\documentclass[%
 onecolumn,
 amsmath,amssymb,
 aps,
longbibliography
]{revtex4-1}

\usepackage{graphicx}
\usepackage{dcolumn}
\usepackage{bm}
\usepackage{subfigure}
\usepackage[euler]{textgreek}
\usepackage{xcolor}  

\begin{document}


\title{Analysis of second-moments and their budgets for \\
Richtmyer--Meshkov instability and \\
variable-density turbulence induced by re-shock}

\author{Man Long Wong}
 \email{mlwong@alumni.stanford.edu }
 \affiliation{%
  Department of Aeronautics and Astronautics, \& \\
  Center for Turbulence Research, Stanford University \\
  Stanford, CA 94305, USA
 }%

\author{Jon R. Baltzer}
 \affiliation{%
  XTD-IDA, Los Alamos National Laboratory \\
  Los Alamos, NM 87545, USA
 }%

\author{Daniel Livescu}
 \affiliation{%
  CCS-2, Los Alamos National Laboratory \\
  Los Alamos, NM 87545, USA
 }%

\author{Sanjiva K. Lele}
 \affiliation{%
  Department of Aeronautics and Astronautics, \\
  Department of Mechanical Engineering, \& \\
  Center for Turbulence Research, Stanford University \\
  Stanford, CA 94305, USA
 }%

\date{\today}

\begin{abstract}
Nonlinear Richtmyer--Meshkov instability and mixing transition induced by a Mach 1.45 shock and subsequent re-shock at an interface between two ideal gases (sulfur hexafluoride and air) with high Atwood number are studied with second-moment analysis using data from high-resolution compressible Navier--Stokes simulations. The analysis first addresses the importance of two second-order moments: turbulent mass flux and density-specific-volume covariance, together with their transport equations. These quantities play an essential role in the development of Favre-averaged Reynolds stress and turbulent kinetic energy in this variable-density flow. Then, grid sensitivities and the time evolution of the turbulent quantities which include the second-moments are investigated, followed by a detailed study of the transport equations for the second-moments including the Reynolds stress and the turbulent kinetic energy with well-resolved data before re-shock. After re-shock, budgets of the same but large-scale turbulent quantities are studied with the effects of the subfilter-scale stress taken into account. The budgets of these large-scale quantities are shown to have an insignificant influence from the numerical regularization. Finally, the effects of the subfilter-scale stress on the budgets of the large-scale turbulent quantities with different degrees of filtering are also examined.

\end{abstract}

\pacs{Valid PACS appear here}
\maketitle


\section{\label{sec:introduction} Introduction}

Richtmyer--Meshkov (RM) instability, or RMI~\cite{richtmyer1960taylor, meshkov1969instability}, arises in many natural phenomena and engineering applications when a shock wave traverses an interface separating two materials of different densities~\cite{zhou2017arayleigh}. RMI is used by astrophysicists to explain the cause of turbulent mixing during supernova explosion~\cite{kifonidis2006non, guzman2009non, hammer2010three} and is also taken into account in many stellar models~\cite{arnett2000role}. In inertial confinement fusion (ICF), it is a common belief that there exists mixing between the capsule material and fuel due to RMI and this prohibits a useful yield obtained from a fusion reaction for power generation~\cite{haan1995design, haan2011point, raman2014flight}. RMI is also employed in some proposed combustion systems since it can enhance the mixing of fuel and oxidizer in supersonic and hypersonic air-breathing engines~\cite{yang1993applications, yang2014richtmyer}. RMI is similar to Rayleigh--Taylor (RT) instability, or RTI, which appears when there is a gravitational acceleration pointing in the opposite direction of the density gradient across an interface. In contrast to RTI, RMI occurs because of impulsive acceleration and is unstable regardless of the direction of the acceleration. Turbulent mixing induced from RMI and RTI with high density variations at the interfaces falls into the category of variable-density turbulence~\cite{livescu2020arfm}, where the Atwood number, which is defined as the difference in the fluid densities divided by their sum, is high. Variable-density turbulent mixing can also be triggered by other types of instabilities at an interface, such as Kelvin--Helmholtz (KH) instability, or KHI, at a shear layer~\cite{livescu2020arfm}.

Direct numerical simulation (DNS)~\cite{moin1998direct}, which resolves all flow scales using a mesh with grid spacing of at least an order of magnitude of the Kolmogorov scale, is a powerful tool for studying turbulent flows, especially flows with a laminar to turbulent transition.
However, its requirements of computational resources is tremendous for high Reynolds number flows and it is computationally too expensive for many engineering applications, even on the largest supercomputers to date. As a result, turbulence modeling approaches are commonly adopted to avoid resolving all spatial and temporal scales in simulations of complex turbulent flows. Large-eddy simulation (LES) and Reynolds-averaged Navier--Stokes (RANS) methodologies are two popular modeling strategies~\cite{lesieur2005large,pope2000turbulent}. LES consists of modeling of small-scales that are assumed to be more universal and self-similar, while the larger scales are resolved on the grid. On the other hand, the entire flow structure is modeled based on statistical averaging in the RANS approach. In general, LES has higher fidelity than RANS-based simulations for turbulent flows where unsteady large scales play critical roles, since the large-scale features are resolved in LES. However, LES also has a larger demand on computational resources due to constraints on grid spacing and time step size for representing the motions of the scales captured. Often, the hybrid RANS-LES approach is chosen as a compromise between computational cost and accuracy~\cite{chaouat2017state}. 

The Besnard--Harlow--Rauenzahn (BHR) family of models based on second-moment closure represent a popular RANS-based approach for variable-density turbulence. The first version of the BHR model was proposed by~\citet{besnard1992turbulence}, in which the unclosed Reynolds stress tensor in the multi-species Favre-averaged (density-weighted-averaged) Navier--Stokes (FANS) equations is closed with the aid
of additional modeled transport equations.
These include modeled equations of decay rate of turbulent kinetic energy and other second-moment quantities, such as turbulent mass flux and density-specific-volume covariance.
These second-moments play important roles in variable-density turbulence; in particular, the turbulent mass flux directly affects the development of Favre-averaged Reynolds stress. The modeling assumptions of the first BHR model were not tested against different types of variable-density flows until the work by~\citet{banerjee2010development}, where simplifications 
of the original BHR model
were also introduced. In their model (BHR $k$-$S$-$a$), the Favre-averaged equations are closed with the turbulent kinetic energy transport equation instead of the equation of the Reynolds stress tensor. The transport equation of the decay rate of turbulent kinetic energy is also replaced with a more physically interpretable transport equation of turbulent length scale. Their model was validated 
with experimental data, but the model coefficients are tuned from flow to flow. Later, the BHR-2 model by~\citet{stalsberg2011bhr2} was proposed. BHR-2 re-adopts the modeled transport equation of density-specific-volume covariance instead of an algebraic model in the BHR $k$-$S$-$a$ model, which was only strictly valid for immiscible fluids. An improved BHR-3 model with modeled transport equations of the Reynolds stress tensor and density-specific-volume covariance was proposed by~\citet{schwarzkopf2011application} and was shown to be capable of capturing
the Reynolds normal stress anisotropy and density-specific-volume covariance well in various variable-density flows, without varying model coefficients. The BHR-3 model was further improved by ~\citet{schwarzkopf2016two} with two length scales to capture the difference between the transport and dissipation turbulent scales in RTI-induced turbulence.
A two-point spectral closure model~\cite{steinkamp1999atwo,steinkamp1999btwo} modified from the constant-density BHRZ model~\cite{besnard1996spectral} for variable-density flows was analyzed for the buoyancy-driven variable-density homogeneous turbulence~\cite{pal2018two}. The model with minimal augmentation was further assessed for the RTI turbulence~\cite{pal2021local}.
In addition to the BHR family of models, there are also similar models for turbulent mixing such as the second-moment model by~\citet{gregoire2005second} with a Boussinesq approximation and the $k$-$L$-$a$ model by~\citet{morgan2015three} extended from the $k$-$L$ model~\cite{dimonte2006k}.
A literature review of different RANS-based models for RMI and other types of variable-density turbulence is provided by~\citet{zhou2017brayleigh}.

Modeling based on the transport of second-moments 
is more popular in the RANS-based approach for variable-density flows, and most proposed LES models for the subfilter-scale (SFS) or subgrid-scale (SGS) terms are based on first-order closures~\cite{chassaing2001modeling}. These include the eddy-viscosity type SFS/SGS closure~\cite{mellado2005large,wang2008large,bai2010investigation} and the stretched-vortex approach, such as~\cite{hill2006large, sidharth2015stretched}. 
There is still a lack of research on the application of second-moments for the closure of SFS/SGS terms in LES. Besides, the role of SFS/SGS terms on the large-scale turbulent quantities, especially the resolved turbulent kinetic energy in variable-density flows, is still unclear. In this paper, we have performed high-resolution RMI simulations with re-shock to provide high-fidelity data for analyzing the 
physical mechanisms underlying the evolution of
second-moments. The set-up of the numerical experiment follows the highest Reynolds number three-dimensional (3D) case in our previous paper~\cite{wong2019high}. Before re-shock, the instability induced at the interface grows nonlinearly but does not achieve the mixing transition. After re-shock, the flow inside the mixing layer transitions and remains turbulent with a wide span of scales until the end of the simulation. 
The 3D simulation presented in this work is advanced to higher grid resolution compared to the cases in the previous work, with the number of grid cells exceeding 4.5 billion.
Grid sensitivity tests show that the second-moments required for closing the FANS equations are well grid-converged during the simulations. We also examine the budgets of the second-moment transport equations before and after re-shock. The budgets analyzed after re-shock are based on large-scale contributions to
second-moments under the influence of the SFS stress. The budgets of the large-scale second-moments are not affected by the numerical regularization, and the effects of SFS stress in
the evolution of large-scale second-moments are studied.
Finally, we also analyze the large-scale second-moment budgets at different filtering scales.


\section{\label{sec:governing_equations} Governing equations}

The conservative multi-component Navier--Stokes equations are solved in this study:
\begin{align}
	\frac{\partial \rho Y_i}{\partial t} + \nabla \cdot \left( \rho \bm{u} Y_i \right) &= - \nabla \cdot \bm{J_i}, \label{eq:species_continuity_eqn} \\
    \frac{\partial \rho \bm{u}}{\partial t} + \nabla \cdot \left( \rho \bm{uu} + p \bm{\delta} \right) &= \nabla \cdot \bm{\tau}, 
    \label{eq:mixture_momentum_eqn} \\
    \frac{\partial E}{\partial t} + \nabla \cdot \left[ \left( E + p \right) \bm{u} \right] &= \nabla \cdot \left( \bm{\tau} \cdot \bm{u} - \bm{q_c} - \bm{q_d} \right), \label{eq:mixture_energy_eqn}
\end{align}
\noindent where $\rho$, $\bm{u}=[u,v,w]^{T}=[u_1,u_2,u_3]^{T}$, $p$, and $E$ are the density, velocity vector, pressure, and total energy of the fluid mixture respectively. $Y_i$ is the mass fraction of species $i \in \left[ 1,2,...,N \right]$, with $N$ the total number of species. $\bm{J_i}$ is diffusive mass flux for species $i$. $\bm{\tau}$, $\bm{q_c}$, and $\bm{q_d}$ are the viscous stress tensor, conductive heat flux, and inter-species diffusional enthalpy flux, respectively. $\bm{\delta}$ is the identity tensor. Since all $Y_i$'s sum up to 1 by definition, the continuity equation for the mixture density can be derived by summing up the continuity equations of all species given by equation~\eqref{eq:species_continuity_eqn} as:
\begin{equation}
	\frac{\partial \rho}{\partial t} + \nabla \cdot \left( \rho \bm{u} \right) = 0. \label{eq:mixture_continuity_eqn}
\end{equation}

The mixture is assumed to be ideal and calorically perfect, with:
\begin{align}
    E = \rho \left( e + \frac{1}{2} \bm{u} \cdot \bm{u} \right), \\
    p = \left( \gamma - 1 \right) \rho e, \quad  e = c_v T,
\end{align}
\noindent where $e$ and $T$ are respectively specific internal energy and temperature of the mixture. $\gamma$ and $c_v$ are the ratio of specific heats and the specific heat at a constant volume of the mixture respectively.

The multi-component diffusive mass flux of species $i$ is given by \cite{hirschfelder1954molecular}:
\begin{equation}
    \bm{J_i} = \rho \frac{M_i}{M^2} \sum^{N}_{j=1} M_j \tilde{D}_{ij} \nabla X_j,
    \label{eqn:multicomponent_diffusive_flux}
\end{equation}

\noindent where $M_i$ and $X_i$ are respectively the molecular weight and the mole fraction of species $i$. $M$ is the molecular weight of the mixture and $\tilde{D}_{ij}$ is the $ij$th 
element of the matrix of ordinary multi-component diffusion coefficients $\tilde{\bm{D}}$. The mole fraction of species $i$ is given by:
\begin{equation}
    X_i = \frac{M}{M_i} Y_i.
\end{equation}
\noindent The multi-component diffusive mass flux is reduced to the Fick's law for a binary mixture:
\begin{equation}
	\bm{J_i} = -\rho D_i \nabla Y_i,\ i=1, 2,
\end{equation}
\noindent where $D_1 = D_2$ is the binary diffusion coefficient. Note that the Fick's law is sufficient in this work since only a binary mixture is studied.

The viscous stress tensor $\bm{\tau}$ for a Newtonian mixture is:
\begin{equation}
	\bm{\tau} = 2 \mu \bm{S} + \left( \mu_v - \frac{2}{3} \mu \right) \bm{\delta} \left( \nabla \cdot \mathbf{u} \right),
\end{equation}
\noindent where $\mu$ and $\mu_v$ are the shear viscosity and bulk viscosity respectively of the mixture. $\bm{S}$ is the strain-rate tensor given by:
\begin{equation}
    \bm{S} = \frac{1}{2} \left[ \nabla \bm{u} + \left( \nabla \bm{u} \right) ^{T} \right].
\end{equation}

The conductive flux and the inter-species diffusional enthalpy flux~\cite{williams2018combustion} are given by:
\begin{align}
	\bm{q_c} = - \kappa \nabla T, \\
	\bm{q_d} = \sum^{N}_{i=1} h_i \bm{J_i},
\end{align}
\noindent where $\kappa$ is the thermal conductivity of the mixture. $h_i$ is the specific enthalpy of species $i$:
\begin{equation}
    h_i = c_{p,i} T,
\end{equation}
\noindent where $c_{p,i}$ is the specific heat capacity at constant pressure of species $i$.

The equations and mixing rules for the fluid properties $\gamma$, $c_v$, $c_{p,i}$,  $\mu$, $\mu_v$, $\kappa$, and $D_i$ are given in the appendices~\ref{sec:appendix_TC} and \ref{sec:appendix_mixing_rules}.


\section{\label{sec:numerical_methods} Numerical methods}

3D numerical experiments with adaptive mesh refinement (AMR) were conducted with the Hydrodynamics Adaptive Mesh Refinement Simulator (HAMeRS) \cite{wong2019thesis} supported with the Structured Adaptive Mesh Refinement Application Infrastructure (SAMRAI) library \cite{gunney2016advances, gunney2006parallel, hornung2006managing, hornung2002managing, wissink2001large} from Lawrence Livermore National Laboratory (LLNL). The convective fluxes of the governing equations are discretized with the explicit form of the sixth-order localized dissipation weighted compact nonlinear scheme (WCNS)~\cite{wong2017high} for shock-capturing and stabilization of solutions. The accuracy and robustness of the WCNS family for compressible multi-fluid flows have been demonstrated in previous works~\cite{nonomura2012numerical,wong2017high,wong2021positivity}.
Derivatives of diffusive and viscous fluxes are computed with explicit sixth-order finite difference schemes in non-conservative form. A third order total variation diminishing Runge--Kutta (RK-TVD) scheme~\cite{shu1989efficient} is employed for the time advancement with a convective Courant--Friedrichs--Lewy (CFL) number of 0.5 and a diffusive CFL number of 0.25. The regions for adaptive mesh refinement are identified with a gradient sensor on the pressure field and a wavelet sensor~\cite{wong2016multiresolution} on the density field to detect shock waves and mixing regions, respectively. An additional sensor based on mass fractions is also used to assist the detection of mixing regions.


\section{\label{sec:IC_comp_domain} Initial conditions and computational domain}

The 3D case set-up in our previous paper~\cite{wong2019high}, with physical transport coefficients for the gases considered, is chosen in this work. In this set-up, the shock-induced mixing problem is simulated in a numerical shock tube with a cross-sectional area of $2.5\ {\mathrm{cm}} \times 2.5\ {\mathrm{cm}}$. A planar shock wave of Mach number $Ma = 1.45$ is initialized in a sulfur hexafluoride ($\mathrm{SF_6}$) region, with the Rankine--Hugoniot jump conditions to interact with a diffuse interface between $\mathrm{SF_6}$ and air. A multi-mode perturbation expressed in the following equation is imposed on the interface:
\begin{equation}
S(y,z) = A \sum_{m} \cos \left( \frac{2\pi m}{L_{yz}} y + \phi_m \right)\cos \left( \frac{2\pi m}{L_{yz}} z + \psi_m \right),
\label{eqn:egg_crate_perturbation}
\end{equation}

\noindent where $L_{yz} = 2.5\ \mathrm{cm}$. The perturbation has 11 modes with wavenumber $m$ between 20 and 30 in each transverse direction. Constant amplitude $A=\sqrt{2} \times 0.01\ \mathrm{mm}$ is used for each mode and random phase shifts $\phi_m$ and $\psi_m$ between 0 and $2\pi$ are introduced to each mode to prevent summing up of harmonic modes. $\phi_m$ and $\psi_m$ of each mode are given in the Supplemental Material~\cite{supple2022wong}.

\begin{figure}[!ht]
\centering
\includegraphics[width=0.5\textwidth]{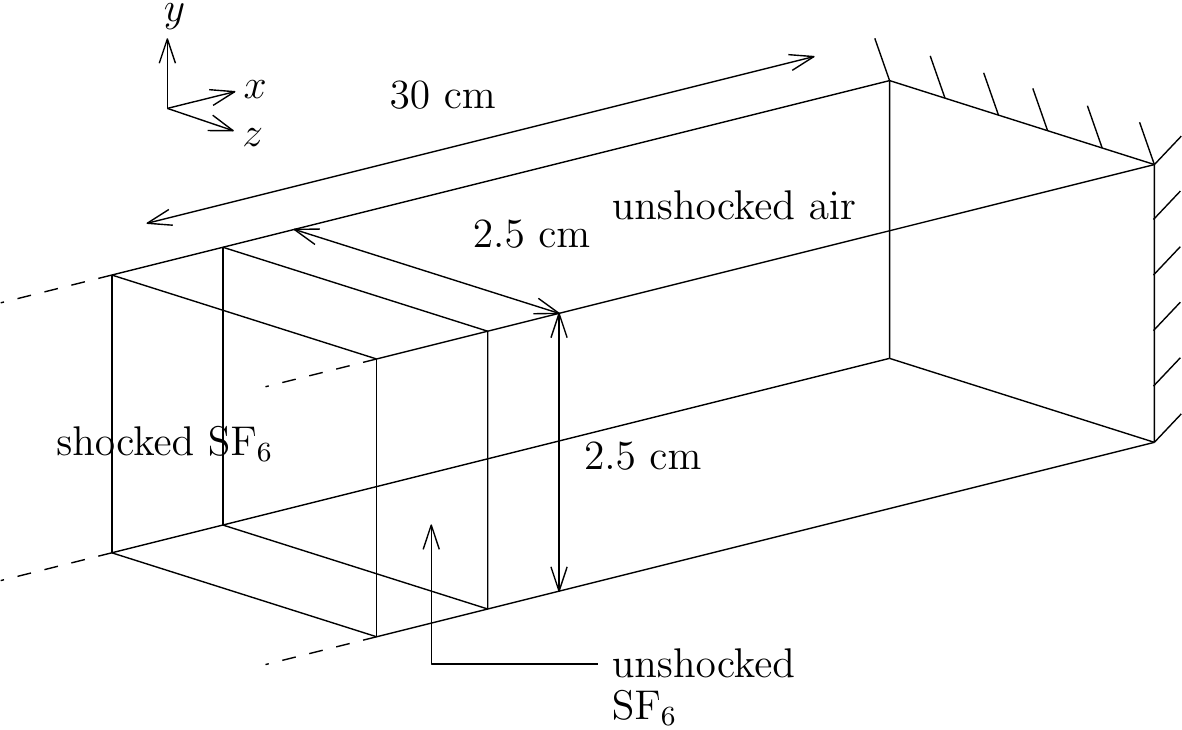}
\caption{Schematic diagram of initial flow field and computational domain.}
\label{fig:configuration}
\end{figure}

The computational domain and initial conditions are shown in figure~\ref{fig:configuration}. Boundaries are periodic in the transverse directions and reflective boundary conditions are applied at the end wall. The length of the domain is chosen to be large enough such that no waves leave the open-sided boundary during the simulations. The pre-shocked gases are stationary initially and have temperature $T=298\ {\mathrm{K}}$ and pressure $p=101325\ {\mathrm{Pa}}$. Table~\ref{tab:IC} shows the initial conditions of gases in different portions of the domain. The initial Atwood number $At = ( \rho_{\mathrm{SF_6}} - \rho_{\mathrm{air}} ) / ( \rho_{\mathrm{SF_6}} + \rho_{\mathrm{air}} )$ across the interface is 0.68.

\begin{table}[!ht]
\caption{\label{tab:IC}%
Initial  conditions  of  the  post-shock  state  and  the  pre-shock  states  of  the  light-  and heavy-gas  sides.}
\begin{ruledtabular}
\begin{tabular}{ c c c c }
 Quantity & Post-shock $\mathrm{SF_6}$ & Pre-shock $\mathrm{SF_6}$ & Air \\ 
 \hline
 $\rho\ ({\mathrm{kg\ m^{-3}}})$ & 11.97082 & 5.972866 & 1.145601 \\
 $p\ ({\mathrm{Pa}})$ & 218005.4 & 101325.0 & 101325.0 \\ 
 $T\ ({\mathrm{K}})$ & 319.9084 & 298.0 & 298.0 \\
 $u\ ({\mathrm{m\ s^{-1}}})$ & 98.93441 & 0 & 0 \\
\end{tabular}
\end{ruledtabular}
\end{table}

\begin{figure}[!ht]
\centering
\includegraphics[width=0.4\textwidth]{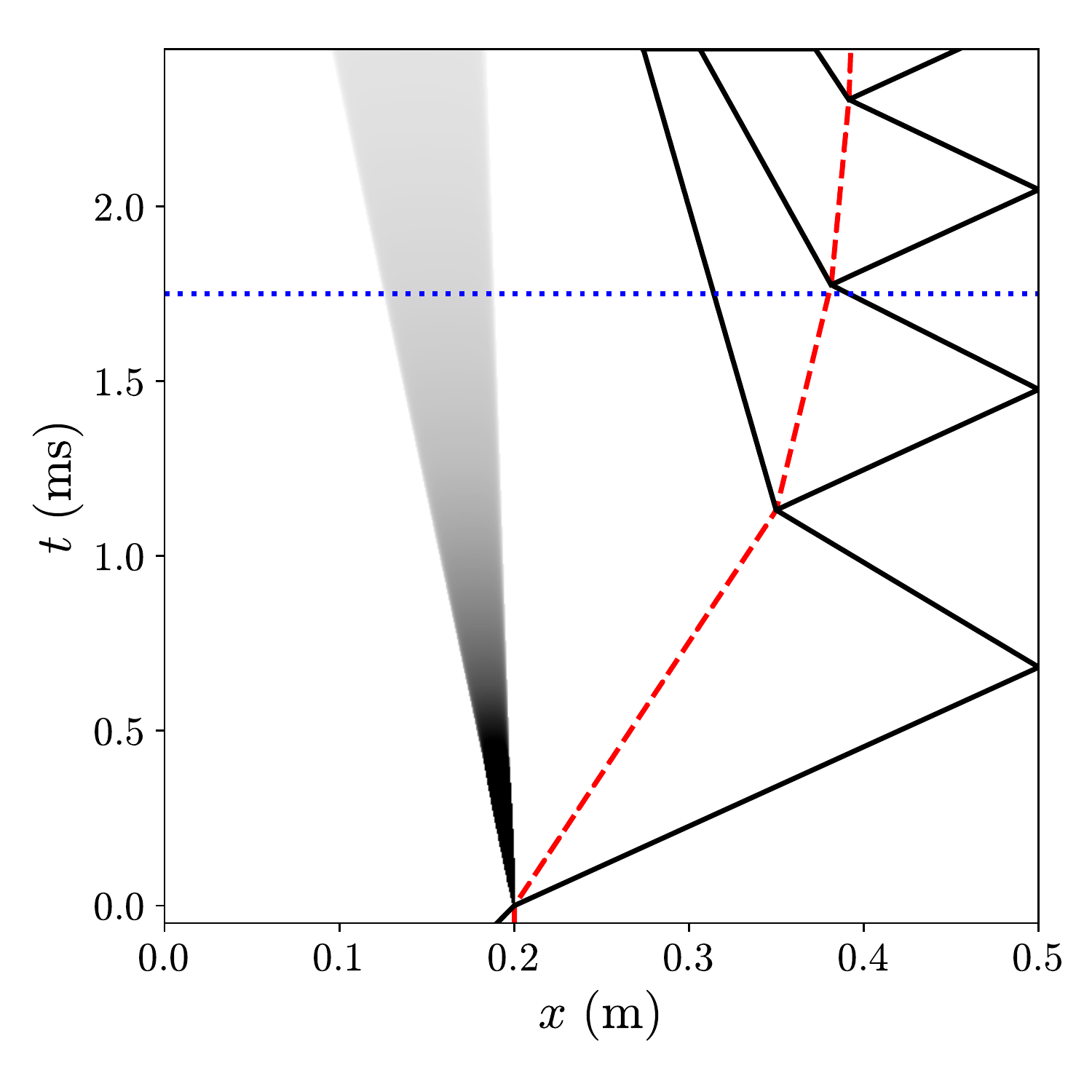}
\caption{$x$-$t$ diagram showing the propagation of material interface, shock waves, and rarefaction. Red dashed line: material interface; black lines: shock waves; gray region: rarefaction. The blue dotted line indicates the end time of the simulations.}
\label{fig:x_t_diagram}
\end{figure}

All simulations start at $t=-0.05\ \mathrm{ms}$, and the shock wave is initially positioned at a location such that the shock-interface interaction first happens at $t=0$.
Since the simulations are initiated in the heavy-light gas setting, the shock wave is transmitted to the light-fluid side and a rarefaction wave is reflected back to the heavy-fluid side. After hitting the wall, the transmitted shock is reflected back towards the interface when it hits the end wall and this causes the re-shock of the interface. Since the shock arrives at the interface from the light-fluid side this time, a transmitted shock and a reflected shock are generated. The reflected shock leads to a second re-shock. The end time of the simulations is chosen at $t=1.75\ \mathrm{ms}$, when the second re-shock is just about to happen, as the grid resolution requirements become too large to accurately capture this flow stage. Figure~\ref{fig:x_t_diagram} shows the space-time ($x$-$t$) diagram for different features in a one-dimensional (1D) flow representation.
This problem was studied in the previous work~\cite{wong2019high} with both two-dimensional (2D) and 3D simulations. In this work, results from a higher resolution 3D AMR simulation are studied for the second-moment analysis of the shock-induced variable-density instability and turbulence.

\section{\label{sec:second_order_moments_transport_equations} Transport equations of the second-moments}

To get a statistical view of a chaotic or turbulent field, it is a common practice to ensemble average the governing equations. The conserved variables are decomposed into ensemble means and fluctuations through Reynolds decomposition. The Reynolds decomposition of an arbitrary variable, $f$, rewrites the variable as:
\begin{equation}
    f = \bar{f} + f^{\prime},
\end{equation}
where $\bar{f}$ and $f^{\prime}$ are the mean and fluctuation of $f$ respectively. If the flow has homogeneous directions and the widths of the domain in the homogeneous directions are sufficiently larger than the length scales of turbulent features, one can estimate the ensemble mean with the mean over all homogeneous directions. For variable-density  flows, after averaging the conserved variables of the governing equations it is natural to see the Favre-averaged (density-weighted-averaged) quantities. The Favre decomposition is given by:
\begin{equation}
    f = \tilde{f} + f^{\prime\prime},
\end{equation}
where $\tilde{f} = \overline{\rho f} / \bar{\rho}$. The Reynolds and Favre averages of the velocity are related with:
\begin{equation}
    \tilde{u}_i = \bar{u}_i + a_i, \label{eq:means_relation}
\end{equation}
where $a_i = \overline{\rho^{\prime} u_i^{\prime}} / \bar{\rho}$ is the velocity associated with the turbulent mass flux $\bar{\rho} a_i$. The fluctuation, $u_i^{\prime}$, and the Favre fluctuation, $u_i^{\prime\prime}$, have similar relation as the averages:
\begin{equation}
    u_i^{\prime\prime} = u_i^{\prime} - a_i. \label{eq:fluctuations_relation}
\end{equation}

If we apply averaging on the continuity equation and the conservative transport equation of momentum given by equations~\eqref{eq:mixture_continuity_eqn} and \eqref{eq:mixture_momentum_eqn}, respectively, we obtain:
\begin{align}
    \frac{\partial \bar{\rho}}{\partial t} + \frac{\partial \left( \bar{\rho} \tilde{u}_k \right)}{\partial x_k} &= 0, \label{eq:Favre_mixture_continuity_eqn} \\
    \frac{\partial \left( \bar{\rho} \tilde{u}_i \right)}{\partial t} + \frac{\partial \left( \bar{\rho} \tilde{u}_k \tilde{u}_i \right)}{\partial x_k} &= - \frac{\partial \left( \bar{p} \delta_{ki} \right)}{\partial x_k} + \frac{\partial \bar{\tau}_{ki}}{\partial x_k} - \frac{\partial \left( \bar{\rho} \tilde{R}_{ki} \right)}{\partial x_k}, \label{eq:Favre_momentum_eqn}
\end{align}
where $\tilde{R}_{ij}$ is the Favre-averaged Reynolds stress tensor given by:
\begin{equation}
    \tilde{R}_{ij} = \frac{\overline{\rho u_i^{\prime\prime} u_j^{\prime\prime}}}{\bar{\rho}}.
\end{equation}

\noindent The Favre-averaged Reynolds stress tensor appears as an unclosed term in the Favre-averaged transport equation of momentum. The development of the Favre-averaged Reynolds stress can be studied through its transport equations given by~\citet{besnard1992turbulence}:
\begin{equation}
\begin{split}
	\underbrace{ \frac{\partial \bar{\rho} \tilde{R}_{ij}}{\partial t} }_{ \text{term (I)} }
	\ \underbrace{ + \frac{\partial \left( \bar{\rho} \tilde{u}_k \tilde{R}_{ij} \right)}{\partial x_k}  }_{ \text{term (II)} } =
	\underbrace{ a_i\left( \frac{\partial \bar{p}}{\partial x_j}
	- \frac{\partial \bar{\tau}_{jk}}{\partial x_k} \right)
	+ a_j\left( \frac{\partial \bar{p}}{\partial x_i}
	- \frac{\partial \bar{\tau}_{ik}}{\partial x_k} \right)  
    - \bar{\rho} \tilde{R}_{ik} \frac{\partial \tilde{u}_j}{\partial x_k}
    - \bar{\rho} \tilde{R}_{jk} \frac{\partial \tilde{u}_i}{\partial x_k} }_{ \text{term (III)} } \\
    \underbrace{ - \frac{\partial \left( \overline{\rho u^{\prime \prime}_{i} u^{\prime \prime}_{j} u^{\prime \prime}_{k} } \right)}{\partial x_k}
    - \frac{\partial \left( \overline{u^{\prime}_{i} p^{\prime}} \right)}{\partial x_j}
    - \frac{\partial \left( \overline{u^{\prime}_{j} p^{\prime}} \right)}{\partial x_i} + \frac{\partial \left( \overline{u^{\prime}_{i} \tau^{\prime}_{jk}} \right)}{\partial x_k}
    + \frac{\partial \left( \overline{u^{\prime}_{j} \tau^{\prime}_{ik}} \right)}{\partial x_k} }_{ \text{term (IV)} }
    \ \underbrace{ + \overline{p^{\prime} \frac{\partial u^{\prime}_{i}}{\partial x_j}}
    + \overline{p^{\prime} \frac{\partial u^{\prime}_{j}}{\partial x_i}} }_{ \text{term (V)} }
    \ \underbrace{ - \overline{\tau^{\prime}_{jk} \frac{\partial u^{\prime}_{i}}{\partial x_k}}
    - \overline{\tau^{\prime}_{ik} \frac{\partial u^{\prime}_{j}}{\partial x_k}} }_{ \text{term (VI)} },
\end{split} \label{eq:Reynolds_stress_transport_eqn}
\end{equation}

\noindent where the left hand side (LHS) consists of rate of change [term (I)] and convection [term (II)]. The right hand side (RHS) consists of production [term (III)], turbulent transport [term (IV)], pressure-strain redistribution [term (V)], and dissipation [term (VI)].

In the paper by~\citet{schwarzkopf2011application}, Favre decomposition is used for the viscous stress in the Favre-averaged Reynolds stress transport equation instead of Reynolds decomposition. However, we follow the original work by~\citet{besnard1992turbulence} to use Reynolds decomposition for the viscous stress as we believe Favre decomposition should only be applied in the advective or convective terms. The Reynolds decomposition of the viscous stress was also employed in the DNS analysis of~\citet{livescu2009rti}, which was later used to refine the model by~\citet{schwarzkopf2016two}. Besides, the dissipation term with Reynolds decomposition on viscous stress in the turbulent kinetic energy transport equation can be proved to be strictly negative if both shear and bulk viscosities are uniform in the domain, while that with Favre decomposition on viscous stress cannot be proved to be strictly negative. Also, note that with the relation given by equation~\eqref{eq:fluctuations_relation}:
\begin{align}
    \frac{\partial \left( \overline{u^{\prime}_{i} p^{\prime}} \right)}{\partial x_j} &= \frac{\partial \left( \overline{u^{\prime \prime}_{i} p^{\prime}} \right)}{\partial x_j}, \\
    \frac{\partial \left( \overline{u^{\prime}_{i} \tau^{\prime}_{jk}} \right)}{\partial x_k} &=
    \frac{\partial \left( \overline{u^{\prime \prime}_{i} \tau^{\prime}_{jk}} \right)}{\partial x_k}, \\
    \overline{p^{\prime} \frac{\partial u^{\prime}_{i}}{\partial x_j}} &=
    \overline{p^{\prime} \frac{\partial u^{\prime \prime}_{i}}{\partial x_j}}, \\
    \overline{\tau^{\prime}_{jk} \frac{\partial u^{\prime}_{i}}{\partial x_k}} &= \overline{\tau^{\prime}_{jk} \frac{\partial u^{\prime \prime}_{i}}{\partial x_k}}.
\end{align}

\noindent The relations above are commonly used to interchange terms in the transport equations of $\tilde{R}_{ij}$ and $k$ in many previous studies.

In flows where the mean is 1D, such as the numerical experiment being studied in this work (where the $y$ and $z$ directions are homogeneous), the transport equation of $\tilde{R}_{11}$ can be simplified to:
\begin{equation}
\begin{split}
	\underbrace{ \frac{\partial \bar{\rho} \tilde{R}_{11}}{\partial t} }_{ \text{term (I)} }
	\ \underbrace{ + \frac{\partial \left( \bar{\rho} \tilde{u} \tilde{R}_{11} \right)}{\partial x} }_{ \text{term (II)} } = 
    \ \underbrace{ 2a_1\left( \frac{\partial \bar{p}}{\partial x} - \frac{\partial \bar{\tau}_{11}}{\partial x} \right) - 2\bar{\rho} \tilde{R}_{11} \frac{\partial \tilde{u}}{\partial x} }_{ \text{term (III)} }
    \ \underbrace{ - \frac{\partial \left( \overline{\rho u^{\prime \prime} u^{\prime \prime} u^{\prime \prime} } \right)}{\partial x}
    - 2\frac{\partial \left( \overline{u^{\prime} p^{\prime}} \right)}{\partial x}
    + 2\frac{\partial \left( \overline{u^{\prime} \tau^{\prime}_{11}} \right)}{\partial x} }_{ \text{term (IV)} } \\
    \underbrace{ + 2\overline{p^{\prime} \frac{\partial u^{\prime}}{\partial x}} }_{ \text{term (V)} }
    \ \underbrace{ - 2\left( \overline{\tau^{\prime}_{11} \frac{\partial u^{\prime}}{\partial x}}
    + \overline{\tau^{\prime}_{12} \frac{\partial u^{\prime}}{\partial y}}
    + \overline{\tau^{\prime}_{13} \frac{\partial u^{\prime}}{\partial z}} \right) }_{ \text{term (VI)} }.
\end{split} \label{eq:R11_transport_eqn_1D}
\end{equation}

The transport equation of $\tilde{R}_{22}$ for 1D mean flow can be reduced to:
\begin{equation}
	\underbrace{ \frac{\partial \bar{\rho} \tilde{R}_{22}}{\partial t} }_{ \text{term (I)} }
	\ \underbrace{ + \frac{\partial \left( \bar{\rho} \tilde{u} \tilde{R}_{22} \right)}{\partial x} }_{ \text{term (II)} } = 
	\ \underbrace{ - \frac{\partial \left( \overline{\rho v^{\prime \prime} v^{\prime \prime} u^{\prime \prime} } \right)}{\partial x}
	+ 2\frac{\partial \left( \overline{v^{\prime} \tau^{\prime}_{21}} \right)}{\partial x} }_{ \text{term (IV)} }
    \ \underbrace{ + 2\overline{p^{\prime} \frac{\partial v^{\prime}}{\partial y}} }_{ \text{term (V)} }
    \ \underbrace{ - 2\left( \overline{\tau^{\prime}_{21} \frac{\partial v^{\prime}}{\partial x}} 
    + \overline{\tau^{\prime}_{22} \frac{\partial v^{\prime}}{\partial y}}
    + \overline{\tau^{\prime}_{23} \frac{\partial v^{\prime}}{\partial z}} \right) }_{ \text{term (VI)} }. \label{eq:R22_transport_eqn_1D}
\end{equation}

\noindent Note that there is no production term [term (III)] in the transport equation of $\tilde{R}_{22}$. The transport equation of $\tilde{R}_{33}$ is similar. In the present flow, the Reynolds shear stress components, $\tilde{R}_{12}$, $\tilde{R}_{13}$, and $\tilde{R}_{23}$, are statistically zero due to the homogeneity of the problem in the transverse directions.

The transport equation of the turbulent kinetic energy per unit mass, $k=\tilde{R}_{ii}/2$, can be simply obtained by taking half of the trace of the Reynolds stress tensor transport equation. For 1D mean flow, it has the following form:
\begin{equation}
\begin{split}
	\underbrace{ \frac{\partial \bar{\rho} k}{\partial t} }_{ \text{term (I)} }
	\ \underbrace{ + \frac{\partial \left( \bar{\rho} \tilde{u} k \right)}{\partial x} }_{ \text{term (II)} } = 
    \ \underbrace{ a_1\left( \frac{\partial \bar{p}}{\partial x} - \frac{\partial \bar{\tau}_{11}}{\partial x} \right) - \bar{\rho} \tilde{R}_{11} \frac{\partial \tilde{u}}{\partial x} }_{ \text{term (III)} }
    \ \underbrace{ - \frac{1}{2} \frac{\partial \left( \overline{\rho u^{\prime \prime}_{i} u^{\prime \prime}_{i} u^{\prime \prime} } \right)}{\partial x}
    - \frac{\partial \left( \overline{u^{\prime} p^{\prime}} \right)}{\partial x}
    + \frac{\partial \left( \overline{u^{\prime}_{i} \tau^{\prime}_{i1}} \right)}{\partial x} }_{ \text{term (IV)} } \\
    \underbrace{ + \overline{p^{\prime} \frac{\partial u^{\prime}_i}{\partial x_i}} }_{ \text{term (V)} }
    \ \underbrace{ - \overline{\tau^{\prime}_{ij} \frac{\partial u^{\prime}_i}{\partial x_j}}}_{ \text{term (VI)} },
\end{split} \label{eq:k_transport_eqn_1D}
\end{equation}
\noindent where the LHS consists of rate of change [term (I)] and convection [term (II)]. The RHS consists of production [term (III)], turbulent transport [term (IV)], pressure-dilatation [term (V)], and dissipation [term (VI)]. Note that the production term represents the energy transfer rate between the mean kinetic energy and the turbulent kinetic energy and can have negative sign.

The velocity associated with the turbulent mass flux, $a_i$, in the mean flow pressure gradient terms only appears in variable-density or/and compressible flows. It is an important term for understanding the energetics in these kinds of flows, as the mean pressure gradient multiplied by it is the agent for the transfer of mean kinetic energy into turbulent kinetic energy. Correct modeling of $a_i$ can also help us close the Reynolds stress transport equations. The transport equation of turbulent mass flux $\bar{\rho}a_i$ is given by~\cite{besnard1992turbulence}:
\begin{equation}
\begin{split}
	\underbrace{ \frac{\partial \left( \bar{\rho}a_i \right)}{\partial t}  }_{ \text{term (I)} }
	\ \underbrace{ + \frac{ \partial \left( \bar{\rho} \tilde{u}_k a_i \right) }{\partial x_k} }_{ \text{term (II)} }
	= \underbrace{ b\left( \frac{\partial \bar{p}}{\partial x_i} - \frac{\partial \bar{\tau}_{ki}}{\partial x_k} \right) - \tilde{R}_{ik} \frac{\partial \bar{\rho}}{\partial x_k}  }_{ \text{term (III)} }
	\ \underbrace{ + \bar{\rho} \frac{\partial \left( a_k a_i \right)}{\partial x_k}
	- \bar{\rho} a_k \frac{\partial \bar{u}_i}{\partial x_k} }_{ \text{term (IV)} }
	\ \underbrace{ - \bar{\rho} \frac{\partial \left( \overline{\rho^{\prime} u^{\prime}_i u^{\prime}_k}/\bar{\rho} \right)}{\partial x_k} }_{ \text{term (V)} } \\ 
	\underbrace{ + \bar{\rho} \overline{\left( \frac{1}{\rho} \right)^{\prime} \left( \frac{\partial p^{\prime}}{\partial x_i} - \frac{\partial \tau^{\prime}_{ik}}{\partial x_k} \right)}
	+ \bar{\rho} \varepsilon_{a_i} }_{ \text{term (VI)} } \label{eq:turb_mass_flux_transport_eqn},
\end{split}
\end{equation}

\noindent where the LHS consists of rate of change [term (I)] and convection [term (II)]. The RHS contains production [term (III)], redistribution [term (IV)], turbulent transport [term (V)], and destruction [term (VI)]. Also,
\begin{equation}
    \varepsilon_{a_i} = - \overline{u_i^{\prime} \frac{\partial u_k^{\prime}}{\partial x_k}}.
\end{equation}
Note that $\varepsilon_{a_i}$ is ignored in the work by~\citet{besnard1992turbulence} and in many turbulence models. However, $\varepsilon_{a_i}$ was shown to be non-negligible at early times in the evolution of constant acceleration RTI \cite{livescu2009rti}. Here, it is also found that $\varepsilon_{a_i}$ is significant in the budgets at different times before re-shock for the flow being studied in this work. 

For 1D mean flow, the transport equation of $\bar{\rho}a_1$ can be simplified to:
\begin{equation}
\begin{split}
	\underbrace{ \frac{\partial \left( \bar{\rho}a_1 \right)}{\partial t} }_{ \text{term (I)} }
	\ \underbrace{ + \frac{\partial \left( \bar{\rho} \tilde{u} a_1 \right)}{\partial x} }_{ \text{term (II)} } =
	\underbrace{ b\left(\frac{\partial \bar{p}}{\partial x} - \frac{\partial \bar{\tau}_{11}}{\partial x} \right) - \tilde{R}_{11} \frac{\partial \bar{\rho}}{\partial x} }_{ \text{term (III)} }
	\ \underbrace{ + \bar{\rho} \frac{\partial \left( a_1 a_1 \right)}{\partial x} - \bar{\rho} a_1 \frac{\partial \bar{u}}{\partial x } }_{ \text{term (IV)} }
	\ \underbrace{ - \bar{\rho} \frac{\partial \left( \overline{\rho^{\prime} u^{\prime} u^{\prime}} / \bar{\rho} \right)}{\partial x} }_{ \text{term (V)} } \\
    \underbrace{ + \bar{\rho} \overline{\left( \frac{1}{\rho} \right)^{\prime} \left( \frac{\partial p^{\prime}}{\partial x} - \frac{\partial \tau^{\prime}_{11}}{\partial x} - \frac{\partial \tau^{\prime}_{12}}{\partial y} - \frac{\partial \tau^{\prime}_{13}}{\partial z} \right)} + \bar{\rho} \varepsilon_{a_1} }_{ \text{term (VI)} }.
\end{split} \label{eq:a1_transport_eqn_1D}
\end{equation}

\noindent Note that $a_2$ and $a_3$ for 1D mean flow are statistically equal to zero. The density-specific-volume covariance, $b=-\overline{\rho^{\prime} (1 / \rho)^{\prime}}$, mediates the turbulent mass flux production mechanism. The component of the production term, $b\bar{p}_{,1}$, is crucial to the prediction of the rate of change of turbulent mass flux and requires the modeling of $b$.

The transport equation of $b$ was first derived by~\citet{besnard1992turbulence} in the following advection form with the Reynolds-averaged velocity:
\begin{equation}
    \frac{\partial b}{\partial t} + \bar{u}_k b_{,k} = -\frac{b+1}{\bar{\rho}} \left( \bar{\rho} a_k \right)_{,k} - \bar{\rho} \left( \overline{\left( \frac{1}{\rho}\right)^{\prime} u_k^{\prime}} \right)_{,k}
    - 2 \bar{\rho} \varepsilon_b, \label{eq:b_advection_eqn}
\end{equation}

\noindent where
\begin{equation}
    \varepsilon_b = \overline{\left(\frac{1}{\rho}\right)^{\prime} \frac{\partial u_k^{\prime}}{\partial x_k}}.
\end{equation}

In~\citet{schwarzkopf2011application}, the transport equation of $\bar{\rho} b$ in the conservative form is derived from equation~\eqref{eq:b_advection_eqn} with the averaged mixture continuity equation (equation~\eqref{eq:mixture_continuity_eqn}) as:
\begin{equation}
	\underbrace{ \frac{\partial \bar{\rho}b}{\partial t} }_{ \text{term (I)} }
	\ \underbrace{ + \frac{\partial \left( \bar{\rho} \tilde{u}_k b \right)}{\partial x_k} }_{ \text{term (II)} } =
	\underbrace{ -2\left( b + 1 \right) a_k \frac{\partial \bar{\rho}}{\partial x_k} }_{ \text{term (III)} }
    \ \underbrace{ + 2 \bar{\rho} a_k \frac{\partial b}{\partial x_k} }_{ \text{term (IV)} }
    \ \underbrace{ + \bar{\rho}^2 \frac{\partial \left( {\overline{\rho^{\prime} \left( 1/\rho \right)^{\prime} u^{\prime}_{k} } / \bar{\rho}} \right)}{\partial x_k} }_{ \text{term (V)} }
    \ \underbrace{ + 2\bar{\rho}^2 \varepsilon_b }_{ \text{term (VI)} }, \label{eq:rho_b_transport_eqn}
\end{equation}

\noindent where the LHS consists of rate of change [term (I)] and convection [term (II)]. The RHS consists of production [term (III)], redistribution [term (IV)], turbulent transport [term (V)], and destruction [term (VI)]. For 1D mean flow, the transport equation of $\bar{\rho}b$ can be simplified to:
\begin{equation}
	\underbrace{ \frac{\partial \left( \bar{\rho}b \right)}{\partial t} }_{ \text{term (I)} }
	\ \underbrace{ + \frac{\partial \left( \bar{\rho} \tilde{u} b \right)}{\partial x} }_{ \text{term (II)} } =
	\underbrace{ -2\left( b + 1 \right) a_1 \frac{\partial \bar{\rho}}{\partial x} }_{ \text{term (III)} }
	\ \underbrace{ + 2 \bar{\rho} a_1 \frac{\partial b}{\partial x} }_{ \text{term (IV)} }
    \ \underbrace{ + \bar{\rho}^2 \frac{ \partial \left( \overline{\rho^{\prime} (1/\rho)^{\prime} u^{\prime} }/\bar{\rho} \right)}{\partial x} }_{ \text{term (V)} }
    \ \underbrace{ + 2\bar{\rho}^2 \varepsilon_b }_{ \text{term (VI)} }. \label{eq:b_transport_eqn_1D}
\end{equation}

\noindent The transport equation of $\bar{\rho} b$ in conservative form shown above instead of that in advection form is studied in this work. Advection form of the $b$ transport equation was considered in the DNS analysis of~\citet{livescu2009rti}.


\section{\label{sec:grid_sensitivity} Grid sensitivity analysis}

In this section, the quality of the simulations is studied through a grid sensitivity analysis. Table~\ref{tab:grids} shows the grid settings used for the problem. There are totally three levels of grids with two levels of mesh refinement in all grid settings. The refinement ratios in each direction from the base level to second level and from second level to the finest level are 1:2 and 1:4 respectively. Four different grid settings are tested, with number of grid points in the transverse directions increasing from 32 points (grid B) to 256 points (grid E) on the base level. The finest level for the largest mesh resolution case has a grid spacing of 12.2 {\textmu}m. With this grid spacing, there are around 68 grid points across the smallest wavelength among the initial modes.
The 3D simulations with grids B-D were first presented in~\cite{wong2019high}, but the new simulation using the grid E settings presented here is a higher grid resolution compared to those runs and provides more accurate statistical results. This ultra-high resolution simulation has cell counts surpassing 4.5 billion, as shown in the appendix~\ref{sec:cells_count}.
Figure~\ref{fig:3D_plot} presents visualizations of the mixing layer at different times with grid E. 

\begin{table}[!ht]
\caption{\label{tab:grids}%
Different grids used for the grid sensitivity study. Three levels of grids with 1:8 overall refinement ratio are used in all cases.}
\begin{ruledtabular}
\begin{tabular}{ c c c c }
Grid & Base grid resolution & Refinement ratios & Finest grid spacing ({\textmu}m) \\ 
\hline
B & $640  \times 32  \times 32$  & 1:2, 1:4 & 97.7  \\
C & $1280 \times 64  \times 64$  & 1:2, 1:4 & 48.8  \\
D & $2560 \times 128 \times 128$ & 1:2, 1:4 & 24.4  \\
E & $5120 \times 256 \times 256$ & 1:2, 1:4 & 12.2  \\
\end{tabular}
\end{ruledtabular}
\end{table}

\begin{figure*}[!ht]
\centering
  \subfigure[$\ t=1.10\ \mathrm{ms}$]{%
\includegraphics[width=0.45\textwidth]{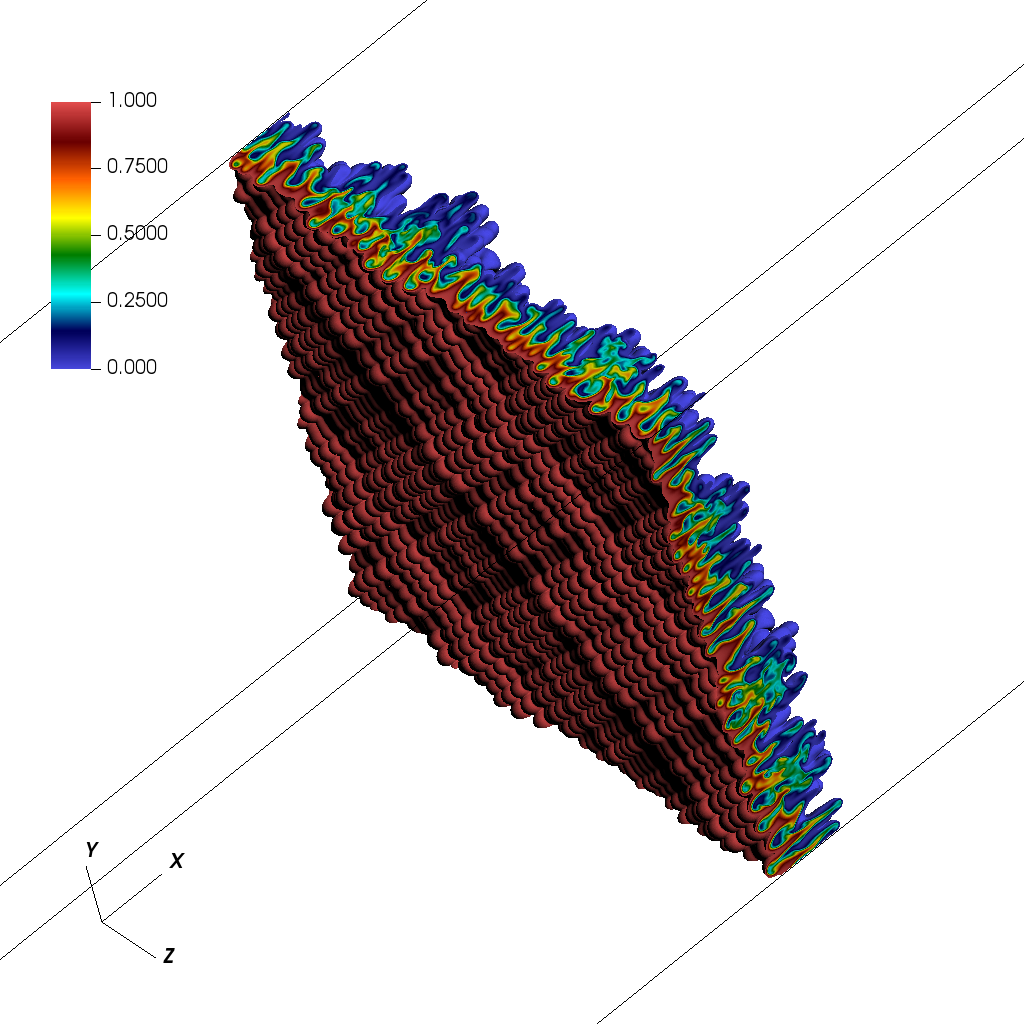}\label{fig:visit_t_1_10}}
  \subfigure[$\ t=1.20\ \mathrm{ms}$]{%
  \includegraphics[width=0.45\textwidth]{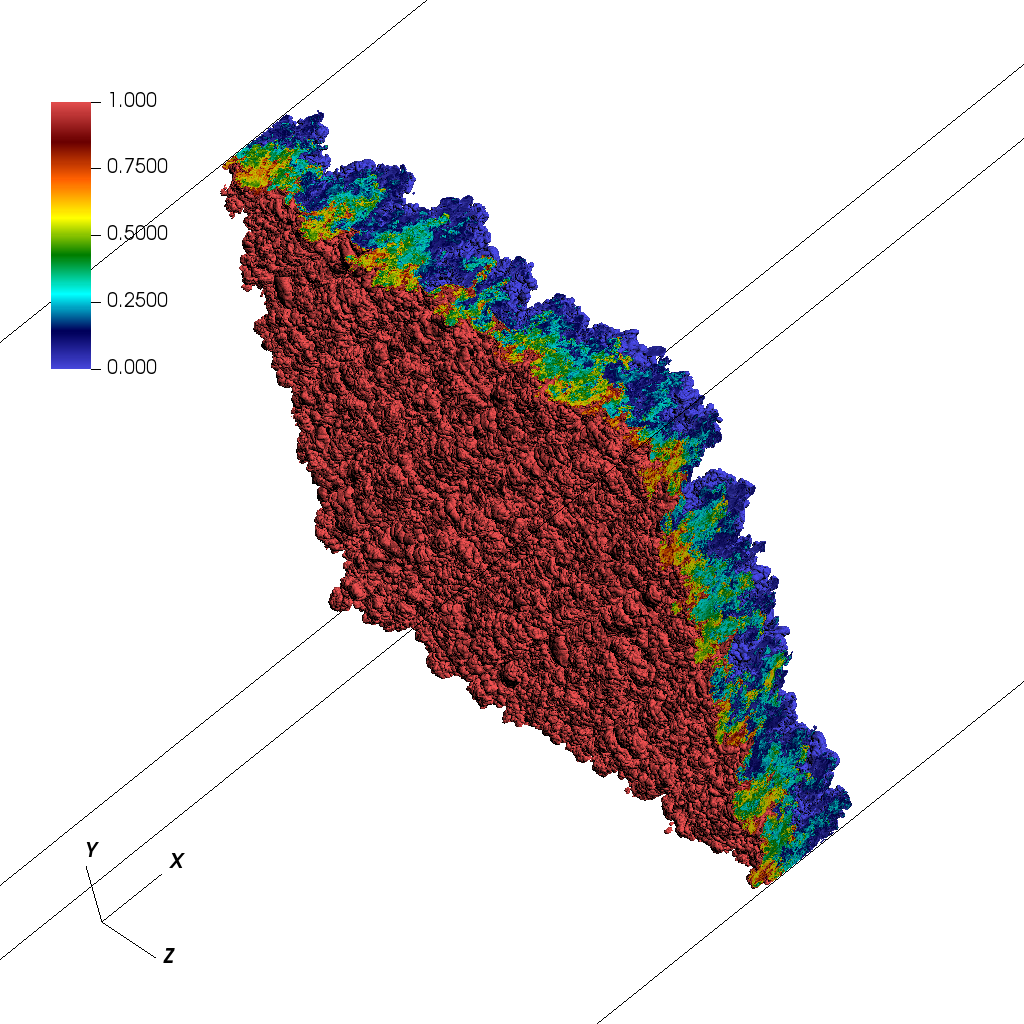}\label{fig:visit_t_1_20}}
  \subfigure[$\ t=1.40\ \mathrm{ms}$]{%
  \includegraphics[width=0.45\textwidth]{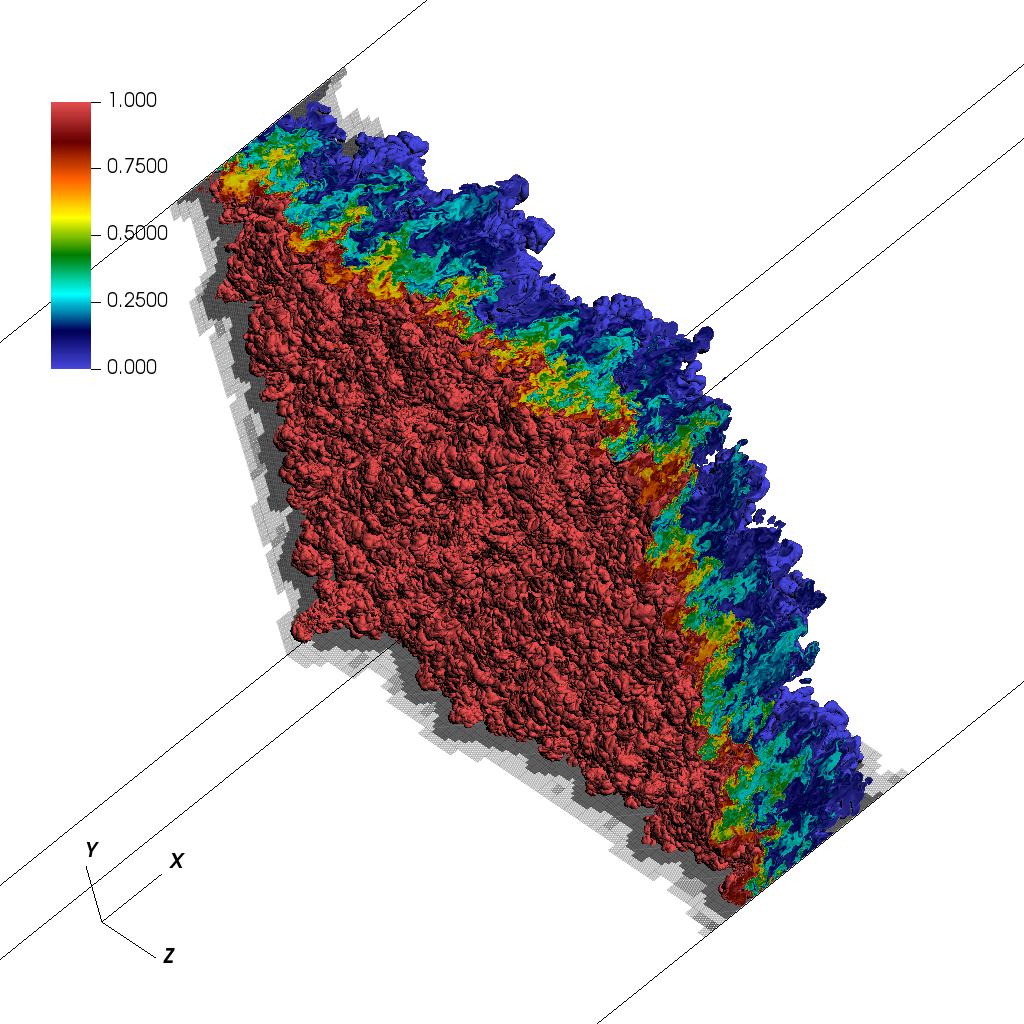}\label{fig:visit_t_1_40}}
  \subfigure[$\ t=1.75\ \mathrm{ms}$]{%
  \includegraphics[width=0.45\textwidth]{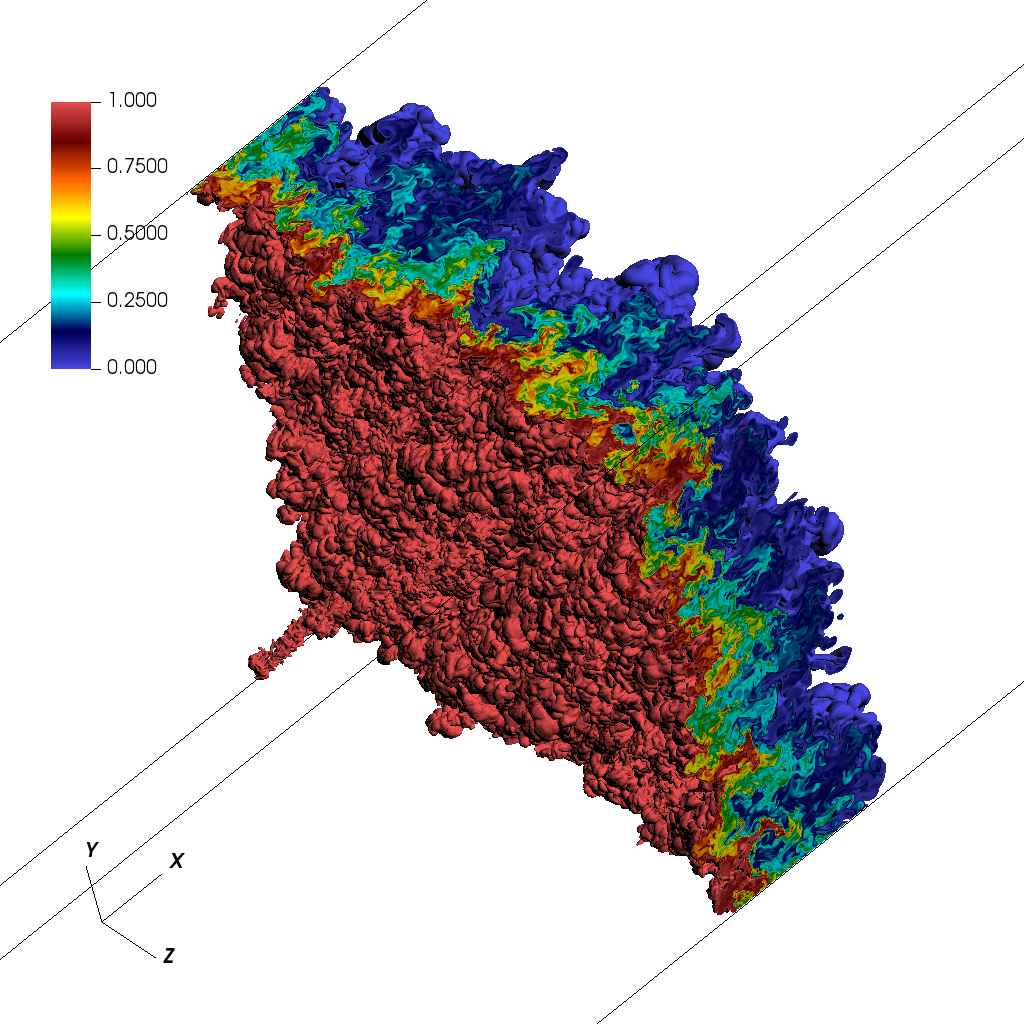}\label{fig:visit_t_1_75}}
  \caption{Isovolumes of the $\mathrm{SF_6}$ mole fraction, $X_{\mathrm{SF_6}}$, at different times in the numerical shock tube with grid E. The colorbar indicates the value of $X_{\mathrm{SF_6}}$. The first and second (last) refinement levels of the AMR grid are shown on the side walls of the domain for the plot at $t=1.40\ \mathrm{ms}$.
  }
  \label{fig:3D_plot}
\end{figure*}

The grid sensitivities of the integral mixing width $W$ and the domain-integrated quantities of interests ($\bar{\rho} a_1$, $\bar{\rho}b$, $\bar{\rho}\tilde{R}_{11}$, $\bar{\rho}\tilde{R}_{22}$, $\bar{\rho}\tilde{R}_{33}$, and $\bar{\rho} k$) in the transport equations of second moment quantities are examined in this section. The mixing width is defined as:
\begin{equation}
    W = \int 4 \bar{X}_{\mathrm{SF_6}} \left( 1 - \bar{X}_{\mathrm{SF_6}} \right) dx. \label{eqn:mixing_width_definition}
\end{equation}
\noindent The mixing width estimates the characteristic length of the mixing layer due to the entrainment of the fluids. Note that since $\bar{\rho}\tilde{R}_{22}$ and $\bar{\rho}\tilde{R}_{33}$ are statistically identical, the grid sensitivity of $\bar{\rho} ( \tilde{R}_{22} + \tilde{R}_{33} ) / 2$ is studied instead.

\begin{figure*}[!ht]
\centering
\subfigure[$\ $Mixing width]{%
\includegraphics[width=0.32\textwidth]{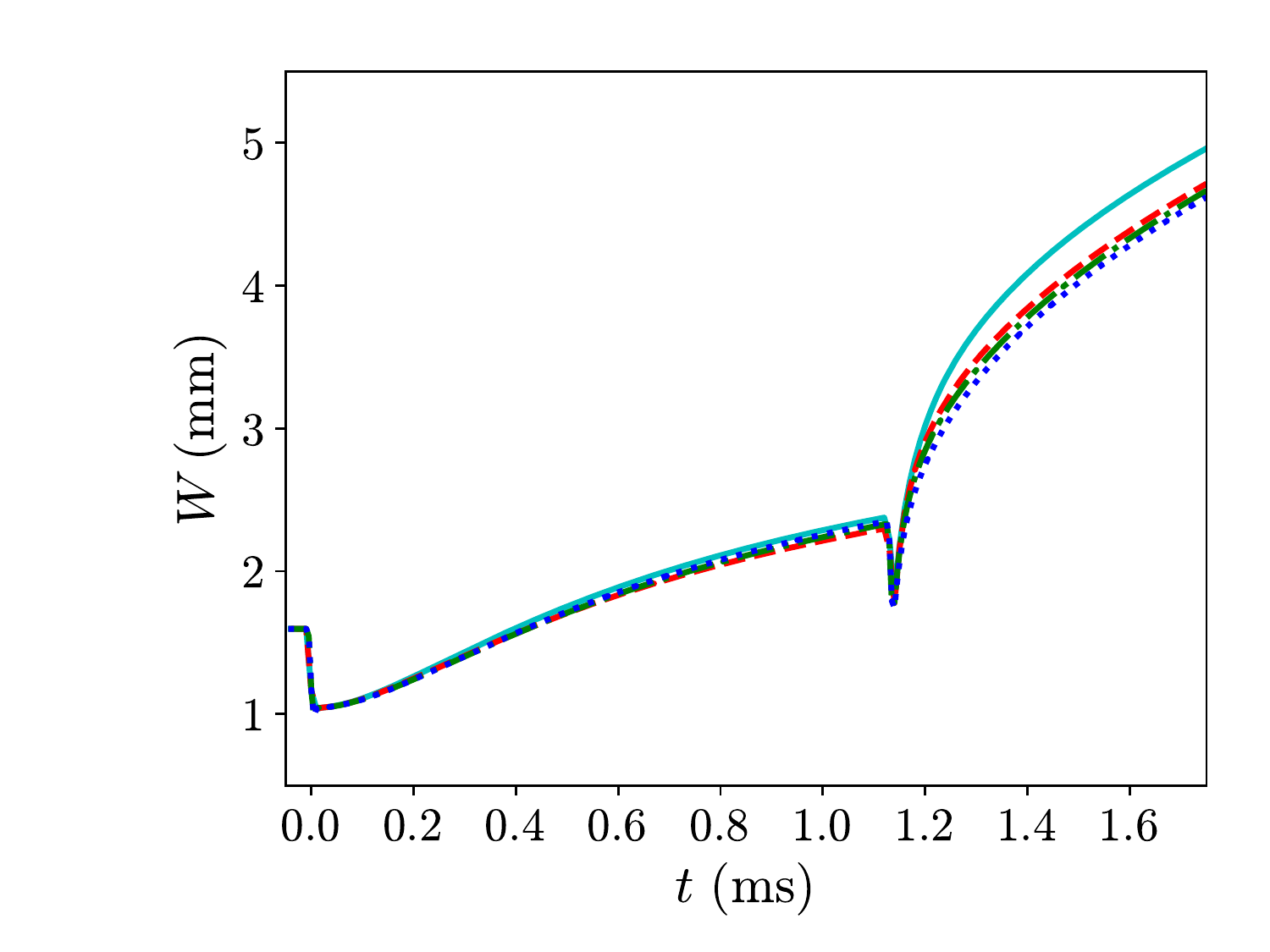}\label{fig:grid_sensitivity_mixing_width}}
\subfigure[$\ $Integrated $\bar{\rho}a_1$]{%
\includegraphics[width=0.32\textwidth]{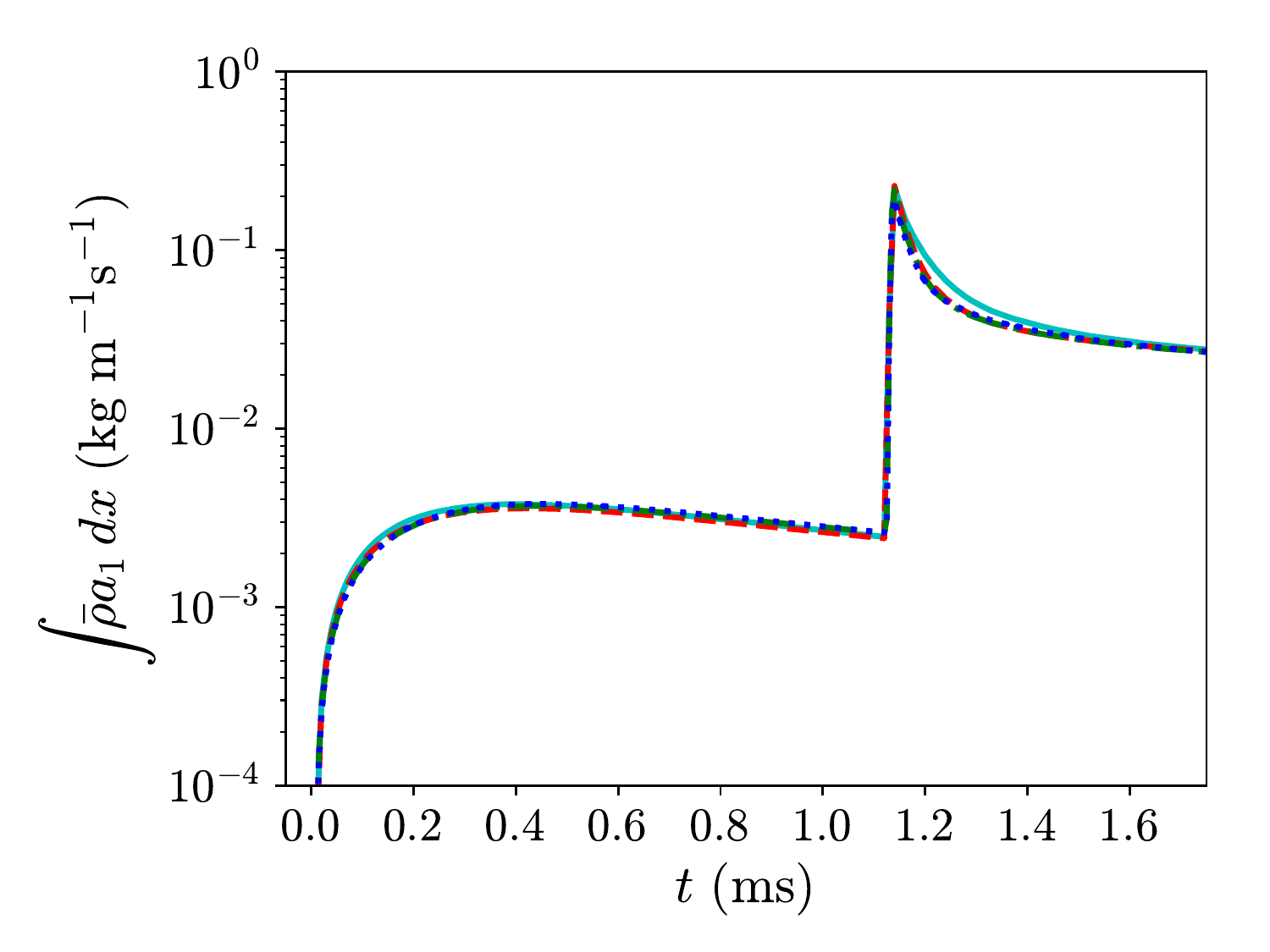}\label{fig:grid_sensitivity_rho_a1}}
\subfigure[$\ $Integrated $\bar{\rho}b$]{%
\includegraphics[width=0.32\textwidth]{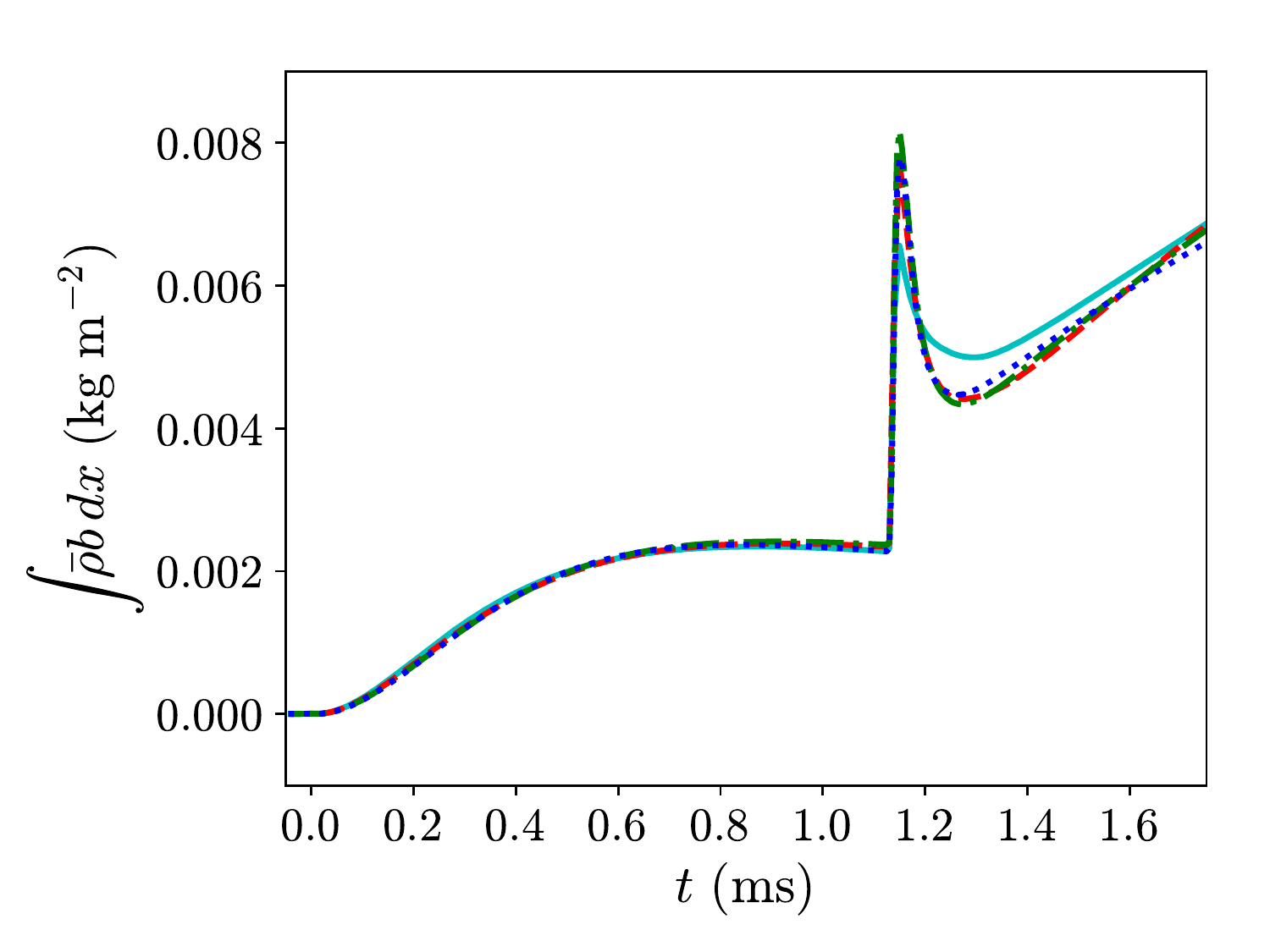}\label{fig:grid_sensitivity_rho_b}}
\subfigure[$\ $Integrated $\bar{\rho}\tilde{R}_{11}$]{%
\includegraphics[width=0.32\textwidth]{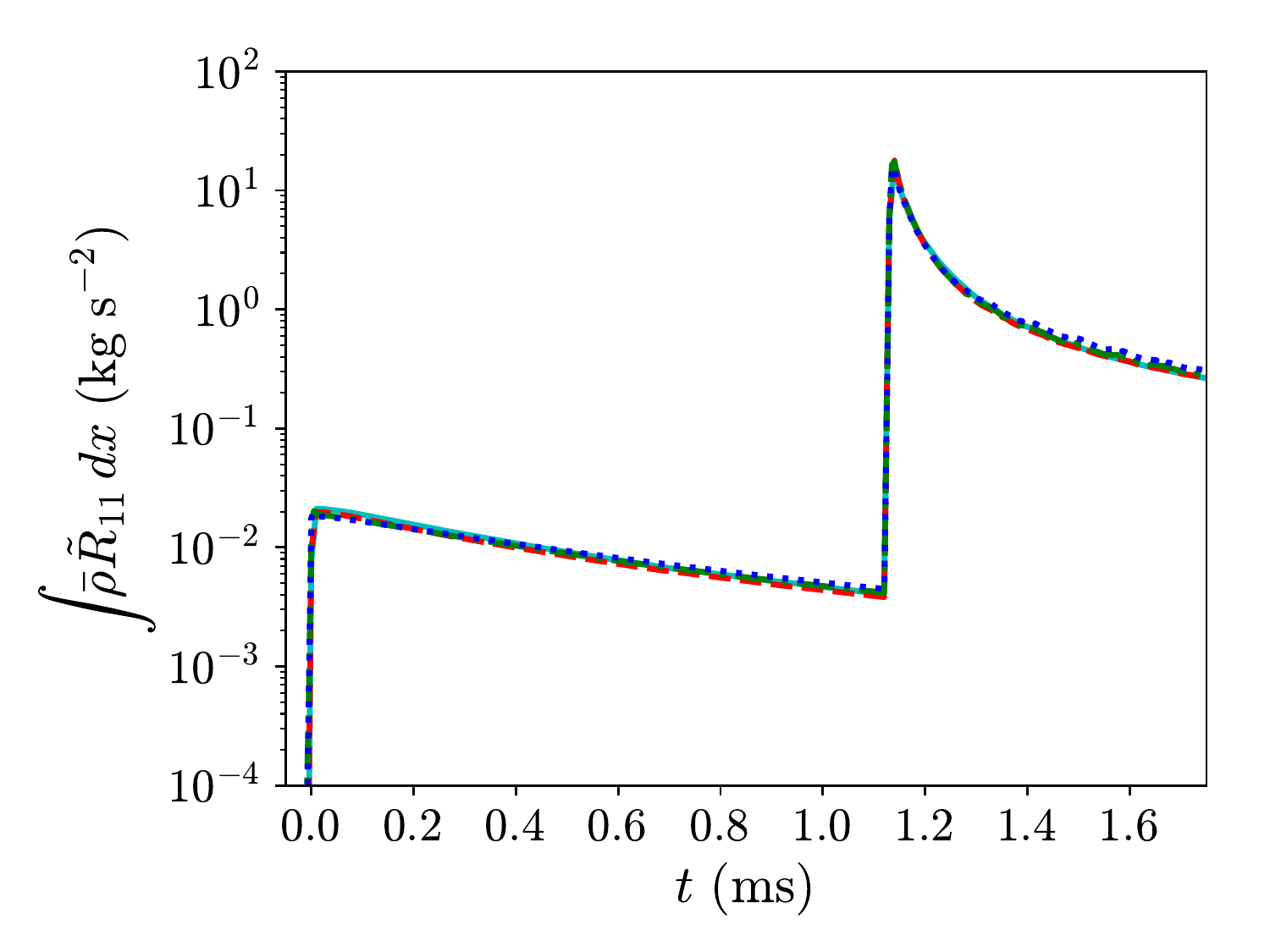}\label{fig:grid_sensitivity_rho_R11}}
\subfigure[$\ $Integrated $\bar{\rho} ( \tilde{R}_{22} + \tilde{R}_{33} ) / 2$]{%
\includegraphics[width=0.32\textwidth]{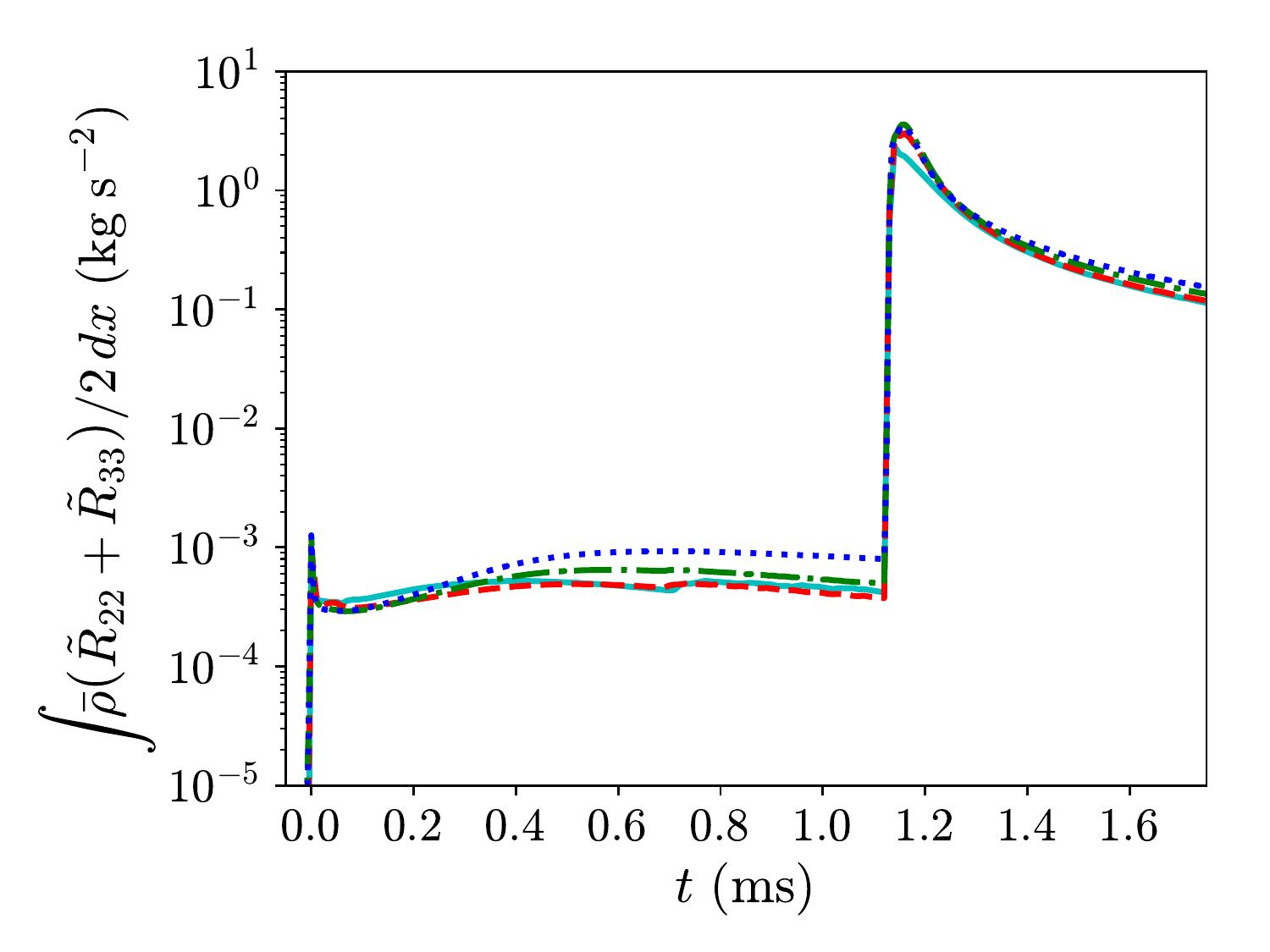}\label{fig:grid_sensitivity_rho_R22_R33}}
\subfigure[$\ $Integrated $\bar{\rho} k$]{%
\includegraphics[width=0.32\textwidth]{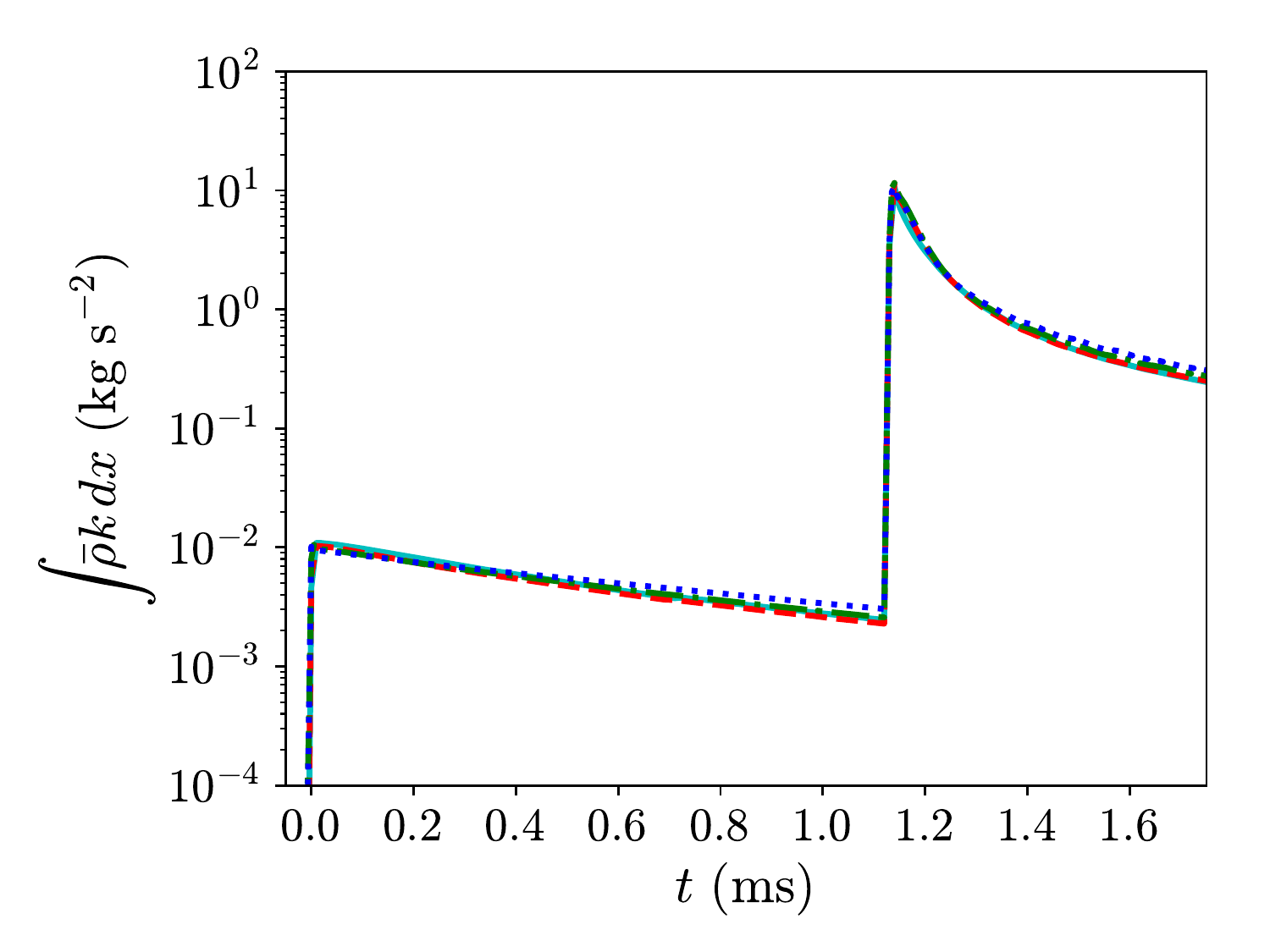}\label{fig:grid_sensitivity_rho_k}}
\caption{Grid sensitivities of mixing width and second moment statistics. Cyan solid line: grid B; red dashed line: grid C; green dash-dotted line: grid D; blue dotted line: grid E.}
\label{fig:grid_sensitivity}
\end{figure*}

Figure~\ref{fig:grid_sensitivity} compares the time evolution of the statistical quantities computed on different grids. From the figure, it can be seen that mixing width, integrals of $\bar{\rho} a_1$, $\bar{\rho}b$, and $\bar{\rho}\tilde{R}_{11}$ are well grid-converged for the entire simulation with the highest resolution grid. The grid sensitivity of the integral of $\bar{\rho} ( \tilde{R}_{22} + \tilde{R}_{33} ) / 2$ is higher than other quantities before re-shock but its contribution to the integral of turbulent kinetic energy, $\bar{\rho}k$, is an order of magnitude smaller than that of $\bar{\rho}\tilde{R}_{11}$. Thus, the integral of $\bar{\rho}k$ is also grid-converged reasonably well at all times. The grid sensitivities of the spatial profiles of these second-moments including the turbulent kinetic energy at different times are also observed to be small between the grid D and the grid E, which are shown in the appendix~\ref{sec:spat_conv_second_moments}.

As the statistical quantities of interests computed on the finest resolution grid (grid E) show very small grid sensitivity throughout the simulation when compared with those from the next finest resolution grid (grid D), only results from grid E are presented and discussed in the remaining sections.

\section{\label{sec:moments} Analysis of the second-moments}

The importance of the second-moments: $\bar{\rho} a_1$, $\bar{\rho} b$, $\bar{\rho} \tilde{R}_{11}$, $\bar{\rho} \tilde{R}_{22}$, and $\bar{\rho} \tilde{R}_{33}$ to close the Favre-averaged momentum equation for the mixture is discussed earlier.
In this section, the time evolution of the spatial profiles of different second-moments including the Favre-averaged Reynolds stress and the turbulent kinetic energy is studied in details, with an examination on their asymmetry due to the variable-density, or non-Boussinesq effects.

At each impulsive acceleration such as at first shock and re-shock, the advection velocity of the mixing layer changes abruptly. However, the advection speed of the mixing layer between impulsive accelerations is essentially constant in time and is close to that given by the solutions of the 1D flow representation, $U_i$. Besides, the mean velocity across the mixing layer is observed to be quite uniform. Therefore, in a moving reference frame with speed $U_i$ relative to the simulation reference frame, $\bar{u} \approx 0$ and $\tilde{u} \approx a_1$ statistically. All of the 1D spatial profiles of the second-moments discussed in this section are plotted in the moving frame of the mixing layer with the $\tilde{x}$ coordinate system. In other words, the $x$ coordinate is shifted as:
\begin{equation}
    \tilde{x}(x,t) = x - x_i(t),
\end{equation}
\noindent where $x_i$ is the location of the interface from the solutions of the 1D flow representation. 

\subsection{Mean density and turbulent mass flux}

\begin{figure*}[!ht]
\centering
\subfigure[$\ $Before re-shock]{%
\includegraphics[width = 0.4\textwidth]{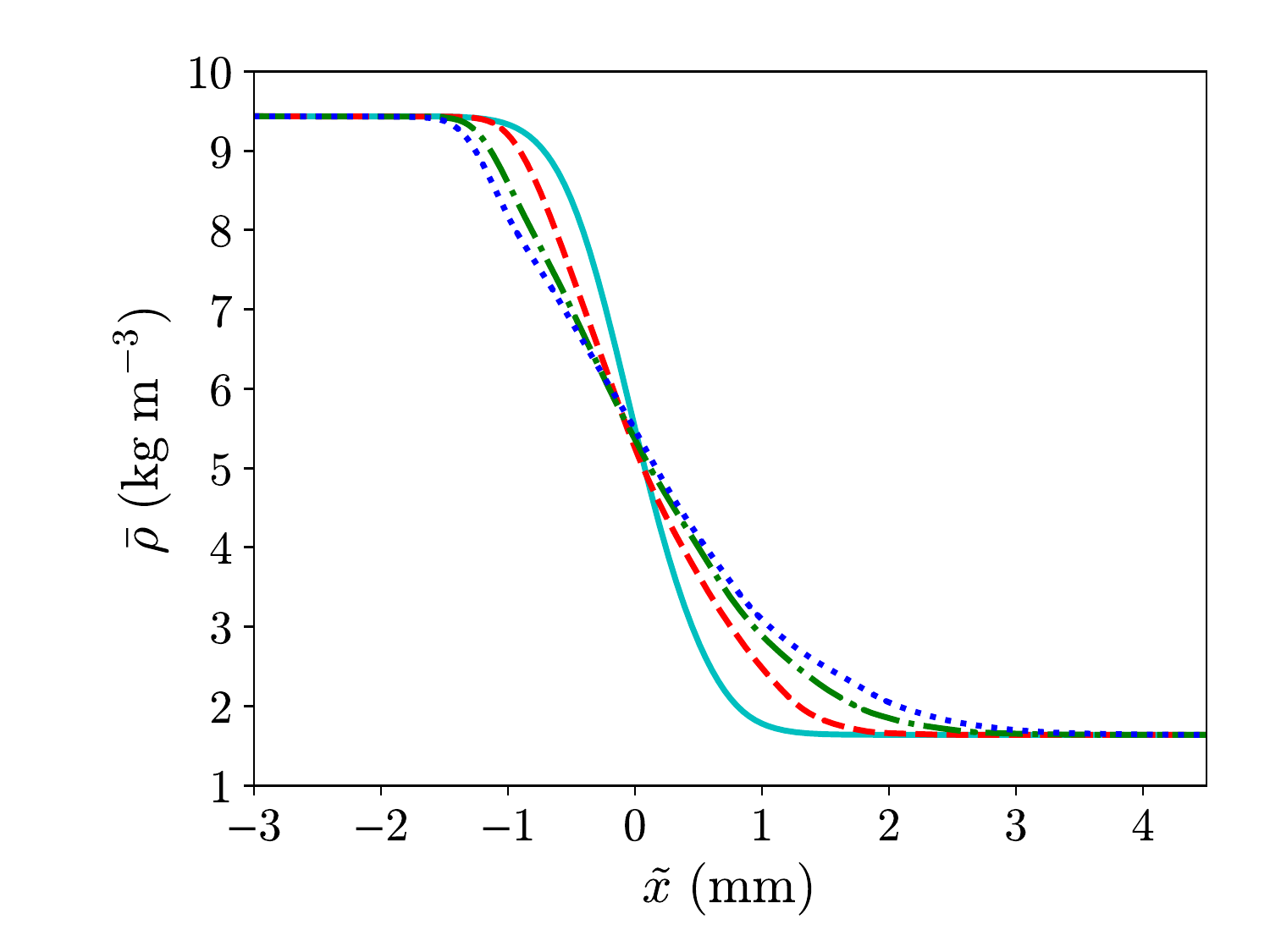}}
\subfigure[$\ $After re-shock]{%
\includegraphics[width = 0.4\textwidth]{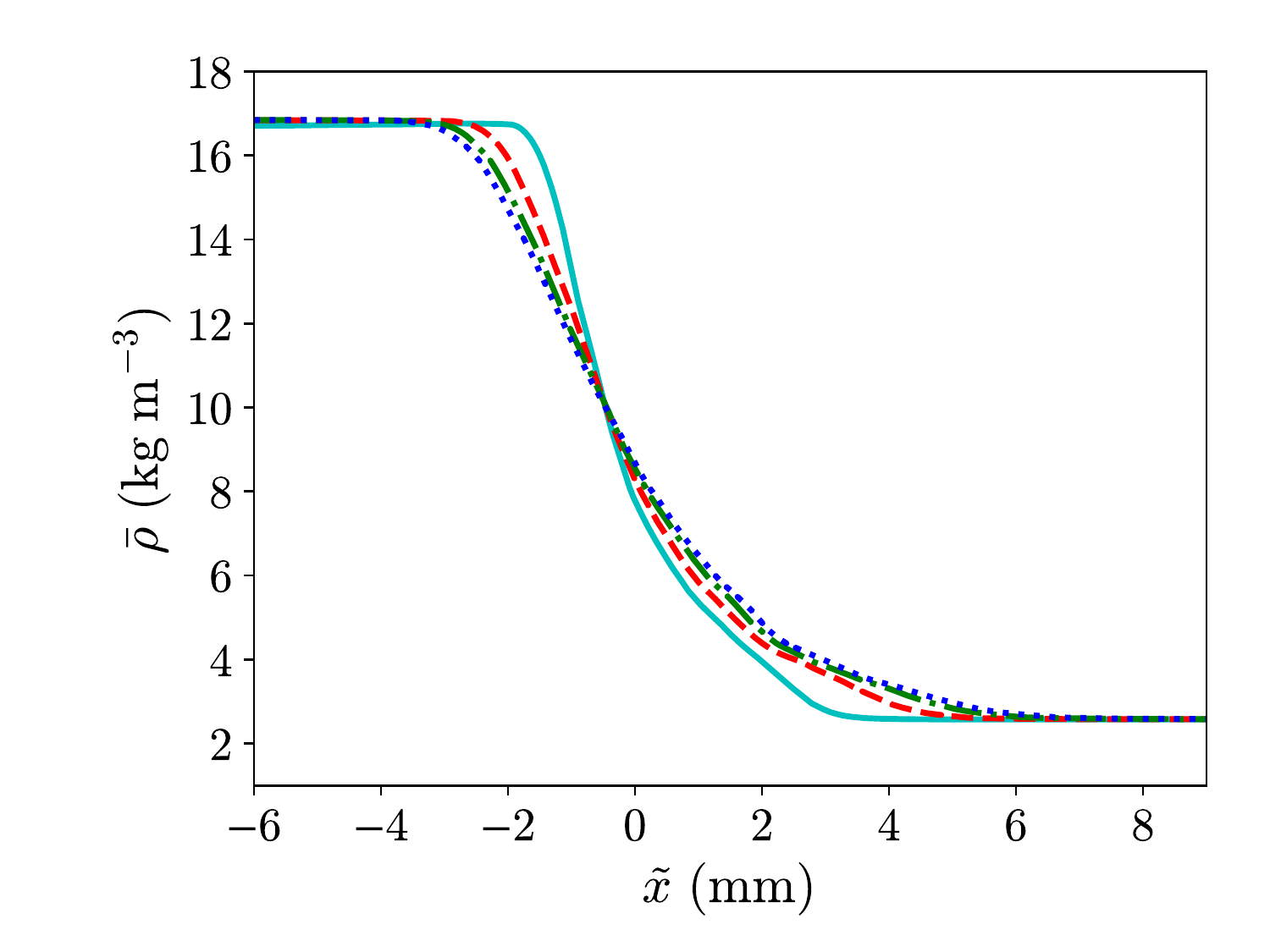}}
\caption{Profiles of the mean density, $\bar{\rho}$, at different times. Cyan solid line in (a): $t=0.05\ \mathrm{ms}$; red dashed line in (a): $t=0.40\ \mathrm{ms}$; green dash-dotted line in (a): $t=0.75\ \mathrm{ms}$; blue dotted line in (a): $t=1.10\ \mathrm{ms}$. Cyan solid line in (b): $t=1.20\ \mathrm{ms}$; red dashed line in (b): $t=1.40\ \mathrm{ms}$; green dash-dotted line in (b): $t=1.60\ \mathrm{ms}$; blue dotted line in (b): $t=1.75\ \mathrm{ms}$.}
\label{fig:rho_profiles}
\end{figure*}

The mean density profiles at different times in the moving frame of the mixing layer are shown in figure~\ref{fig:rho_profiles}.
The density profiles are asymmetric where the spikes penetrate into the lighter fluid more than the bubbles into the heavier fluid due to variable-density effects that are also observed in RTI~\cite{livescu2009rti,livescu2010new}.
The density profiles become wider over time after first shock and re-shock due to the mixing caused by the RMI. While not shown here, the density profiles collapse reasonably well at late times after both first shock and re-shock when they are normalized with the mixing width $W$, similar to RTI~\cite{livescu2009rti}. A similar collapse for the mole fraction profiles was also reported in our previous RMI work~\cite{wong2019high}.

\begin{figure*}[!ht]
\centering
\subfigure[$\ $Before re-shock]{%
\includegraphics[width = 0.4\textwidth]{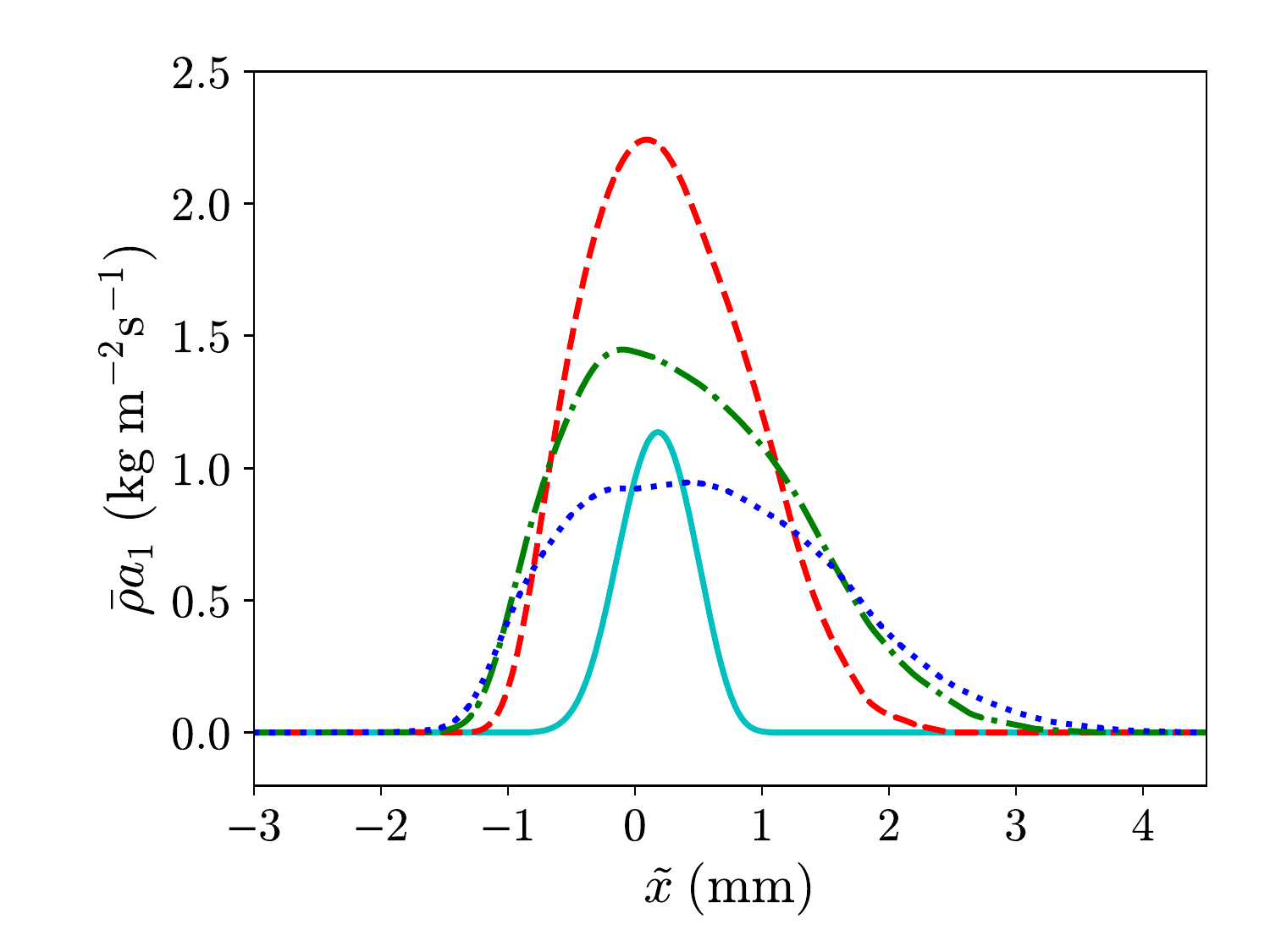}}
\subfigure[$\ $After re-shock]{%
\includegraphics[width = 0.4\textwidth]{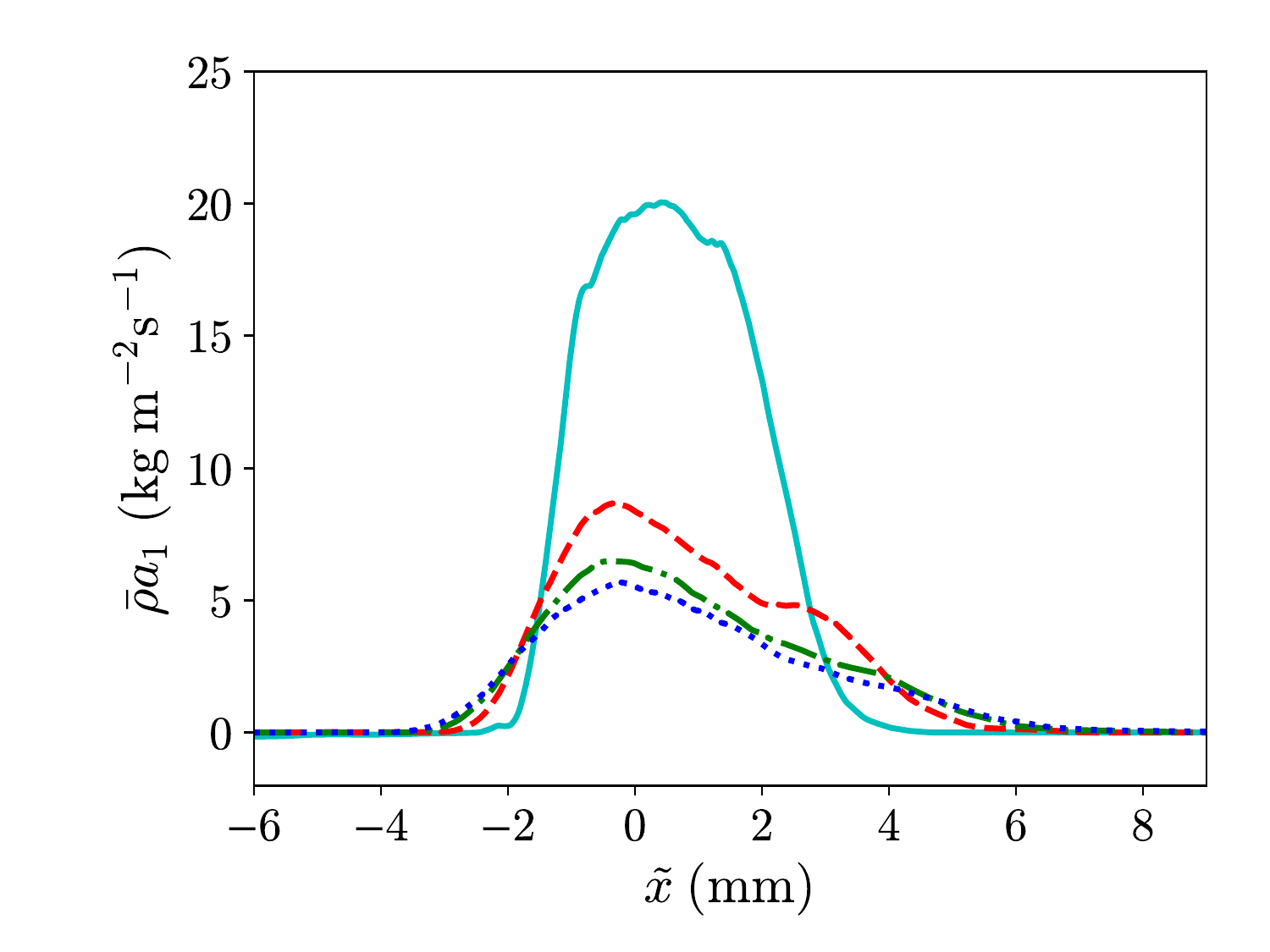}}
\caption{Profiles of the turbulent mass flux component in the streamwise direction, $\bar{\rho}a_1$, at different times. Cyan solid line in (a): $t=0.05\ \mathrm{ms}$; red dashed line in (a): $t=0.40\ \mathrm{ms}$; green dash-dotted line in (a): $t=0.75\ \mathrm{ms}$; blue dotted line in (a): $t=1.10\ \mathrm{ms}$. Cyan solid line in (b): $t=1.20\ \mathrm{ms}$; red dashed line in (b): $t=1.40\ \mathrm{ms}$; green dash-dotted line in (b): $t=1.60\ \mathrm{ms}$; blue dotted line in (b): $t=1.75\ \mathrm{ms}$.}
\label{fig:rho_a1_profiles}
\end{figure*}

Figure~\ref{fig:rho_a1_profiles} compares the profiles of $\bar{\rho} a_1$ at different times before and after re-shock.
The study of turbulent mass flux, $\bar{\rho} a_1$, and the velocity associated with turbulent mass flux, $a_1$, is very important for understanding variable-density effects in the current problem and modeling similar types of flows.
The turbulent mass flux determines the growth of the Favre-averaged Reynolds stress and turbulent kinetic energy in variable-density flows and is studied in previous works on RMI~\cite{balakumar2012turbulent, mohaghar2017evaluation,reese2018simultaneous}, RTI~\cite{livescu2009rti,livescu2010new,aslangil2022study} and buoyancy-driven variable-density turbulence~\cite{livescu2007buoyancy}.
From the figure, it can be seen that there is a sudden rise in $\bar{\rho} a_1$, followed by its decay after each shock event. The jump in the magnitude of $\bar{\rho} a_1$ is caused by the large amount of energy injected at the mixing layer at each impulsive acceleration.
The profiles of $\bar{\rho} a_1$ are asymmetric and have longer tails on the light fluid side.
It can also be noticed that at late times after first-shock and re-shock, $\bar{\rho} a_1$ peaks at a position slightly towards the heavier fluid side (slightly negative $\tilde{x}$). This suggests that there is a fixed point at the same location in the mean density profiles at late times, which can be deduced from equation~\eqref{eq:Favre_mixture_continuity_eqn}. The fixed point can be seen and verified from the mean density profiles shown in figure~\ref{fig:rho_profiles}. A fixed point in mean density profiles was also observed in RTI~\cite{livescu2009rti}.

\subsection{Density-specific-volume covariance}

\begin{figure*}[!ht]
\centering
\subfigure[$\ $Before re-shock]{%
\includegraphics[width=0.4\textwidth]{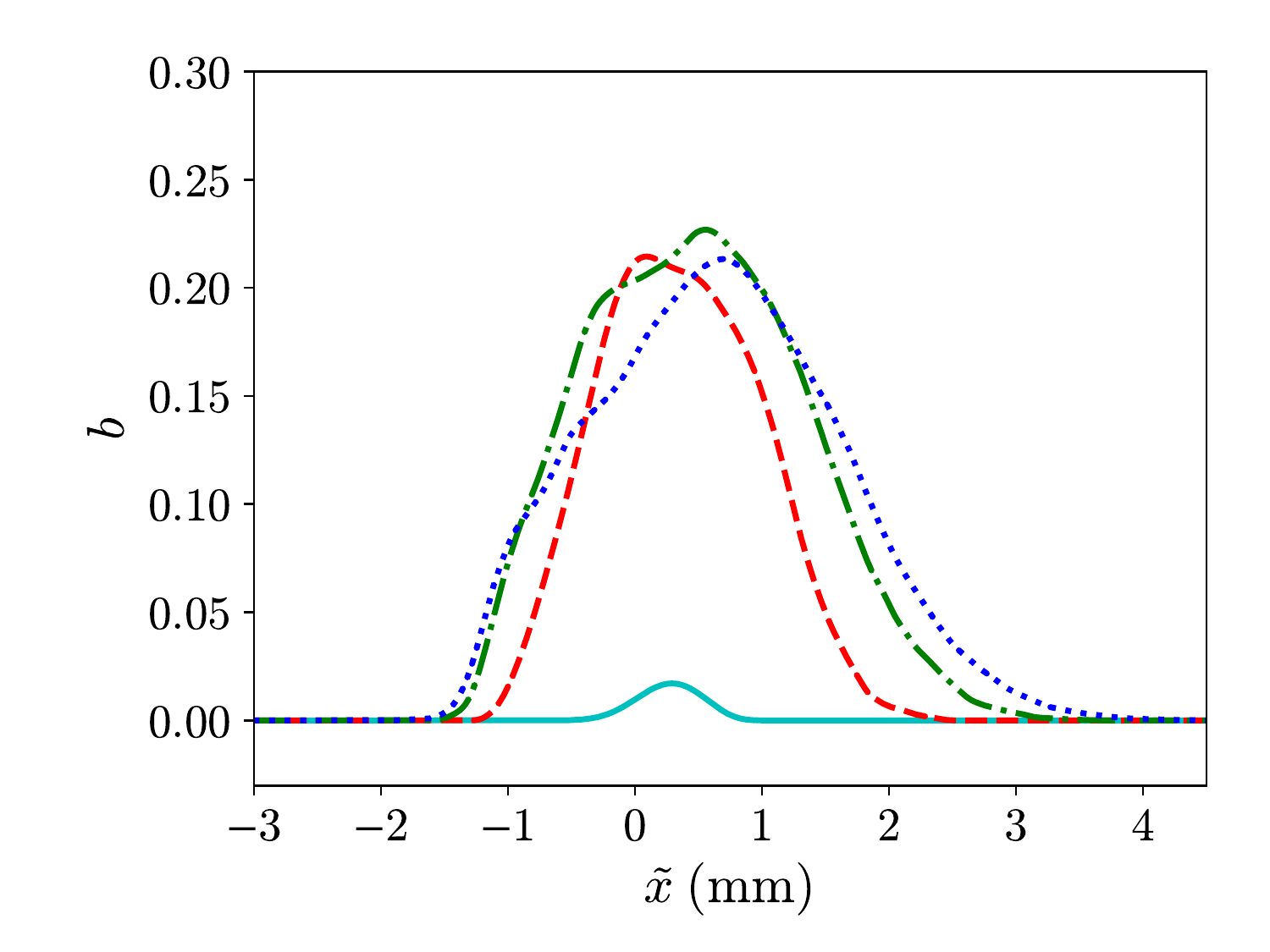}}
\subfigure[$\ $After re-shock]{%
\includegraphics[width=0.4\textwidth]{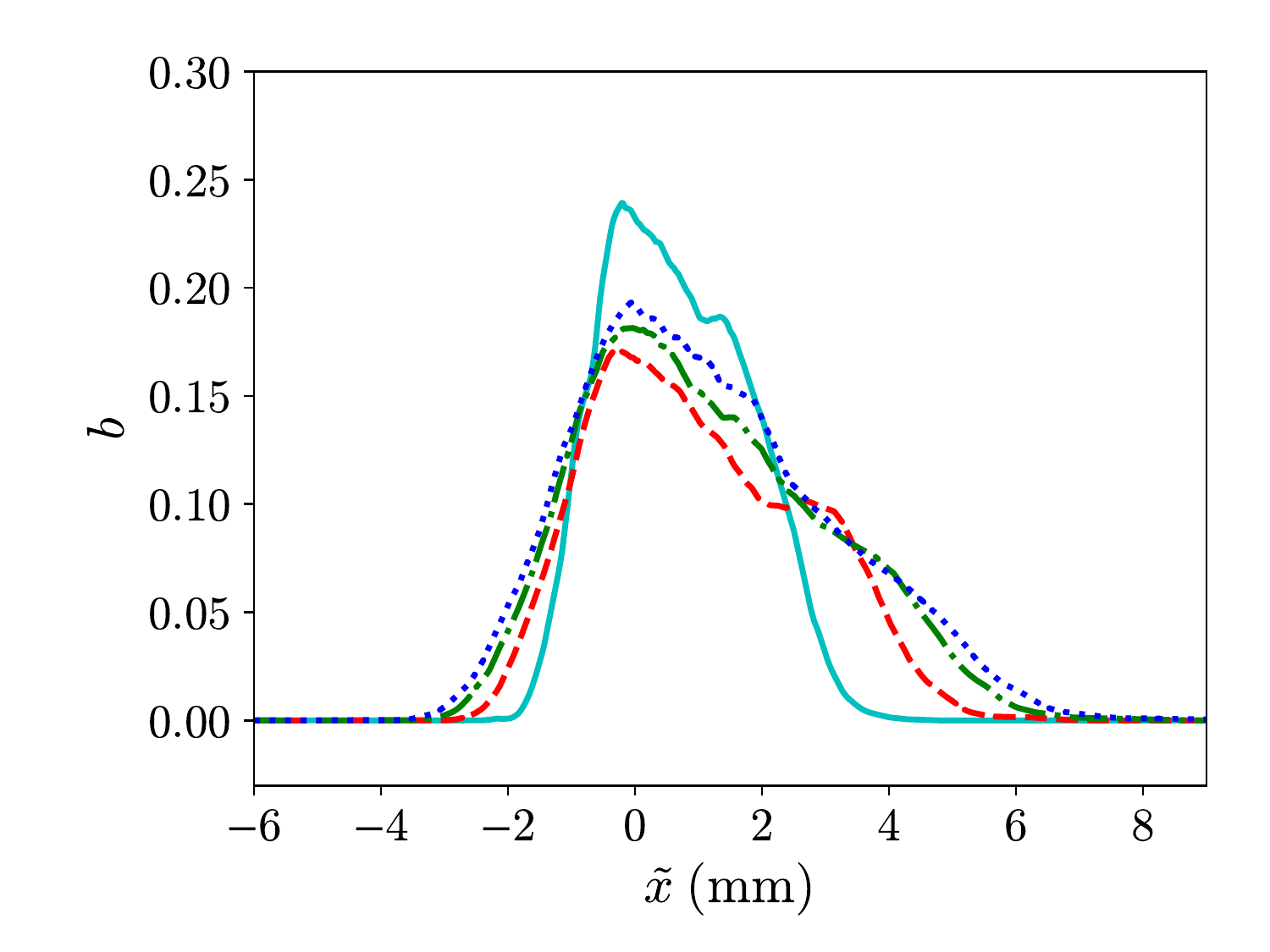}}
\caption{Profiles of the density-specific-volume covariance, $b$, at different times. Cyan solid line in (a): $t=0.05\ \mathrm{ms}$; red dashed line in (a): $t=0.40\ \mathrm{ms}$; green dash-dotted line in (a): $t=0.75\ \mathrm{ms}$; blue dotted line in (a): $t=1.10\ \mathrm{ms}$. Cyan solid line in (b): $t=1.20\ \mathrm{ms}$; red dashed line in (b): $t=1.40\ \mathrm{ms}$; green dash-dotted line in (b): $t=1.60\ \mathrm{ms}$; blue dotted line in (b): $t=1.75\ \mathrm{ms}$.}
\label{fig:b_profiles}
\end{figure*}

The density-specific-volume covariance, $b$, mediates the turbulent mass flux production mechanism. It can also be viewed as a metric for the homogeneity of mixing. $b$ is a non-negative quantity, and $b=0$ corresponds to fluids that are homogeneously mixed. On the contrary, a high value of $b$ indicates inhomogeneous mixing of the fluids. 
This statistical quantity was extensively studied in many previous investigations on RMI~\cite{balasubramanian2013experimental,orlicz2013incident, tomkins2013evolution, weber2014experimental, tritschler2014richtmyer, lombardini2014turbulent,mohaghar2017evaluation,mohaghar2019transition} and also RTI~\cite{livescu2009rti}.
Figure~\ref{fig:b_profiles} displays the profiles of $b$ at different times before and after re-shock.
The shapes of $b$ have a single peak and are asymmetric at different times due to the variable-density, or non-Boussinesq effects. The shapes have longer tails on the lighter fluid side at all times.
Before re-shock, the peak appears to be on the lighter fluid side at late times, but the peak shifts to the heavier fluid side after re-shock.
It can be seen that the peak of $b$ at late times after first shock and re-shock remains quite stationary.
This slow rate of change in the magnitude of $b$ at late times was also observed in the RMI experiments by~\citet{balasubramanian2013experimental} and~\citet{tomkins2013evolution} after incident shock and re-shock respectively, and RMI simulations by~\citet{tritschler2014richtmyer} after re-shock. Similar late-time behavior was also seen in the RTI simulations by~\citet{livescu2009rti}.

$b$ can also be expressed as a sum of a series of density probability density function (PDF) moments~\cite{livescu2008variable}:
\begin{equation}
   b = \frac{\overline{{\rho^{\prime}}^{2}}}{\bar{\rho}^2}
      \left[ 1 - i_{\rho} \frac{\overline{{\rho^{\prime}}^{3} } }{( \overline{ {\rho^{\prime}}^{2} } )^{3/2} }
      + i_{\rho}^{2} \frac{\overline{{\rho^{\prime}}^{4} } }{( \overline{ {\rho^{\prime}}^{2} } )^{2} }
      - i_{\rho}^{3} \frac{\overline{{\rho^{\prime}}^{5} } }{( \overline{ {\rho^{\prime}}^{2} } )^{5/2} }
      + ...
      \right],
\label{eqn:b_taylor_expansion}
\end{equation}

\noindent where $i_{\rho}=(\overline{{\rho^{\prime}}^{2}})^{1/2} / \bar{\rho}$. If $i_{\rho}$ is very small, the equation reduces to the Boussinesq relation:
\begin{equation}
    b \approx \frac{\overline{{\rho^{\prime}}^{2}}}{\bar{\rho}^2}.
\label{eqn:Boussinesq_relation}
\end{equation}

\begin{figure*}[!ht]
\centering
\subfigure[$\ $Before re-shock]{%
\includegraphics[width=0.4\textwidth]{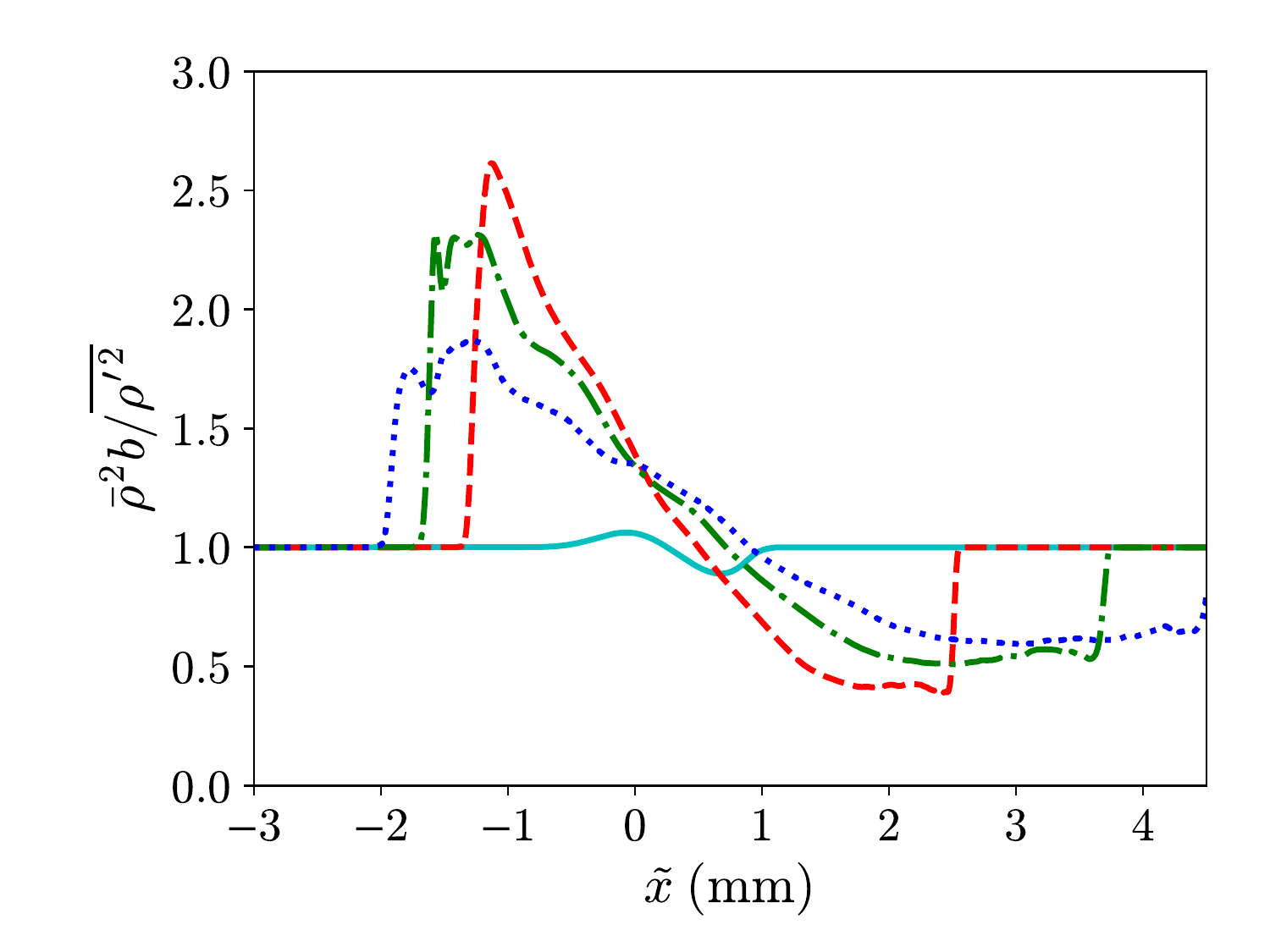}}
\subfigure[$\ $After re-shock]{%
\includegraphics[width=0.4\textwidth]{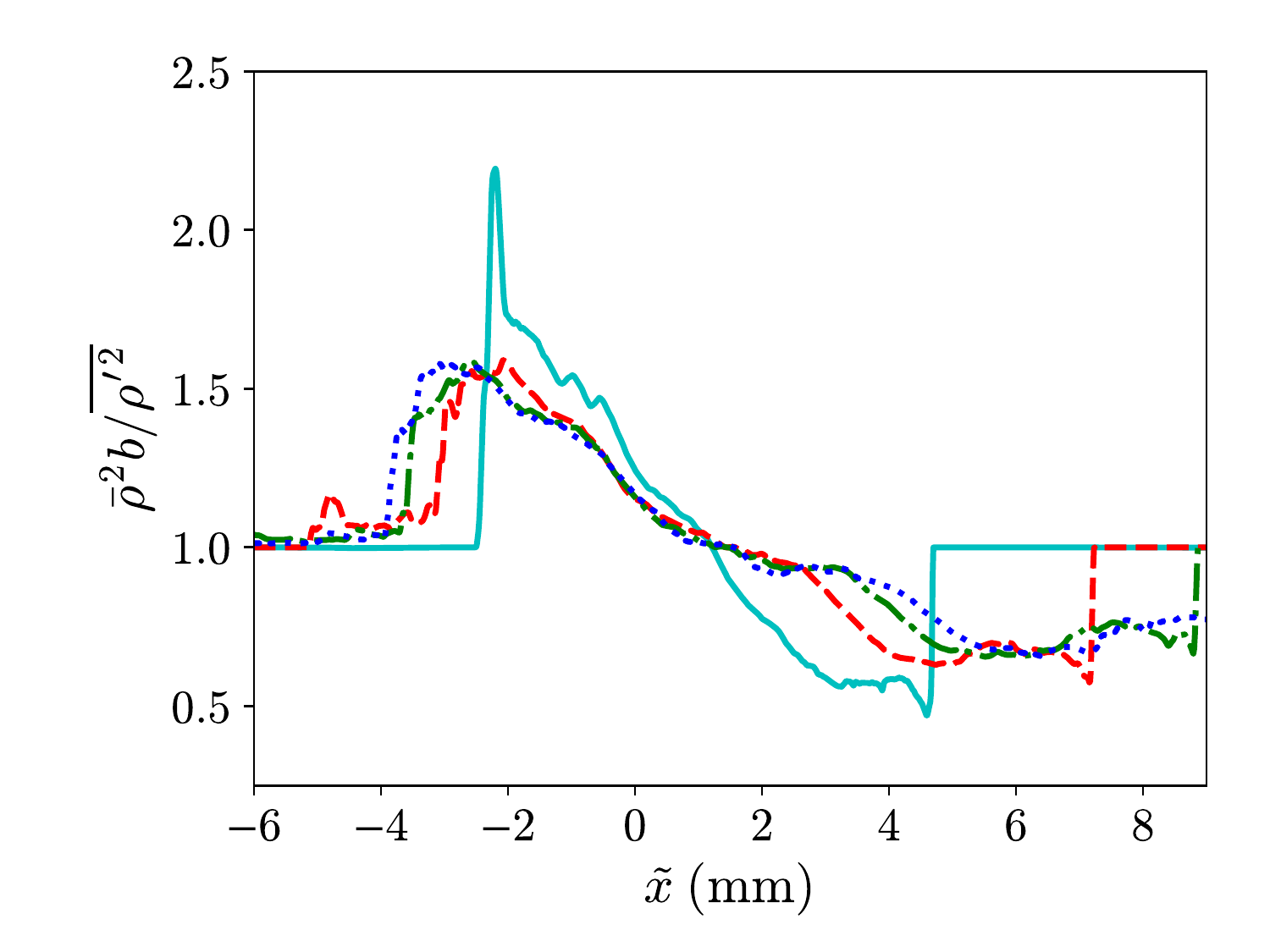}}
\caption{Profiles of the ratio of the density-specific-volume covariance to the square of density intensity at different times. Cyan solid line in (a): $t=0.05\ \mathrm{ms}$; red dashed line in (a): $t=0.40\ \mathrm{ms}$; green dash-dotted line in (a): $t=0.75\ \mathrm{ms}$; blue dotted line in (a): $t=1.10\ \mathrm{ms}$. Cyan solid line in (b): $t=1.20\ \mathrm{ms}$; red dashed line in (b): $t=1.40\ \mathrm{ms}$; green dash-dotted line in (b): $t=1.60\ \mathrm{ms}$; blue dotted line in (b): $t=1.75\ \mathrm{ms}$.}
\label{fig:Bouss_profiles}
\end{figure*}

\noindent The ratio of the left hand side (density-specific-volume covariance) and the right hand side (square of density intensity) of equation~\eqref{eqn:Boussinesq_relation} can be used to test the Boussinesq approximation, where the corresponding component in turbulent mass flux production can be approximated with density variance instead of $b$. Boussinesq approximation is valid when the ratio is close to 1. Figure~\ref{fig:Bouss_profiles} shows the variations in the ratio across the mixing region at different times. It can be seen that the ratio varies from 0.5 to 2.5. The ratio is, in general, larger than 1 on the heavier fluid side and smaller than 1 on the lighter fluid side, because of the skewness of the density field. The peaks are located at the edges of the mixing layers, which indicates that variable-density effects are larger at the edges than at the central part of the mixing layer. The same behavior is also observed in the spherical RMI simulations by~\citet{lombardini2014turbulent} with essentially the same Atwood number and the planar RTI simulations by~\citet{livescu2009rti} with slightly smaller Atwood number ($At = 0.5$). As as result, Boussinesq equations would lead to an underestimation of the energy conversion rate on the heavier fluid side and an overestimation on the lighter fluid side for high Atwood number flows.
A grid sensitivity analysis of the profiles of the ratio at different times is given in the Supplemental Material~\cite{supple2022wong}. The analysis shows that the differences of the profiles between the grid D and the grid E at different times are minor.

\subsection{Favre-averaged Reynolds stress and turbulent kinetic energy}

\begin{figure*}[!ht]
\centering
\subfigure[$\ $Before re-shock]{%
\includegraphics[width=0.4\textwidth]{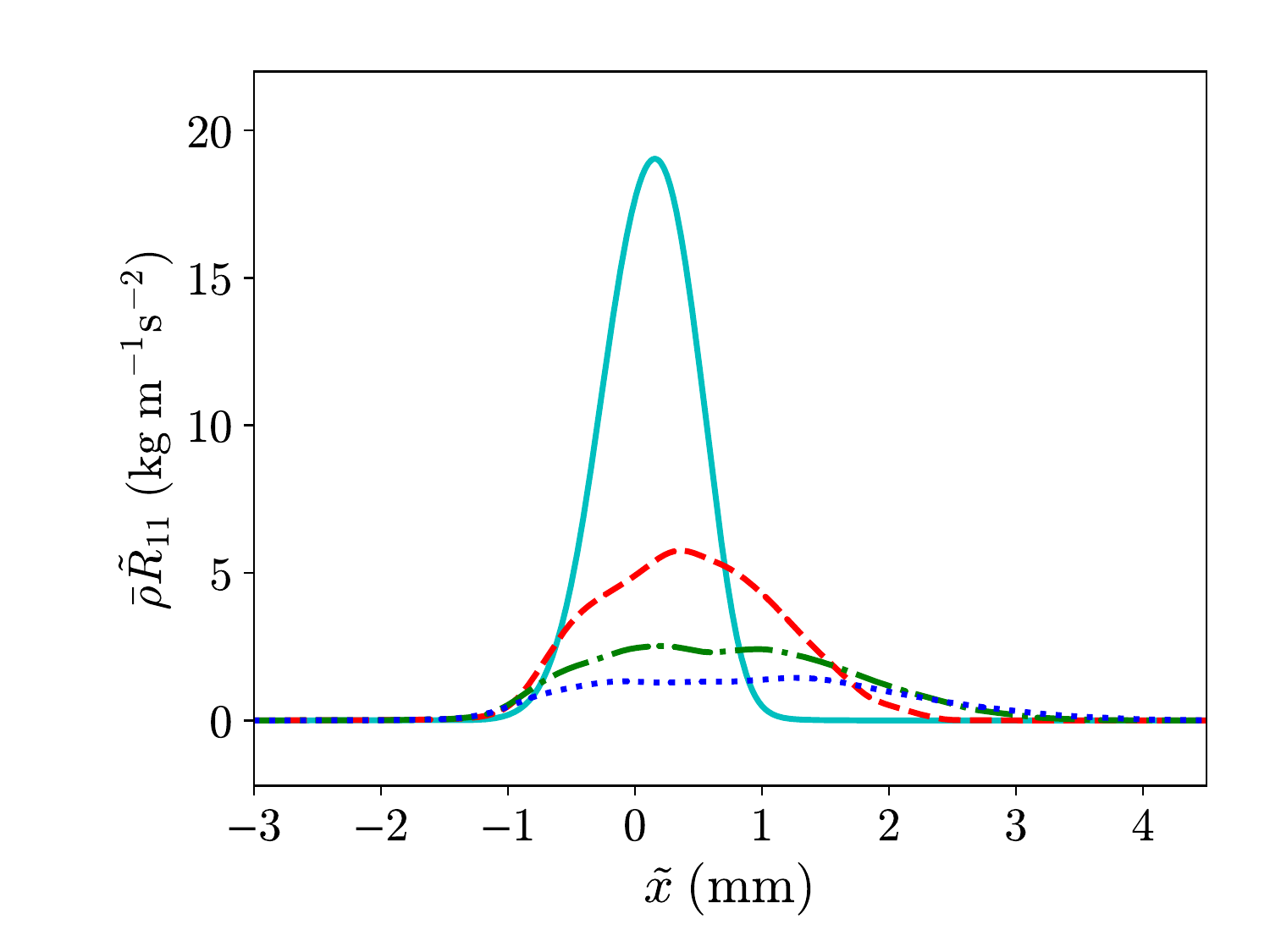}}
\subfigure[$\ $After re-shock]{%
\includegraphics[width=0.4\textwidth]{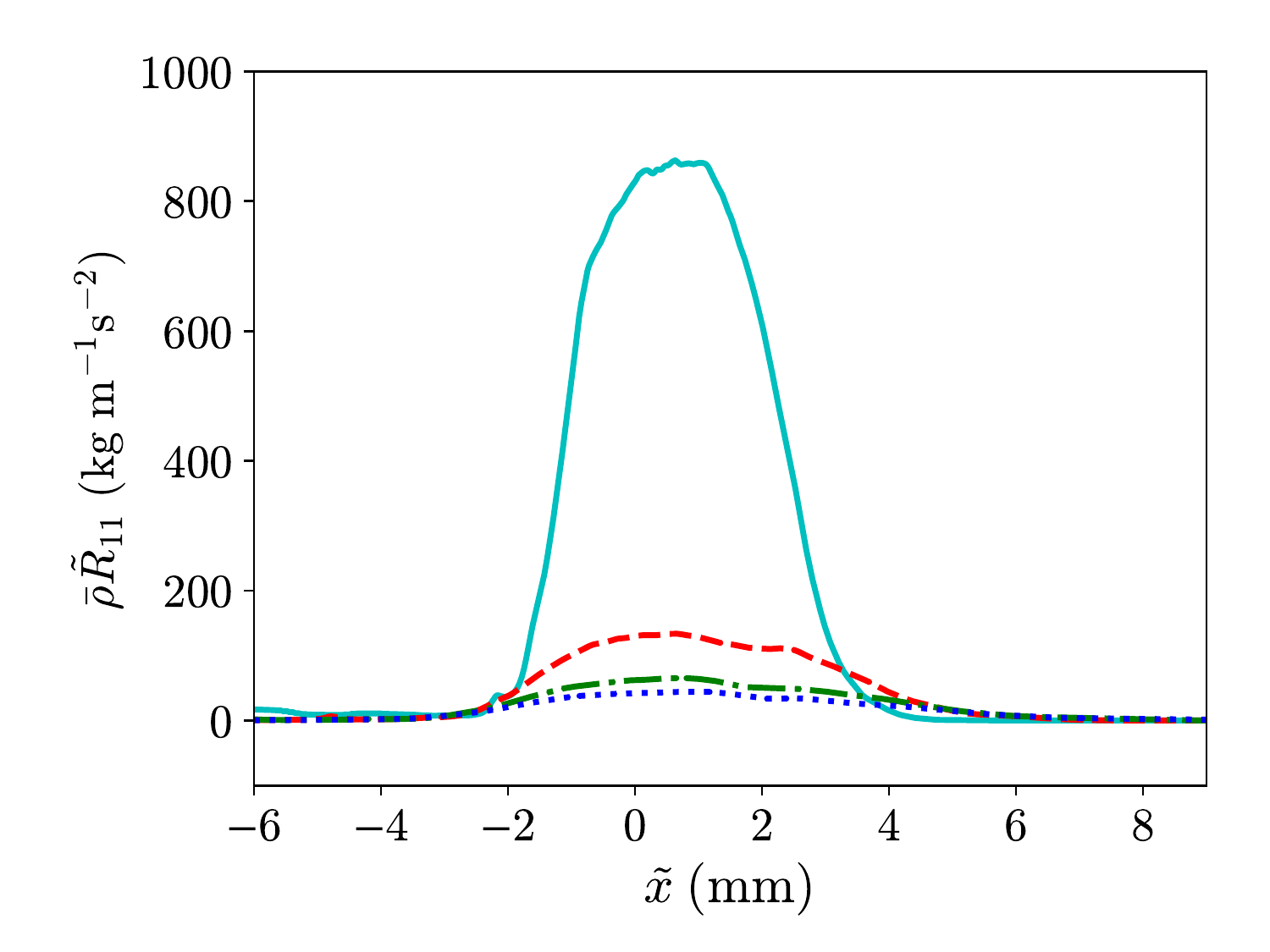}}
\caption{Profiles of the Reynolds normal stress component in the streamwise direction multiplied by the mean density, $\bar{\rho} \tilde{R}_{11}$, at different times. Cyan solid line in (a): $t=0.05\ \mathrm{ms}$; red dashed line in (a): $t=0.40\ \mathrm{ms}$; green dash-dotted line in (a): $t=0.75\ \mathrm{ms}$; blue dotted line in (a): $t=1.10\ \mathrm{ms}$. Cyan solid line in (b): $t=1.20\ \mathrm{ms}$; red dashed line in (b): $t=1.40\ \mathrm{ms}$; green dash-dotted line in (b): $t=1.60\ \mathrm{ms}$; blue dotted line in (b): $t=1.75\ \mathrm{ms}$.}
\label{fig:rho_R11_profiles}
\end{figure*}

\begin{figure*}[!ht]
\centering
\subfigure[$\ $Before re-shock]{%
\includegraphics[width=0.4\textwidth]{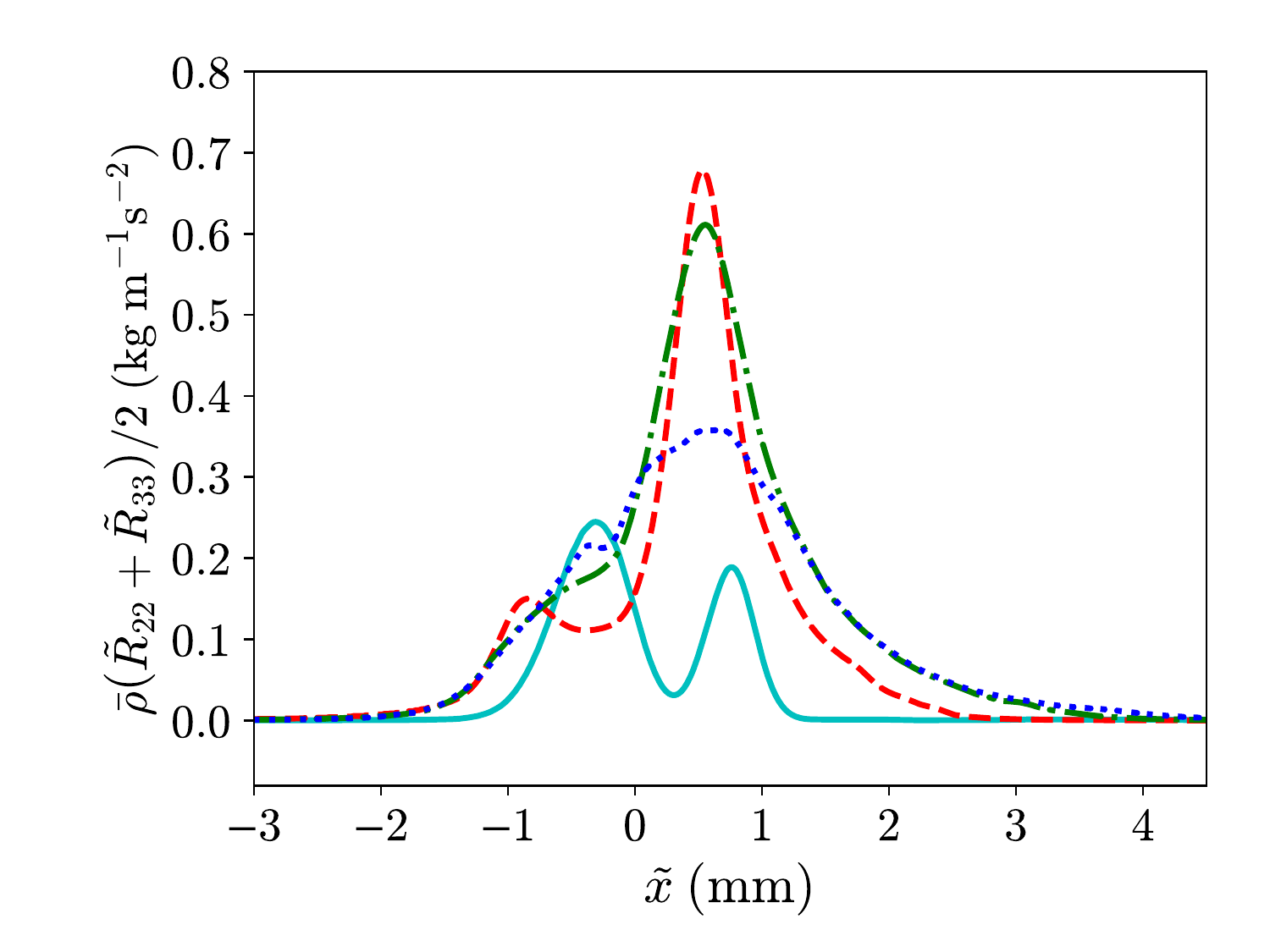}}
\subfigure[$\ $After re-shock]{%
\includegraphics[width=0.4\textwidth]{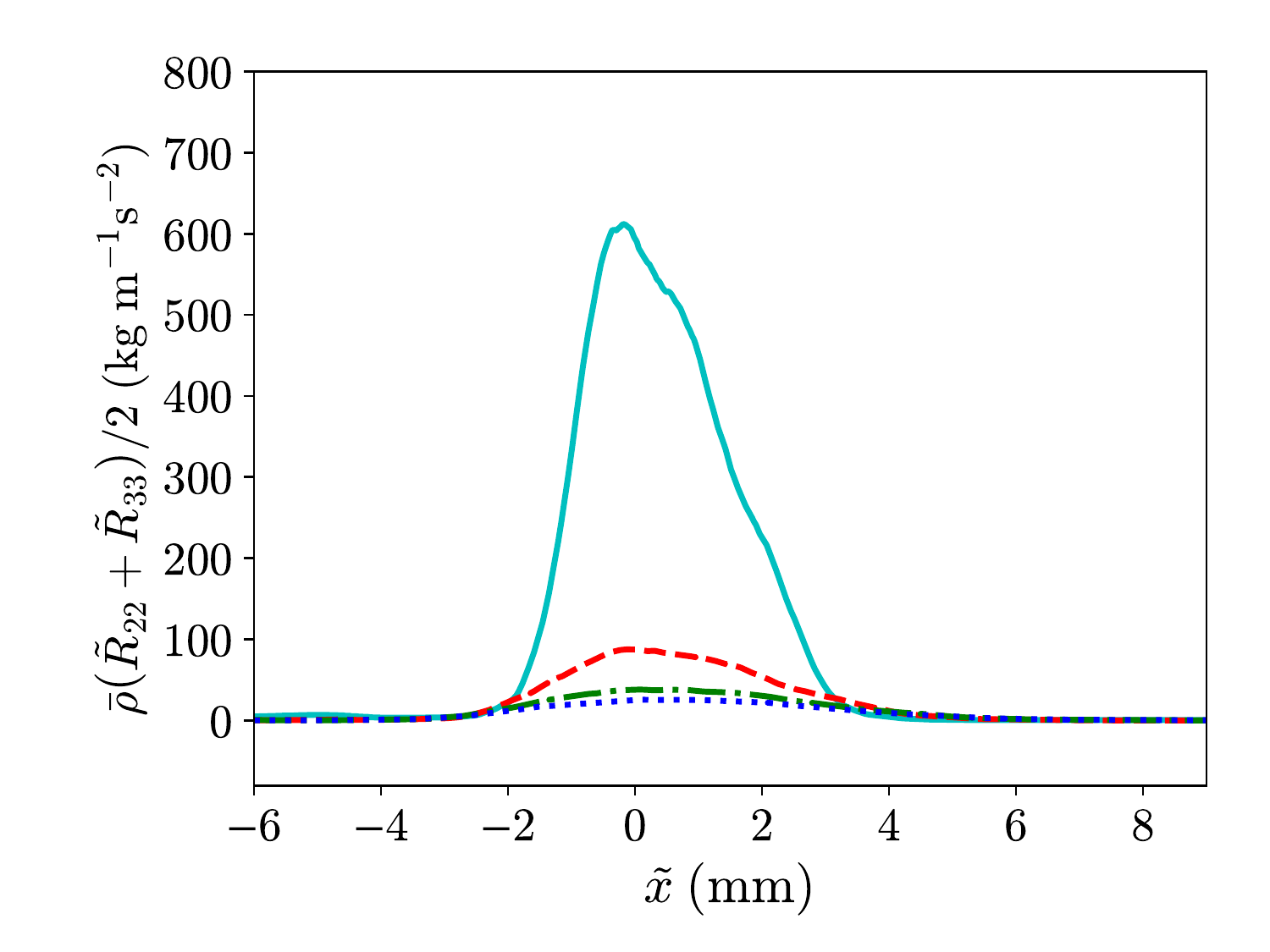}}
\caption{Profiles of the Reynolds normal stress component in the transverse directions multiplied by the mean density, $\bar{\rho} ( \tilde{R}_{11} + \tilde{R}_{22} )/2$, at different times. Cyan solid line in (a): $t=0.05\ \mathrm{ms}$; red dashed line in (a): $t=0.40\ \mathrm{ms}$; green dash-dotted line in (a): $t=0.75\ \mathrm{ms}$; blue dotted line in (a): $t=1.10\ \mathrm{ms}$. Cyan solid line in (b): $t=1.20\ \mathrm{ms}$; red dashed line in (b): $t=1.40\ \mathrm{ms}$; green dash-dotted line in (b): $t=1.60\ \mathrm{ms}$; blue dotted line in (b): $t=1.75\ \mathrm{ms}$.}
\label{fig:rho_R22_33_profiles}
\end{figure*}

\begin{figure*}[!ht]
\centering
\subfigure[$\ t=0.40\ \mathrm{ms}$]{%
\includegraphics[height = 0.33\textwidth]{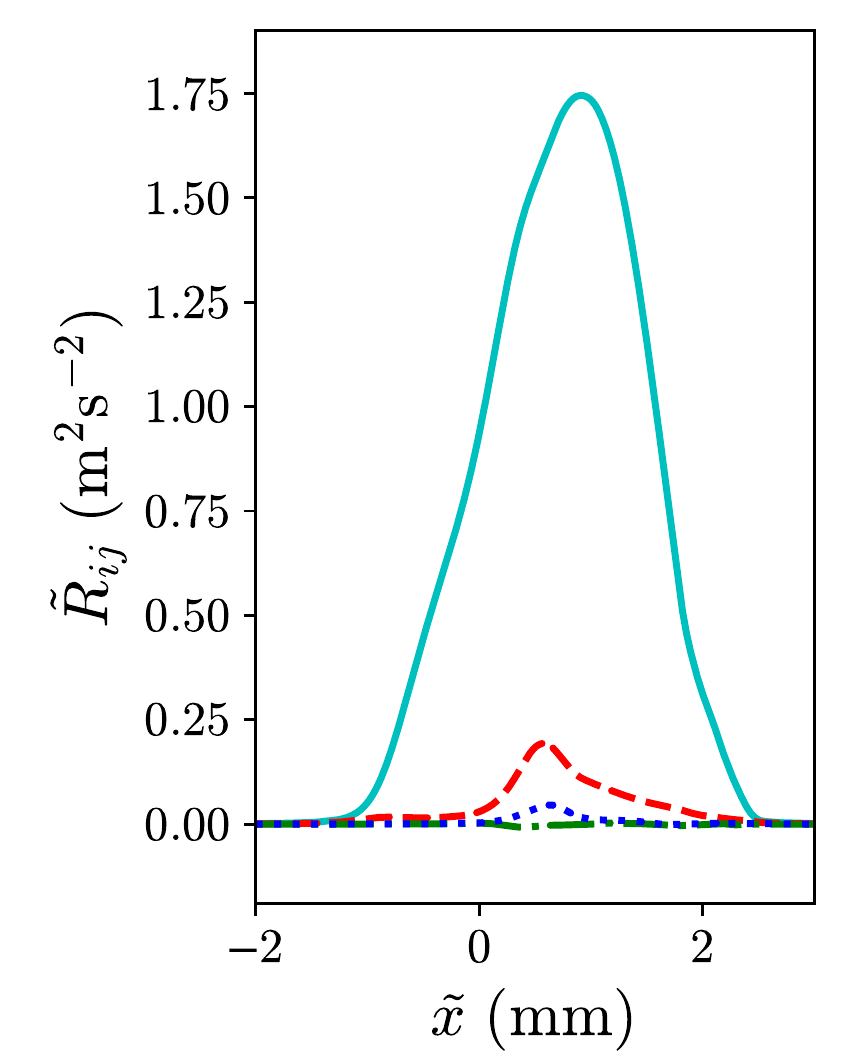}}
\subfigure[$\ t=1.10\ \mathrm{ms}$]{%
\includegraphics[height = 0.33\textwidth]{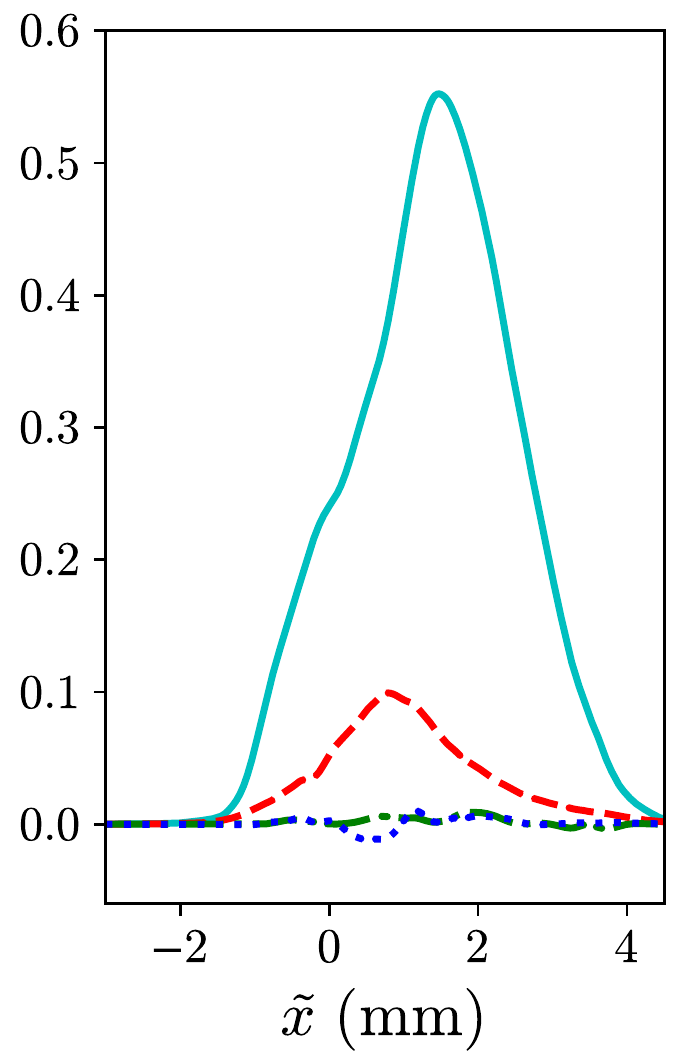}}
\subfigure[$\ t=1.20\ \mathrm{ms}$]{%
\includegraphics[height = 0.33\textwidth]{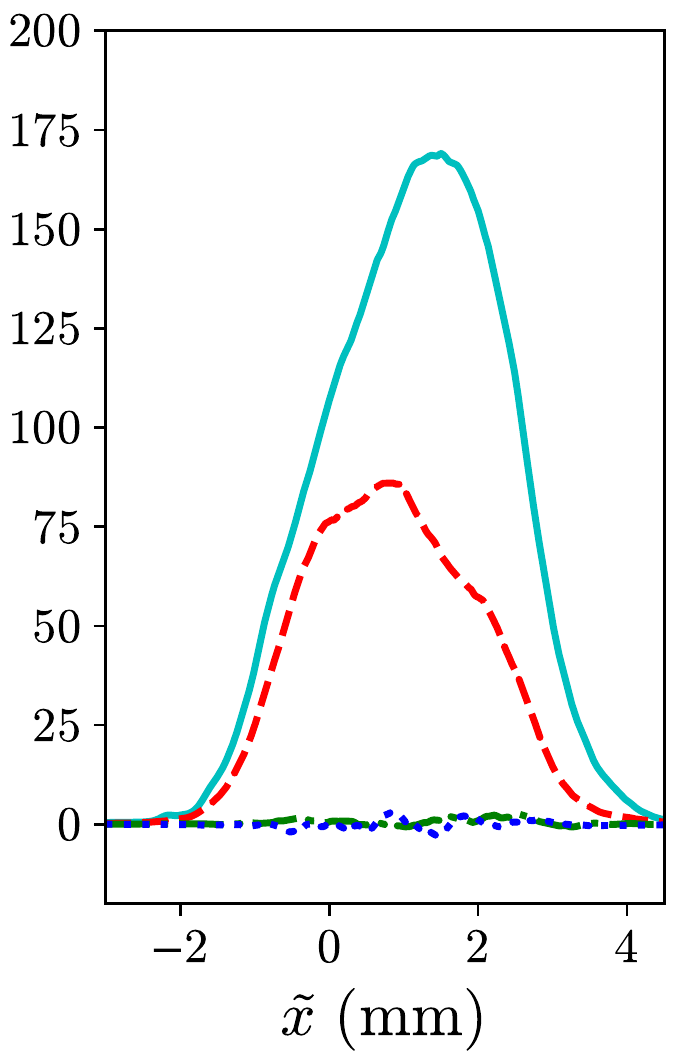}}
\subfigure[$\ t=1.75\ \mathrm{ms}$]{%
\includegraphics[height = 0.33\textwidth]{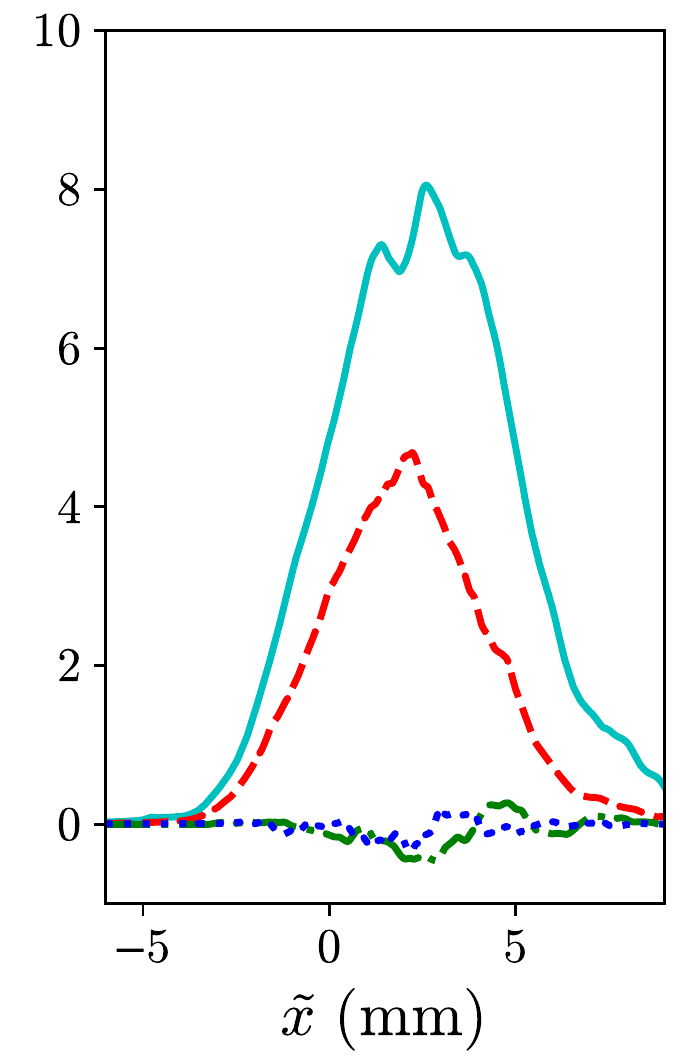}}
\caption{Comparison of the Reynolds stress components multiplied by the mean density at different times. Cyan solid line: $\tilde{R}_{11}$; red dashed line: $(\tilde{R}_{22}+\tilde{R}_{33})/2$; green dash-dotted line: $(\tilde{R}_{12}+\tilde{R}_{13})/2$; blue dotted line: $\tilde{R}_{23}$.}
\label{fig:R_ij_profiles}
\end{figure*}

The Favre-averaged Reynolds stress tensor, $\tilde{R}_{ij}$, appears as an unclosed term in the averaged transport equation of momentum given by equation~\eqref{eq:Favre_momentum_eqn}. Figures~\ref{fig:rho_R11_profiles} and \ref{fig:rho_R22_33_profiles} respectively show the profiles of Favre-averaged Reynolds normal stress components in the streamwise and transverse directions at different times. Immediately after first shock, there is generation of the Favre-averaged Reynolds normal stress in the mixing region. However, the Favre-averaged Reynolds normal stress component in the streamwise direction is much larger than those in the transverse directions at that instance. As time advances, the ratios of the component in the streamwise direction to those in the transverse directions decreases, but the Reynolds normal stress fields are still very anisotropic at the moment just before re-shock. The streamwise Favre-averaged Reynolds normal stress component peaks at the lighter fluid side because of smaller inertia to entrain the fluid from nonlinear convection. After re-shock, the Favre-averaged Reynolds normal stress fields become more isotropic but there is still more contribution to the turbulent kinetic energy from the streamwise Reynolds normal stress component until the end of simulation. The comparison of different Favre-averaged Reynolds stress components is shown in figure~\ref{fig:R_ij_profiles}. All Reynolds shear stress components should be statistically zero but figure~\ref{fig:R_ij_profiles} shows that the Reynolds shear stress components are not absolutely zero. This is due to some lack of full statistical convergence, but the values are all negligible compared to the Reynolds normal stress components.

\begin{figure*}[!ht]
\centering
\subfigure[$\ t=0.40\ \mathrm{ms}$]{%
\includegraphics[height = 0.33\textwidth]{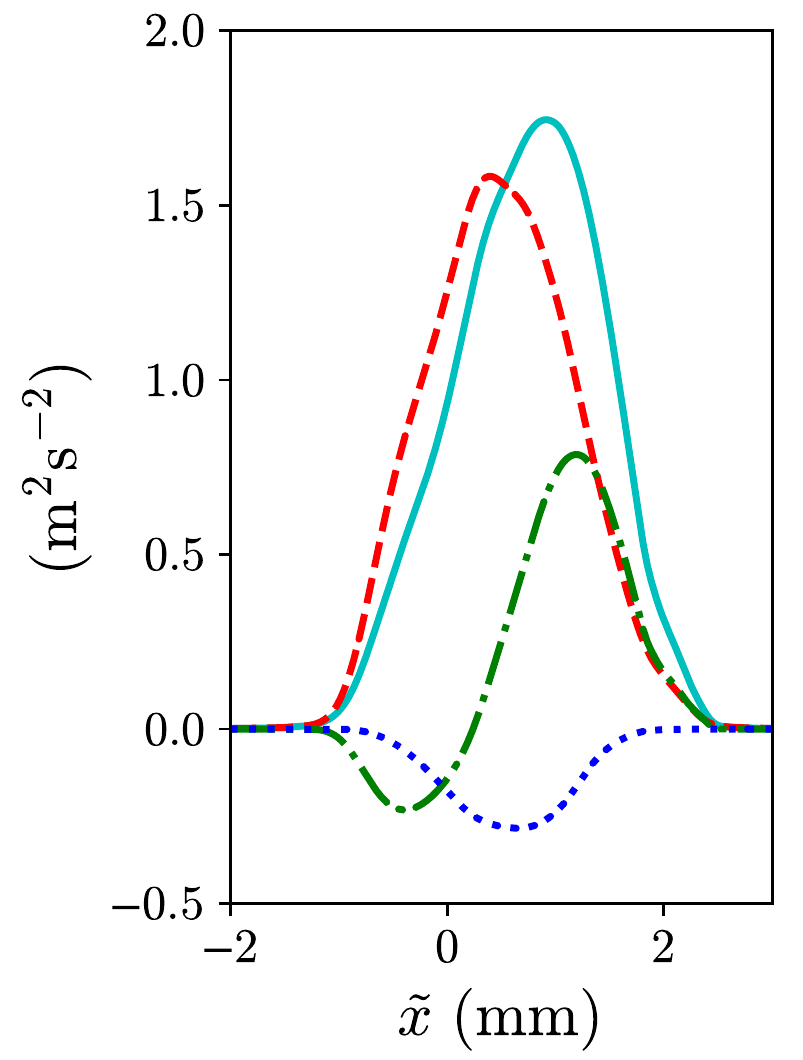}}
\subfigure[$\ t=1.10\ \mathrm{ms}$]{%
\includegraphics[height = 0.33\textwidth]{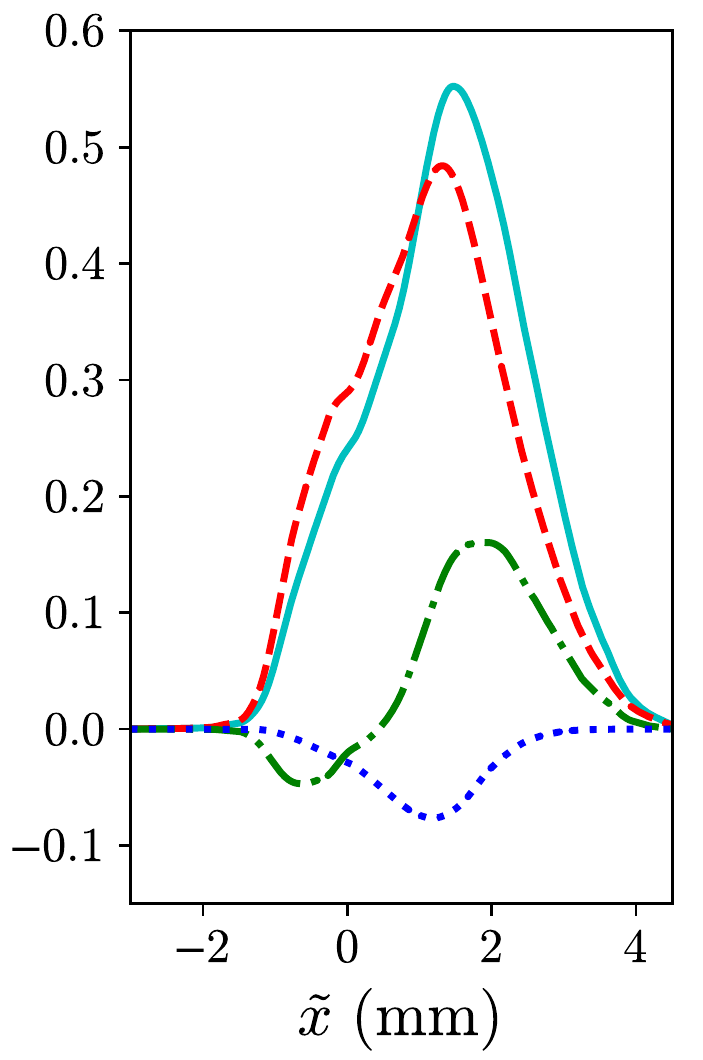}}
\subfigure[$\ t=1.20\ \mathrm{ms}$]{%
\includegraphics[height = 0.33\textwidth]{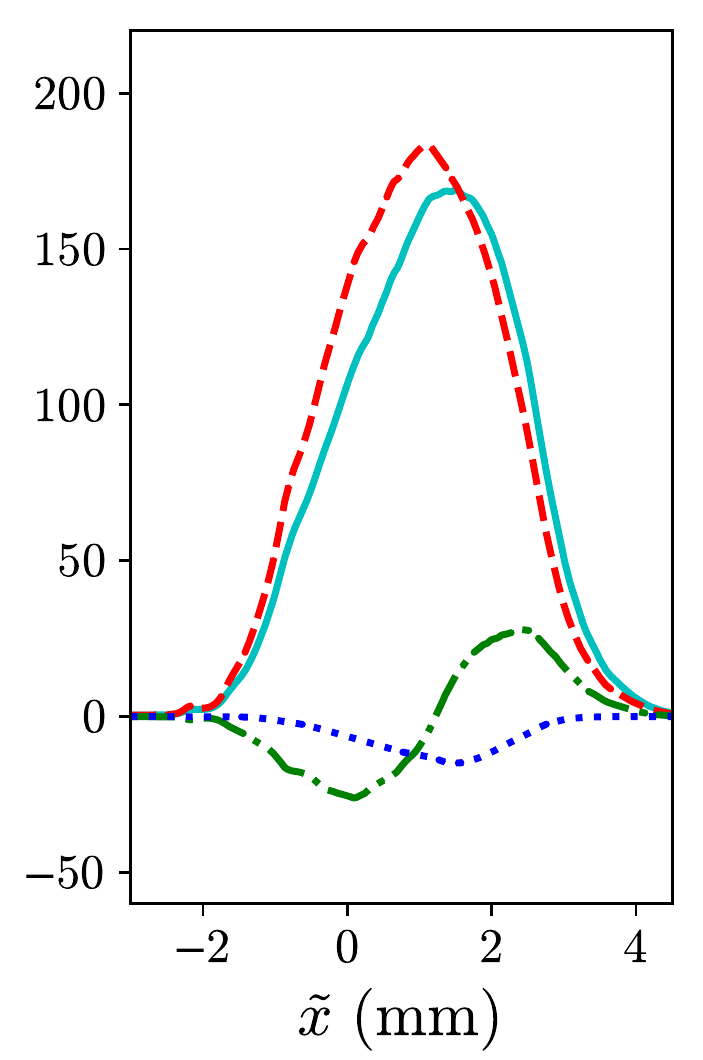}}
\subfigure[$\ t=1.75\ \mathrm{ms}$]{%
\includegraphics[height = 0.33\textwidth]{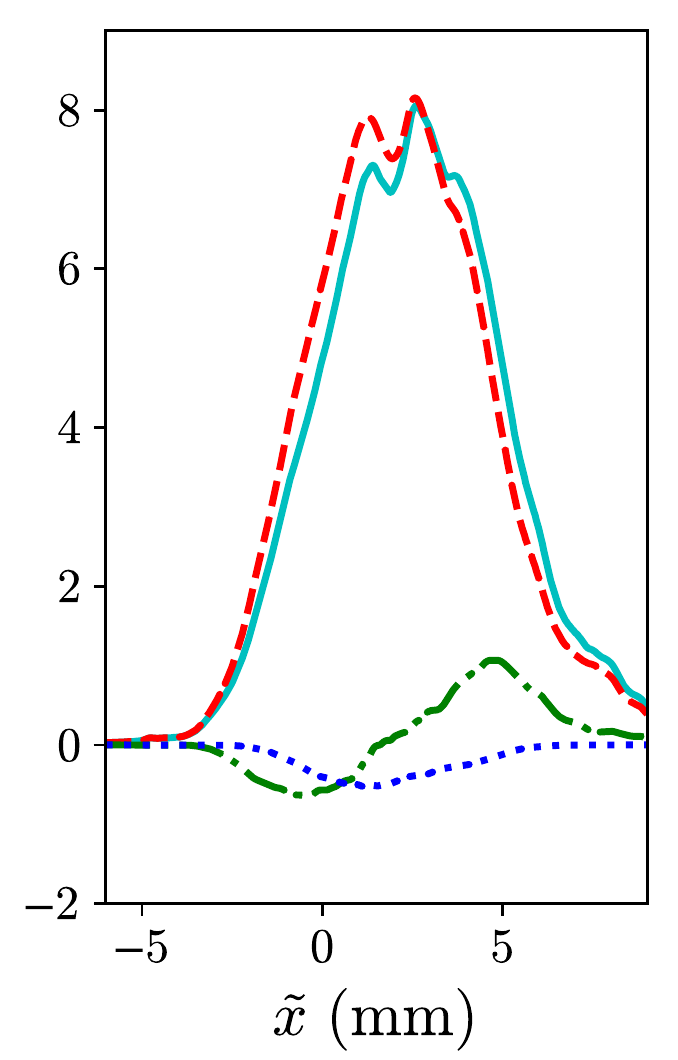}}
\caption{Decomposition of the Reynolds normal stress component in the streamwise direction multiplied by the mean density, $\bar{\rho} \tilde{R}_{11}$, at different times. Cyan solid line: $\tilde{R}_{11}$; red dashed line: $\overline{u^{\prime}u^{\prime}}$ [term (I)]; green dash-dotted line: $\overline{\rho^{\prime}u^{\prime}u^{\prime}} / \bar{\rho}$ [term (II)]; blue dotted line: $-a_1^2$ [term (III)].}
\label{fig:R11_decomposition}
\end{figure*}

\begin{figure*}[!ht]
\centering
\subfigure[$\ $Approximation (i) of $\tilde{R}_{11}$]{%
\includegraphics[width=0.4\textwidth]{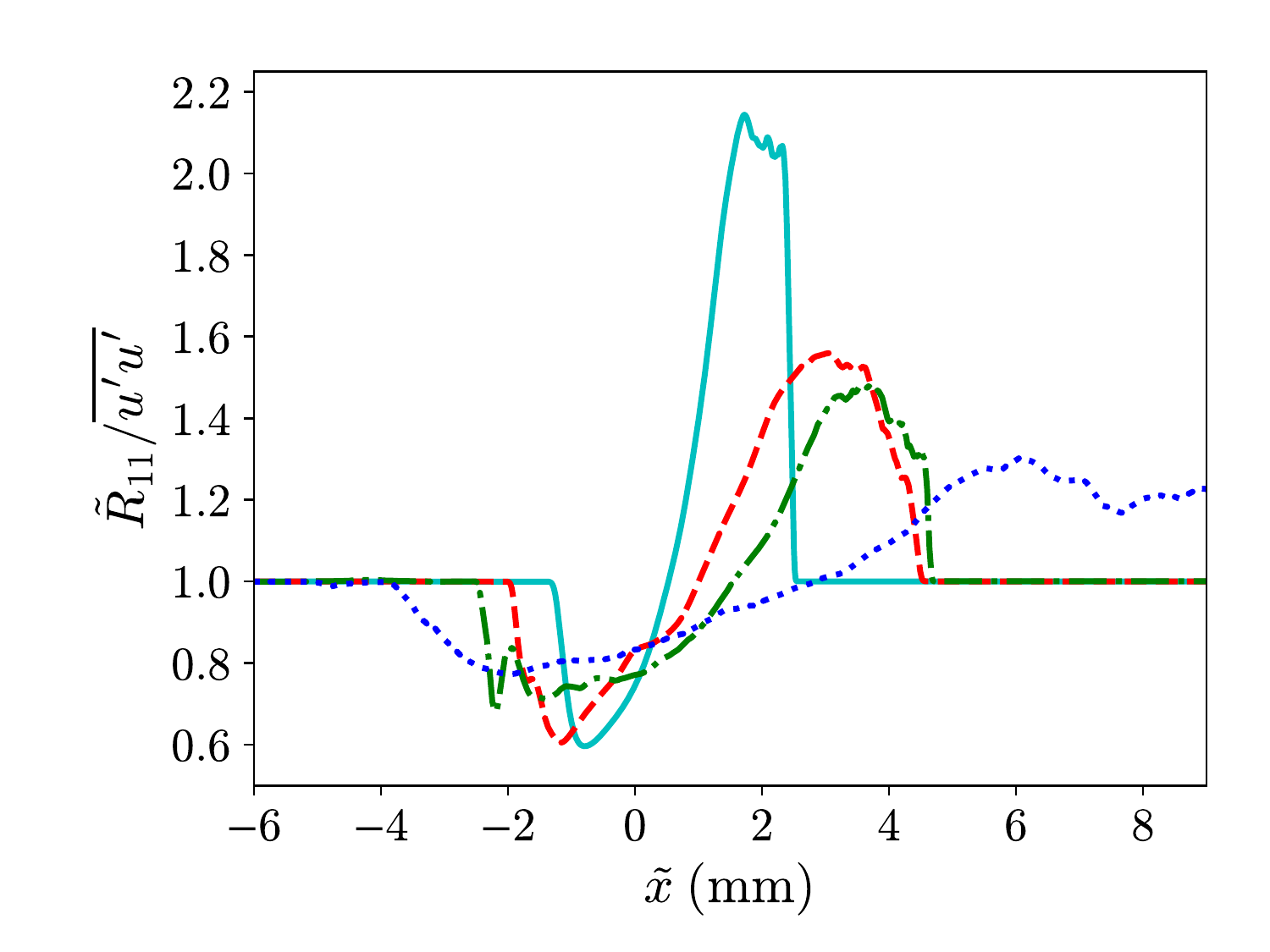}}
\subfigure[$\ $Approximation (ii) of $\tilde{R}_{11}$]{%
\includegraphics[width=0.4\textwidth]{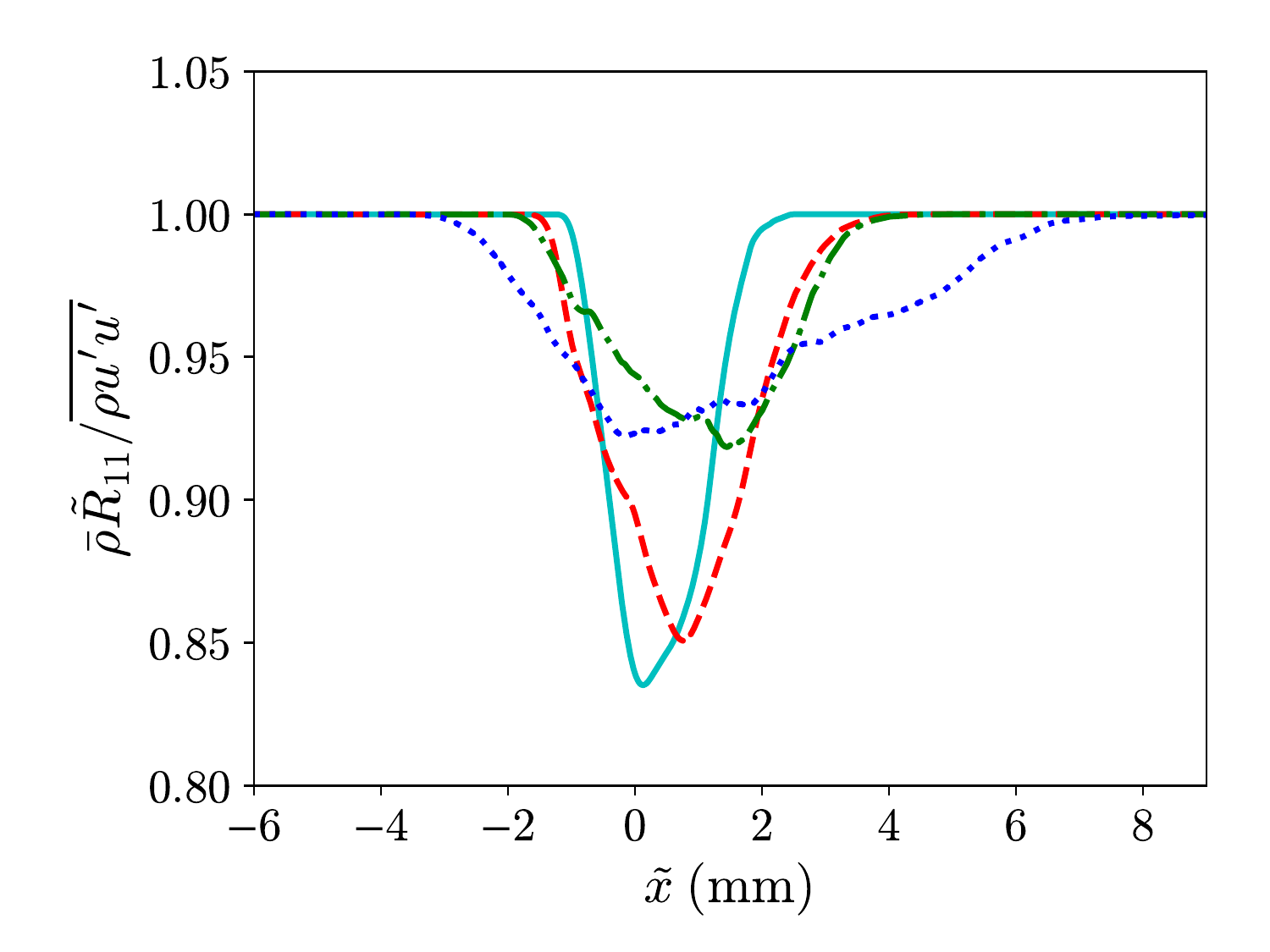}}
\caption{Ratios of the Reynolds normal stress component in the streamwise direction multiplied by the mean density, $\bar{\rho} \tilde{R}_{11}$, to different approximations at different times. Cyan solid line: $t=0.40\ \mathrm{ms}$; red dashed line: $t=1.10\ \mathrm{ms}$; green dash-dotted line: $t=1.20\ \mathrm{ms}$; blue dotted line: $t=1.75\ \mathrm{ms}$.}
\label{fig:R11_approximations}
\end{figure*}

The Favre-averaged Reynolds stress $\tilde{R}_{ij}$ can be decomposed as:
\begin{equation*}
    \tilde{R}_{ij} = \frac{\overline{\rho u_i^{\prime}u_j^{\prime}}}{\bar{\rho}} - a_i a_j
    = \underbrace{ \overline{u_i^{\prime}u_j^{\prime}} }_{ \text{term (I)} } + \underbrace{  \frac{\overline{\rho^{\prime}  u_i^{\prime}u_j^{\prime}}}{\bar{\rho}} }_{ \text{term (II)} } - \underbrace{ a_i a_j }_{ \text{term (III)} },
\end{equation*}
\noindent where term (I), $\overline{u_i^{\prime}u_j^{\prime}}$, is the definition of the Reynolds stress tensor for single-species incompressible flows.
This decomposition is commonly found in previous papers on RMI, such as~\cite{balakumar2012turbulent,shankar2014numerical,mohaghar2019transition}.
Figure~\ref{fig:R11_decomposition} compares the contributions of different terms to $\tilde{R}_{11}$ at different times. We can see from the plots that, although the profile of $\tilde{R}_{11}$ is very similar to $\overline{u^{\prime}u^{\prime}}$ [term (I)], the contributions of the other two terms, especially $\overline{\rho^{\prime}u^{\prime}u^{\prime}} / \bar{\rho}$ [term (II)], are not negligible.
Term (II) is around 20\% to 40\% of 
term (I) within the mixing region at late times after first shock and different times after re-shock. $-a_1^2$ [term (III)] is around one order of magnitude smaller than term (I). This is different from the observations in~\cite{balakumar2012turbulent,mohaghar2019transition}, where terms (II) and (III) are at least 100 and 1000 times smaller, respectively, than term (I).
Figure~\ref{fig:R11_approximations} shows the discrepancies between $\tilde{R}_{11}$ and the two different approximations: (i) $\overline{u^{\prime}u^{\prime}}$ and (ii) $\overline{\rho u^{\prime}u^{\prime}} / \bar{\rho}=\overline{u^{\prime}u^{\prime}} + \overline{\rho^{\prime}  u^{\prime}u^{\prime}} / \bar{\rho}$, through the ratios of $\tilde{R}_{11}$ to the approximations. It can be seen that $\tilde{R}_{11}$ cannot be well represented by $\overline{u^{\prime}u^{\prime}}$ alone, as the ratio can vary from 0.6 to 2.2. This is associated with the strong variable-density effects of the flow. If $\overline{\rho^{\prime}  u^{\prime}u^{\prime}} / \bar{\rho}$ is included to approximate $\tilde{R}_{11}$, there is a huge improvement in the approximation, as the ratio now only varies from 0.85 to 1. Although this suggests that $a_1$ has a small contribution to the decomposition of $\tilde{R}_{11}$, this does not mean that $a_1$ has an insignificant effect on the time evolution of $\tilde{R}_{11}$. Thus, it is shown in the next few sections that $a_1$ plays an important role in the transport equation of $\tilde{R}_{11}$ through the component of the production term, $2a_1 \bar{p}_{,1}$.
A grid sensitivity analysis of the spatial profiles of different contributions to $\tilde{R}_{11}$ is also provided in the Supplemental Material~\cite{supple2022wong}. The profiles only show small grid sensitivities between the grid D and the grid E, and the discussion above is not much affected by the grid sensitivities.

\begin{figure*}[!ht]
\centering
\subfigure[$\ $Before re-shock]{%
\includegraphics[width=0.4\textwidth]{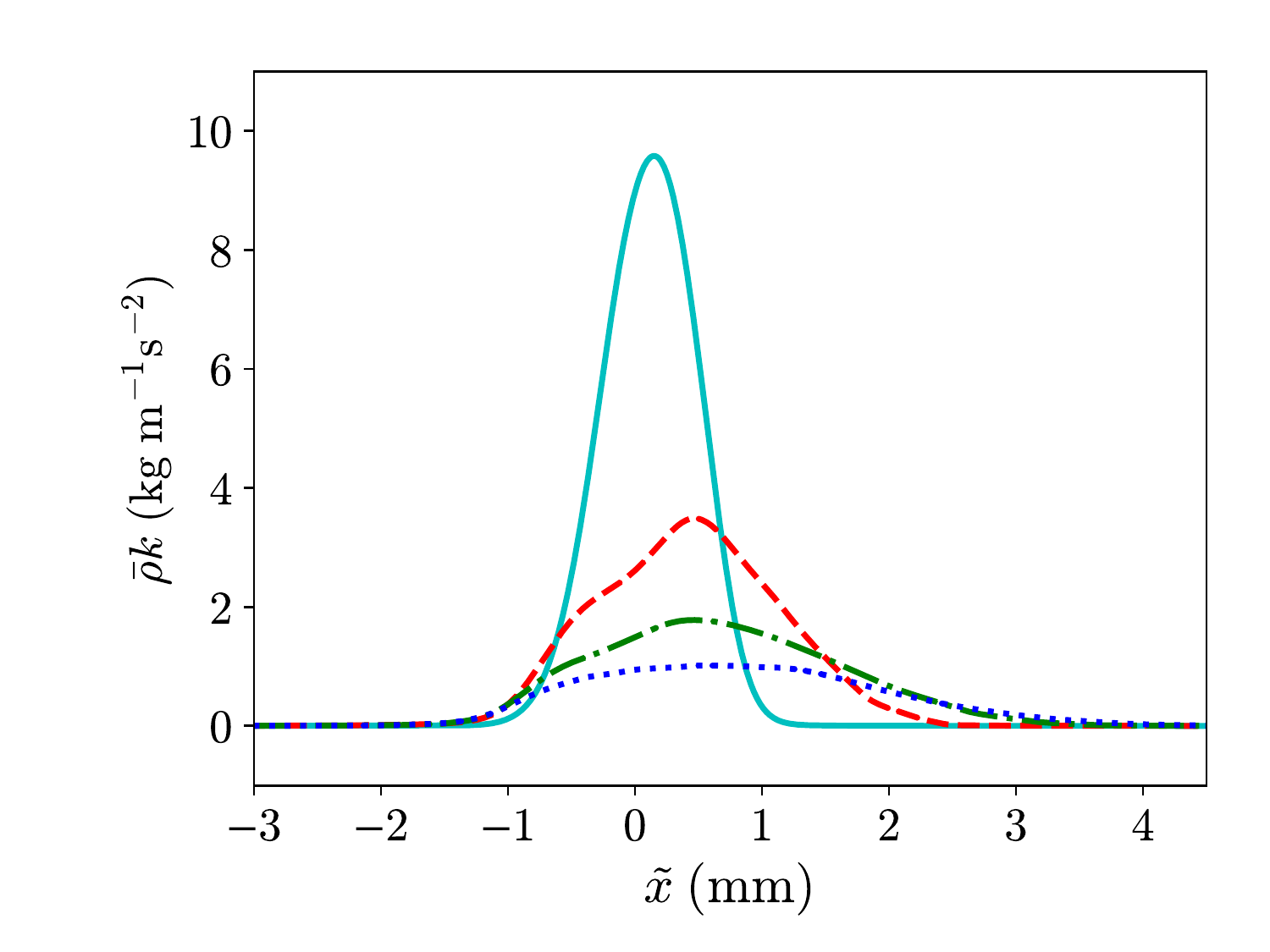}}
\subfigure[$\ $After re-shock]{%
\includegraphics[width=0.4\textwidth]{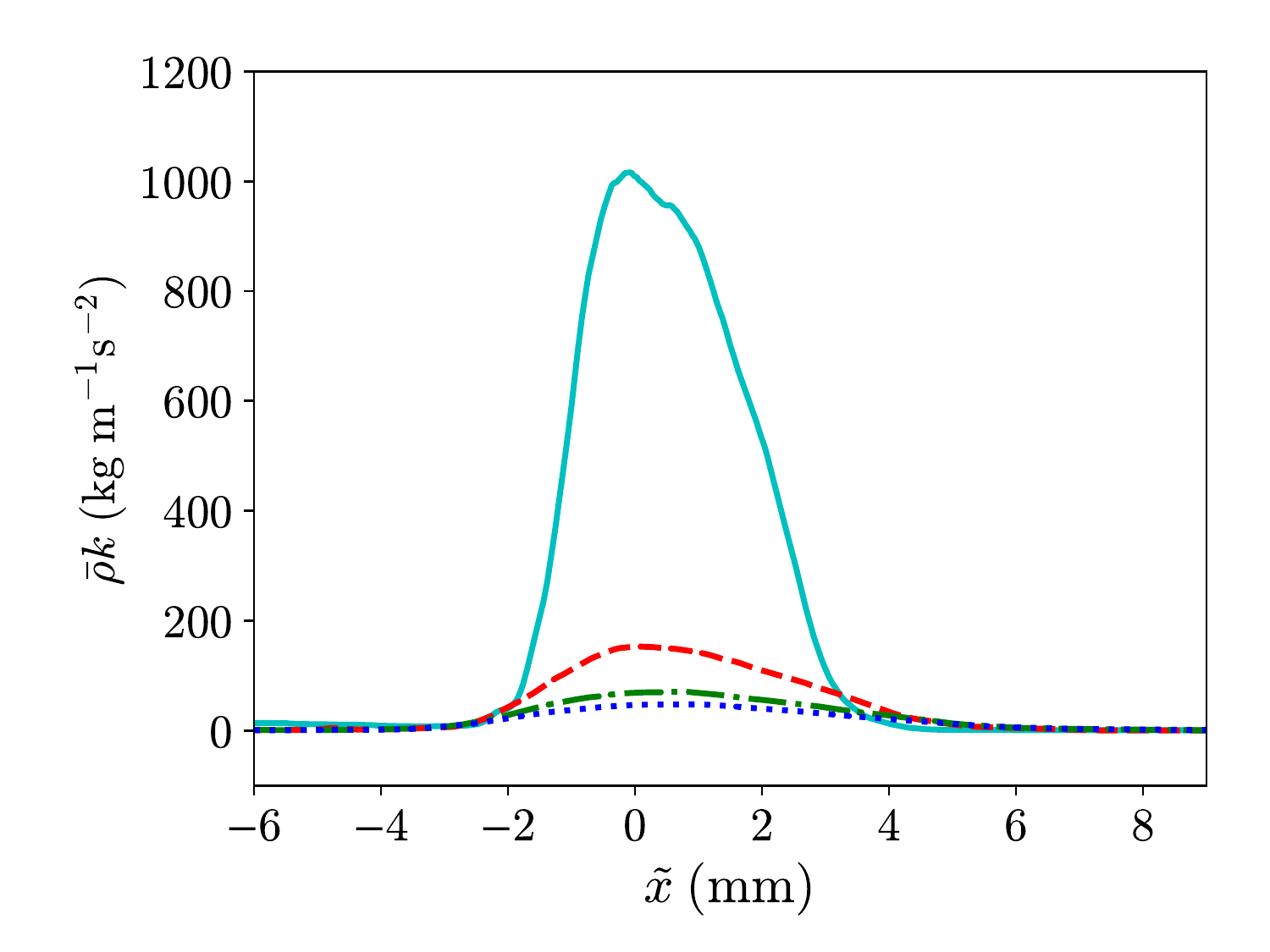}}
\caption{Profiles of the turbulent kinetic energy, $\bar{\rho} k$, at different times. Cyan solid line in (a): $t=0.05\ \mathrm{ms}$; red dashed line in (a): $t=0.40\ \mathrm{ms}$; green dash-dotted line in (a): $t=0.75\ \mathrm{ms}$; blue dotted line in (a): $t=1.10\ \mathrm{ms}$. Cyan solid line in (b): $t=1.20\ \mathrm{ms}$; red dashed line in (b): $t=1.40\ \mathrm{ms}$; green dash-dotted line in (b): $t=1.60\ \mathrm{ms}$; blue dotted line in (b): $t=1.75\ \mathrm{ms}$.}
\label{fig:rho_k_profiles}
\end{figure*}

Figure~\ref{fig:rho_k_profiles} shows the profiles of the turbulent kinetic energy at different times. Before re-shock, the profiles look similar to those of $\bar{\rho} \tilde{R}_{11}$ as most of the turbulent kinetic energy is contributed by the Reynolds normal stress component in the streamwise direction. At re-shock, the turbulent kinetic energy is amplified by three orders of magnitude. However, it decays rapidly due to large viscous dissipation over time.

\section{\label{sec:budgets_before_reshock} Budgets of the second-moments before re-shock}

In this section, the budgets of second-moments: $\bar{\rho} a_1$, $\bar{\rho} b$, and $\bar{\rho} \tilde{R}_{11}$, together with $\bar{\rho} k$ across the mixing layer before re-shock are studied.
All budgets are computed with the results from the highest resolution (grid E) simulation, for which the flow fields are well-resolved.
A grid sensitivity analysis of the budgets is also given in the Supplemental Material~\cite{supple2022wong}.
The budgets are studied in the $\tilde{x}$ coordinate system, equivalent to studying the budgets in the moving reference frame of the mixing layer. The convective terms in all of the transport equations of second-moments for 1D mean flow have the common form of $[ \bar{\rho} \tilde{u} ( \cdot ) ]_{,1}$, where $( \cdot )$ represents any of the second-moments ($a_1$, $b$, or $R_{ij}$). Using equation~\eqref{eq:means_relation}, the convective terms can be rewritten as:
\begin{equation}
    \frac{\partial \bar{\rho} \tilde{u} \left( \cdot \right) }{\partial x} = \underbrace{ \frac{\partial \bar{\rho} \bar{u} \left( \cdot \right) }{\partial x} }_{ \text{term (I)} } + \underbrace{ \frac{\partial \bar{\rho} a_1 \left( \cdot \right) }{\partial x} }_{ \text{term (II)} },
\end{equation}
\noindent where term (I) is the convection due to mean velocity and term (II) is the convection due to velocity associated with turbulent mass flux. In the moving reference frame of the mixing layer, it is observed that $\bar{u}$ is quite uniformly close to zero compared to $a_1$.
Hence term (I) can be ignored.
The convective term in this section is assumed to be fully represented by $[ \bar{\rho} a_1 ( \cdot ) ]_{,1}$.

\subsection{Turbulent mass flux}

\begin{figure*}[!ht]
\centering
\subfigure[$\ t=0.40\ \mathrm{ms}$]{%
\includegraphics[width=0.4\textwidth]{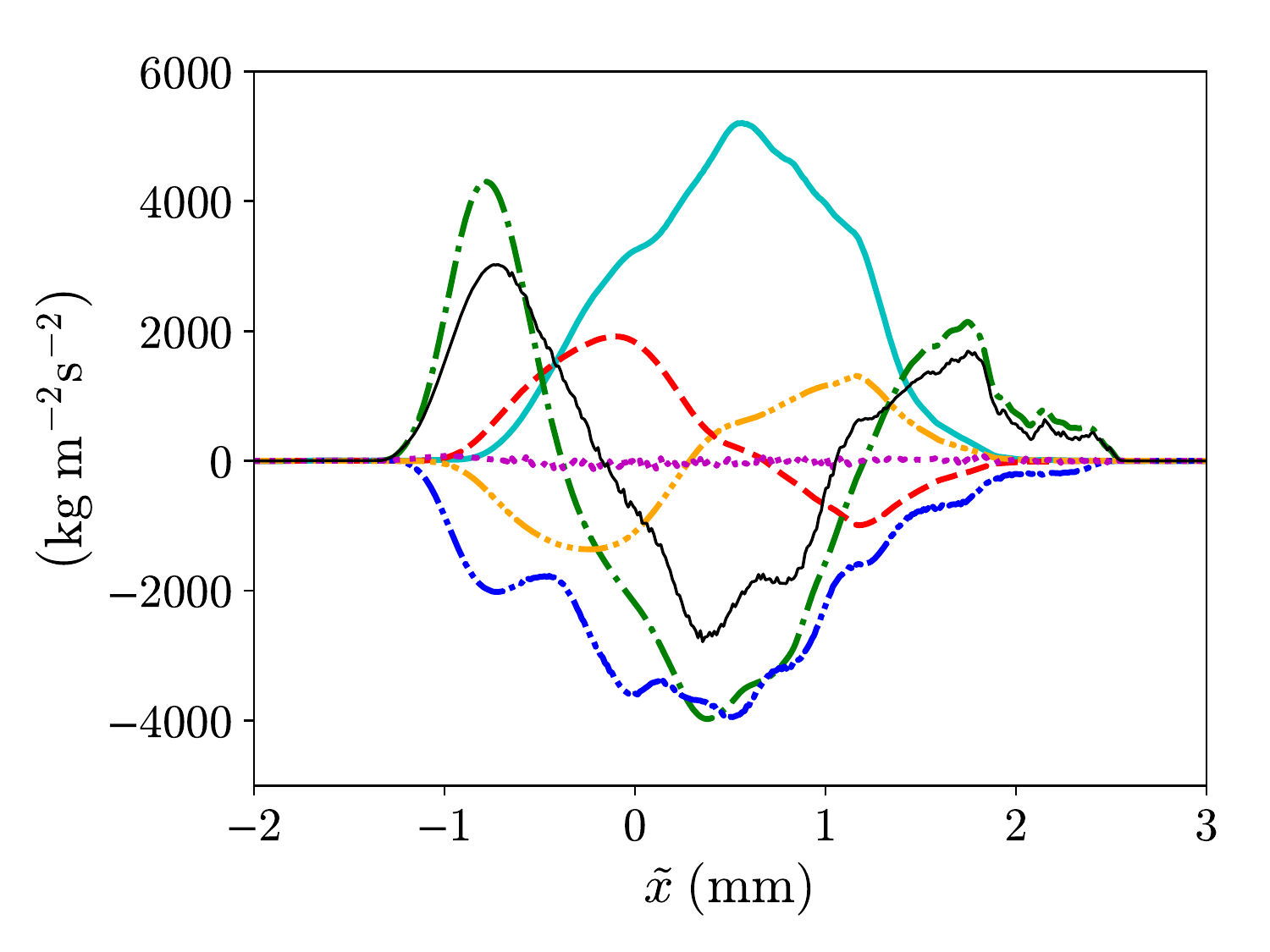}\label{fig:rho_a1_budget_t_0_40}}
\subfigure[$\ t=1.10\ \mathrm{ms}$]{%
\includegraphics[width=0.4\textwidth]{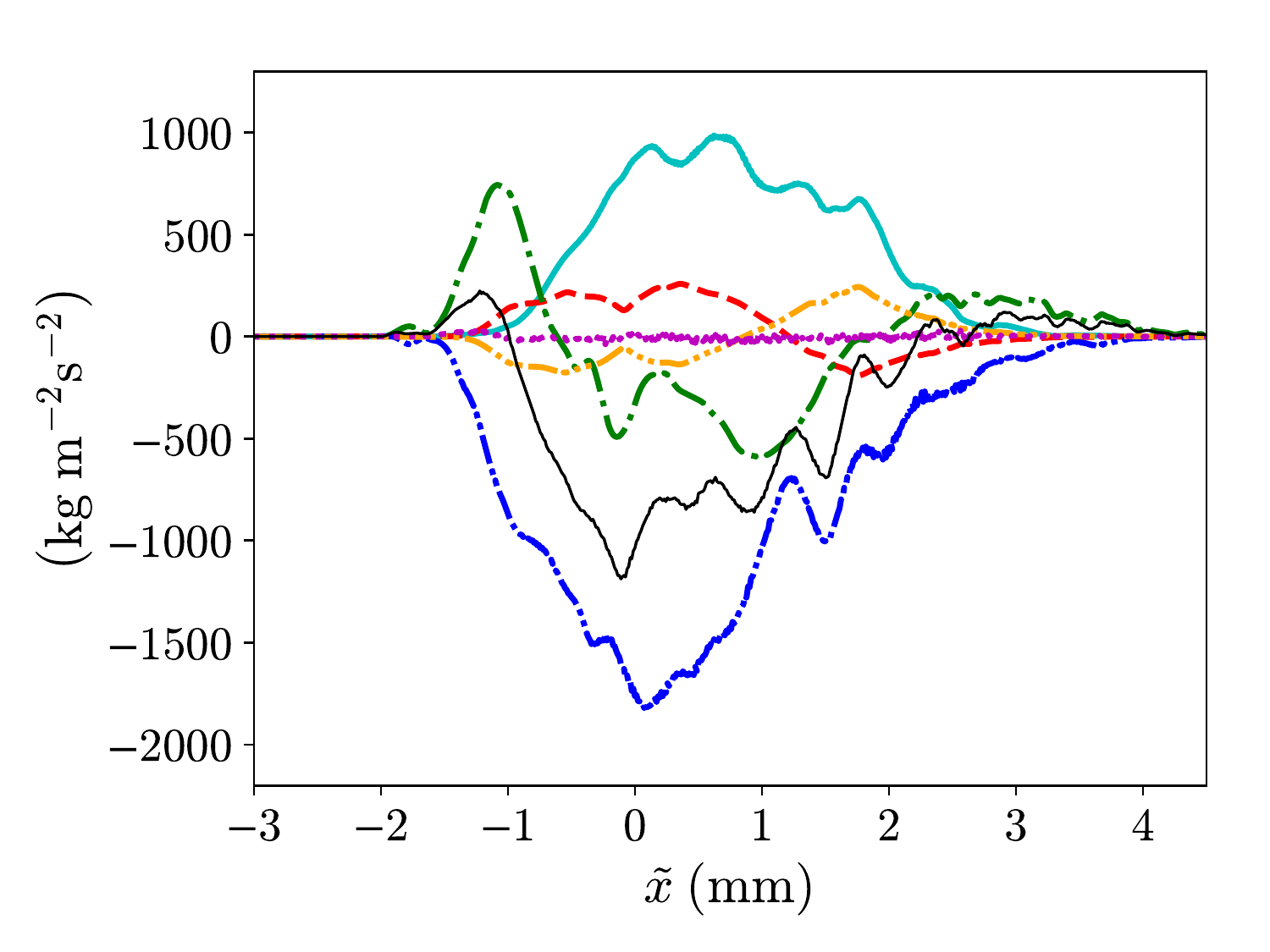}\label{fig:rho_a1_budget_before_reshock}}
\caption{Budgets of the turbulent mass flux component in the streamwise direction, $\bar{\rho} a_1$, given by equation~\eqref{eq:a1_transport_eqn_1D}, at different times before re-shock. Cyan solid line: production [term (III)]; red dashed line: redistribution [term (IV)]; green dash-dotted line: turbulent transport [term (V)]; blue dash-dot-dotted line: destruction [term (VI)]; orange dash-triple-dotted line: negative of convection due to streamwise velocity associated with turbulent mass flux; magenta dotted line: residue; thin black solid line: summation of all terms (rate of change in the moving frame).}
\label{fig:rho_a1_budget}
\end{figure*}

Figure~\ref{fig:rho_a1_budget} shows the spatial profiles of different terms in the RHS of the transport equation of the streamwise component of turbulent mass flux, $\bar{\rho} {a_1}$, given by equation~\eqref{eq:a1_transport_eqn_1D} at different times before re-shock across the mixing layer. The negative of the convective term of the same equation due to $a_1$ is also shown in the figure. The rate of change term on the LHS of the transport equation is computed by restarting the simulation at different checkpoints. In each plot, the magenta dotted line shows the profile of the residue, which is defined as the subtraction of the net RHS term from the net LHS term in the simulation frame. Therefore, the residue represents the numerical effect or the SGS effect on the rate of change of the conserved variable, i.e. $\bar{\rho} a_1$ here. From both figures~\ref{fig:rho_a1_budget_t_0_40} and \ref{fig:rho_a1_budget_before_reshock}, it can be seen that the residue is virtually zero across the mixing layer at different times before re-shock. Note that the thin black solid line in each plot is the sum of all of the RHS terms including the residue and the negative of the convection term due to $a_1$, thus it represents the rate of change of $\bar{\rho} {a_1}$ in the moving frame of the mixing layer.

From figure~\ref{fig:rho_a1_budget}, we can see that production [term (III)], destruction [term (VI)], and turbulent transport [term (V)] terms play important roles in the budget equation at the chosen times before re-shock: $t = 0.40\ \mathrm{ms}$ and $t = 1.10\ \mathrm{ms}$. At the two chosen times, the instability is in the nonlinear growth regime. The production and destruction terms are not symmetric as they are skewed to the lighter fluid side with peaks also slightly positioned at that side. In the interior part of the the mixing layer, the production, destruction, and turbulent transport terms are the dominant terms. The production term is strictly positive in the mixing region and peaks around the middle part of the mixing region. Nonetheless, both destruction and turbulent transport terms are negative in the interior part of mixing layer to offset the effect from production. Overall, the combined effect of the destruction and turbulent transport terms is larger than that of the production and hence the peak of turbulent mass flux reduces over time.  At the edges of the mixing layer, all RHS terms are small except the turbulent transport term, which is positive and responsible for the spreading of the turbulent mass flux. The magnitudes of redistribution [term (IV)] and convective terms are smaller than those of other terms but still have significant effects at different times before re-shock. The two terms have similar magnitudes but opposite signs. The convective term decreases the turbulent mass flux on the heavier fluid side and increases that on the lighter fluid side. The redistribution term has the opposite effect of bringing the turbulent mass flux from the light fluid side back to the heavy fluid side. Opposite sign but close magnitude for the two corresponding transport terms in the budgets of $\bar{\rho} {a_1}$ is also noticed in the RTI turbulence by~\citet{livescu2009rti}.

\begin{figure*}[!ht]
\centering
\subfigure[$\ t=0.40\ \mathrm{ms}$]{%
\includegraphics[width=0.4\textwidth]{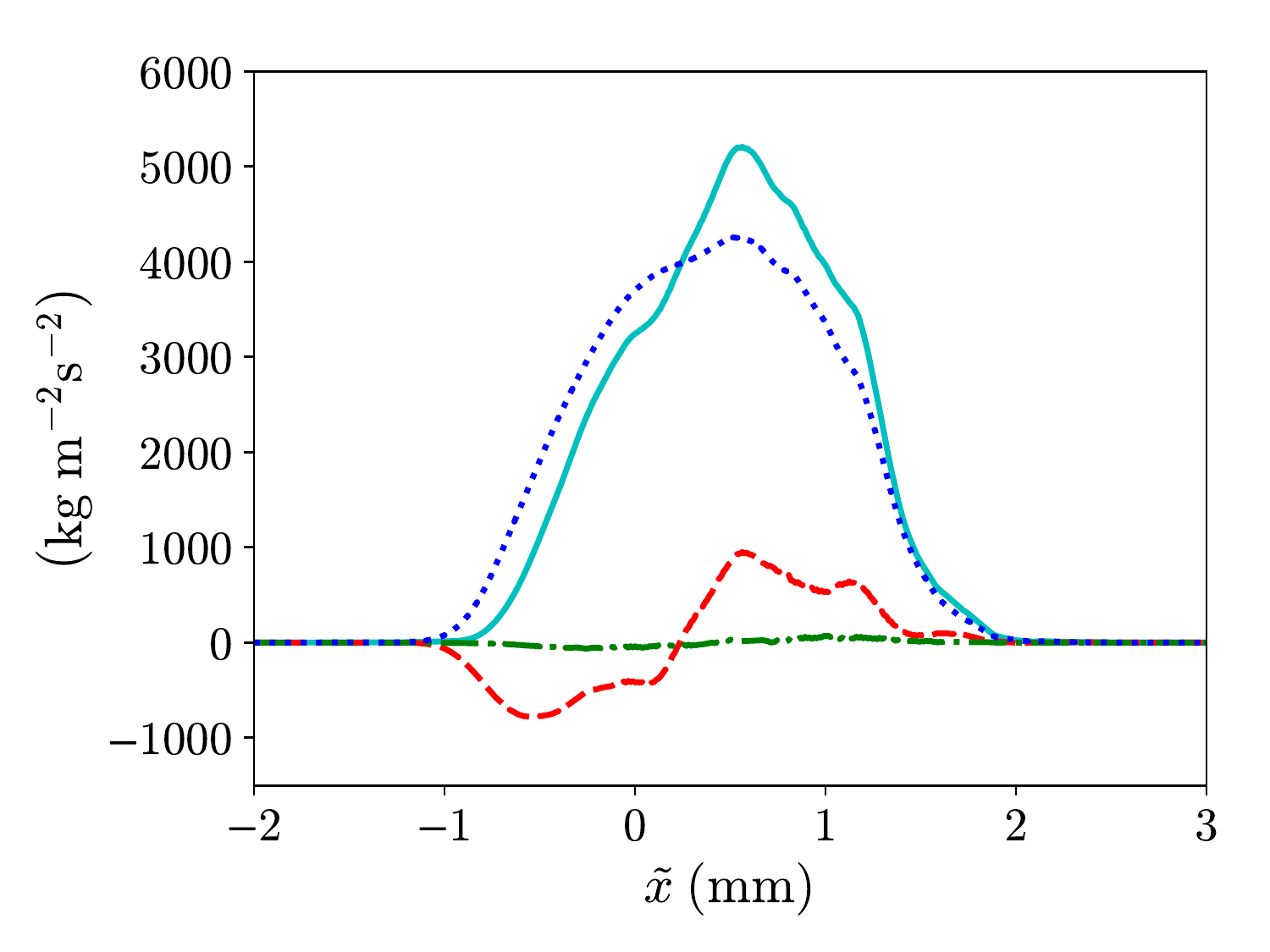}\label{fig:rho_a1_budget_production_terms_t_0_40}}
\subfigure[$\ t=1.10\ \mathrm{ms}$]{%
\includegraphics[width=0.4\textwidth]{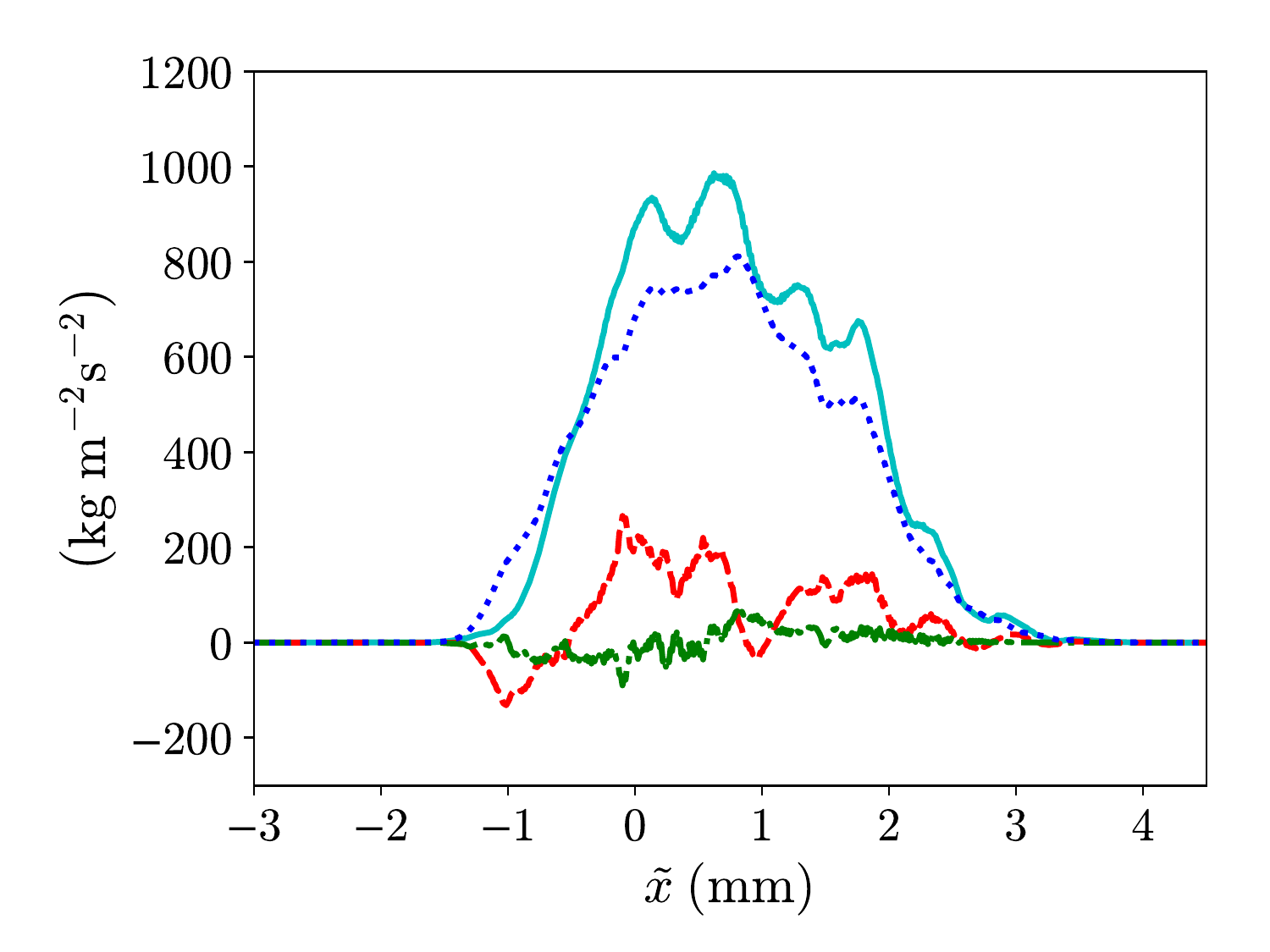}\label{fig:rho_a1_budget_production_terms_before_reshock}}
\caption{Compositions of the production term [term (III)] in the transport equation for the turbulent mass flux component in the streamwise direction, $\bar{\rho} a_1$, at different times before re-shock. Cyan solid line: overall production; red dashed line: $b \bar{p}_{,1}$; green dash-dotted line: $-b \bar{\tau}_{11_{,1}}$; blue dotted line: $-\tilde{R}_{11} \bar{\rho}_{,1}$.}
\label{fig:rho_a1_budget_production_terms}
\end{figure*}

\begin{figure*}[!ht]
\centering
\subfigure[$\ t=0.40\ \mathrm{ms}$]{%
\includegraphics[width=0.4\textwidth]{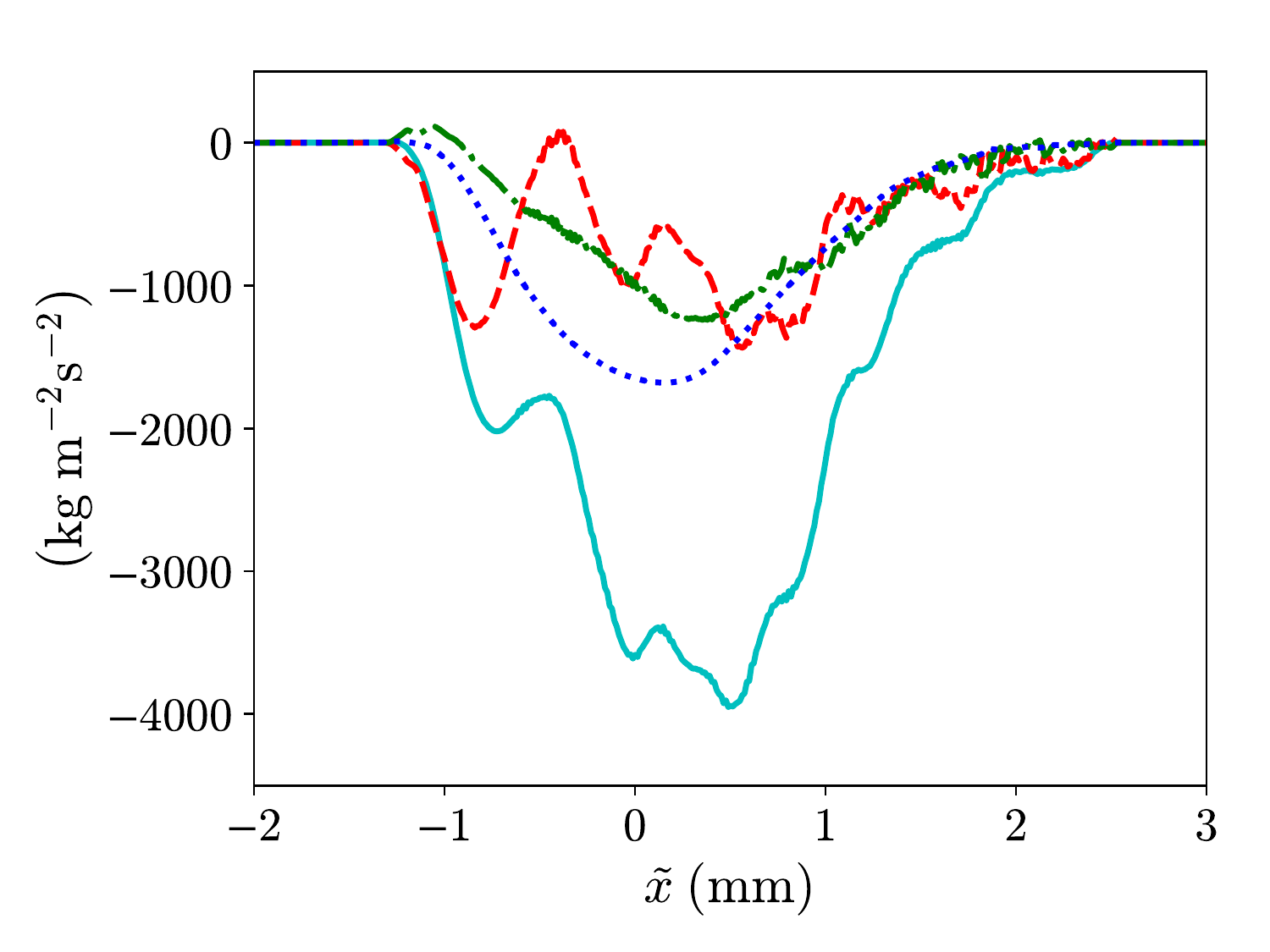}\label{fig:rho_a1_budget_destruction_terms_t_0_40}}
\subfigure[$\ t=1.10\ \mathrm{ms}$]{%
\includegraphics[width=0.4\textwidth]{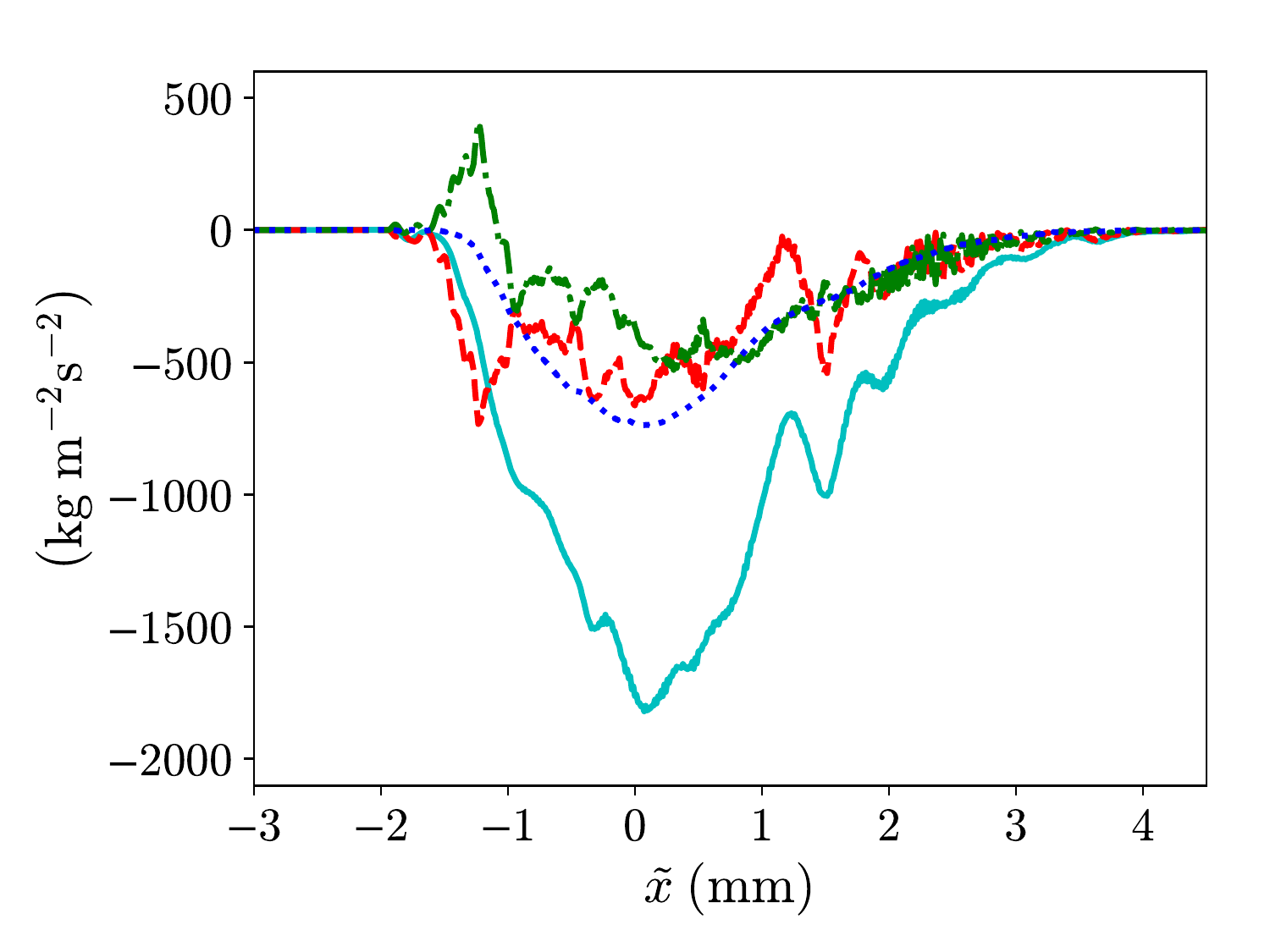}\label{fig:rho_a1_budget_destruction_terms_before_reshock}}
\caption{Compositions of the destruction term [term (VI)] in the transport equation for the turbulent mass flux component in the streamwise direction, $\bar{\rho} a_1$, at different times before re-shock. Cyan solid line: overall destruction; red dashed line: $\bar{\rho} \overline{ ( 1/\rho )^{\prime} p^{\prime}_{,1} }$; green dash-dotted line: $-\bar{\rho} \overline{ ( 1/\rho )^{\prime} \tau^{\prime}_{1i_{,i}} }$; blue dotted line: $\bar{\rho} \varepsilon_{a_1}$.}
\label{fig:rho_a1_budget_destruction_terms}
\end{figure*}

The composition of the production term [term (III)] is shown in figure~\ref{fig:rho_a1_budget_production_terms}. It can be seen that at the chosen times before re-shock, the production term is mainly contributed from the component, $-\tilde{R}_{11} \bar{\rho}_{,1}$, which is observed to be strictly positive. Another constituent, $b \bar{p}_{,1}$, has a smaller contribution to the overall term and $-b \bar{\tau}_{11_{,1}}$ is negligible. $b \bar{p}_{,1}$ transfers the turbulent mass flux from the heavier fluid side to the lighter fluid side, although this effect is hidden in the overall production term. As for the destruction term [term (VI)], figure~\ref{fig:rho_a1_budget_destruction_terms} shows that all three constituents ($\bar{\rho} \overline{ ( 1/\rho )^{\prime} p^{\prime}_{,1} }$, $-\bar{\rho} \overline{ ( 1/\rho )^{\prime} \tau^{\prime}_{1i_{,i}} }$, and $\bar{\rho} \varepsilon_{a_1}$) have similar magnitudes and are generally negative at the two times before re-shock.

\subsection{Density-specific-volume covariance}

\begin{figure*}[!ht]
\centering
\subfigure[$\ t=0.40\ \mathrm{ms}$]{%
\includegraphics[width=0.4\textwidth]{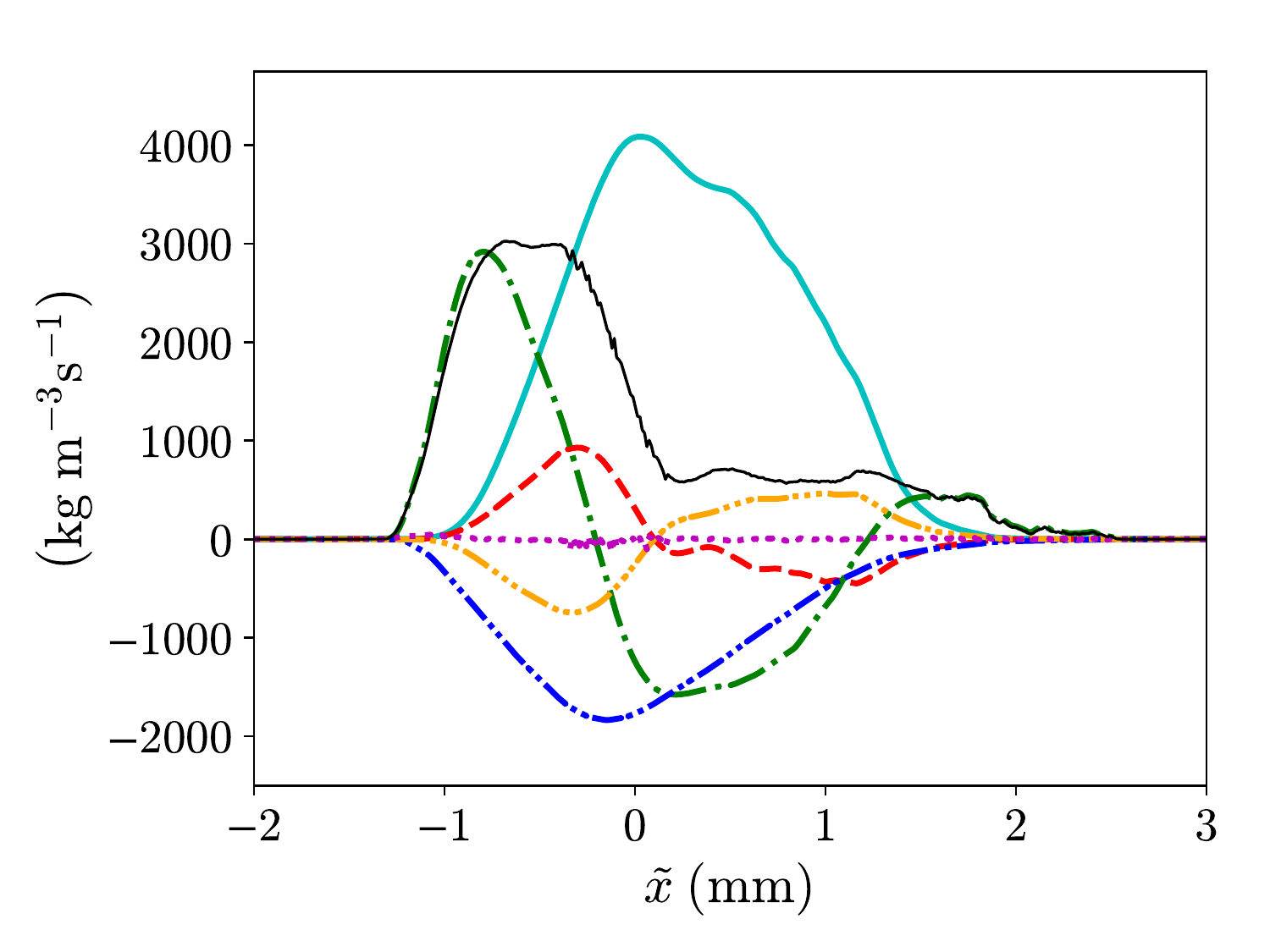}\label{fig:rho_b_budget_t_0_40}}
\subfigure[$\ t=1.10\ \mathrm{ms}$]{%
\includegraphics[width=0.4\textwidth]{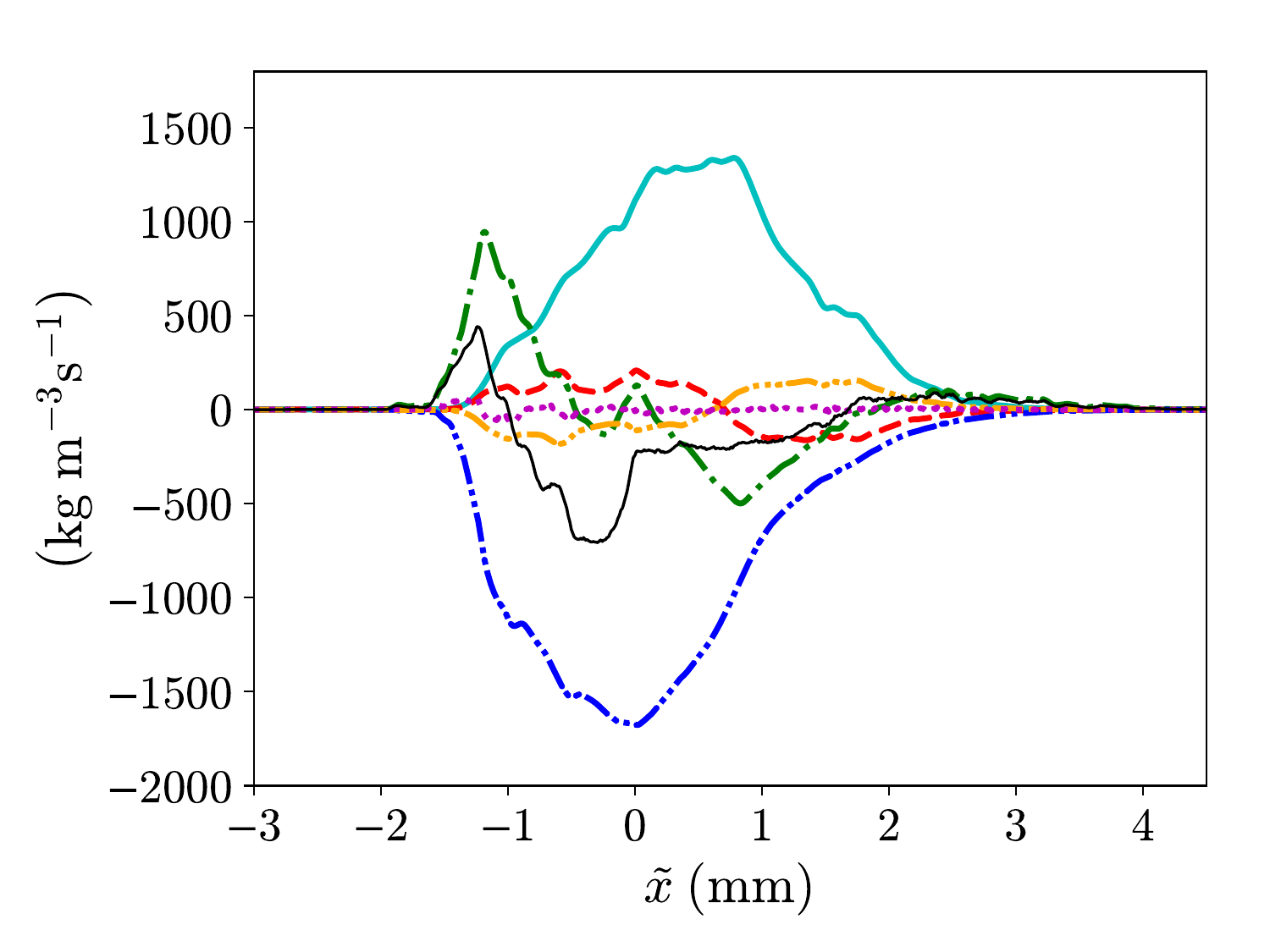}\label{fig:rho_b_budget_before_reshock}}
\caption{Budgets of the density-specific-volume covariance multiplied by the mean density, $\bar{\rho} b$, given by equation~\eqref{eq:b_transport_eqn_1D}, at different times before re-shock. Cyan solid line: production [term (III)]; red dashed line: redistribution [term (IV)]; green dash-dotted line: turbulent transport [term (V)]; blue dash-dot-dotted line: destruction [term (VI)]; orange dash-triple-dotted line: negative of convection due to streamwise velocity associated with turbulent mass flux; magenta dotted line: residue; thin black solid line: summation of all terms (rate of change in the moving frame).}
\label{fig:rho_b_budget}
\end{figure*}

Figure~\ref{fig:rho_b_budget} shows the spatial profiles of different budget terms that appear in the transport equation of $\bar{\rho} b$ given by equation~\eqref{eq:b_transport_eqn_1D} before re-shock. Similar to the plots for budgets of the turbulent mass flux, the magenta dotted line represents the residue, which is the difference between the net LHS and net RHS terms. As seen in figures~\ref{fig:rho_b_budget_t_0_40} and \ref{fig:rho_b_budget_before_reshock}, the residue is basically zero. This means that there is a negligible numerical effect due to insufficient spatial grid spacing on the time evolution of $\bar{\rho} b$ before re-shock. Before re-shock, the production [term (III)], turbulent transport [term (V)], and destruction [term (VI)] terms are dominant, but the redistribution [term (IV)] and convective terms cannot be neglected in the transport equation of $\bar{\rho} b$ either. Similar to the budgets of $\bar{\rho} {a_1}$, both production and destruction terms are asymmetric and skewed to the light fluid side. Although there is a positive effect in the interior part of the mixing layer from the production term to increase $\bar{\rho} b$, the effect is offset by both turbulent transport and destruction terms. The net rate of change of $\bar{\rho} b$ around the peak of $b$ is small so the peak of $b$ (similarly for $\bar{\rho} b$) remains relatively constant in time compared to peaks of other second-moments, which is shown earlier. At the edges of the mixing layer, most terms are small except the turbulent transport term, which is positive, which leads $\bar{\rho} b$ and $b$ to spread. Both the redistribution term and the convection term due to $a_1$ redistribute $\bar{\rho} b$ across the layer but they have exactly opposite effects (the redistribution term brings $\bar{\rho}$ from the lighter fluid side to the heavy fluid side and vice versa the convective term).
They also have similar shapes and thus roughly cancel effects from each other. A similar cancellation is also shown earlier for the corresponding terms in the budgets of $\bar{\rho} {a_1}$.

\subsection{Favre-averaged Reynolds stress and turbulent kinetic energy}

\begin{figure*}[!ht]
\centering
\subfigure[$\ t=0.40\ \mathrm{ms}$]{%
\includegraphics[width=0.4\textwidth]{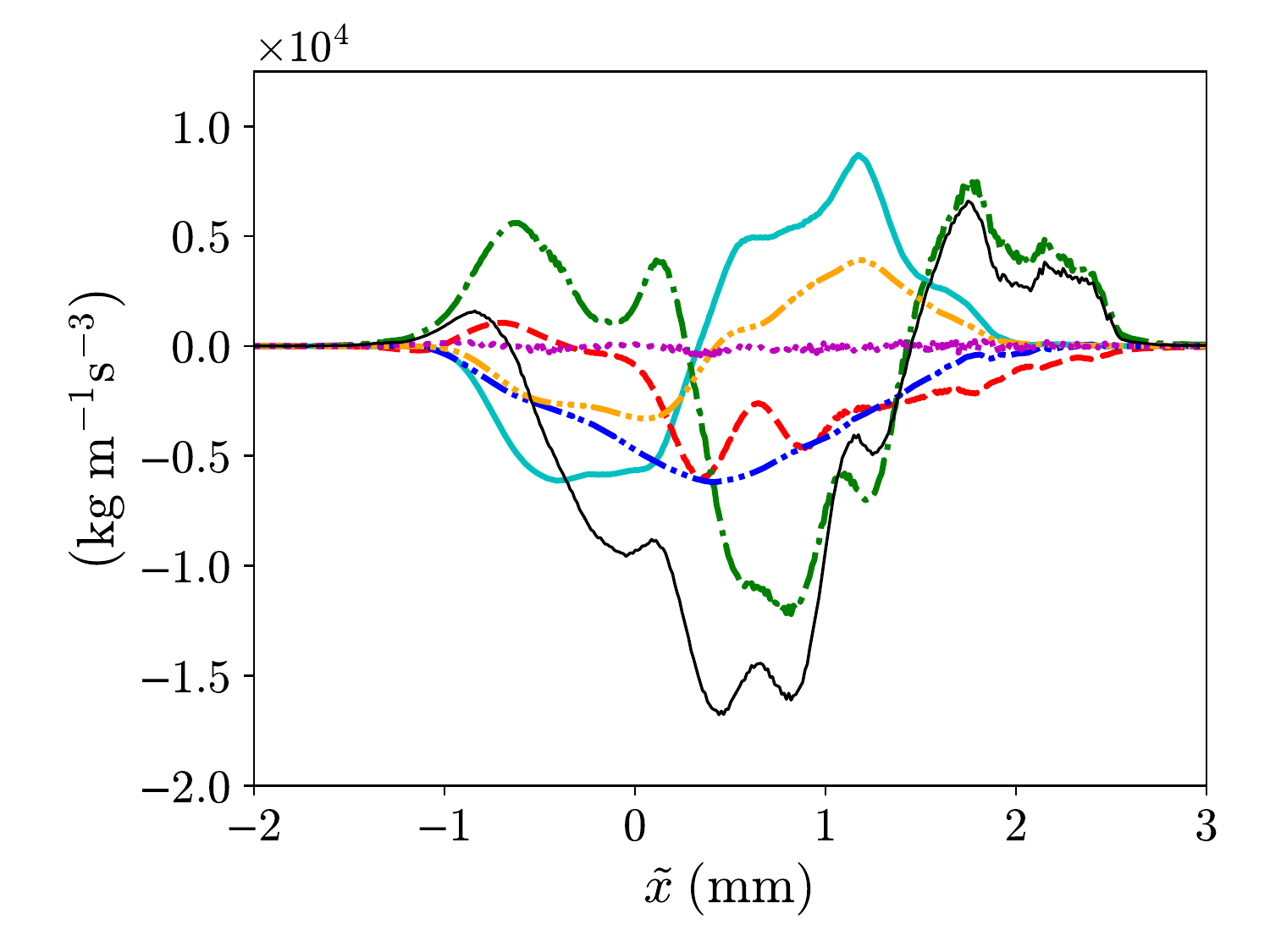}}
\subfigure[$\ t=1.10\ \mathrm{ms}$]{%
\includegraphics[width=0.4\textwidth]{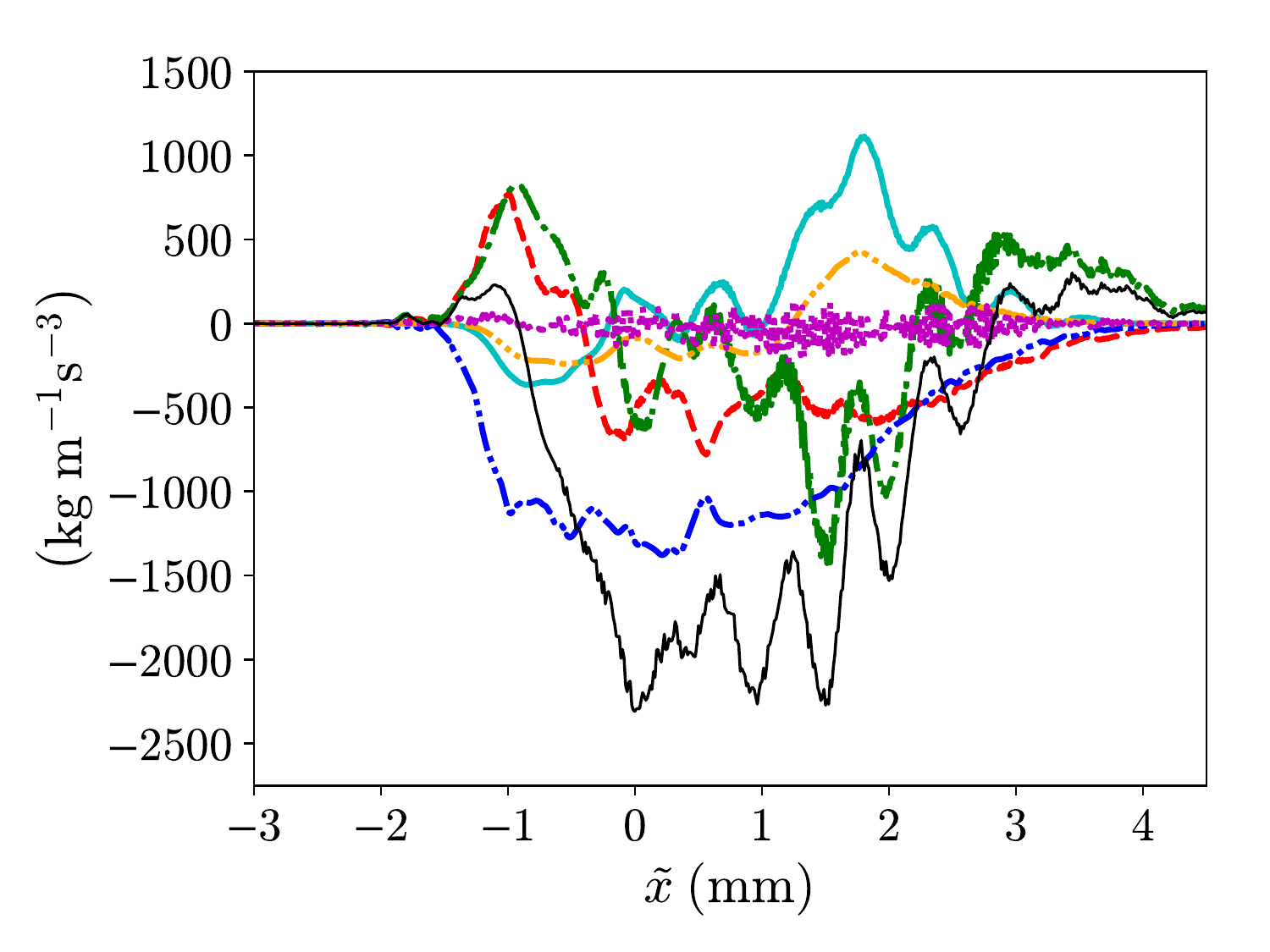}}
\caption{Budgets of the Reynolds normal stress component in the streamwise direction multiplied by the mean density, $\bar{\rho} \tilde{R}_{11}$, given by equation~\eqref{eq:R11_transport_eqn_1D}, at different times before re-shock. Cyan solid line: production [term (III)]; red dashed line: press-strain redistribution [term (V)]; green dash-dotted line: turbulent transport [term (IV)]; blue dash-dot-dotted line: dissipation [term (VI)]; orange dash-triple-dotted line: negative of convection due to streamwise velocity associated with turbulent mass flux; magenta dotted line: residue; thin black solid line: summation of all terms (rate of change in the moving frame).}
\label{fig:rho_R11_budget}
\end{figure*}

In figure~\ref{fig:rho_R11_budget}, the spatial profiles of different terms in the transport equation of $\bar{\rho} \tilde{R}_{11}$ given by equation~\eqref{eq:R11_transport_eqn_1D} at different times before re-shock are compared. Similar to the budgets of other second-moments, the residue due to spatial discretization is negligible before re-shock. The critical terms in the interior mixing region that cause the peak of $\bar{\rho} \tilde{R}_{11}$ (slightly inclined towards the lighter fluid side) to decrease before re-shock are the pressure-strain redistribution [term (V)], turbulent transport [term (IV)], and dissipation [term (VI)] terms. The production [term (III)] and convection terms are quite positive there but their combined effect is smaller than that from the negative terms.
In general, both production and convection terms are positive on the lighter fluid side and negative on the heavier fluid side. These two terms transport $\bar{\rho} \tilde{R}_{11}$ from the heavier fluid side to the lighter fluid side.
On the other hand, the turbulent transport term helps bring $\bar{\rho} \tilde{R}_{11}$ from the lighter fluid side to the heavier fluid side and more importantly it is also responsible for the spreading of the statistical quantity at the edges of the mixing layer.

\begin{figure*}[!ht]
\centering
\subfigure[$\ t=0.40\ \mathrm{ms}$]{%
\includegraphics[width=0.4\textwidth]{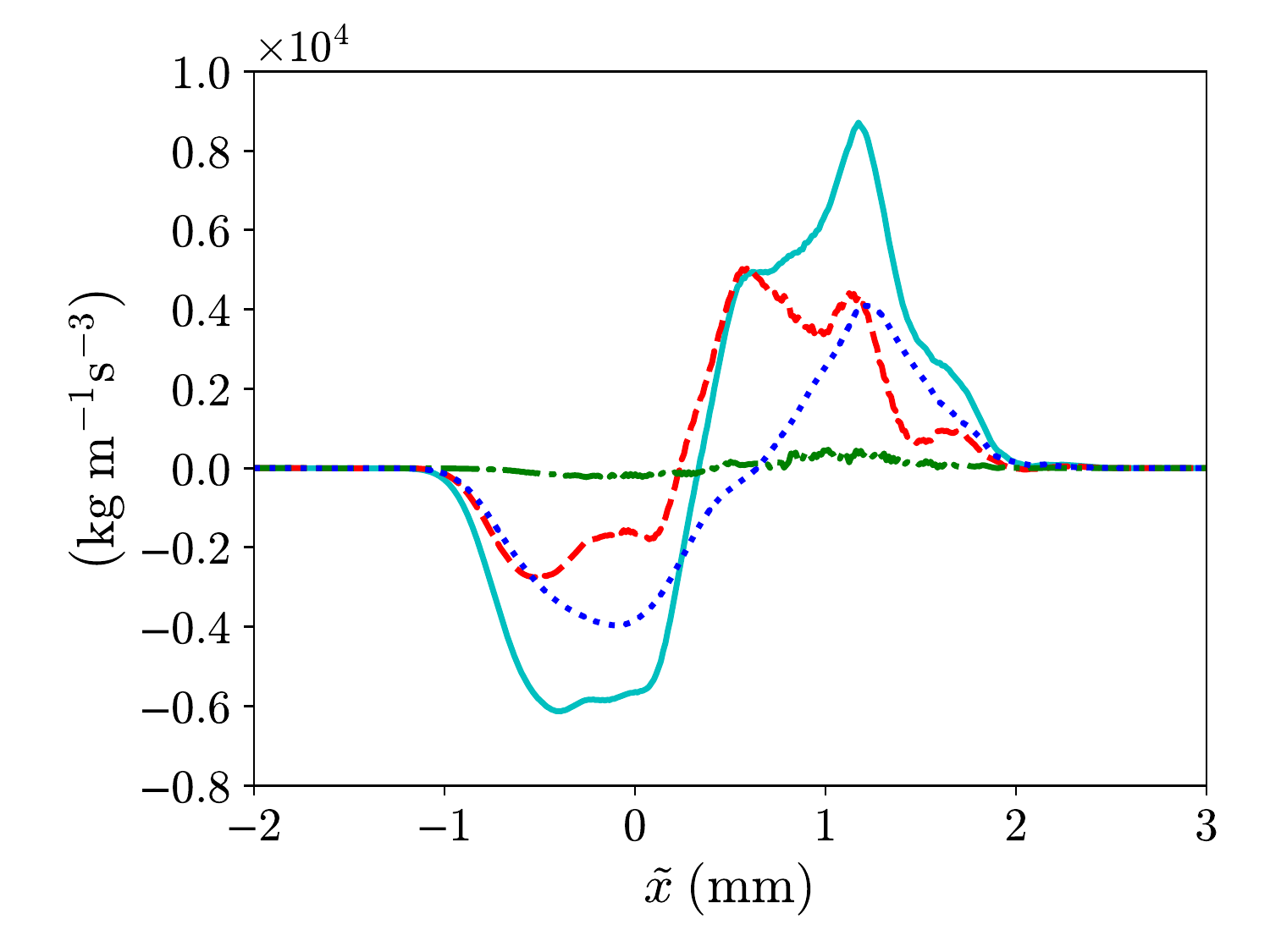}\label{fig:rho_R11_budget_production_terms_t_0_40}}
\subfigure[$\ t=1.10\ \mathrm{ms}$]{%
\includegraphics[width=0.4\textwidth]{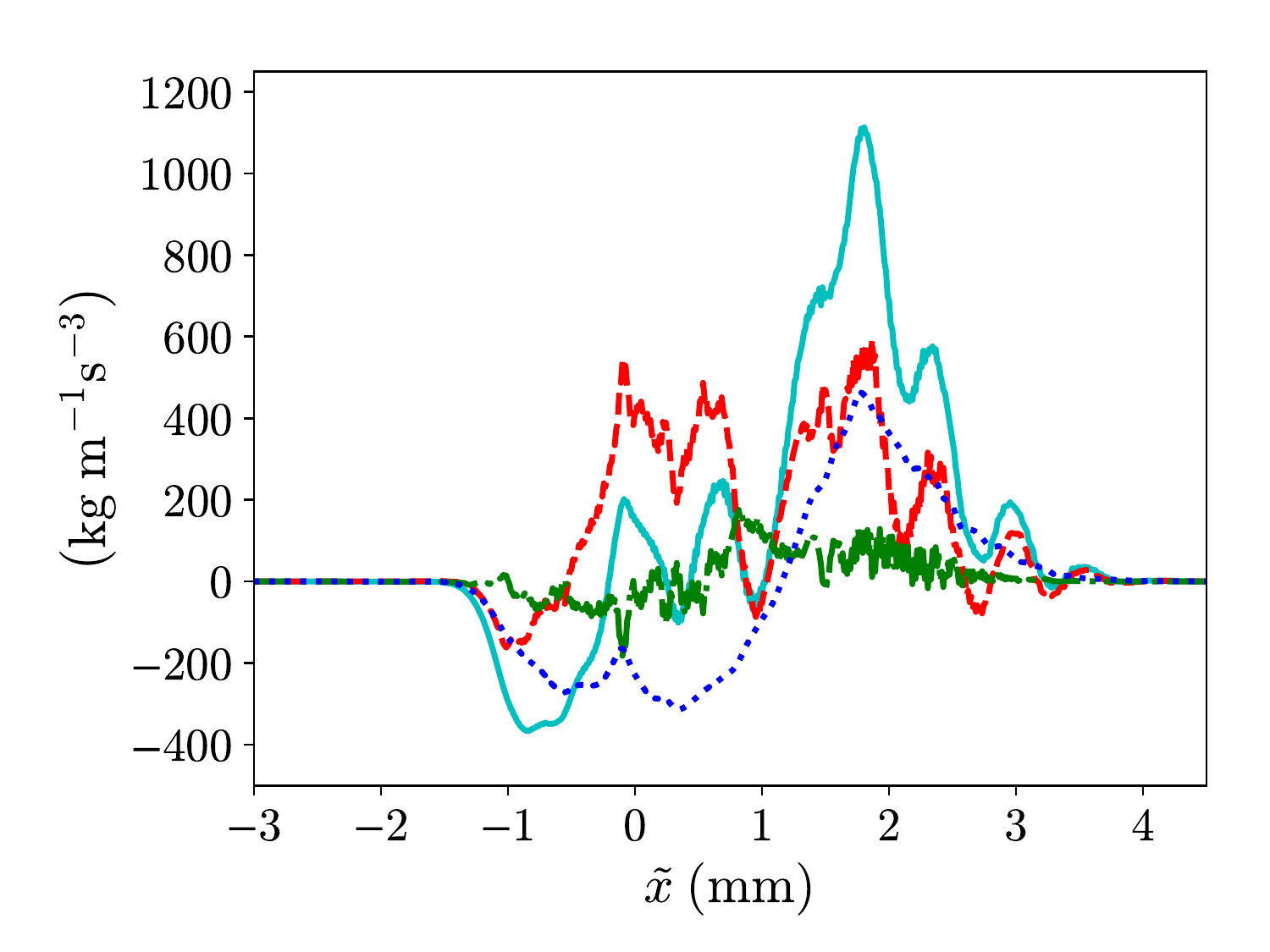}\label{fig:rho_R11_budget_production_terms_before_reshock}}
\caption{Compositions of the production term [term (III)] in the transport equation for the Reynolds normal stress component in the streamwise direction multiplied by the mean density, $\bar{\rho} \tilde{R}_{11}$, at different times before re-shock. Cyan solid line: overall production; red dashed line: $2a_1 \bar{p}_{,1}$; green dash-dotted line: $-2a_1 \bar{\tau}_{11_{,1}}$; blue dotted line: $-2\bar{\rho}\tilde{R}_{11} \tilde{u}_{,1}$.}
\label{fig:rho_R11_budget_production_terms}
\end{figure*}

\begin{figure*}[!ht]
\centering
\subfigure[$\ t=0.40\ \mathrm{ms}$]{%
\includegraphics[width=0.4\textwidth]{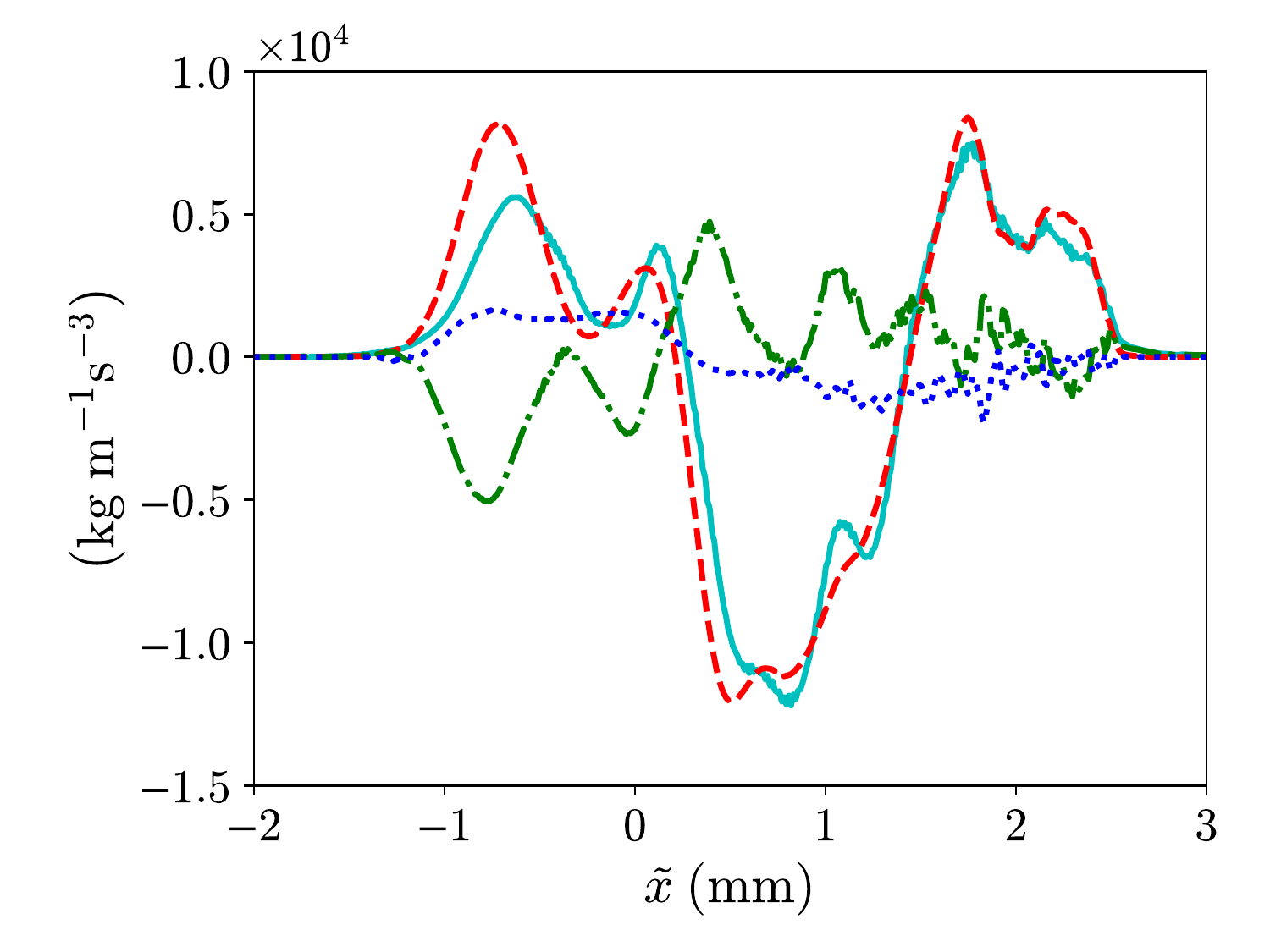}\label{fig:rho_R11_budget_turb_transport_terms_t_0_40}}
\subfigure[$\ t=1.10\ \mathrm{ms}$]{%
\includegraphics[width=0.4\textwidth]{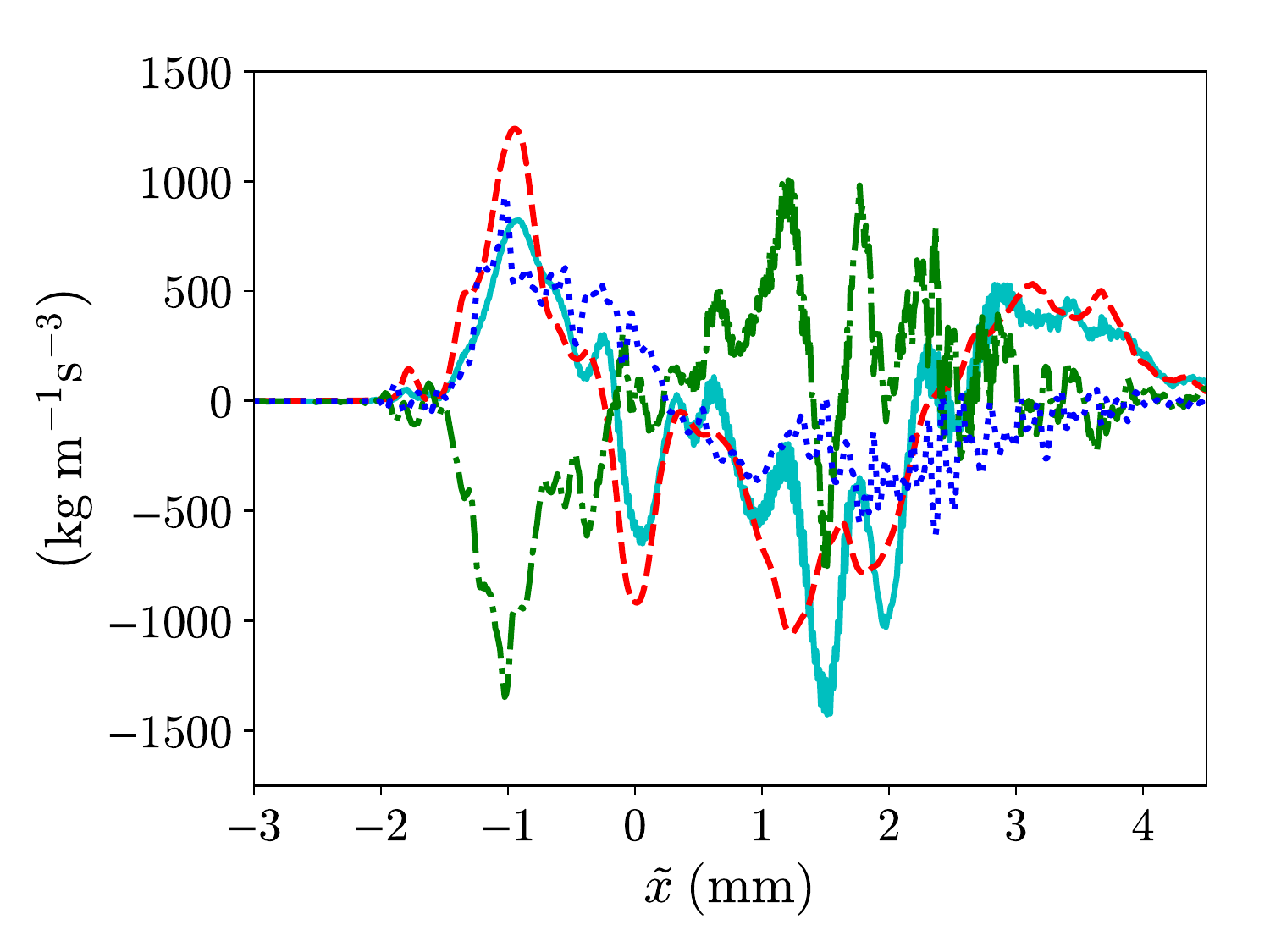}\label{fig:rho_R11_budget_turb_transport_terms_before_reshock}}
\caption{Compositions of the turbulent transport term [term (IV)] in the transport equation for the Reynolds normal stress component in the streamwise direction multiplied by the mean density, $\bar{\rho} \tilde{R}_{11}$, at different times before re-shock. Cyan solid line: overall turbulent transport; red dashed line: $-( \overline{\rho u^{\prime \prime} u^{\prime \prime} u^{\prime \prime} } )_{,1}$; green dash-dotted line: $-2 ( \overline{u^{\prime} p^{\prime}} )_{,1}$; blue dotted line: $2 ( \overline{u^{\prime} \tau^{\prime}_{11}})_{,1}$.}
\label{fig:rho_R11_budget_turb_transport_terms}
\end{figure*}

Figure~\ref{fig:rho_R11_budget_production_terms} shows the composition of production term [term (III)] before re-shock. It can be seen that both $2a_1 \bar{p}_{,1}$ and $-2\bar{\rho}\tilde{R}_{11} \tilde{u}_{,1}$ have large contributions to the production term, while the remaining component, $-2a_1 \bar{\tau}_{11_{,1}}$, is negligible. 
The composition of the turbulent transport term [term (IV)] is shown in figure~\ref{fig:rho_R11_budget_turb_transport_terms}. All three constituents: $-( \overline{\rho u^{\prime \prime} u^{\prime \prime} u^{\prime \prime} } )_{,1}$, $-2( \overline{u^{\prime} p^{\prime}} )_{,1}$, and $2 ( \overline{u^{\prime} \tau^{\prime}_{11}} )_{,1}$ have significant contributions to the term before re-shock. The triple correlation component, $-( \overline{\rho u^{\prime \prime} u^{\prime \prime} u^{\prime \prime} } )_{,1}$, is the root of the spreading effect while $-2( \overline{u^{\prime} p^{\prime}} )_{,1}$ and $2( \overline{u^{\prime} \tau^{\prime}_{11}} )_{,1}$ have opposite effects for the transfer of $\bar{\rho} \tilde{R}_{11}$ between heavy and light fluid regions. $-2 ( \overline{u^{\prime} p^{\prime}} )_{,1}$ transports $\bar{\rho} \tilde{R}_{11}$ from the heavier fluid side to the lighter fluid side and vice versa for $2 ( \overline{u^{\prime} \tau^{\prime}_{11}} )_{,1}$.

\begin{figure*}[!ht]
\centering
\subfigure[$\ t=0.40\ \mathrm{ms}$]{%
\includegraphics[width=0.4\textwidth]{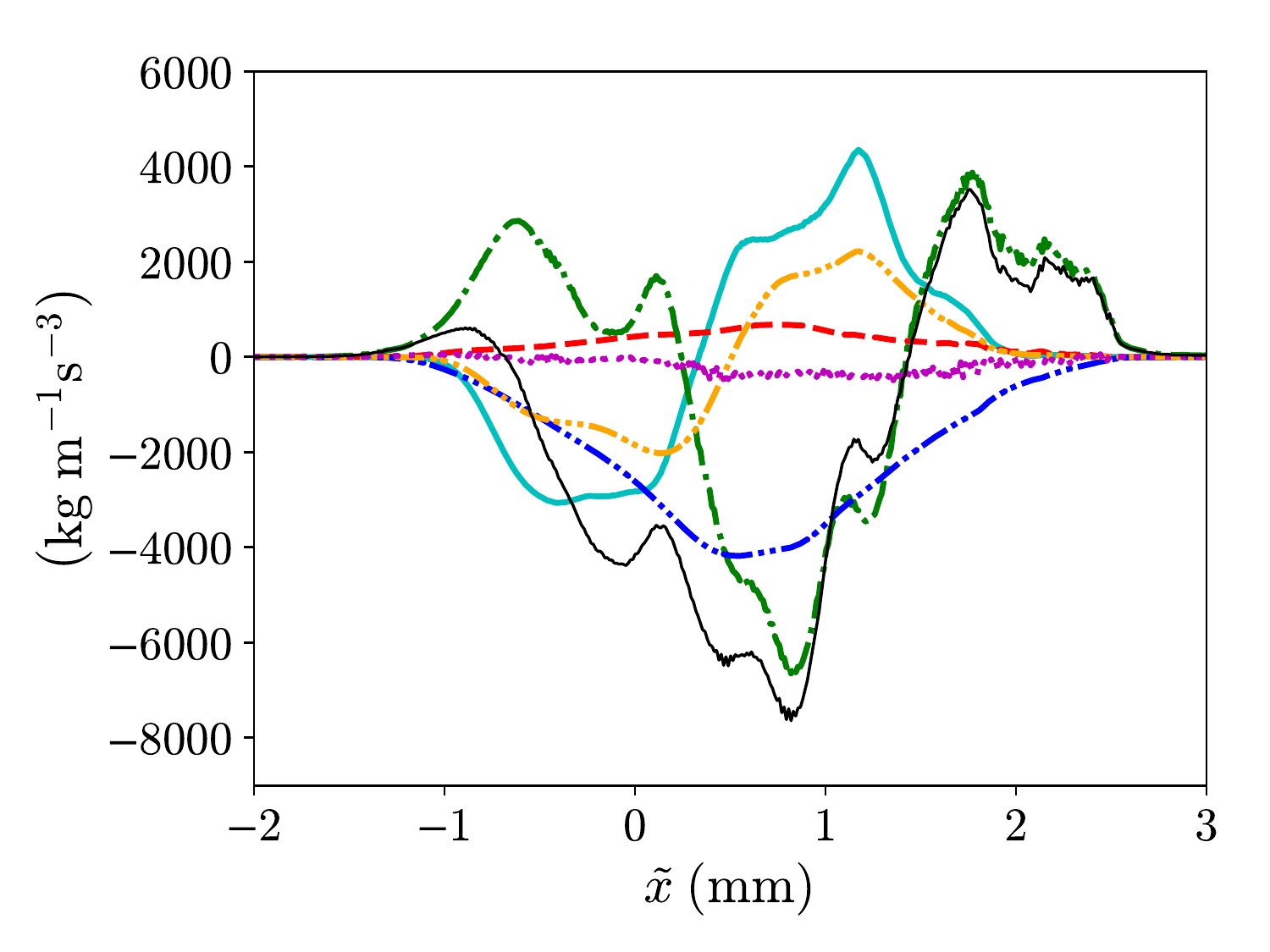}}
\subfigure[$\ t=1.10\ \mathrm{ms}$]{%
\includegraphics[width=0.4\textwidth]{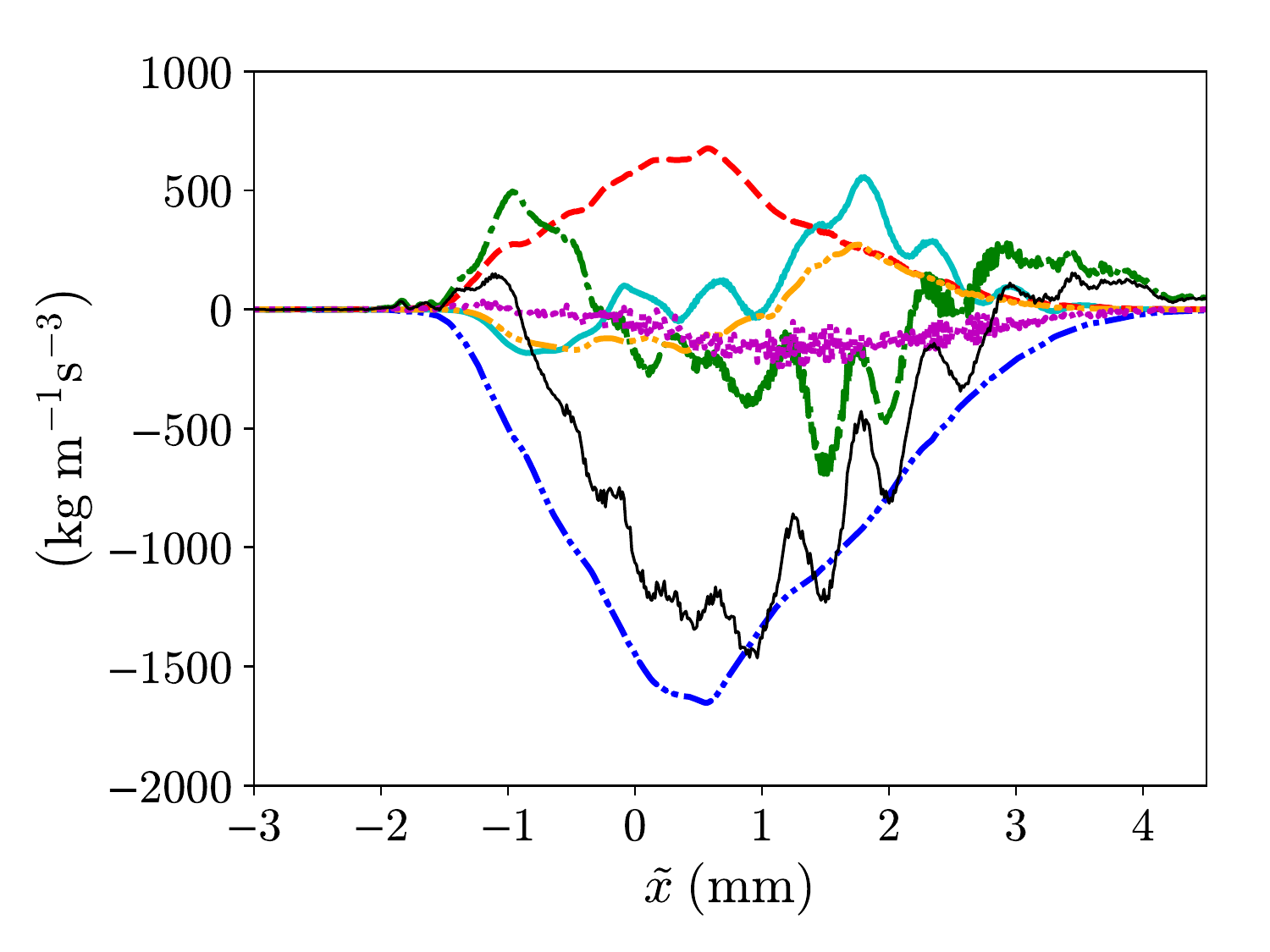}}
\caption{
Budgets of the turbulent kinetic energy, $\bar{\rho} k$, given by equation~\eqref{eq:k_transport_eqn_1D}, at different times before re-shock. Cyan solid line: production [term (III)]; red dashed line: pressure-dilatation [term (V)]; green dash-dotted line: turbulent transport [term (IV)]; blue dash-dot-dotted line: dissipation [term (VI)]; orange dash-triple-dotted line: negative of convection due to streamwise velocity associated with turbulent mass flux; magenta dotted line: residue; thin black solid line: summation of all terms (rate of change in the moving frame).
}
\label{fig:rho_k_budget}
\end{figure*}

Finally, the budget terms in the transport equation for the turbulent kinetic energy, $\bar{\rho} k$, given by equation~\eqref{eq:k_transport_eqn_1D} are compared at different times before re-shock in figure~\ref{fig:rho_k_budget}. The residue due to numerical discretization is negligible at $t=0.40\ \mathrm{ms}$. At later times, it becomes slightly larger relative to other budget terms, although it is still small in the budgets. At late times, the major terms in the interior part of the mixing layer are the pressure-dilatation [term (V)] and dissipation [term (VI)] terms. In single-species incompressible flows, the pressure-dilatation term is absent, but this term plays a large role to reduce the effects of dissipation term in this variable-density decaying flow before re-shock. At the edges of the mixing layer, the turbulent transport term [term (IV)] is relatively more important and is responsible for the spread of the turbulent kinetic energy.

\section{\label{sec:filtered_equations} Filtered Navier--Stokes equations and transport equations of the large-scale second-moments}

In the present flow, 
the mixing transition 
follows after the mixing layer is traversed by the reflected shock. This re-shock deposits baroclinic vorticity at both large and small scales, and rapid breakdown to fully-developed turbulence ensues.
The eddies span a wide range of length scales, where the largest and smallest eddies are estimated to be at scales of $O(1000)$ and $O(1)$ {\textmu}m respectively~\cite{wong2019high}. The small scales of the turbulent flow after re-shock are not well-resolved even in the highest resolution simulation. Therefore, it is more appropriate to study the transport equations of large-scale second-moments derived from the filtered Navier--Stokes equations at times after re-shock. The idea is that numerical regularization is assumed to have negligible effects on large-scale second-moments that only contain scales from zero wavenumber to a cut-off wavenumber imposed by a filter that is considerably larger than the grid cut-off wavenumber. The analysis of the transport equations of the large-scale second-moments is useful for (i) studying the mechanisms of the generation, destruction and spreading of the large-scale turbulent features in shock-induced variable-density turbulence, (ii) examining the self-similarity of the turbulent flow, and (iii) understanding how the subfilter-scale stress can affect the resolved large-scale turbulent features in LES.

While the effects of filtering on the buoyancy-driven variable-density turbulence have been studied in~\cite{saenz2021filtering}, the focus in the current work is different.
The analysis proposed here mainly focuses on the effects of SFS stress on large-scale statistical quantities resolved on a lower resolution grid, and the aim is to gain insight into the suitability of using LES data for analyzing RANS-based models. Spherical surface-averaged transport equations of different statistical quantities were also derived with the SGS stress in~\cite{lombardini2014turbulent}. Here, we present the planar surface-averaged transport equations of the second-moments, including the Favre-averaged Reynolds stress and turbulent kinetic energy, with the effects of SFS stress included.

The filtering operation of a variable, $f = f\left(x_i, t\right)$, with filter width, $\ell$, can be defined as:
\begin{equation}
    \left< f\left(x_i, t\right) \right>_{\ell} = \int_{-\infty}^{\infty} f\left(x_i^{\prime}, t\right) G\left(x_i^{\prime}, x_i \right) d {x_i^{\prime}},
\end{equation}
where $\left< f \right>_{\ell}$ is the filtered value and $G\left(x_i^{\prime}, x_i \right)$ denotes a filter function. In variable-density flows, it is also convenient to define the Favre-filtered value, $\left< f \right>_{L}$, as:
\begin{equation}
    \left< f \right>_{L} = \frac{ \left< \rho f \right>_{\ell} }{ \left< \rho \right>_{\ell} }.
\end{equation}

If we apply the filter on the mixture continuity equation and transport equation of momentum given by equations~\eqref{eq:mixture_continuity_eqn} and \eqref{eq:mixture_momentum_eqn} respectively, we can obtain the filtered Navier--Stokes equations:
\begin{align}
    \frac{\partial \left< \rho \right>_{\ell}}{\partial t} + \frac{\partial \left( \left< \rho \right>_{\ell} \left< u_k \right>_{L} \right)}{\partial x_k} &= 0, \label{eq:filtered_mixture_continuity_eqn} \\
    \frac{\partial \left( \left< \rho \right>_{\ell} \left< u_i \right>_{L} \right)}{\partial t} + \frac{\partial \left( \left< \rho \right>_{\ell} \left< u_k \right>_{L} \left< u_i 
    \right>_L \right)}{\partial x_k} &= - \frac{\partial \left( \left< p \right>_{\ell} \delta_{ki} \right)}{\partial x_k} + \frac{\partial \left< \tau_{ki} \right>_{\ell}}{\partial x_k} - \frac{\partial \tau_{ki}^{SFS}}{\partial x_k}, \label{eq:filtered_momentum_eqn}
\end{align}
where commutation terms are assumed to be negligible. $\tau_{ij}^{SFS}$ is the SFS stress tensor given by:
\begin{equation}
    \tau_{ij}^{SFS} = \left< \rho u_i u_j \right>_{\ell} - \left< \rho \right>_{\ell} \left< u_i \right>_{L} \left< u_j \right>_{L}  .
\end{equation}

If averaging is further applied on the filtered continuity equation and transport equation of momentum given by equations~\eqref{eq:filtered_mixture_continuity_eqn} and \eqref{eq:filtered_momentum_eqn} respectively, the following  Favre-averaged filtered Navier--Stokes equations are obtained:
\begin{align}
    \frac{\partial \overline{\left< \rho \right>}_{\ell}}{\partial t}
    + \frac{\partial \left( \overline{\left< \rho \right>}_{\ell} \widetilde{\left< u_k \right>}_{L} \right)}{\partial x_k} &= 0, \label{eq:averaged_filtered_mixture_continuity_eqn} \\
    \frac{\partial \left( \overline{\left< \rho \right>}_{\ell} \widetilde{\left< u_i \right>}_{L} \right)}{\partial t} + \frac{\partial \left( \overline{\left< \rho \right>}_{\ell} \widetilde{\left< u_k \right>}_{L} \widetilde{\left< u_i \right>}_{L} \right)}{\partial x_k} &= - \frac{\partial \left( \overline{\left< p \right>}_{\ell} \delta_{ki} \right)}{\partial x_k} + \frac{\partial \overline{\left< \tau_{ki} \right>}_{\ell}}{\partial x_k}  - \frac{\partial \overline{\tau_{ki}^{SFS}}}{\partial x_k} - \frac{\partial \left( \overline{\left< \rho \right>}_{\ell} \widetilde{R}_{L,ki} \right)}{\partial x_k}, \label{eq:averaged_filtered_momentum_eqn}
\end{align}
where Reynolds and Favre decompositions on the filtered variables ($\left< f \right>_{\ell}$ or $\left< f \right>_{L}$) are involved:
\begin{equation}
    \left< f \right>_{\ell/L} =  \overline{\left< f \right>}_{\ell/L} + \left< f \right>_{\ell/L}^{\prime} = \widetilde{\left< f \right>}_{\ell/L} + \left< f \right>_{\ell/L}^{\prime\prime},
\end{equation}
and $\widetilde{R}_{L,ij}$ is the large-scale Favre-averaged Reynolds stress tensor computed with the filtered density and velocity fields and is given by:
\begin{equation}
    \widetilde{R}_{L,ij} = \frac{\overline{ \left< \rho \right>_{\ell} \left< u_i \right>_{L}^{\prime\prime} \left< u_j \right>_{L}^{\prime\prime}}}{\overline{\left< \rho \right>}_{\ell}}.
\end{equation}

In a 1D mean flow, the transport equation of $\overline{\left< \rho \right>}_{\ell} \widetilde{R}_{L,11}$ is given by:
\begin{equation}
\begin{split}
	\underbrace{ \frac{\partial \left( \overline{\left< \rho \right>}_{\ell} \widetilde{R}_{L,11} \right)}{\partial t} }_{ \text{term (I)} }
	\ \underbrace{ + \frac{\partial \left( \overline{\left< \rho \right>}_{\ell} \widetilde{\left< u \right>}_{L} \widetilde{R}_{L,11} \right)}{\partial x} }_{ \text{term (II)} } = 
    \ \underbrace{ 2a_{L,1} \left( \frac{\partial \overline{\left< p \right>}_{\ell} }{\partial x} - \frac{\partial \overline{\left< \tau_{11} \right>}_{\ell}}{\partial x} + \frac{\partial \overline{ \tau_{11}^{SFS} }}{\partial x} \right) - 2 \overline{\left< \rho \right>}_{\ell} \widetilde{R}_{L,11} \frac{\partial \widetilde{\left< u \right>}_{L} }{\partial x} }_{ \text{term (III)} } \\
    \underbrace{ - \frac{\partial \left( \overline{ \left< \rho \right>_{\ell} \left< u \right>_{L}^{\prime\prime} \left< u \right>_{L}^{\prime\prime} \left< u \right>_{L}^{\prime\prime} } \right)}{\partial x} 
    - 2 \frac{\partial \left( \overline{\left< u \right>_{L}^{\prime} \left< p \right>_{\ell}^{\prime}} \right)}{\partial x}
    + 2 \frac{\partial \left( \overline{ \left< u \right>_{L}^{\prime} \left< \tau_{11} \right>_{\ell}^{\prime} } \right)}{\partial x} - 2 \frac{\partial \left( \overline{ \left< u \right>_{L}^{\prime} {\tau_{11}^{SFS}}^{\prime} } \right)}{\partial x} }_{ \text{term (IV)} }
    \
    \underbrace{ + 2 \overline{\left< p \right>_{\ell}^{\prime} \frac{\partial \left< u \right>_{L}^{\prime}}{\partial x}} }_{ \text{term (V)} } \\
    \underbrace{ - 2\left( \overline{\left< \tau_{11} \right>_{\ell}^{\prime} \frac{\partial \left< u \right>_{L}^{\prime}}{\partial x}}
    + \overline{\left< \tau_{12} \right>_{\ell}^{\prime} \frac{\partial \left< u \right>_{L}^{\prime}}{\partial y}}
    + \overline{\left< \tau_{13} \right>_{\ell}^{\prime} \frac{\partial \left< u \right>_{L}^{\prime}}{\partial z}} \right) + 2\left( \overline{ {\tau_{11}^{SFS}}^{\prime} \frac{\partial \left< u \right>_{L}^{\prime}}{\partial x}}
    + \overline{ {\tau_{12}^{SFS}}^{\prime} \frac{\partial \left< u \right>_{L}^{\prime}}{\partial y}}
    + \overline{ {\tau_{13}^{SFS}}^{\prime} \frac{\partial \left< u \right>_{L}^{\prime}}{\partial z}} \right) }_{ \text{term (VI)} },
\end{split} \label{eq:RL11_transport_eqn_1D}
\end{equation}
\noindent where the LHS consists of the rate of change [term (I)] and convection [term (II)]. The RHS consists of production [term (III)], turbulent transport [term (IV)], pressure-strain redistribution [term (V)], and dissipation [term (VI)]. $a_{L,i} = \overline{ \left< \rho \right>_{\ell}^{\prime} \left< u_i \right>_{L}^{\prime} } / \overline{\left< \rho \right>}_{\ell}$ is the velocity associated with the large-scale turbulent mass flux $\overline{\left< \rho \right>}_{\ell} a_{L,i} = \overline{ \left< \rho \right>_{\ell}^{\prime} \left< u_i \right>_{L}^{\prime} }$ computed on filtered fields.

The transport equation of $\overline{\left< \rho \right>}_{\ell} \widetilde{R}_{L,22}$ for 1D mean flow can be reduced to:
\begin{equation}
\begin{split}
	\underbrace{ \frac{\partial \left( \overline{\left< \rho \right>}_{\ell} \widetilde{R}_{L,22} \right)}{\partial t} }_{ \text{term (I)} }
	\ \underbrace{ + \frac{\partial \left( \overline{\left< \rho \right>}_{\ell} \widetilde{\left< u \right>}_{L} \widetilde{R}_{L,22} \right)}{\partial x} }_{ \text{term (II)} } = 
	\ \underbrace{ - \frac{\partial \left( \overline{ \left< \rho \right>_{\ell} \left< v \right>_{L}^{\prime\prime} \left< v \right>_{L}^{\prime\prime} \left< u \right>_{L}^{\prime\prime} } \right)}{\partial x}
	+ 2\frac{\partial \left( \overline{ \left< v \right>_{L}^{\prime} \left< \tau_{21} \right>_{\ell}^{\prime}} \right)}{\partial x} - 2\frac{\partial \left( \overline{ \left< v \right>_{L}^{\prime} { \tau_{21}^{SFS} }^{\prime}} \right)}{\partial x} }_{ \text{term (IV)} }
    \ \underbrace{ + 2\overline{\left< p \right>_{\ell}^{\prime} \frac{\partial \left< v \right>_{L}^{\prime}}{\partial y}} }_{ \text{term (V)} } \\
    \underbrace{ - 2\left( \overline{\left< \tau_{21} \right>_{\ell}^{\prime} \frac{\partial \left< v \right>_{L}^{\prime}}{\partial x}} 
    + \overline{\left< \tau_{22} \right>_{\ell}^{\prime} \frac{\partial \left< v \right>_{L}^{\prime}}{\partial y}}
    + \overline{\left< \tau_{23} \right>_{\ell}^{\prime} \frac{\partial \left< v \right>_{L}^{\prime}}{\partial z}} \right) + 2\left( \overline{ { \tau_{21}^{SFS} }^{\prime} \frac{\partial \left< v \right>_{L}^{\prime}}{\partial x}} 
    + \overline{ { \tau_{22}^{SFS} }^{\prime} \frac{\partial \left< v \right>_{L}^{\prime}}{\partial y}}
    + \overline { { \tau_{23}^{SFS} }^{\prime} \frac{\partial \left< v \right>_{L}^{\prime}}{\partial z}} \right) }_{ \text{term (VI)} }.
\end{split} \label{eq:RL22_transport_eqn_1D}
\end{equation}

\noindent The transport equation of $\overline{\left< \rho \right>}_{\ell} \widetilde{R}_{L,33}$ is similar.

The large-scale turbulent kinetic energy per unit mass is defined as $k_L = \widetilde{R}_{L,ii}/2$. The transport equation of $\overline{\left< \rho \right>}_{\ell} k_L$ can be obtained by taking half of the trace of the transport equation of $\overline{\left< \rho \right>}_{\ell} \widetilde{R}_{L,ij}$. In 1D mean flow, it has the following form:
\begin{equation}
\begin{split}
	\underbrace{ \frac{\partial \left( \overline{\left< \rho \right>}_{\ell} k_L \right)}{\partial t} }_{ \text{term (I)} }
	\ \underbrace{ + \frac{\partial \left( \overline{\left< \rho \right>}_{\ell} \widetilde{\left< u \right>}_{L} k_L \right)}{\partial x} }_{ \text{term (II)} } = 
    \ \underbrace{ a_{L,1}\left( \frac{\partial \overline{\left< p \right>}_{\ell}}{\partial x} - \frac{\partial \overline{\left< \tau_{11} \right>}_{\ell}}{\partial x} + \frac{\partial \overline{ \tau_{11}^{SFS} }}{\partial x} \right) - \overline{\left< \rho \right>}_{\ell} \widetilde{R}_{L,11} \frac{\partial \widetilde{\left< u \right>}_{L} }{\partial x} }_{ \text{term (III)} } \\
    \underbrace{ - \frac{1}{2} \frac{\partial \left( \overline{ \left< \rho \right>_{\ell} \left< u_i \right>_{L}^{\prime\prime} \left< u_i \right>_{L}^{\prime\prime} \left< u \right>_{L}^{\prime\prime} } \right)}{\partial x}
    - \frac{\partial \left( \overline{\left< u \right>_{L}^{\prime} \left< p \right>_{\ell}^{\prime}} \right)}{\partial x}
    + \frac{\partial \left( \overline{ \left< u_i \right>_{L}^{\prime} \left< \tau_{i1} \right>_{\ell}^{\prime} } \right)}{\partial x} - \frac{\partial \left( \overline{ \left< u_i \right>_{L}^{\prime} {\tau_{i1}^{SFS}}^{\prime} } \right)}{\partial x} }_{ \text{term (IV)} } \\
    \underbrace{ + \overline{\left< p \right>_{\ell}^{\prime} \frac{\partial \left< u_i \right>_{L}^{\prime}}{\partial x_i}} }_{ \text{term (V)} }
    \ \underbrace{ -\overline{\left< \tau_{ij} \right>_{\ell}^{\prime} \frac{\partial \left< u_i \right>_{L}^{\prime}}{\partial x_j}} + \overline{ { \tau_{ij}^{SFS} }^{\prime} \frac{\partial \left< u_i \right>_{L}^{\prime}}{\partial x_j}} }_{ \text{term (VI)} },
\end{split} \label{eq:kL_transport_eqn_1D}
\end{equation}
\noindent where the LHS consists of the rate of change [term (I)] and convection [term (II)]. The RHS consists of production [term (III)], turbulent transport [term (IV)], pressure-dilatation [term (V)], and dissipation [term (VI)]. Note that term (III) represents the transfer of energy between $\overline{\left< \rho \right>}_{\ell} k_L$ and the the mean kinetic energy computed from filtered fields, $K_L = \overline{\left< \rho \right>}_{\ell} \widetilde{\left< u_i \right>}_{L} \widetilde{\left< u_i \right>}_{L} /2$. Besides, the combination of $- ( \overline{ \left< u_i \right>_{L}^{\prime} {\tau_{i1}^{SFS}}^{\prime} } )_{,1}$ and $\overline{ { \tau_{ij}^{SFS} }^{\prime} \partial \left< u_i \right>_{L,j}^{\prime}}$ contributes to the transfer of energy between $\overline{\left< \rho \right>}_{\ell} k_L$ and the mean SFS turbulent kinetic energy, $\overline{\tau_{ii}^{SFS}}/2$.

In 1D mean flow, the transport equation of the large-scale turbulent mass flux component in the streamwise direction, $\overline{\left< \rho \right>}_{\ell} a_{L,1}$, can be simplified to:
\begin{equation}
\begin{split}
	\underbrace{ \frac{\partial \left( \overline{\left< \rho \right>}_{\ell} a_{L,1} \right)}{\partial t} }_{ \text{term (I)} }
	\ \underbrace{ + \frac{\partial \left( \overline{\left< \rho \right>}_{\ell} \widetilde{\left< u \right>}_{L} a_{L,1} \right)}{\partial x} }_{ \text{term (II)} } =
	\underbrace{ b_L \left(\frac{\partial \overline{\left< p \right>}_{\ell}}{\partial x} - \frac{\partial \overline{\left< \tau_{11} \right>}_{\ell} }{\partial x} + \frac{\partial \overline{\tau_{11}^{SFS}} }{\partial x} \right) - \widetilde{R}_{L,11} \frac{\partial \overline{\left< \rho \right>}_{\ell}}{\partial x} }_{ \text{term (III)} } \\
	\underbrace{ + \overline{\left< \rho \right>}_{\ell} \frac{\partial \left( a_{L,1} a_{L,1} \right)}{\partial x} - \overline{\left< \rho \right>}_{\ell} a_{L,1} \frac{\partial \overline{\left< u \right>}_{L}}{\partial x } }_{ \text{term (IV)} }
	\ \underbrace{ - \overline{\left< \rho \right>}_{\ell} \frac{\partial \left( \overline{\left< \rho \right>_{\ell}^{\prime} \left< u \right>_{L}^{\prime} \left< u \right>_{L}^{\prime}} / \overline{ \left< \rho \right>_{\ell} } \right)}{\partial x} }_{ \text{term (V)} } \\
    \underbrace{ + \overline{\left< \rho \right>}_{\ell} \overline{\left( \frac{1}{\left< \rho \right>_{\ell}} \right)^{\prime} \left( \frac{\partial \left< p \right>_{\ell}^{\prime}}{\partial x} - \frac{\partial \left< \tau_{11} \right>_{\ell}^{\prime}}{\partial x} - \frac{\partial \left< \tau_{12} \right>_{\ell}^{\prime}}{\partial y} - \frac{\partial \left< \tau_{13} \right>_{\ell}^{\prime}}{\partial z} + \frac{\partial {\tau_{11}^{SFS}}^{\prime}}{\partial x} + \frac{\partial {\tau_{12}^{SFS}}^{\prime}}{\partial y} + \frac{\partial {\tau_{13}^{SFS}}^{\prime}}{\partial z} \right)}
    + \overline{\left< \rho \right>}_{\ell} \varepsilon_{a_{L,1}} }_{ \text{term (VI)} },
\end{split} \label{eq:aL1_transport_eqn_1D}
\end{equation}

\noindent where the LHS consists of the rate of change [term (I)] and convection [term (II)]. The RHS contains production [term (III)], redistribution [term (IV)], turbulent transport [term (V)], and destruction [term (VI)]. Also,
\begin{equation}
    \varepsilon_{a_{L,i}} = - \overline{\left< u_i \right>_{L}^{\prime} \frac{\partial \left< u_k \right>_{L}^{\prime}}{\partial x_k}}.
\end{equation}

\noindent $b_L$ is the large-scale density-specific-volume covariance computed from the filtered fields and is given by $b_L=-\overline{\left< \rho \right>_{\ell}^{\prime} (1 / \left< \rho \right>_{\ell})^{\prime}}$.

In 1D mean flow, the transport equation of $\overline{\left< \rho \right>}_{\ell} b_L$ is given by:
\begin{equation}
\begin{split}
	\underbrace{ \frac{\partial \left( \overline{\left< \rho \right>}_{\ell} b_L \right) }{\partial t} }_{ \text{term (I)} }
	\ \underbrace{ + \frac{\partial \left( \overline{\left< \rho \right>}_{\ell} \widetilde{\left< u \right>}_{L} b_L \right)}{\partial x} }_{ \text{term (II)} } =
	\underbrace{ -2 \left( b_L + 1 \right) a_{L,1} \frac{\partial \overline{\left< \rho \right>}_{\ell}}{\partial x} }_{ \text{term (III)} }
	\ \underbrace{ + 2 \overline{\left< \rho \right>}_{\ell} a_{L,1} \frac{\partial b_L}{\partial x} }_{ \text{term (IV)} } \\
    \underbrace{ + \overline{\left< \rho \right>}_{\ell}^2 \frac{ \partial \left( \overline{\left< \rho \right>_{\ell}^{\prime} (1/\left< \rho \right>_{\ell})^{\prime} \left< u \right>_{L}^{\prime} }/\overline{\left< \rho \right>}_{\ell} \right)}{\partial x} }_{ \text{term (V)} }
    \ \underbrace{ + 2\overline{\left< \rho \right>}_{\ell}^2 \varepsilon_{b_L} }_{ \text{term (VI)} },
\end{split} \label{eq:bL_transport_eqn_1D}
\end{equation}
\noindent where the LHS consists of the rate of change [term (I)] and convection [term (II)]. The RHS consists of production [term (III)], redistribution [term (IV)], turbulent transport [term (V)], and destruction [term (VI)]. Also,
\begin{equation}
    \varepsilon_{b_L} = \overline{\left(\frac{1}{\left< \rho \right>_{\ell}}\right)^{\prime} \frac{\partial \left< u_k \right>_{L}^{\prime}}{\partial x_k}}.
\end{equation}

A truncated Gaussian filter~\cite{cook2005hyperviscosity} is used. At each filtering operation, 1D filters in the $x$-, $y$-, and $z$-directions are applied successively to the 3D fields. 
The filter in the $x$-direction is given by:
\begin{equation}
\begin{split}
    \left< f_{i,j,k} \right>_{\ell,x} &=
    \frac{3565}{10368} f_{i,j,k} + \frac{3091}{12960} \left( f_{i-1,j,k} + f_{i+1,j,k} \right)
    + \frac{1997}{25920} \left( f_{i-2,j,k} + f_{i+2,j,k} \right) \\
    &\quad + \frac{149}{12960} \left( f_{i-3,j,k} + f_{i+3,j,k} \right)
    + \frac{107}{103680} \left( f_{i-4,j,k} + f_{i+4,j,k} \right),
\end{split}
\end{equation}
where the effective filter width of one filtering operation is $\ell = 4 \Delta$, and $\Delta$ is the grid spacing of the finest grid level. Filtering in the $y$ and $z$ directions is in similar forms.
The Gaussian filtering operation can be applied successively to achieve filtering with an essentially larger filter width. If the filter is applied $N_f$ times repeatedly, the effective filter width is $\ell \approx 4 \sqrt{N_f} \Delta$. The approximated filter widths obtained on the finest grid level of grid E with different numbers of filtering operations are shown in table~\ref{tab:filter_widths}. The truncated Gaussian filter is selected because of its positivity-preserving property for the density field.

\begin{table}[!ht]
\caption{\label{tab:filter_widths}%
Approximated filter widths obtained on the finest grid level of grid E with different numbers of filtering operations.}
\begin{ruledtabular}
\begin{tabular}{ c c c }
 Number of filtering operations & Approximated filter width in $\Delta$ & Approximated filter width in physical unit ($\mathrm{mm}$) \\ 
 \hline
   1 &  4 $\Delta$ & 0.049 \\
   4 &  8 $\Delta$ & 0.098 \\
  16 & 16 $\Delta$ & 0.195 \\ 
  64 & 32 $\Delta$ & 0.391 \\
 256 & 64 $\Delta$ & 0.781 \\
\end{tabular}
\end{ruledtabular}
\end{table}



\section{\label{sec:second_moments_effect_filtering} Effects of the filter width on the large-scale second-moments and the SFS stress}

\begin{figure*}[!ht]
\centering
\subfigure[$\ \overline{\left< \rho \right>}_{\ell} a_{L,1}$]{%
\includegraphics[width=0.4\textwidth]{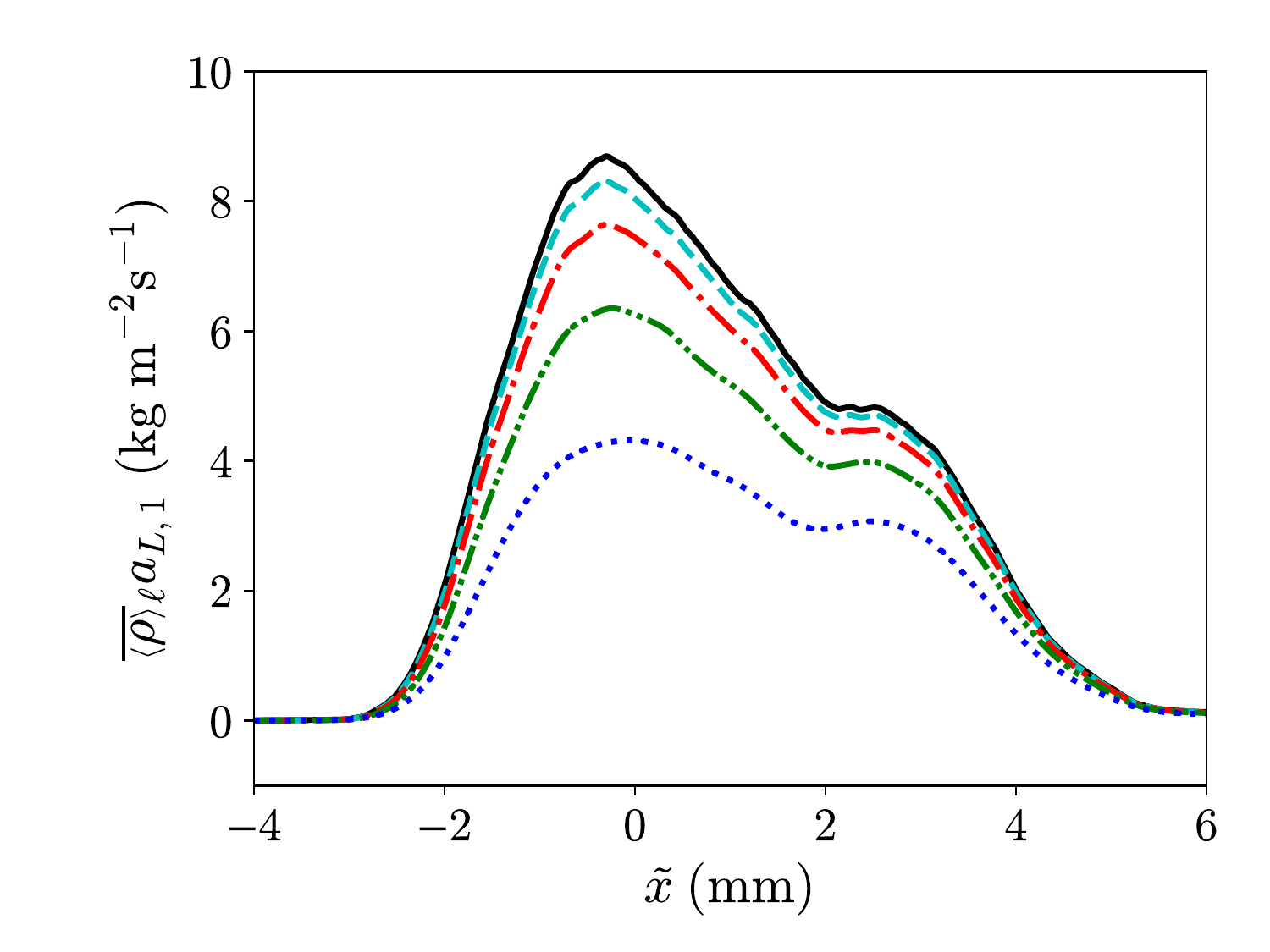}\label{fig:rho_a1_filtered_t_1_40}}
\subfigure[$\ \overline{\left< \rho \right>}_{\ell} b_L$]{%
\includegraphics[width=0.4\textwidth]{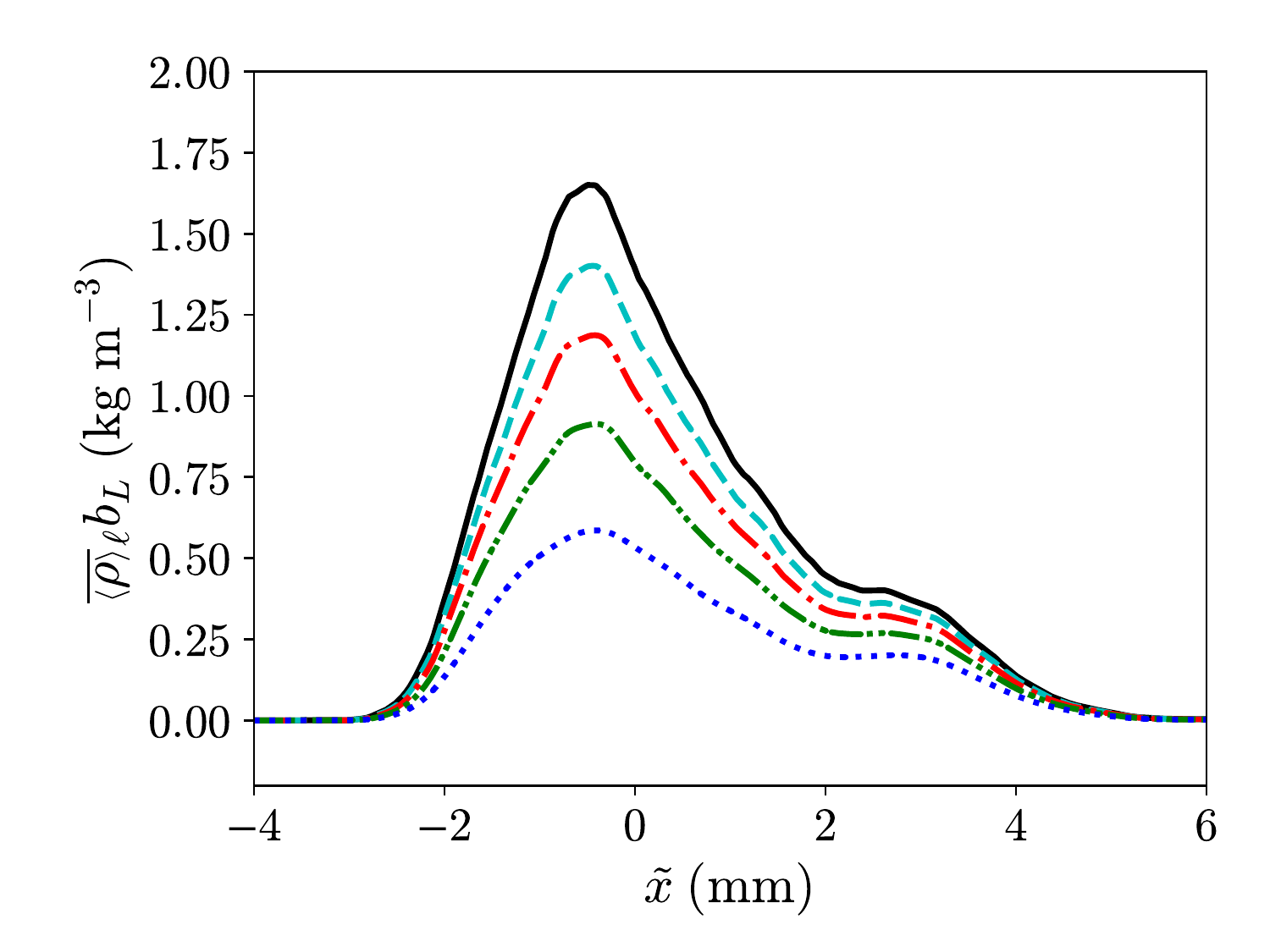}\label{fig:rho_b_filtered_t_1_40}}
\label{fig:rho_a1_filtered}
\subfigure[$\ \overline{\left< \rho \right>}_{\ell} \widetilde{R}_{L,11}$]{%
\includegraphics[width=0.4\textwidth]{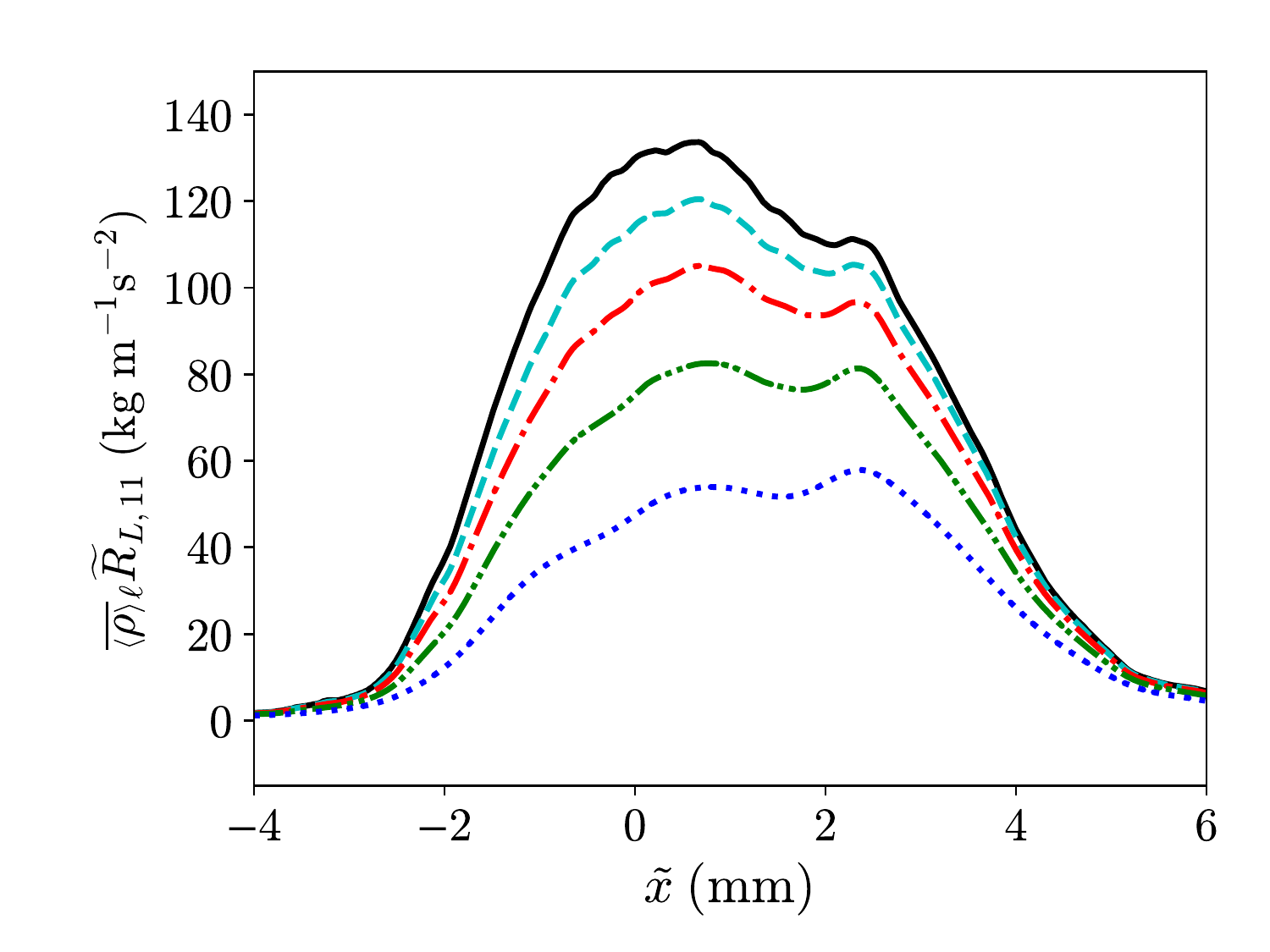}\label{fig:rho_R11_filtered_t_1_40}}
\subfigure[$\ \overline{\left< \rho \right>}_{\ell} k_L$]{%
\includegraphics[width=0.4\textwidth]{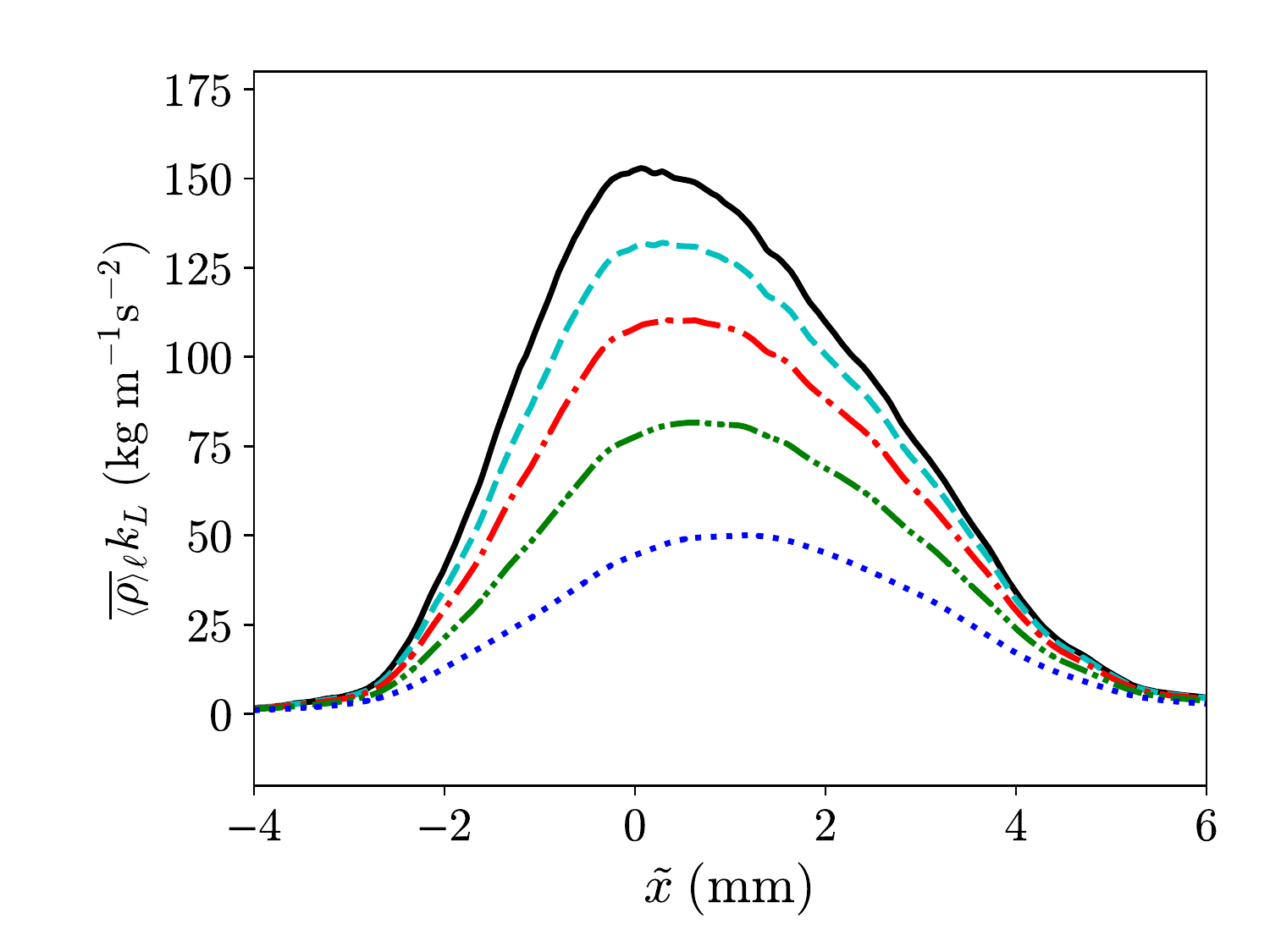}\label{fig:rho_k_filtered_t_1_40}}
\caption{Effect of filtering on the large-scale second-moments and the turbulent kinetic energy, $\overline{\left< \rho \right>}_{\ell} k_L$, at $t=1.40\ \mathrm{ms}$ after re-shock. Black solid line: no filtering; cyan dashed line: $\ell \approx 8 \Delta$; red dash-dotted line: $\ell \approx 16 \Delta$; green dash-dot-dotted line: $\ell \approx 32 \Delta$; blue dotted line: $\ell \approx 64 \Delta$.
}
\end{figure*}

The effects of filter width on the large-scale second-moments including the Reynolds normal stress in the streamwise direction multiplied by the mean filtered density:
$\overline{\left< \rho \right>}_{\ell} a_{L,1}$, $\overline{\left< \rho \right>}_{\ell} b_L$, and $\overline{\left< \rho \right>}_{\ell} \widetilde{R}_{L,11}$
at $t=1.40\ \mathrm{ms}$ are shown in figures~\ref{fig:rho_a1_filtered_t_1_40}, \ref{fig:rho_b_filtered_t_1_40}, and \ref{fig:rho_R11_filtered_t_1_40} respectively. It can be seen that the magnitudes of the large-scale quantities reduce when the essential width of the filter applied to the density and momentum fields is increased because they are composed of scales from zero wavenumber to a larger cut-off wavenumber. The shape of each quantity remains quite self-similar with different filter widths. Especially the location of the peak of each quantity does not move significantly under the effect of filtering. At large filter widths, all large-scale second-moments including $\overline{\left< \rho \right>}_{\ell} \widetilde{R}_{L,11}$ have similar degrees of changes in the magnitudes with the same filter width change.

\begin{figure*}[!ht]
\centering
\subfigure[$\ \overline{\tau_{11}^{SFS}}$]{%
\includegraphics[width=0.4\textwidth]{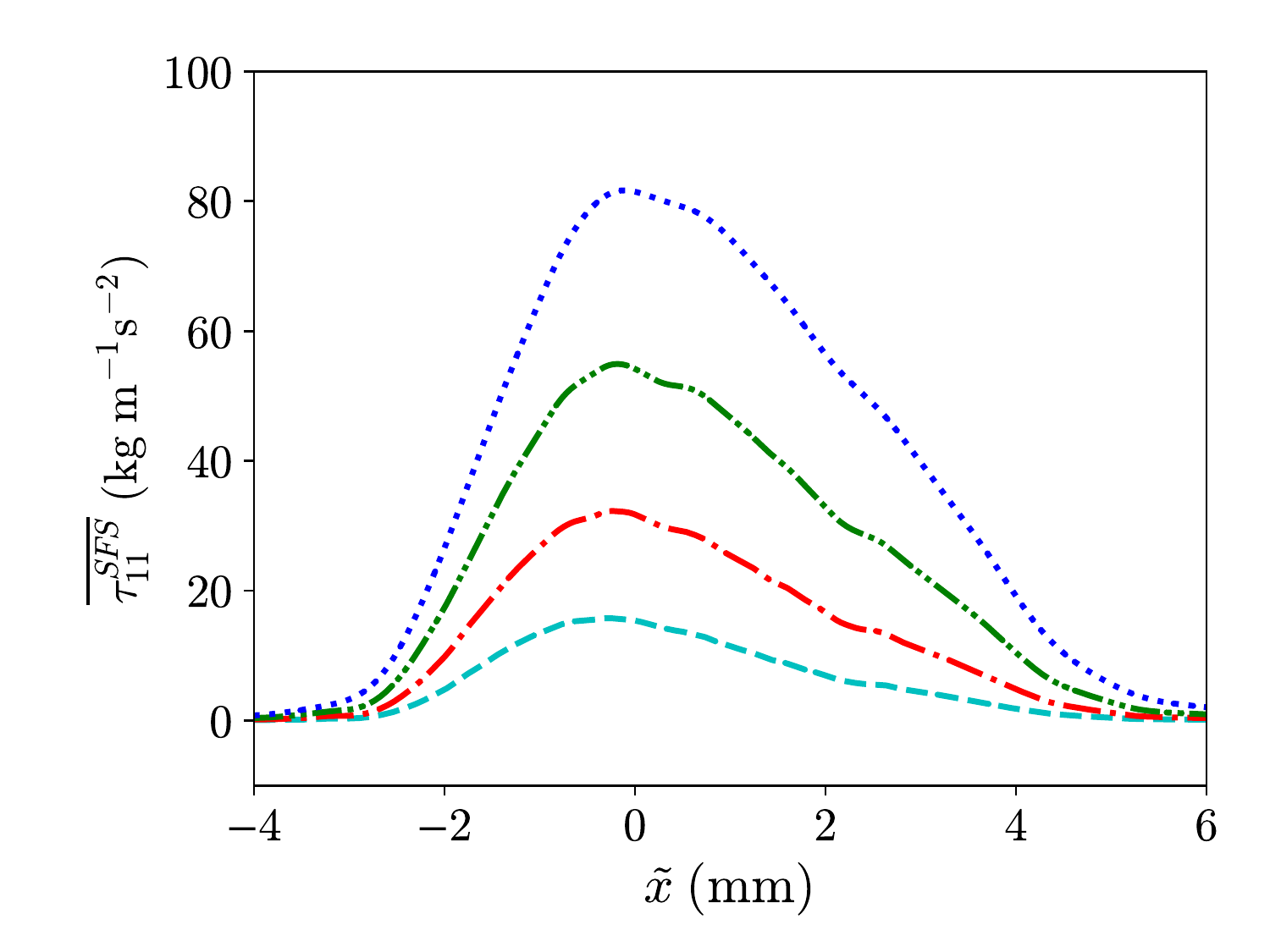}\label{fig:SFS_stress_t_1_40}}
\subfigure[$\ \overline{\tau_{ii}^{SFS}}/2$]{%
\includegraphics[width=0.4\textwidth]{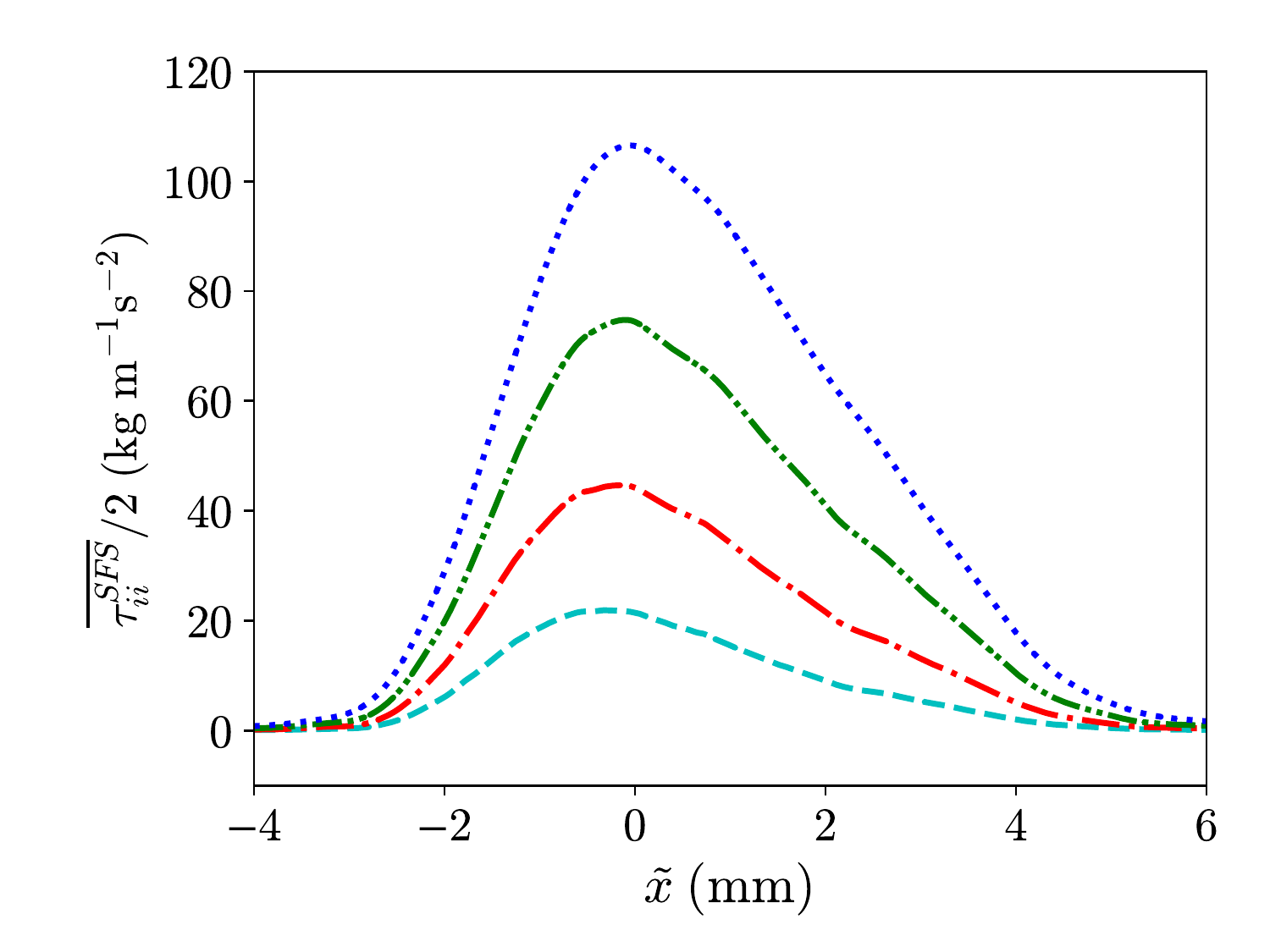}\label{fig:SFS_TKE_t_1_40}}
\caption{Effect of filtering on the mean SFS stress component in the streamwise direction, $\overline{\tau_{11}^{SFS}}$, and the mean SFS turbulent kinetic energy, $\overline{\tau_{ii}^{SFS}}/2$ at $t=1.40\ \mathrm{ms}$ after re-shock. Cyan dashed line: $\ell \approx 8 \Delta$; red dash-dotted line: $\ell \approx 16 \Delta$; green dash-dot-dotted line: $\ell \approx 32 \Delta$; blue dotted line: $\ell \approx 64 \Delta$.
}
\label{fig:SFS_stress}
\end{figure*}

Figure~\ref{fig:SFS_stress_t_1_40} shows the effect of filtering on the mean SFS normal stress component in the streamwise direction, $\overline{\tau_{11}^{SFS}}$. It can be seen that the magnitude of the SFS stress component increases with larger filter width. In fact, it is noticed that the sum of $\overline{\tau_{11}^{SFS}}$ and large-scale $\overline{\left< \rho \right>}_{\ell} \widetilde{R}_{L,11}$ is virtually constant under the filtering effect. The same relation is also observed for the sum of the large-scale turbulent kinetic energy and the mean SFS turbulent kinetic energy, $\overline{\tau_{ii}^{SFS}}/2$. The magnitude of the large-scale turbulent kinetic energy decreases while that of the SFS turbulent kinetic energy rises when more filtering operations are applied, as seen in figures~\ref{fig:rho_k_filtered_t_1_40} and \ref{fig:SFS_TKE_t_1_40} respectively. These suggest that the correlations between the small scales and large scales are negligible compared to the large-scale-large-scale and small-scale-small-scale correlations. The effects of filter width on large-scale second-moments, SFS stress, and SFS turbulent kinetic energy at other times after re-shock are shown in the Supplemental Material~\cite{supple2022wong}.

\section{\label{sec:budgets_after_reshock} Budgets of the large-scale second-moments after re-shock}

In this section, the budgets of large-scale second-moments computed with the filtered density and Favre-filtered velocity fields:
$\overline{\left< \rho \right>}_{\ell} a_{L,1}$, $\overline{\left< \rho \right>}_{\ell} b_L$, and $\overline{\left< \rho \right>}_{\ell} \widetilde{R}_{L,11}$, together with $\overline{\left< \rho \right>}_{\ell} k_L$
across the mixing layer after re-shock are examined. The chosen filter width is $\ell \approx 64 \Delta = 0.781\ \mathrm{mm}$.
A grid sensitivity analysis of the budgets at this filter width is provided in the Supplemental Material~\cite{supple2022wong}.

As with the unfiltered budgets before re-shock, the budgets of large-scale second-moments after re-shock are studied in the $\tilde{x}$ coordinate system, equivalent to studying the budgets in the moving reference frame of the mixing layer. The convective terms in all of the transport equations of large-scale second-moments for 1D mean flow have the common form of $[ \overline{\left< \rho \right>}_{\ell} \widetilde{\left< u \right>}_{L} ( \cdot ) ]_{,1}$, where $( \cdot )$ represents any of the large-scale second-moments ($a_{L,1}$, $b_L$, $\widetilde{R}_{L,ij}$, or $k_L$). Using the relation $\widetilde{\left< u \right>}_{L} = \overline{\left< u \right>}_{L} + a_{L,1}$, the convective terms can be rewritten as:
\begin{equation}
    \frac{\partial \overline{\left< \rho \right>}_{\ell} \widetilde{\left< u \right>}_{L} \left( \cdot \right) }{\partial x} = \underbrace{ \frac{\partial \overline{\left< \rho \right>}_{\ell} \overline{\left< u \right>}_{L} \left( \cdot \right) }{\partial x} }_{ \text{term (I)} } + \underbrace{ \frac{\partial \overline{\left< \rho \right>}_{\ell} a_{L,1} \left( \cdot \right) }{\partial x} }_{ \text{term (II)} },
\end{equation}
\noindent where term (I) is the convection due to mean Favre-filtered velocity and term (II) is the convection due to velocity associated with large-scale turbulent mass flux. Similar to the scenario without filtering, $\overline{\left< u \right>}_{L}$ is observed to be uniformly close to zero in the moving reference frame of the mixing layer and the term (I) can be neglected. Thus, the convective term is thought as fully contributed by $[ \overline{\left< \rho \right>}_{\ell} a_{L,1} ]_{,1}$ in the analysis of this section. While only the budgets of the large-scale second-moments at $t=1.20\ \mathrm{ms}$ and $t=1.60\ \mathrm{ms}$ are shown in this section, the budgets at two other times after re-shock are included in the Supplemental Material~\cite{supple2022wong}.

\subsection{Large-scale turbulent mass flux}

\begin{figure*}[!ht]
\centering
\subfigure[$\ t=1.20\ \mathrm{ms}$]{%
\includegraphics[width=0.4\textwidth]{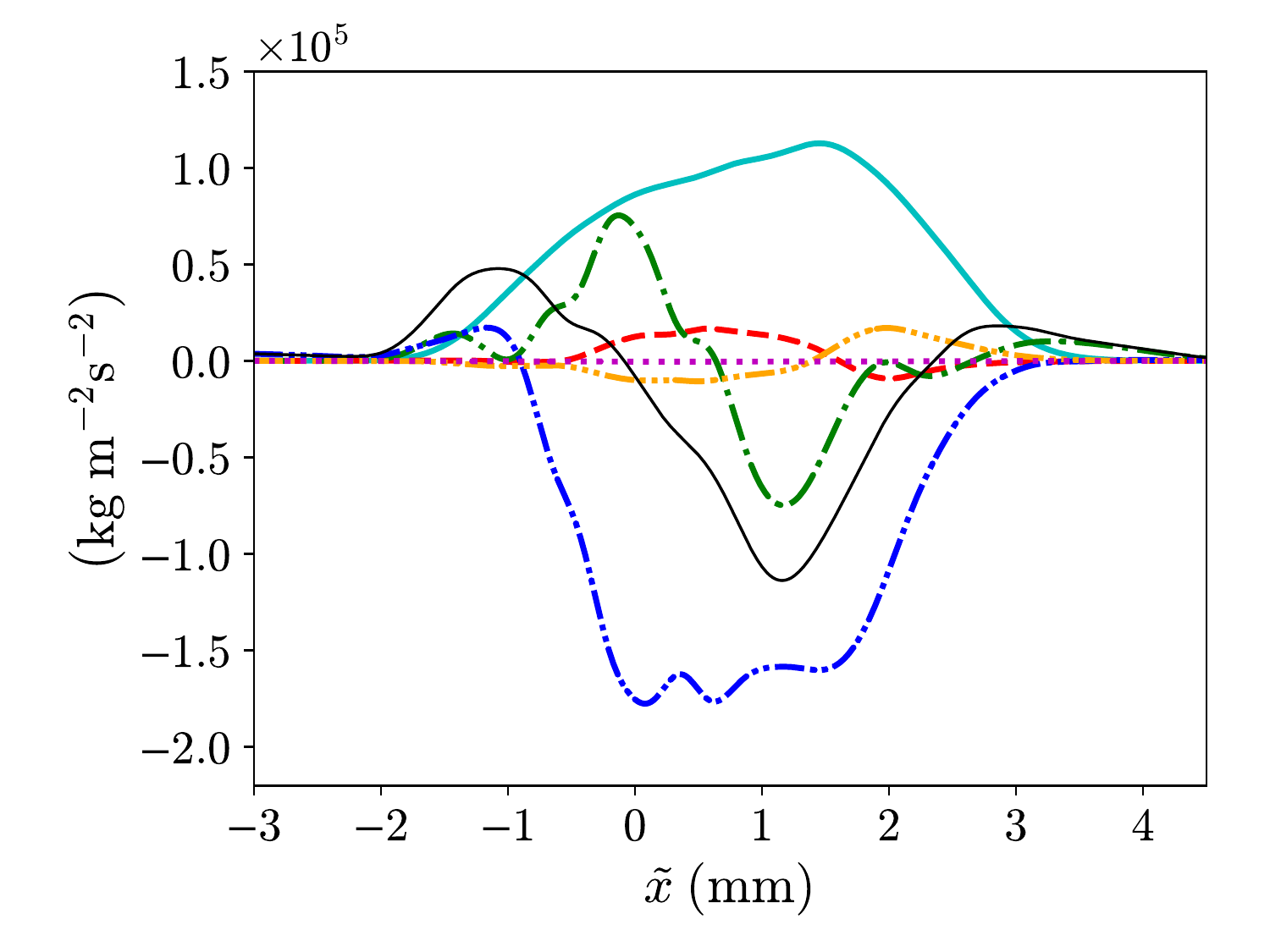}\label{fig:rho_a1_budget_filtered_after_reshock}}
\subfigure[$\ t=1.60\ \mathrm{ms}$]{%
\includegraphics[width=0.4\textwidth]{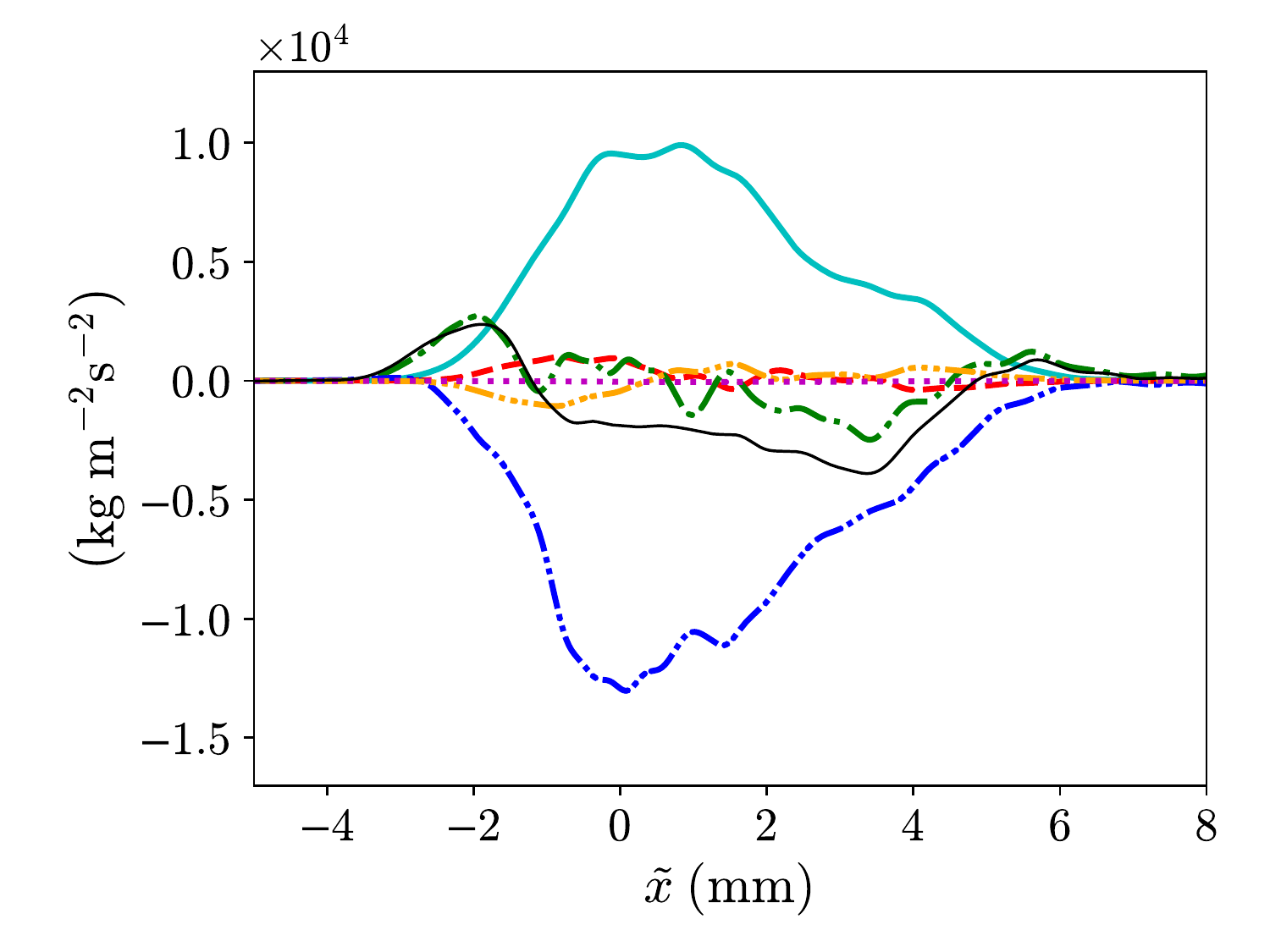}\label{fig:rho_a1_budget_filtered_t_1_60}}
\caption{Budgets of the large-scale turbulent mass flux component in the streamwise direction, $\overline{\left< \rho \right>}_{\ell} a_{L,1}$, given by equation~\eqref{eq:aL1_transport_eqn_1D}, at different times after re-shock. Cyan solid line: production [term (III)]; red dashed line: redistribution [term (IV)]; green dash-dotted line: turbulent transport [term (V)]; blue dash-dot-dotted line: destruction [term (VI)]; orange dash-triple-dotted line: negative of convection due to streamwise velocity associated with turbulent mass flux; magenta dotted line: residue; thin black solid line: summation of all terms (rate of change in the moving frame).}
\label{fig:rho_a1_budget_filtered}
\end{figure*}

Figure~\ref{fig:rho_a1_budget_filtered} shows the spatial profiles of different RHS terms in the transport equation for the large-scale turbulent mass flux component in the streamwise direction, $\overline{\left< \rho \right>}_{\ell} a_{L,1}$, given by equation~\eqref{eq:aL1_transport_eqn_1D}, together with the negative of the convection term due to $a_{L,1}$ after re-shock. Similar to the budgets before re-shock, the magenta dotted line represents the residue which is defined as the subtraction of the net RHS terms from the net LHS term in the simulation frame. The residue here provides a way to verify that the numerical regularization or SGS effect has negligible effects on the budgets of large-scale second-moments at the chosen filter width. The rate of change of $\overline{\left< \rho \right>}_{\ell} a_{L,1}$ in the moving reference frame of the mixing layer is represented by the thin black line, which is the subtraction of the convective term due to $a_{L,1}$ from the summation of the net RHS and the residue. 

\begin{figure*}[!ht]
\centering
\subfigure[$\ t=1.20\ \mathrm{ms}$]{%
\includegraphics[width=0.4\textwidth]{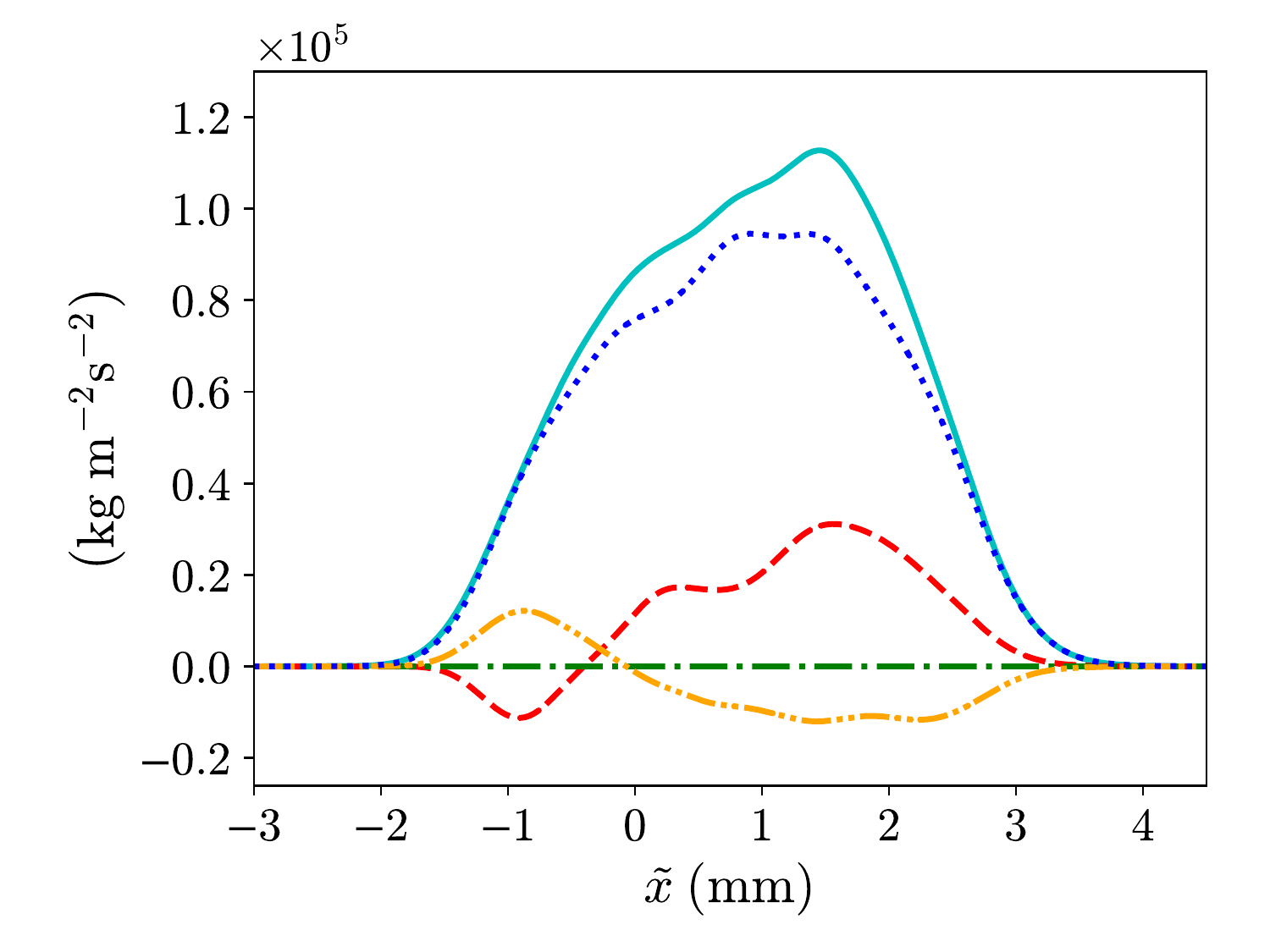}\label{fig:rho_a1_budget_filtered_production_terms_after_reshock}}
\subfigure[$\ t=1.60\ \mathrm{ms}$]{%
\includegraphics[width=0.4\textwidth]{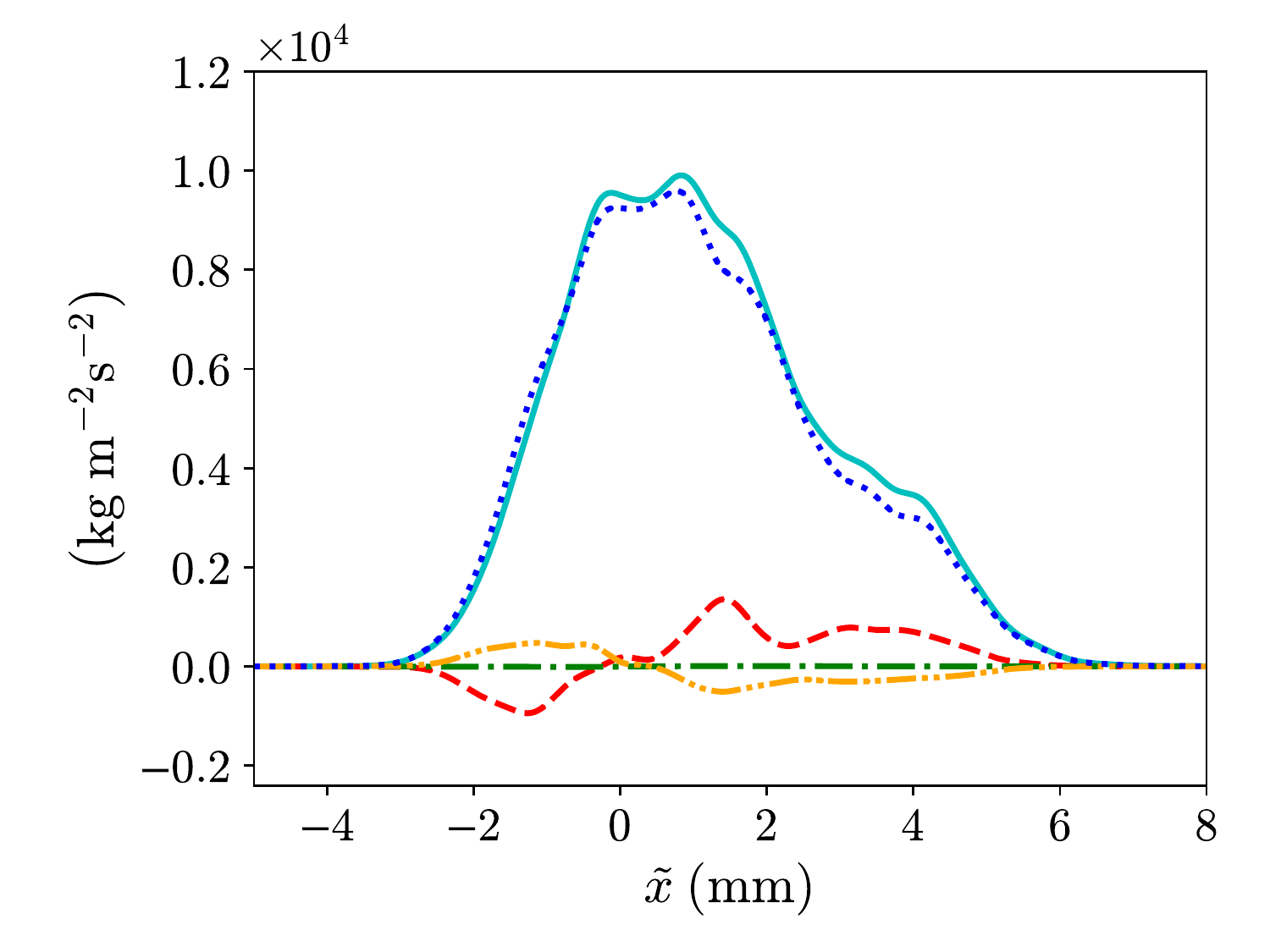}\label{fig:rho_a1_budget_filtered_production_terms_t_1_60}}
\caption{Compositions of the production term [term (III)] in the transport equation for the large-scale turbulent mass flux component in the streamwise direction, $\overline{\left< \rho \right>}_{\ell} a_{L,1}$, at different times after re-shock. Cyan solid line: overall production; red dashed line: $b_L \overline{\left< p \right>}_{\ell,1}$; green dash-dotted line: $-b_L \overline{\left< \tau_{11} \right>}_{\ell,1}$; orange dash-dot-dotted line: $b_L \overline{\tau_{11}^{SFS}}_{,1}$; blue dotted line: $-\widetilde{R}_{L,11} \overline{\left< \rho \right>}_{\ell,1}$.}
\label{fig:rho_a1_budget_filtered_production_terms}
\end{figure*}

\begin{figure*}[!ht]
\centering
\subfigure[$\ t=1.20\ \mathrm{ms}$]{%
\includegraphics[width=0.4\textwidth]{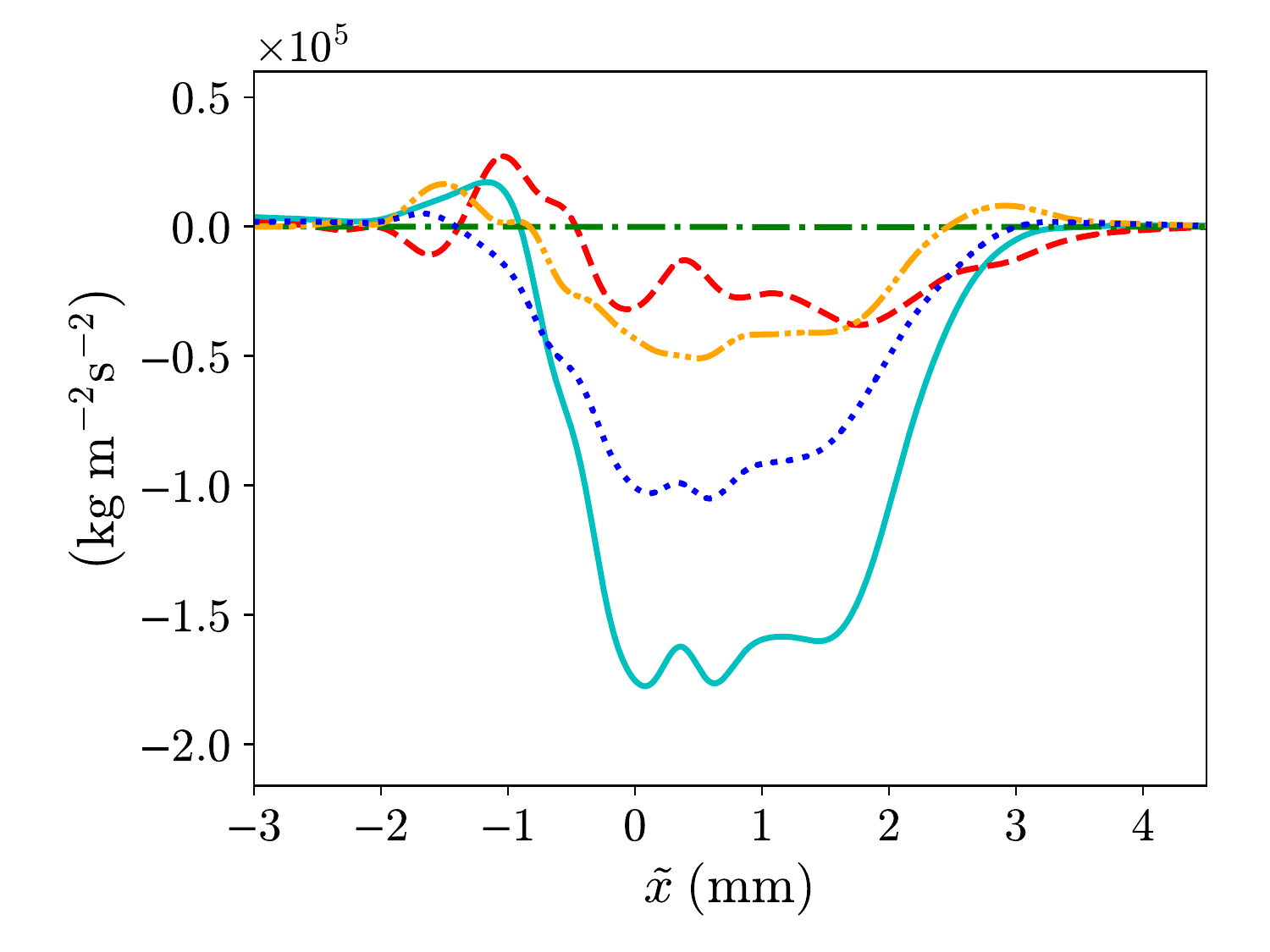}\label{fig:rho_a1_budget_filtered_destruction_terms_after_reshock}}
\subfigure[$\ t=1.60\ \mathrm{ms}$]{%
\includegraphics[width=0.4\textwidth]{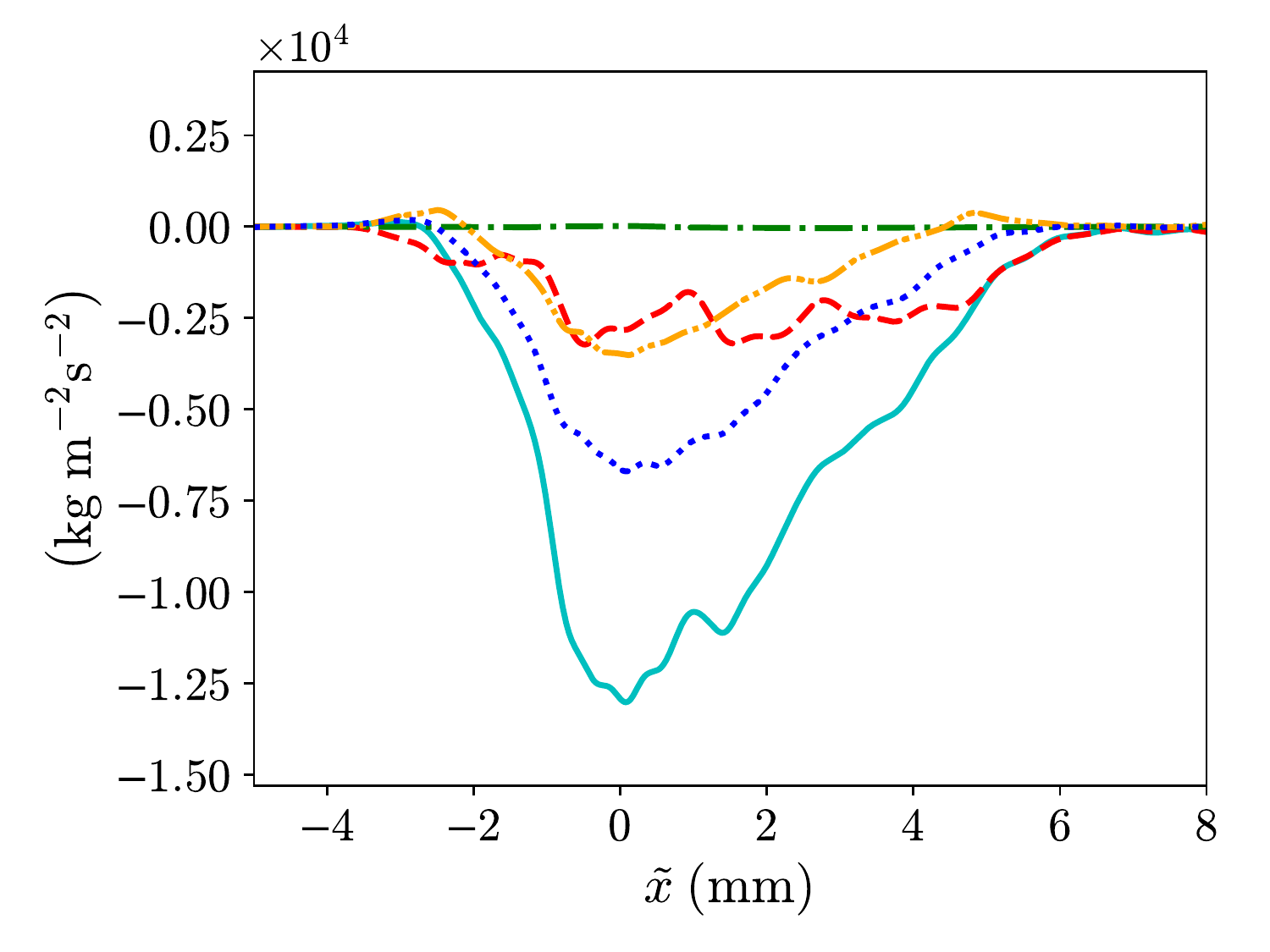}\label{fig:rho_a1_budget_filtered_destruction_terms_t_1_60}}
\caption{Compositions of the destruction term [term (VI)] in the transport equation for the large-scale turbulent mass flux component in the streamwise direction, $\overline{\left< \rho \right>}_{\ell} a_{L,1}$, at different times after re-shock. Cyan solid line: overall destruction; red dashed line: $\overline{\left< \rho \right>}_{\ell} \overline{ \left( 1/\left< \rho \right>_{\ell} \right)^{\prime} \left< p \right>^{\prime}_{\ell,1} }$; green dash-dotted line: $-\overline{\left< \rho \right>}_{\ell} \overline{ \left( 1/\left< \rho \right>_{\ell} \right)^{\prime} \left( \partial \left< \tau_{1i} \right>_{\ell}^{\prime} / \partial x_i \right) }$; orange dash-dot-dotted line: $\overline{\left< \rho \right>}_{\ell} \overline{ \left( 1/\left< \rho \right>_{\ell} \right)^{\prime} \left( \partial {\tau_{1i}^{SFS}}^{\prime} / \partial x_i \right) }$; blue dotted line: $\overline{\left< \rho \right>}_{\ell} \varepsilon_{a_{L,1}}$.}
\label{fig:rho_a1_budget_filtered_destruction_terms}
\end{figure*}

Similar to the times before re-shock, production [term (III)] and destruction [term (VI)] terms are asymmetric after re-shock, as shown in figure~\ref{fig:rho_a1_budget_filtered}. Both terms are skewed towards and have peaks at positions slightly towards the lighter fluid side.
As seen from figure~\ref{fig:rho_a1_budget_filtered}, the production and destruction terms are the the dominant terms among all the RHS terms in the interior mixing region after re-shock. In this region, the magnitude of the destruction is larger than that of the production, and this drives the peak of the large-scale turbulent mass flux to diminish over time after re-shock, as indicated by the negative rate of change at all times. At the edges of the mixing region, the turbulent transport term [term (V)] becomes relatively more important and causes the turbulent mass flux to spread over time. The redistribution [term (IV)] and convective terms are small across the mixing layer compared to the other RHS terms. In fact, the redistribution term is commonly ignored in many turbulent mixing models, such as the BHR $k$-$S$-$a$ model by~\citet{banerjee2010development} and the $k$-$L$-$a$ model by~\citet{morgan2015three}. The compositions of the production and destruction terms are shown in figures~\ref{fig:rho_a1_budget_filtered_production_terms} and \ref{fig:rho_a1_budget_filtered_destruction_terms} respectively. As seen from both figures, the components with filtered molecular shear stress, $-b_L \overline{\left< \tau_{11} \right>}_{\ell,1}$ and $-\overline{\left< \rho \right>}_{\ell} \overline{ \left( 1/\left< \rho \right>_{\ell} \right)^{\prime} \left( \partial \left< \tau_{1i} \right>_{\ell}^{\prime} / \partial x_i \right) }$, are both zero at different times and this indicates that the molecular shear stress has no direct effect on the large-scale turbulent mass flux through its budget.

Examining figure~\ref{fig:rho_a1_budget_filtered_production_terms} for the composition of the production term [term (III)], it can be seen that that the shapes and relative importance of the two components of the production term, $-\widetilde{R}_{L,11} \overline{\left< \rho \right>}_{\ell,1}$ and $b_L \overline{\left< p \right>}_{\ell,1}$, after re-shock are similar to those of the corresponding ones before re-shock. However, there is an additional term with the SFS stress, $b_L \overline{\tau_{11}^{SFS}}_{,1}$, in the composition due to filtering. In general, the component with the large-scale Reynolds stress, $-\widetilde{R}_{L,11} \overline{\left< \rho \right>}_{\ell,1}$, has the largest contribution to the production term and appears strictly positive. Another two constituents, $b_L \overline{\left< p \right>}_{\ell,1}$ and $b_L \overline{\tau_{11}^{SFS}}_{,1}$, have smaller contributions and have conflicting effects. The latter largely reduces the influence of the former on the production term. Therefore, the production term can be regarded as mainly supplied by the component with the large-scale Reynolds stress. As for the destruction [term (VI)], it can be seen in figure~\ref{fig:rho_a1_budget_filtered_destruction_terms} that the contribution of each constituent after re-shock is similar to the corresponding one in the unfiltered budgets before re-shock, except that the role of the component with molecular shear stress is replaced by a component with SFS stress, $\overline{\left< \rho \right>}_{\ell} \overline{ ( 1/\left< \rho \right>_{\ell} )^{\prime} ( \partial {\tau_{1i}^{SFS}}^{\prime} / \partial x_i ) }$. Similar to the corresponding component with $\varepsilon_{a_1}$ in the budgets before re-shock, the component with $\varepsilon_{a_{L,1}}$ also contributes significantly to the destruction term after re-shock.

\subsection{Large-scale density-specific-volume covariance}

\begin{figure*}[!ht]
\centering
\subfigure[$\ t=1.20\ \mathrm{ms}$]{%
\includegraphics[width=0.4\textwidth]{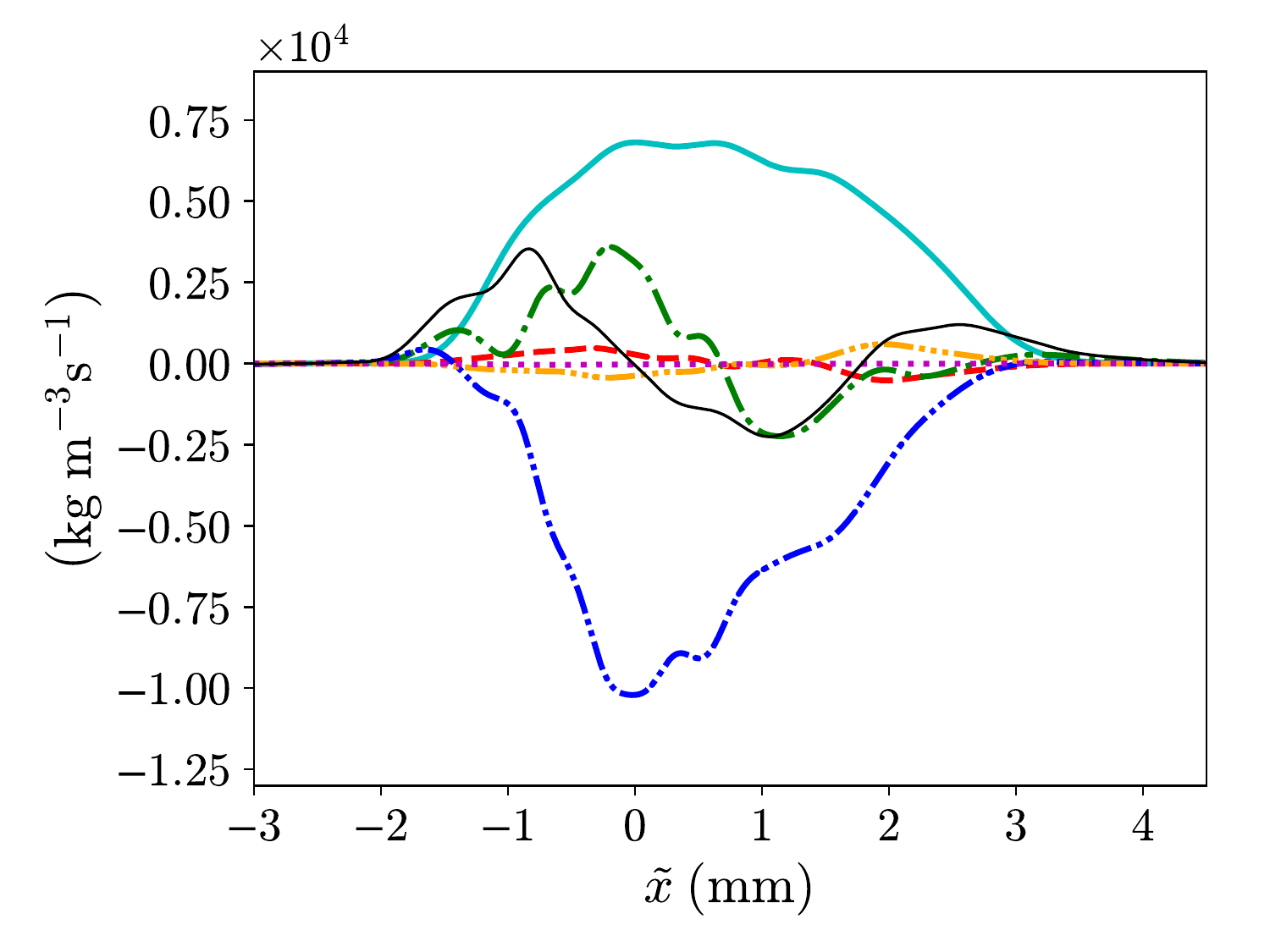}\label{fig:rho_b_budget_filtered_after_reshock}}
\subfigure[$\ t=1.60\ \mathrm{ms}$]{%
\includegraphics[width=0.4\textwidth]{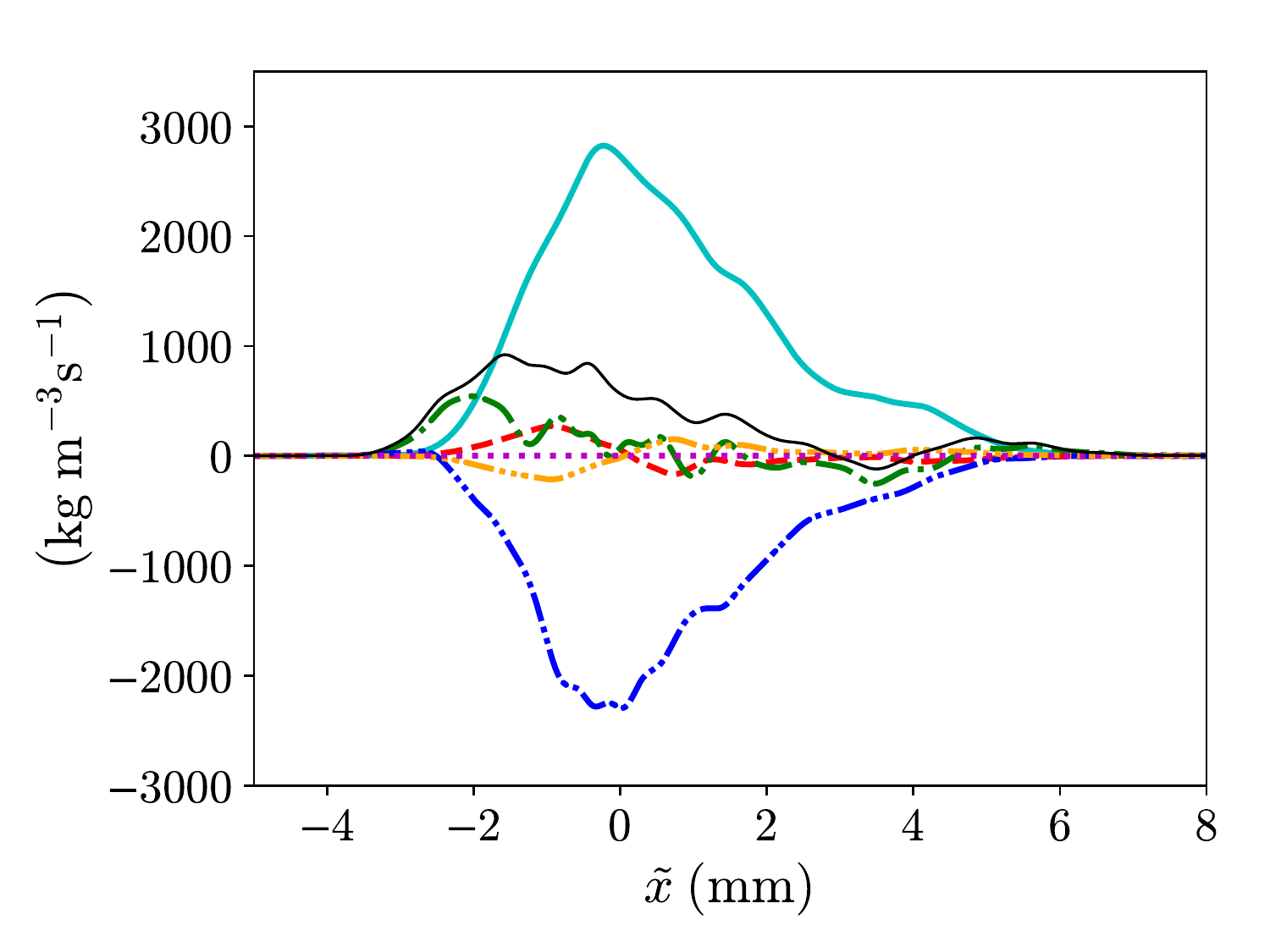}\label{fig:rho_b_budget_filtered_t_1_60}}
\caption{Budgets of the large-scale density-specific-volume covariance multiplied by the mean filtered density, $\overline{\left< \rho \right>}_{\ell} b_L$, given by equation~\eqref{eq:bL_transport_eqn_1D}, at different times after re-shock. Cyan solid line: production [term (III)]; red dashed line: redistribution [term (IV)]; green dash-dotted line: turbulent transport [term (V)]; blue dash-dot-dotted line: destruction [term (VI)]; orange dash-triple-dotted line: negative of convection due to streamwise velocity associated with turbulent mass flux; magenta dotted line: residue; thin black solid line: summation of all terms (rate of change in the moving frame).}
\label{fig:rho_b_budget_filtered}
\end{figure*}

Figure~\ref{fig:rho_b_budget_filtered} shows the spatial profiles of different terms that appear in the transport equation for the large-scale density-specific-volume covariance multiplied by the mean filtered density, $\overline{\left< \rho \right>}_{\ell} b_L$, given by equation~\eqref{eq:bL_transport_eqn_1D} after re-shock. As for the plots for budgets of the large-scale turbulent mass flux, the magenta dotted line represents the residue. As seen in the sub-figures, the residue is virtually zero at different times after re-shock and this means that there is an insignificant effect of numerical regularization on the rate of change of $\overline{\left< \rho \right>}_{\ell} b_L$.

As seen from the figure, the production [term (III)] and destruction [term (VI)] terms are the dominant terms in the interior region of the mixing layer.
In the papers by~\citet{tomkins2013evolution} and~\citet{mohaghar2019transition}, it was also observed in the layer interior that the production term is dominant in the budgets of density-specific-volume covariance.
In the interior part of the mixing layer, the rate of change of $\overline{\left< \rho \right>}_{\ell} b_L$ is negative just after re-shock as the magnitude of the negative destruction term is larger than that of the positive production term. Thus, the amplitude of the large-scale second-moment decreases just after re-shock. Nevertheless, soon after re-shock, the relative magnitude of the production term in the middle part of the mixing layer becomes larger, and even larger than that of the destruction term at late times. As a result, the rate of change of $\overline{\left< \rho \right>}_{\ell} b_L$ at the peak location turns slightly positive at later times.
Overall, the budget terms are quite balanced in the interior part of the mixing layer at late times as the production term roughly cancels the destruction term. 
This is similar to the observations on the budgets of density-specific-volume covariance in the heavy-light case of the spherical RMI~\cite{lombardini2014turbulent} and the planar RTI~\cite{livescu2009rti}. 
As a result, $\overline{\left< \rho \right>}_{\ell} b_L$ and $b_L$ have quite stationary peaks at late times.
Although the turbulent transport term [term (V)] is not small in the central part of the mixing layer, its effect is small compared to the production and destruction terms. However, the turbulent transport term becomes relatively more important at the heavier fluid edge of the mixing region. The term is positive at both edges of the mixing layer and is the vital term at the heavier fluid side for the spreading of $\overline{\left< \rho \right>}_{\ell} b_L$ over time.

\subsection{Large-scale Favre-averaged Reynolds stress and large-scale turbulent kinetic energy}

\begin{figure*}[!ht]
\centering
\subfigure[$\ t=1.20\ \mathrm{ms}$]{%
\includegraphics[width=0.4\textwidth]{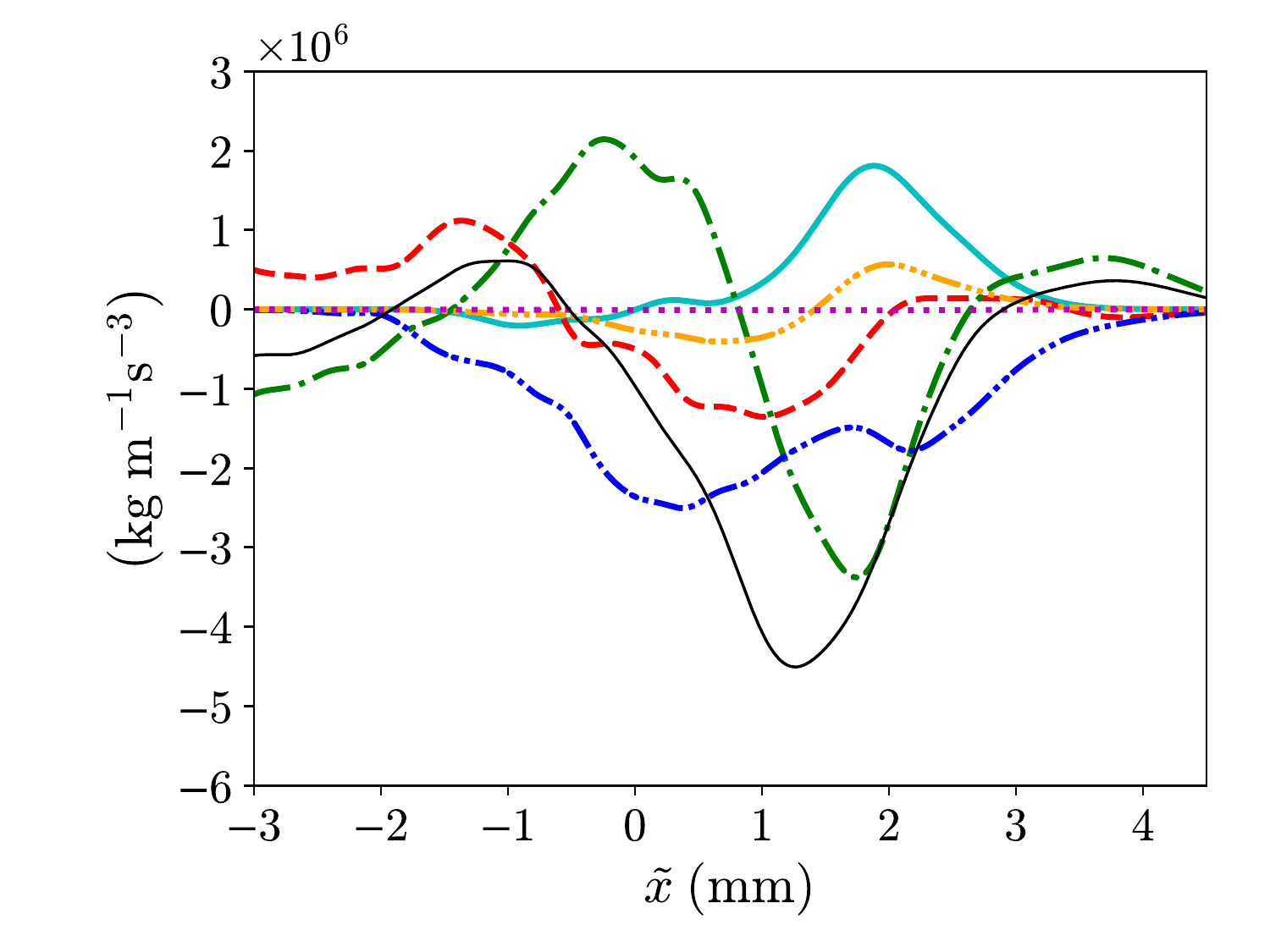}}
\subfigure[$\ t=1.60\ \mathrm{ms}$]{%
\includegraphics[width=0.4\textwidth]{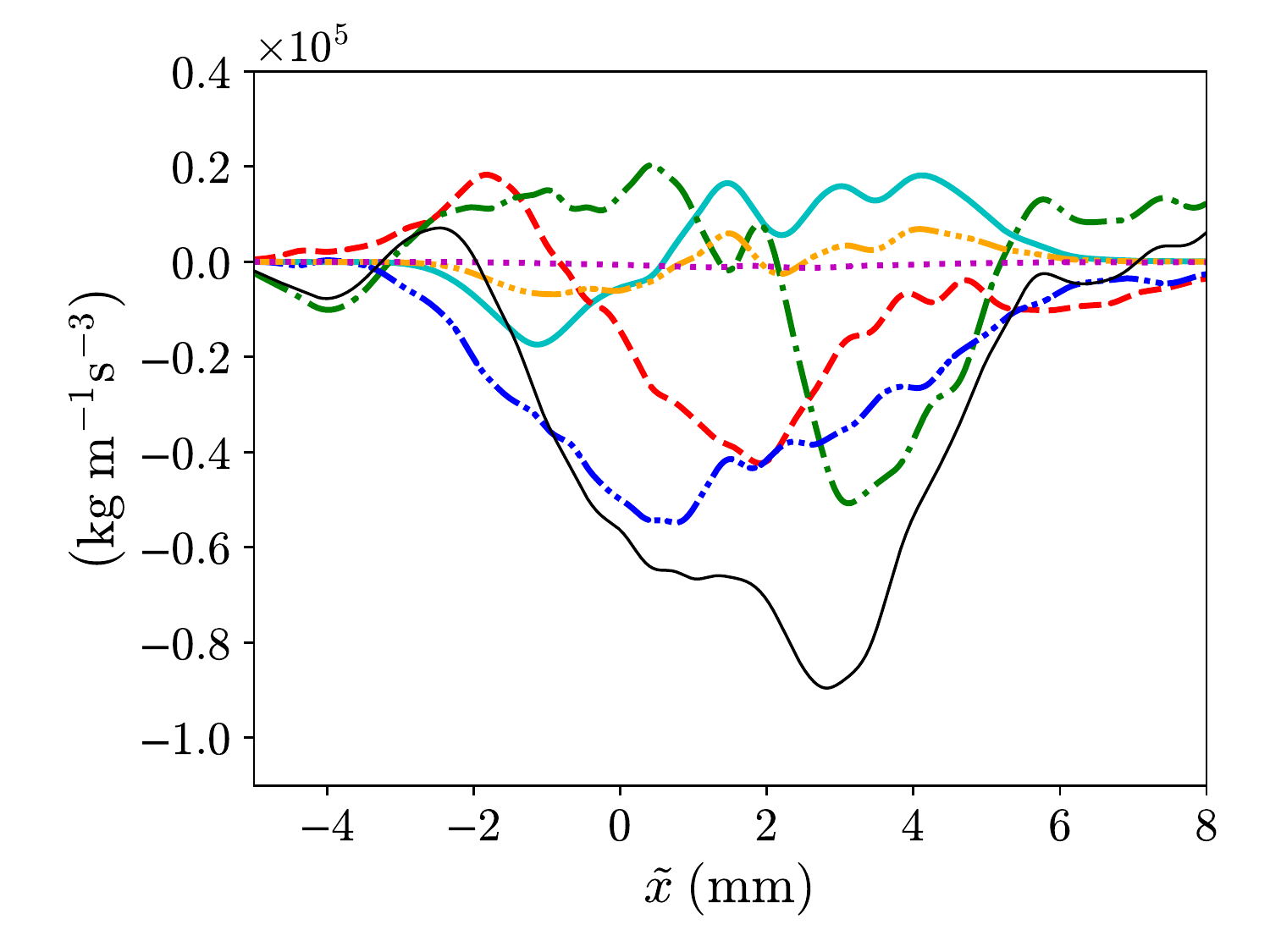}}
\caption{Budgets of the large-scale Favre-averaged Reynolds normal stress component in the streamwise direction multiplied by the mean filtered density, $\overline{\left< \rho \right>}_{\ell} \widetilde{R}_{L,11}$, given by equation~\eqref{eq:RL11_transport_eqn_1D}, at different times after re-shock. Cyan solid line: production [term (III)]; red dashed line: press-strain redistribution [term (V)]; green dash-dotted line: turbulent transport [term (IV)]; blue dash-dot-dotted line: dissipation [term (VI)]; orange dash-triple-dotted line: negative of convection due to streamwise velocity associated with turbulent mass flux; magenta dotted line: residue; thin black solid line: summation of all terms (rate of change in the moving frame).}
\label{fig:rho_R11_budget_filtered}
\end{figure*}

In figure~\ref{fig:rho_R11_budget_filtered}, the spatial profiles of different budget terms of the large-scale Favre-averaged Reynolds stress component in the streamwise direction multiplied by the mean filtered density, $\overline{\left< \rho \right>}_{\ell} \widetilde{R}_{L,11}$, after re-shock are shown. Each budget term in the transport equation for $\overline{\left< \rho \right>}_{\ell} \widetilde{R}_{L,11}$ is given by equation~\eqref{eq:RL11_transport_eqn_1D}. As shown in the figure, the residue, represented by the magenta dotted line, is basically zero at all times. Thus, the effect of numerical regularization on $\overline{\left< \rho \right>}_{\ell} \widetilde{R}_{L,11}$ can be ignored.

From the figure, we can see that all terms except the convection term play significant roles in the rate of change of $\overline{\left< \rho \right>}_{\ell} \widetilde{R}_{L,11}$ in the interior part of the mixing region. Similar to the budgets before re-shock, the production term [term (III)] is positive on the light fluid side and negative on the heavy fluid side to transport $\overline{\left< \rho \right>}_{\ell} \widetilde{R}_{L,11}$ from the heavier fluid side to the lighter fluid side. On the other hand, generally the turbulent transport term [term (IV)] has a larger magnitude but the opposite effect in the interior region of the mixing layer compared to the production term. In the same region, both pressure-strain redistribution [term (V)] and dissipation [term (VI)] terms are negative in general and hence the overall rate of change is negative. At the edges of the mixing layer, only the turbulent transport and pressure-strain redistribution are critical terms. Their combined effect contributes to the spreading of $\overline{\left< \rho \right>}_{\ell} \widetilde{R}_{L,11}$ on the lighter fluid side over time while there is some anti-spreading effect on the heavier fluid side for quite a long period of time after re-shock.

\begin{figure*}[!ht]
\centering
\subfigure[$\ t=1.20\ \mathrm{ms}$]{%
\includegraphics[width=0.4\textwidth]{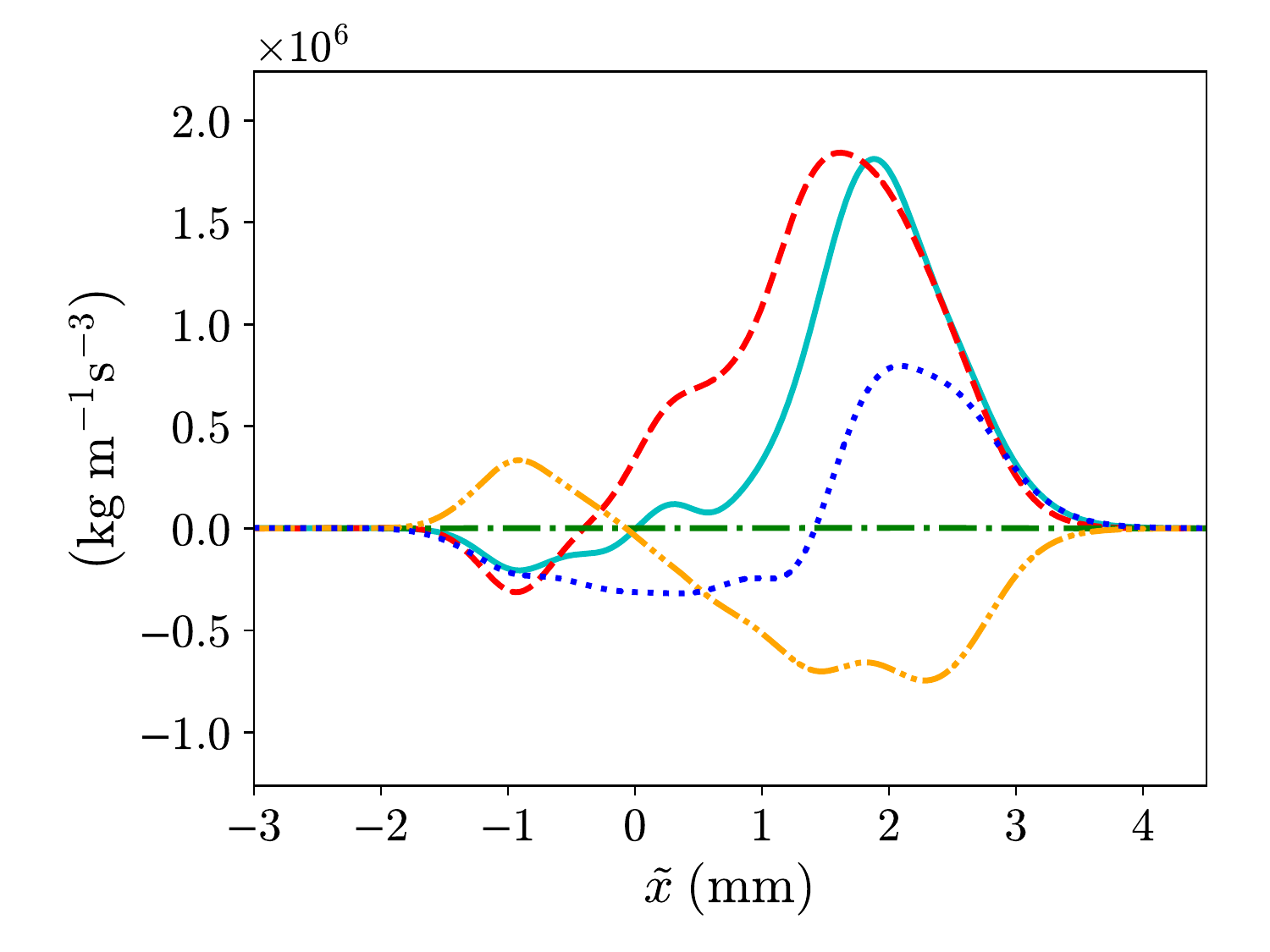}\label{fig:rho_R11_budget_filtered_production_terms_after_reshock}}
\subfigure[$\ t=1.60\ \mathrm{ms}$]{%
\includegraphics[width=0.4\textwidth]{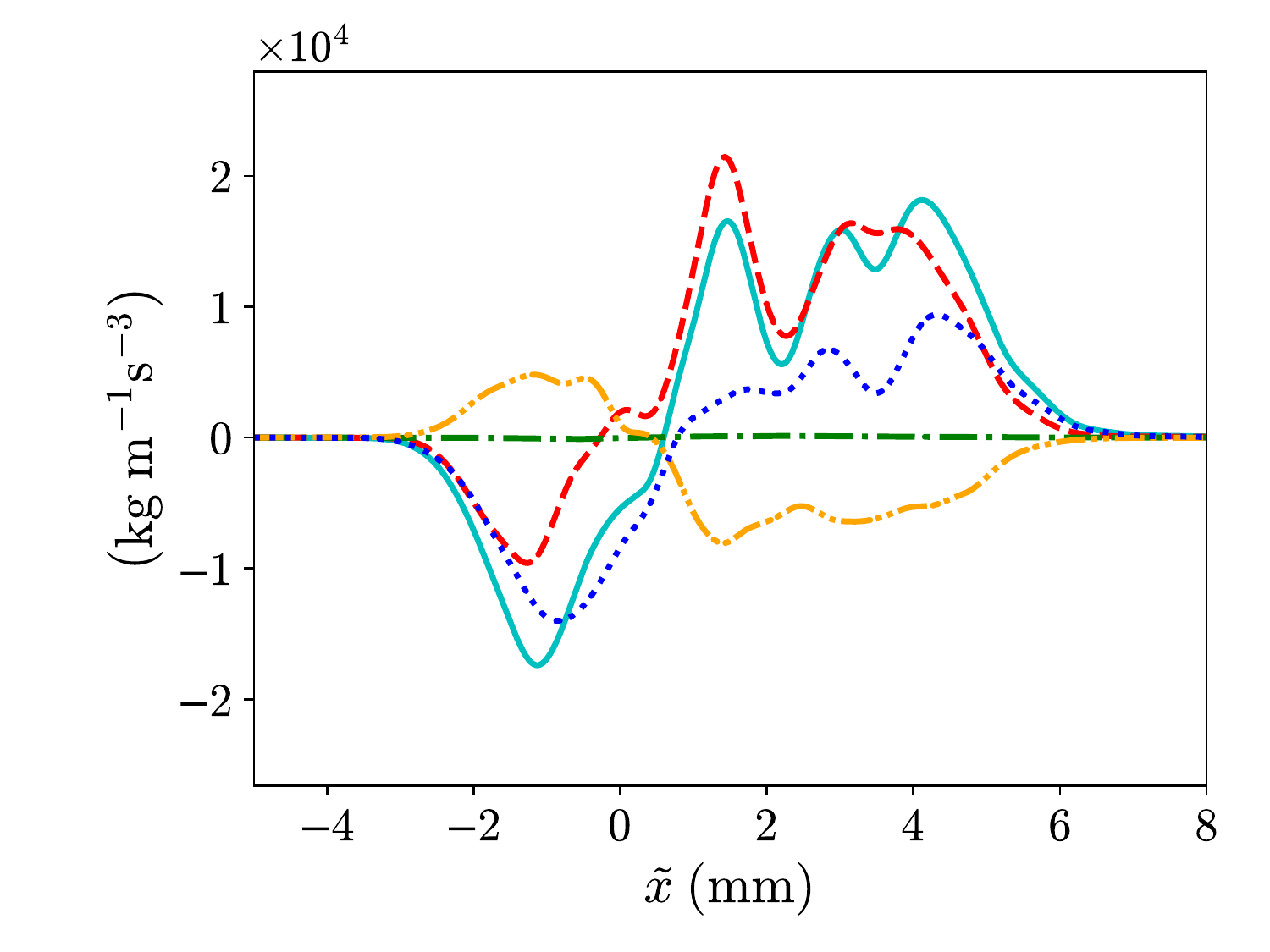}\label{fig:rho_R11_budget_filtered_production_terms_t_1_60}}
\caption{Compositions of the production term [term (III)] in the transport equation for the large-scale Favre-averaged Reynolds normal stress component in the streamwise direction multiplied by the mean filtered density, $\overline{\left< \rho \right>}_{\ell} \widetilde{R}_{L,11}$, at different times after re-shock. Cyan solid line: overall production; red dashed line: $2a_{L,1} \overline{\left< p \right>}_{\ell,1}$; green dash-dotted line: $-2a_{L,1} \overline{\left< \tau_{11} \right>}_{\ell,1}$; orange dash-dot-dotted line: $2a_{L,1} {\overline{ \tau_{11}^{SFS} }}_{,1}$; blue dotted line: $-2\overline{\left< \rho \right>}_{\ell} \widetilde{R}_{L,11} \widetilde{\left< u \right>}_{L,1}$.}
\label{fig:rho_R11_budget_filtered_production_terms}
\end{figure*}

\begin{figure*}[!ht]
\centering
\subfigure[$\ t=1.20\ \mathrm{ms}$]{%
\includegraphics[width=0.4\textwidth]{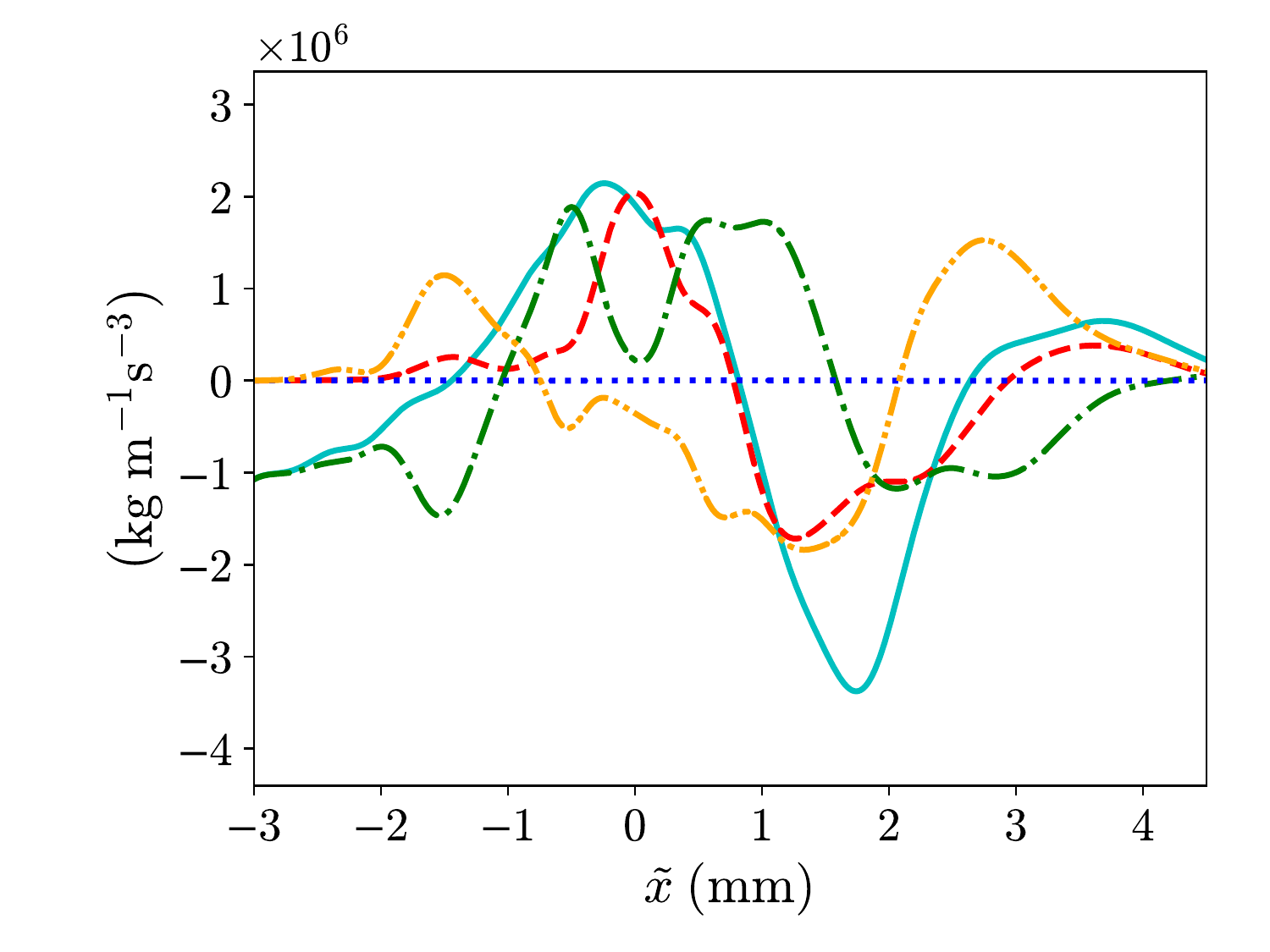}\label{fig:rho_R11_budget_filtered_turb_transport_terms_after_reshock}}
\subfigure[$\ t=1.60\ \mathrm{ms}$]{%
\includegraphics[width=0.4\textwidth]{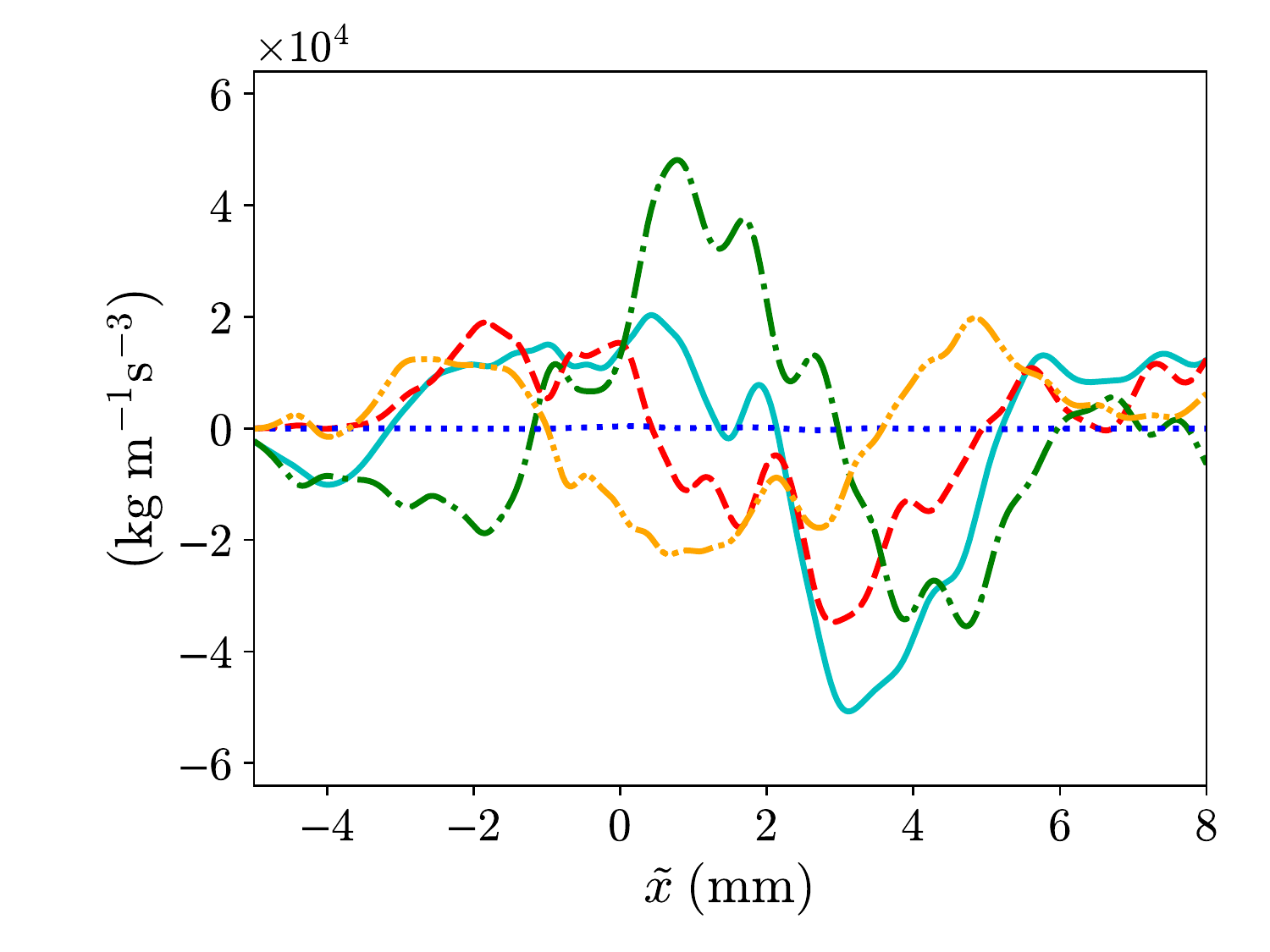}\label{fig:rho_R11_budget_filtered_turb_transport_terms_t_1_60}}
\caption{Compositions of the turbulent transport term [term (IV)] in the transport equation for the large-scale Favre-averaged Reynolds normal stress component in the streamwise direction multiplied by the mean filtered density, $\overline{\left< \rho \right>}_{\ell} \widetilde{R}_{L,11}$, at different times after re-shock. Cyan solid line: overall turbulent transport; red dashed line: $- ( \overline{ \left< \rho \right>_{\ell} \left< u \right>_{L}^{\prime\prime} \left< u \right>_{L}^{\prime\prime} \left< u \right>_{L}^{\prime\prime} } )_{,1}$; green dash-dotted line: $-2 ( \overline{\left< u \right>_{L}^{\prime} \left< p \right>_{\ell}^{\prime}} )_{,1}$; blue dotted line: $2 ( \overline{ \left< u \right>_{L}^{\prime} \left< \tau_{11} \right>_{\ell}^{\prime} } )_{,1}$; orange dash-dot-dotted line: $-2 ( \overline{ \left< u \right>_{L}^{\prime} {\tau_{11}^{SFS}}^{\prime} } )_{,1}$.}
\label{fig:rho_R11_budget_filtered_turb_transport_terms}
\end{figure*}

Figures~\ref{fig:rho_R11_budget_filtered_production_terms} and \ref{fig:rho_R11_budget_filtered_turb_transport_terms} show the compositions of production [term (III)] and turbulent transport [term (IV)] terms, respectively, after re-shock. Both figures show that the components due to filtered molecular shear stress, $-2a_{L,1} \overline{\left< \tau_{11} \right>}_{\ell,1}$ and $2 ( \overline{ \left< u \right>_{L}^{\prime} \left< \tau_{11} \right>_{\ell}^{\prime} } )_{,1}$, are insignificant to the budgets at different times after re-shock. Considering the composition of the production term in figure~\ref{fig:rho_R11_budget_filtered_production_terms}, the component with SFS stress, $2a_{L,1} {\overline{ \tau_{11}^{SFS} }}_{,1}$, appears as a new term compared to the budgets without filtering before re-shock. Both constituents $2a_{L,1} \overline{\left< p \right>}_{\ell,1}$ and $-2\overline{\left< \rho \right>}_{\ell} \widetilde{R}_{L,11} \widetilde{\left< u \right>}_{L,1}$ play similar roles to the production term. They are negative on the heavier fluid side and positive on the lighter fluid side. However, the former has a larger effect on the lighter side while the effect of the latter is stronger on the heavier fluid side. The component with SFS stress has a similar magnitude to $-2\overline{\left< \rho \right>}_{\ell} \widetilde{R}_{L,11} \widetilde{\left< u \right>}_{L,1}$ but the opposite effect that brings $\overline{\left< \rho \right>}_{\ell} \widetilde{R}_{L,11}$ from the lighter fluid side to the heavier fluid side.
Inspecting the composition of the turbulent transport term, the three constituents, $-( \overline{ \left< \rho \right>_{\ell} \left< u \right>_{L}^{\prime\prime} \left< u \right>_{L}^{\prime\prime} \left< u \right>_{L}^{\prime\prime} } )_{,1}$, $-2 ( \overline{\left< u \right>_{L}^{\prime} \left< p \right>_{\ell}^{\prime}} )_{,1}$, and $-2 ( \overline{ \left< u \right>_{L}^{\prime} {\tau_{11}^{SFS}}^{\prime} } )_{,1}$, have significant contributions to the term after re-shock. The triple velocity correlation component, $-( \overline{ \left< \rho \right>_{\ell} \left< u \right>_{L}^{\prime\prime} \left< u \right>_{L}^{\prime\prime} \left< u \right>_{L}^{\prime\prime} } )_{,1}$, and the component arisen from filtering, $-2 ( \overline{ \left< u \right>_{L}^{\prime} {\tau_{11}^{SFS}}^{\prime} } )_{,1}$, are responsible for the spreading of $\overline{\left< \rho \right>}_{\ell} \widetilde{R}_{L,11}$. On the other hand, the constituent $-2 ( \overline{\left< u \right>_{L}^{\prime} \left< p \right>_{\ell}^{\prime}} )_{,1}$ has an anti-spreading effect.

\begin{figure*}[!ht]
\centering
\subfigure[$\ t=1.20\ \mathrm{ms}$]{%
\includegraphics[width=0.4\textwidth]{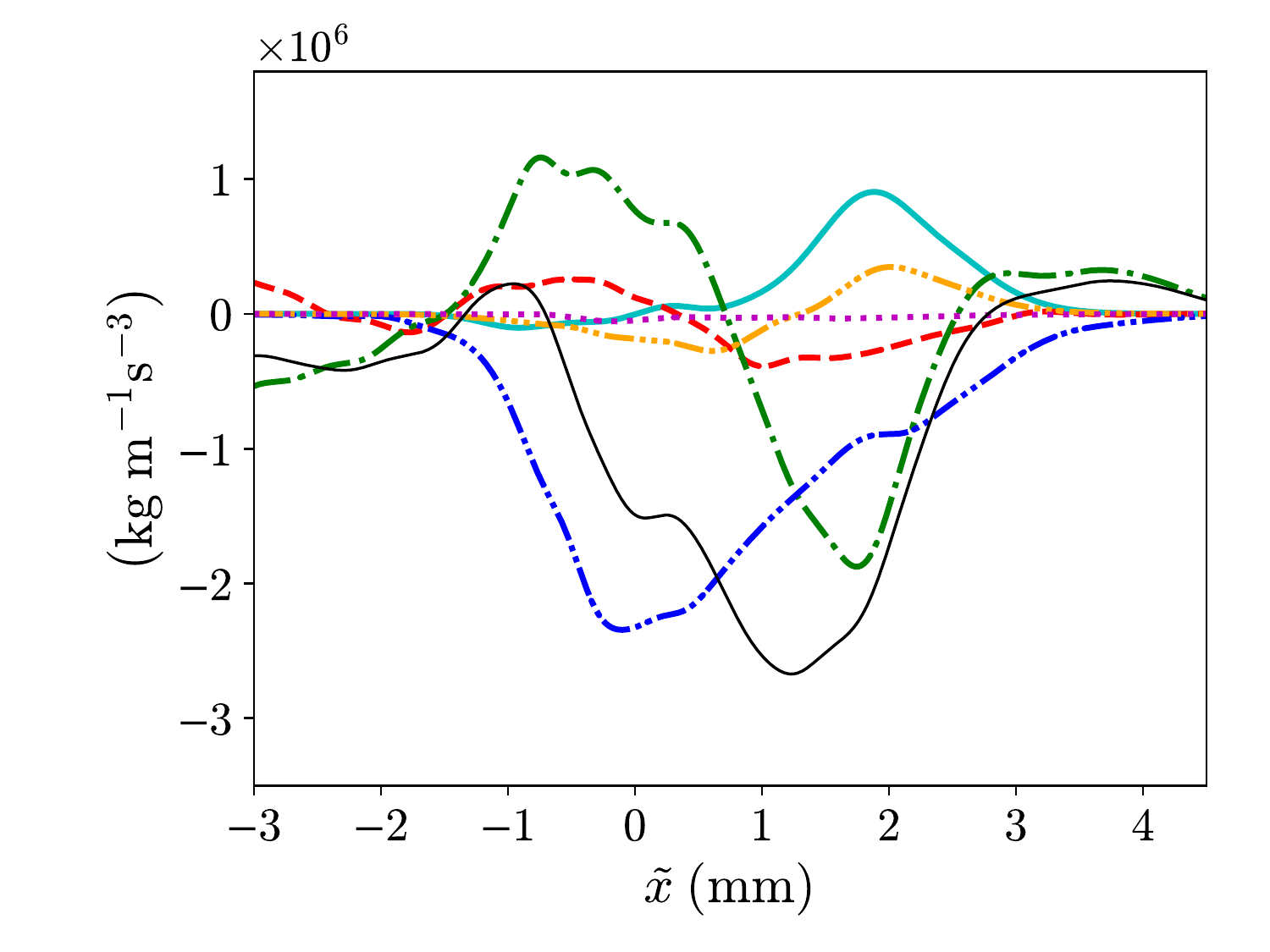}}
\subfigure[$\ t=1.60\ \mathrm{ms}$]{%
\includegraphics[width=0.4\textwidth]{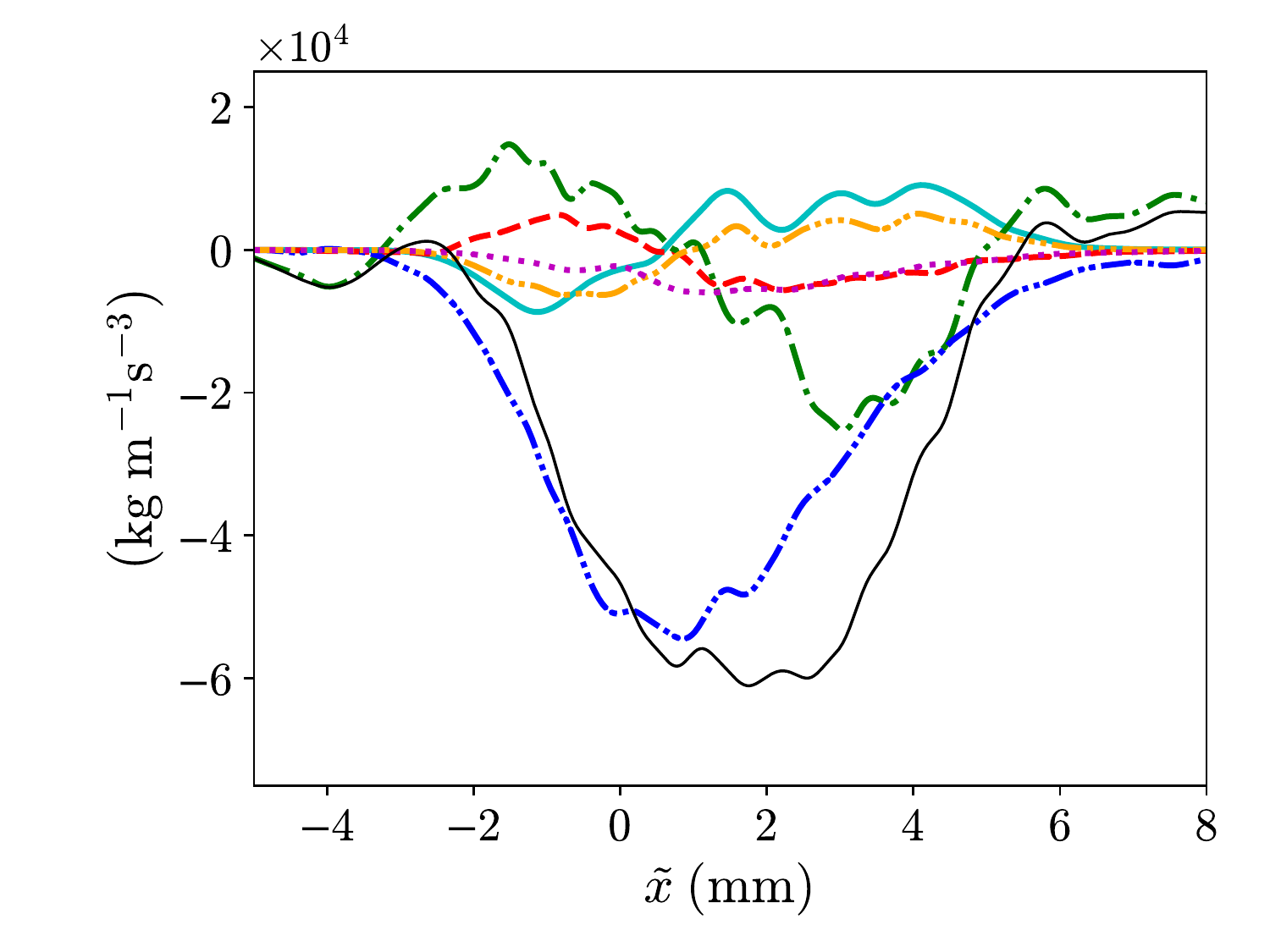}}
\caption{Budgets of the large-scale turbulent kinetic energy, $\overline{\left< \rho \right>}_{\ell} k_L$, given by equation~\eqref{eq:kL_transport_eqn_1D}, at different times after re-shock. Cyan solid line: production [term (III)]; red dashed line: pressure-dilatation [term (V)]; green dash-dotted line: turbulent transport [term (IV)]; blue dash-dot-dotted line: dissipation [term (VI)]; orange dash-triple-dotted line: negative of convection due to streamwise velocity associated with turbulent mass flux; magenta dotted line: residue; thin black solid line: summation of all terms (rate of change in the moving frame).}
\label{fig:rho_k_budget_filtered}
\end{figure*}

Figure~\ref{fig:rho_k_budget_filtered} shows a comparison of different budget terms in the transport equation for the large-scale turbulent kinetic energy, $\overline{\left< \rho \right>}_{\ell} k_L$, given by equation~\eqref{eq:kL_transport_eqn_1D}.
The residue of $\overline{\left< \rho \right>}_{\ell} k_L$ is negligible at early times after re-shock but becomes slightly larger at later times. At $t=1.60\ \mathrm{ms}$, the residue cannot be considered as zero but is still small compared with other budget terms. Through grid sensitivity analysis presented in the Supplemental Material~\cite{supple2022wong}, it is found that the residue computed with the grid E is largely reduced compared to that with the grid D.
It should be noted again that in incompressible single-species flow, the pressure-dilatation term [term (V)] is zero. As seen from the figure, the pressure-dilatation term in the variable-density flow being studied here is not zero. However, its influence is generally very small across the mixing layer and its effect is roughly canceled by the convection term. The effect from pressure-dilatation is commonly ignored in many RANS-based models~\cite{morgan2015three,banerjee2010development} for RMI-induced turbulence. In the interior part of the mixing region, the dissipation dominates the overall rate of change of the large-scale turbulent kinetic energy, and the quantity decays over time. Note that the dissipation term is contributed mainly by the component with SFS stress, $\overline{ { \tau_{ij}^{SFS} }^{\prime} ( \partial \left< u_i \right>_{L}^{\prime} / \partial x_j) }$. In the interior region of RTI~\cite{livescu2009rti}, the production and dissipation terms are equally important in the turbulent kinetic energy budget, while the former has a small contribution for the RMI turbulence studied in this work. The production term is large over time in RTI and buoyancy-driven variable-density turbulence due to the continuous conversion of potential energy into kinetic energy~\cite{livescu2009rti,livescu2007buoyancy,aslangil2020effects}. However, this mechanism does not exist in RMI.

\section{Effects of filtering on the budgets of the large-scale second-moments after re-shock}

\begin{figure*}[!ht]
\centering
\subfigure[$\ \ell \approx 16 \Delta$]{%
\includegraphics[width=0.4\textwidth]{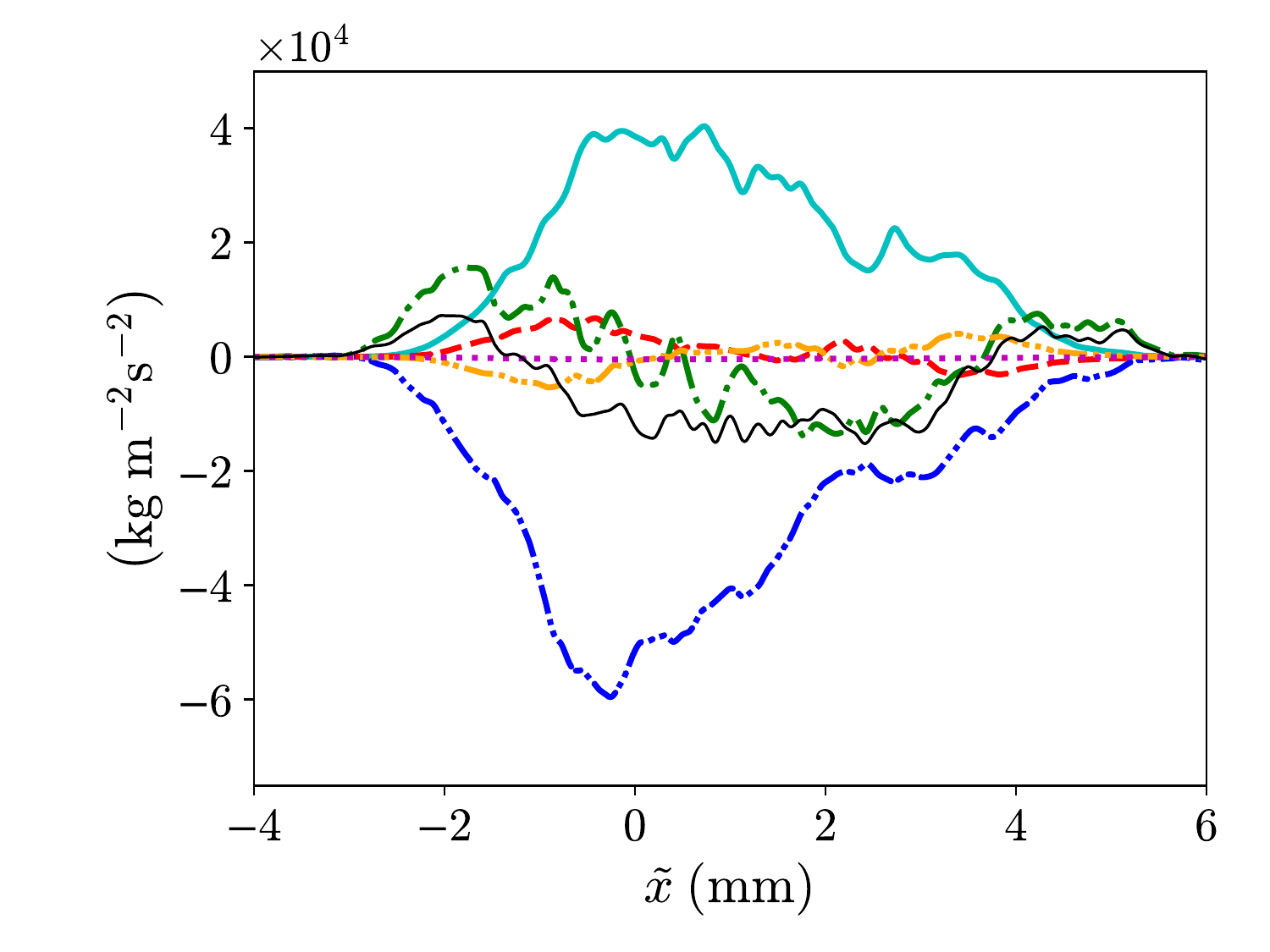}\label{fig:rho_a1_budget_filtered_t_1_40_016x}}
\subfigure[$\ \ell \approx 64 \Delta$]{%
\includegraphics[width=0.4\textwidth]{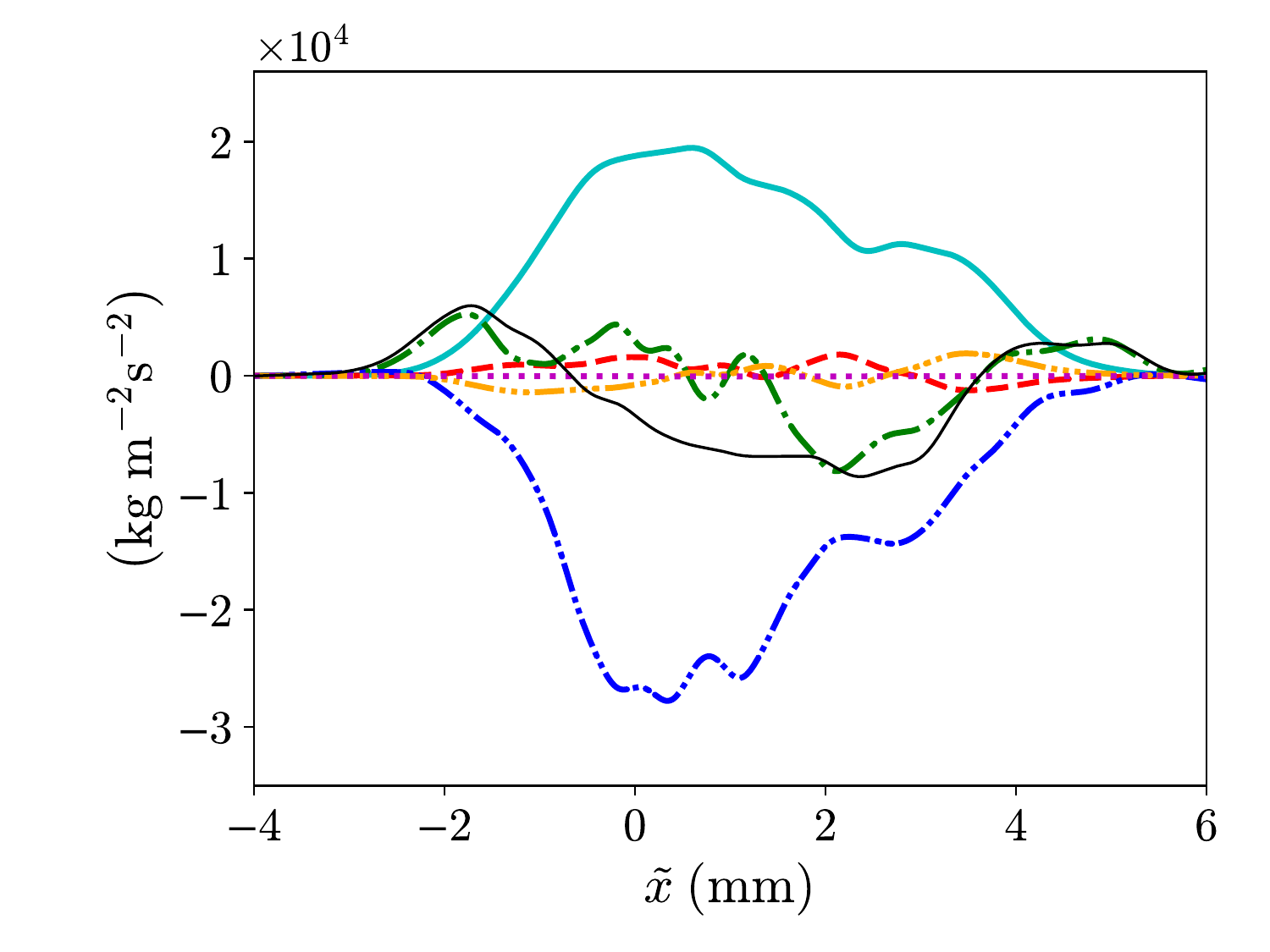}\label{fig:rho_a1_budget_filtered_t_1_40_256x}}
\caption{Effect of filtering on the budgets of the large-scale turbulent mass flux component in the streamwise direction, $\overline{\left< \rho \right>}_{\ell} a_{L,1}$, given by equation~\eqref{eq:aL1_transport_eqn_1D}, at $t =1.40\ \mathrm{ms}$. Cyan solid line: production [term (III)]; red dashed line: redistribution [term (IV)]; green dash-dotted line: turbulent transport [term (V)]; blue dash-dot-dotted line: destruction [term (VI)]; orange dash-triple-dotted line: negative of convection due to streamwise velocity associated with turbulent mass flux; magenta dotted line: residue; thin black solid line: summation of all terms (rate of change in the moving frame).}
\label{fig:rho_a1_budget_filtered_effect_filtering}
\end{figure*}

\begin{figure*}[!ht]
\centering
\subfigure[$\ \ell \approx 16 \Delta$]{%
\includegraphics[width=0.4\textwidth]{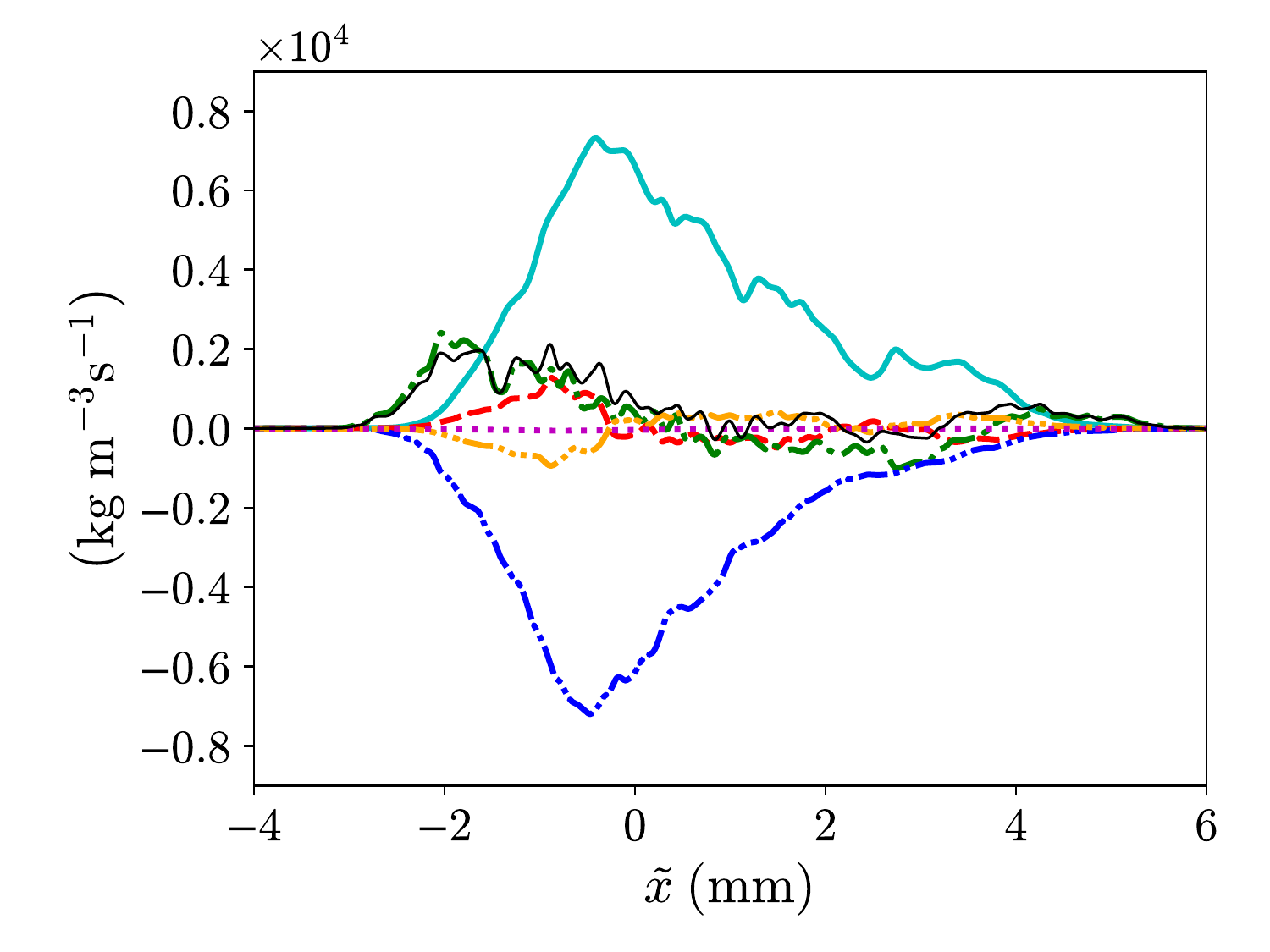}\label{fig:rho_b_budget_filtered_t_1_40_016x}}
\subfigure[$\ \ell \approx 64 \Delta$]{%
\includegraphics[width=0.4\textwidth]{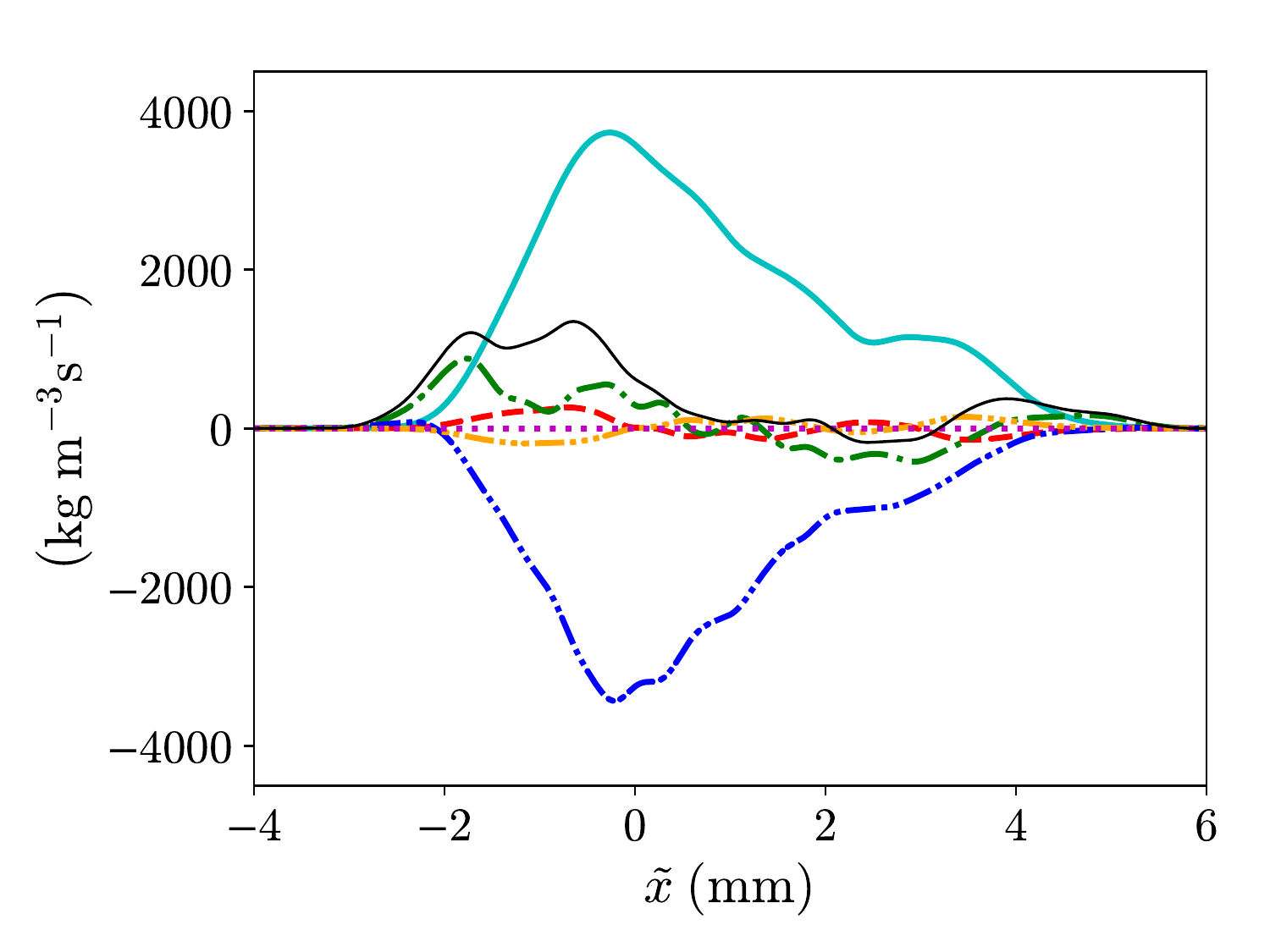}\label{fig:rho_b_budget_filtered_t_1_40_256x}}
\caption{Effect of filtering on the budgets of the large-scale density-specific-volume covariance multiplied by the mean filtered density, $\overline{\left< \rho \right>}_{\ell} b_L$, given by equation~\eqref{eq:bL_transport_eqn_1D}, at $t =1.40\ \mathrm{ms}$. Cyan solid line: production [term (III)]; red dashed line: redistribution [term (IV)]; green dash-dotted line: turbulent transport [term (V)]; blue dash-dot-dotted line: destruction [term (VI)]; orange dash-triple-dotted line: negative of convection due to streamwise velocity associated with turbulent mass flux; magenta dotted line: residue; thin black solid line: summation of all terms (rate of change in the moving frame).}
\label{fig:rho_b_budget_filtered_effect_filtering}
\end{figure*}

\begin{figure*}[!ht]
\centering
\subfigure[$\ \ell \approx 16 \Delta$]{%
\includegraphics[width=0.4\textwidth]{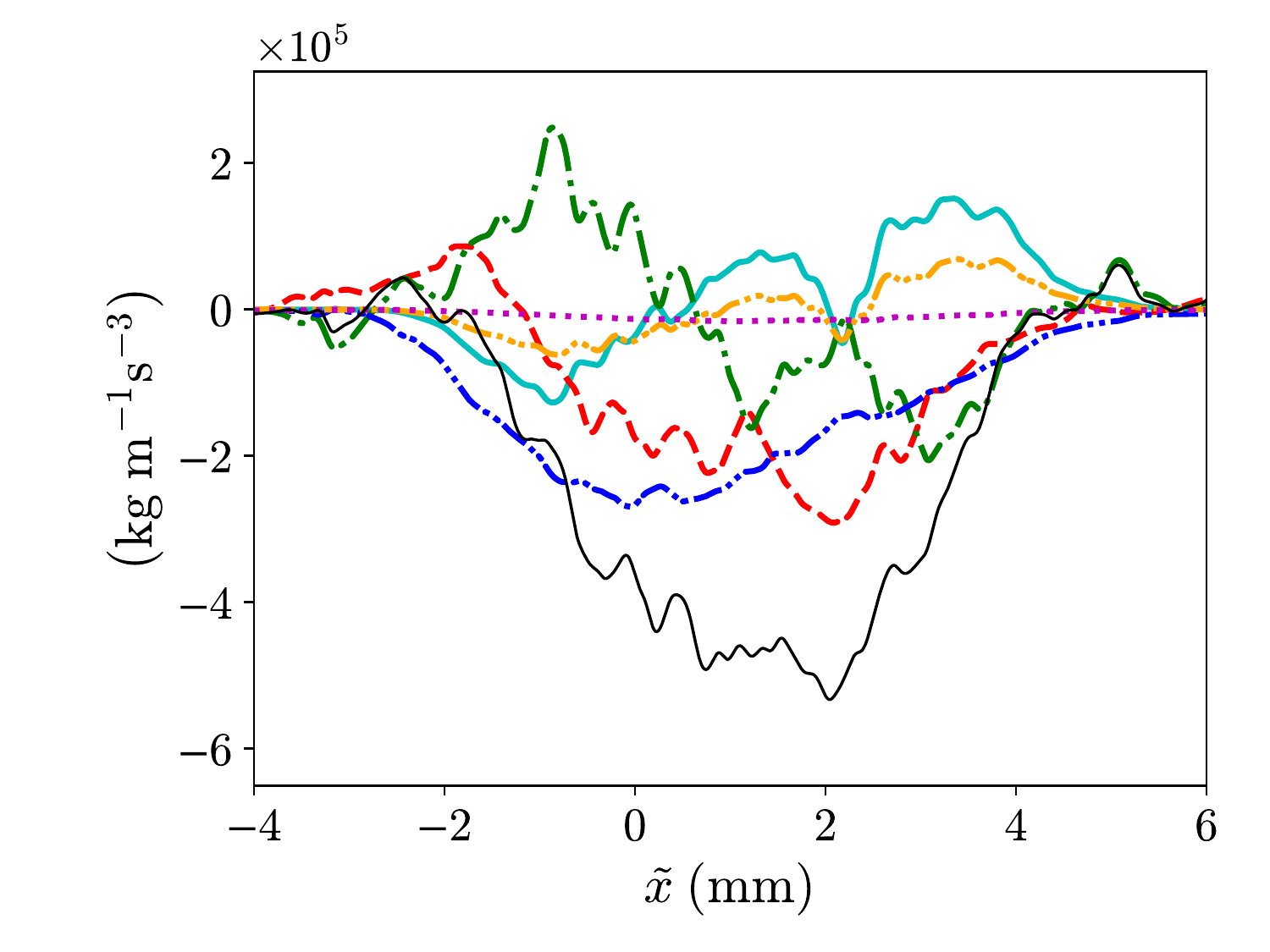}}
\subfigure[$\ \ell \approx 64 \Delta$]{%
\includegraphics[width=0.4\textwidth]{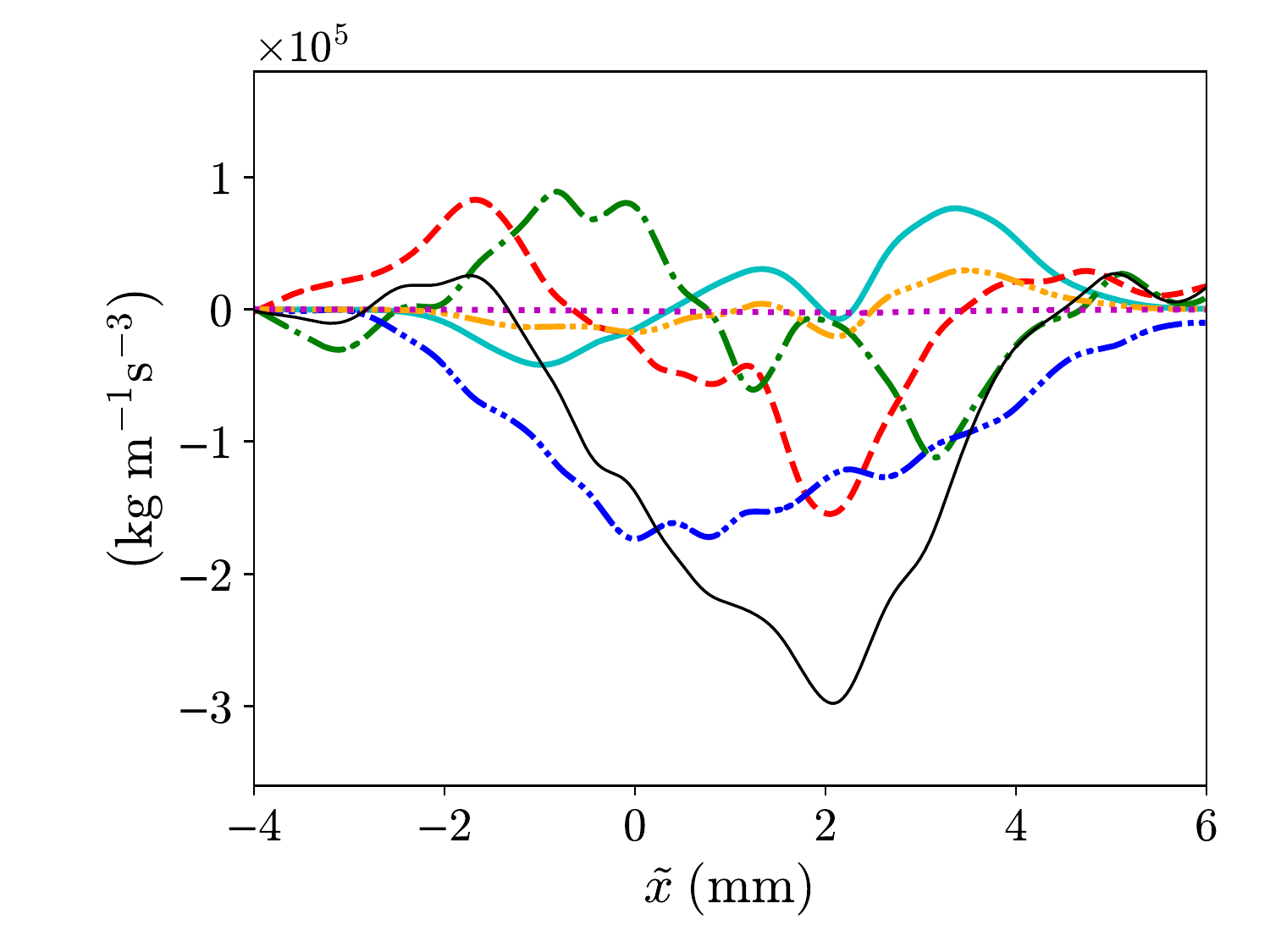}}
\caption{Effect of filtering on the budgets of the large-scale Reynolds normal stress component in the streamwise direction multiplied by the mean filtered density, $\overline{\left< \rho \right>}_{\ell} \widetilde{R}_{L,11}$, given by equation~\eqref{eq:RL11_transport_eqn_1D}, at $t =1.40\ \mathrm{ms}$. Cyan solid line: production [term (III)]; red dashed line: press-strain redistribution [term (V)]; green dash-dotted line: turbulent transport [term (IV)]; blue dash-dot-dotted line: dissipation [term (VI)]; orange dash-triple-dotted line: negative of convection due to streamwise velocity associated with turbulent mass flux; magenta dotted line: residue; thin black solid line: summation of all terms (rate of change in the moving frame).}
\label{fig:rho_R11_budget_filtered_effect_filtering}
\end{figure*}

\begin{figure*}[!ht]
\centering
\subfigure[$\ \ell \approx 16 \Delta$]{%
\includegraphics[width=0.4\textwidth]{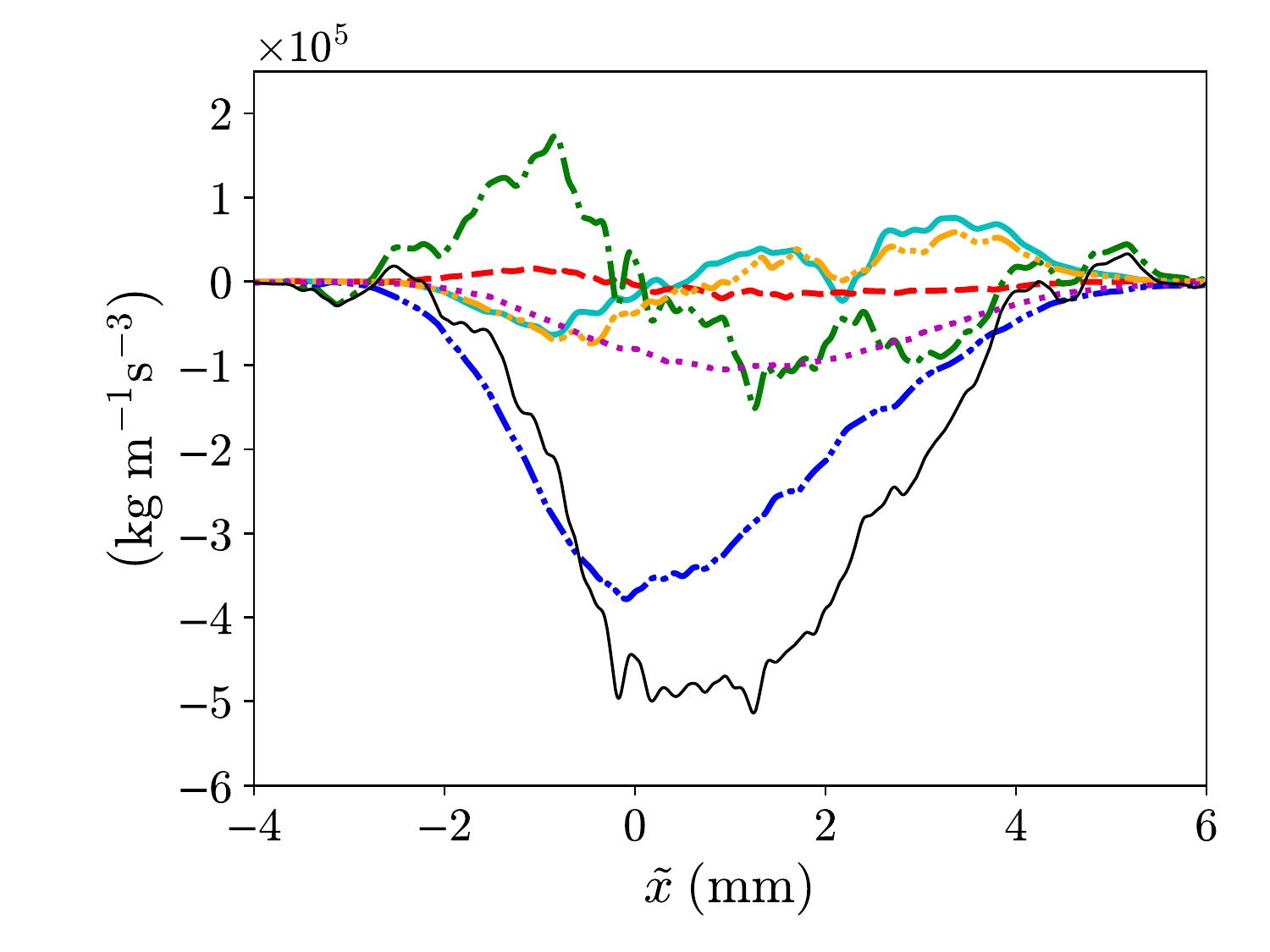}}
\subfigure[$\ \ell \approx 64 \Delta$]{%
\includegraphics[width=0.4\textwidth]{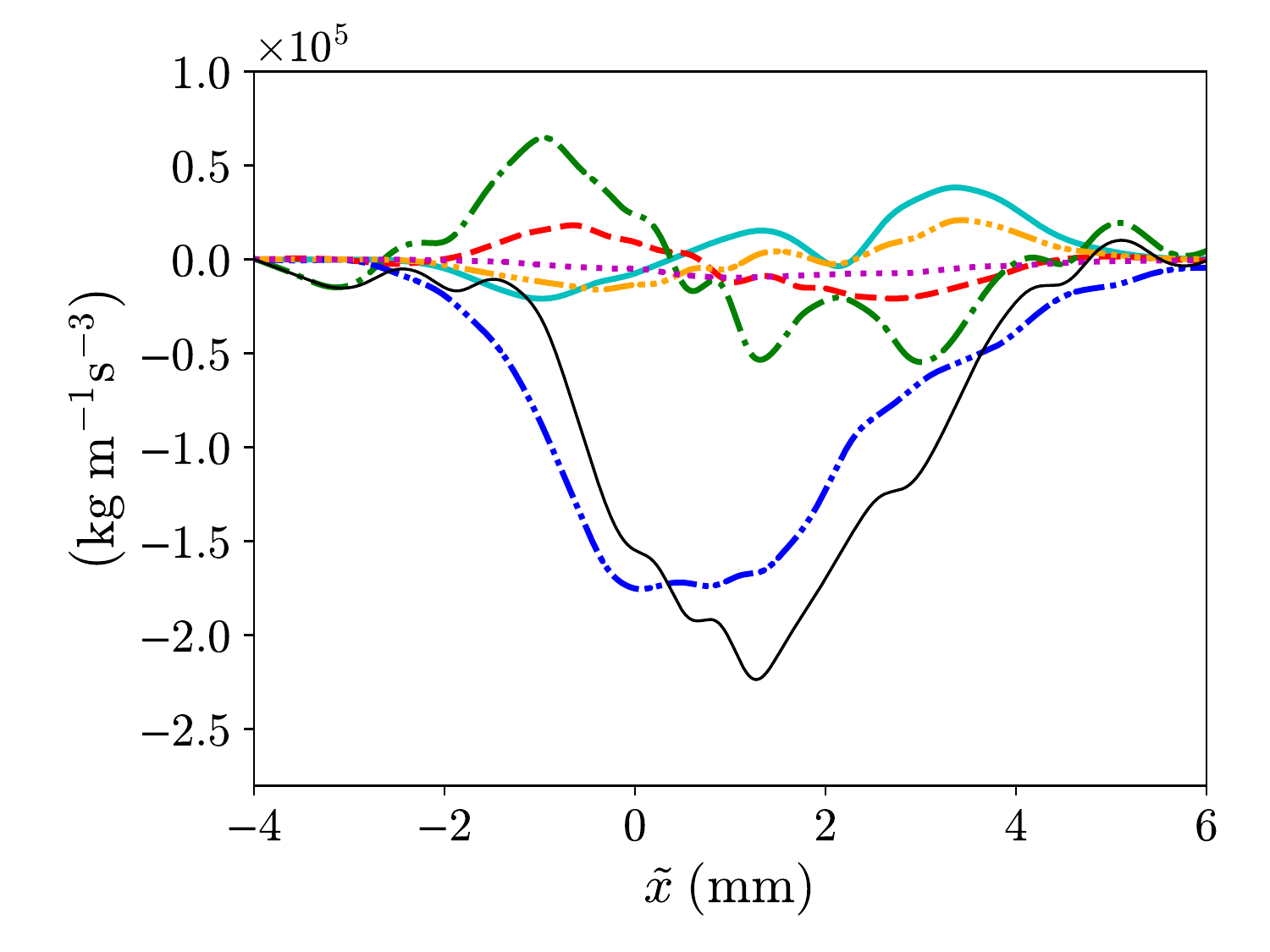}}
\caption{Effect of filtering on the budgets of the large-scale turbulent kinetic energy, $\overline{\left< \rho \right>}_{\ell} k_L$, given by equation~\eqref{eq:kL_transport_eqn_1D}, at $t =1.40\ \mathrm{ms}$. Cyan solid line: production [term (III)]; red dashed line: pressure-dilatation [term (V)]; green dash-dotted line: turbulent transport [term (IV)]; blue dash-dot-dotted line: dissipation [term (VI)]; orange dash-triple-dotted line: negative of convection due to streamwise velocity associated with turbulent mass flux; magenta dotted line: residue; thin black solid line: summation of all terms (rate of change in the moving frame).}
\label{fig:rho_k_budget_filtered_effect_filtering}
\end{figure*}

The effects of filtering on the budgets of different large-scale second-moments and turbulent kinetic energy with $\ell \approx 16 \Delta$ and $\ell \approx 64 \Delta$ at $t = 1.40\ \mathrm{ms}$ after re-shock are shown in figures~\ref{fig:rho_a1_budget_filtered_effect_filtering}, \ref{fig:rho_b_budget_filtered_effect_filtering}, \ref{fig:rho_R11_budget_filtered_effect_filtering}, and \ref{fig:rho_k_budget_filtered_effect_filtering} respectively. Note that the unfiltered budgets and the filtered budgets with another filter width can be found in the Supplemental Material~\cite{supple2022wong}.

It should be mentioned that the budget of $\bar{\rho} a_1$ is already closed when no filtering is used and hence from figure~\ref{fig:rho_a1_budget_filtered_effect_filtering}, it can be seen that the residues in the budgets of the corresponding large-scale turbulent mass flux component with different filter widths are also negligible across the entire mixing region. Similar to the effects of the filter on the large-scale second-moments, the magnitudes of different terms in the transport equation for the large-scale turbulent mass flux component decrease when a larger filter width is applied on the mixture density and momentum equations, but their shapes remain quite similar.
From figures~\ref{fig:rho_b_budget_filtered_effect_filtering} and  \ref{fig:rho_R11_budget_filtered_effect_filtering}, it can be seen that the residues in the budgets of the density-specific-volume covariance and the Reynolds normal stress component (both multiplied by the mean filtered density) are already virtually zero when the Navier--Stokes equations are filtered with filter width $\ell \approx 16 \Delta$. The shapes of different terms in the budgets of the two large-scale second-moments also appear similar and the magnitudes reduce with a larger filter width. As for the budgets of large-scale turbulent kinetic energy, an even larger filter width, or more successive filtering operations are needed for the residue to become negligibly small, which is indicated by figure~\ref{fig:rho_k_budget_filtered_effect_filtering}. Nevertheless, the budget terms of the large-scale turbulent kinetic energy are also quite similar for filter widths $\ell \approx 16 \Delta$ and $\ell \approx 64 \Delta$.
From all of these figures mentioned above, it can also be noticed that the ratios of the magnitudes between the major terms for each budget do not change much with different filter widths. Thus, the budget terms in each transport equation are quite self-similar with different degrees of filtering.

\begin{figure*}[!ht]
\centering
\subfigure[$\ \ell \approx 16 \Delta$]{%
\includegraphics[width=0.4\textwidth]{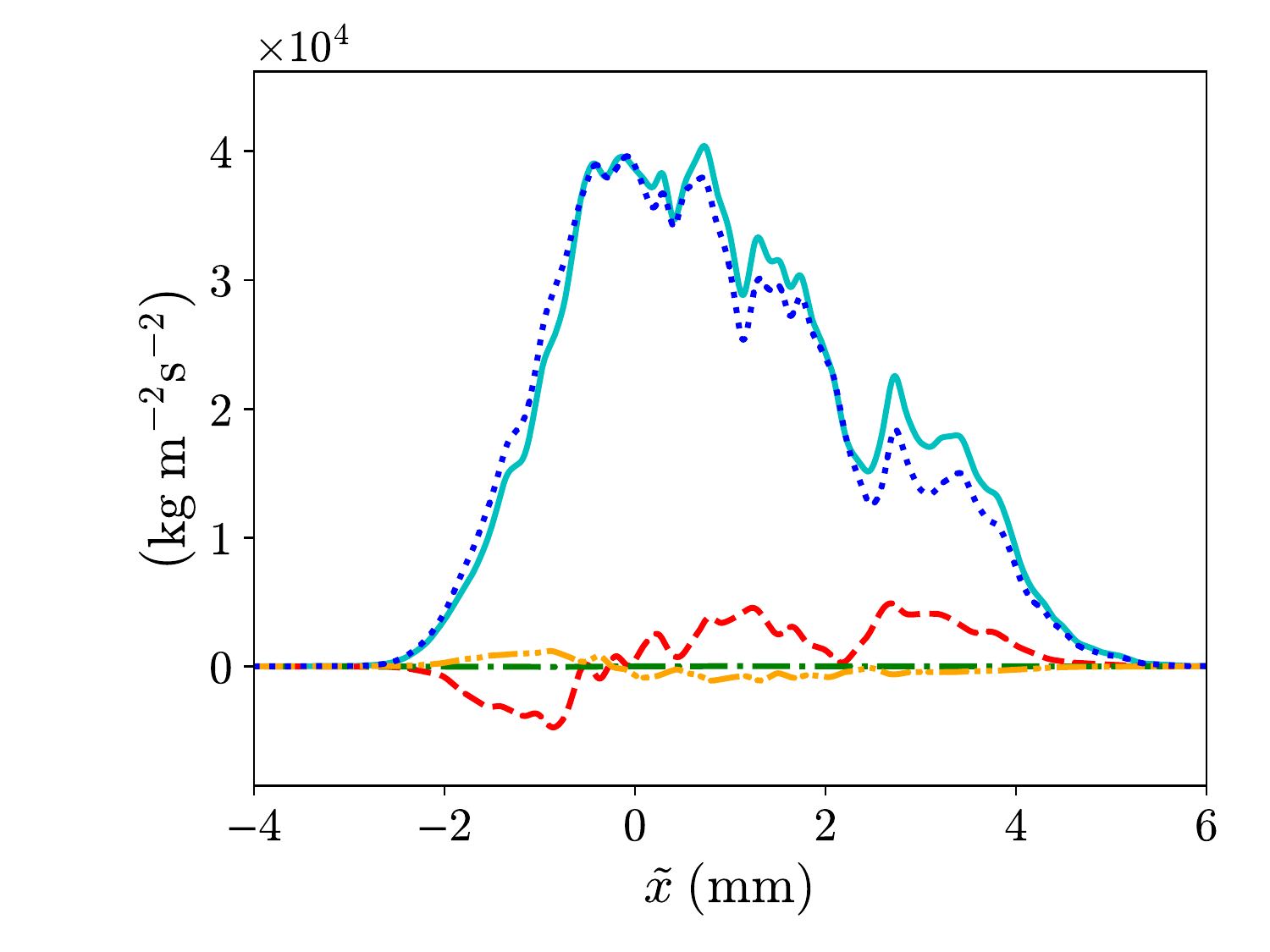}}
\subfigure[$\ \ell \approx 64 \Delta$]{%
\includegraphics[width=0.4\textwidth]{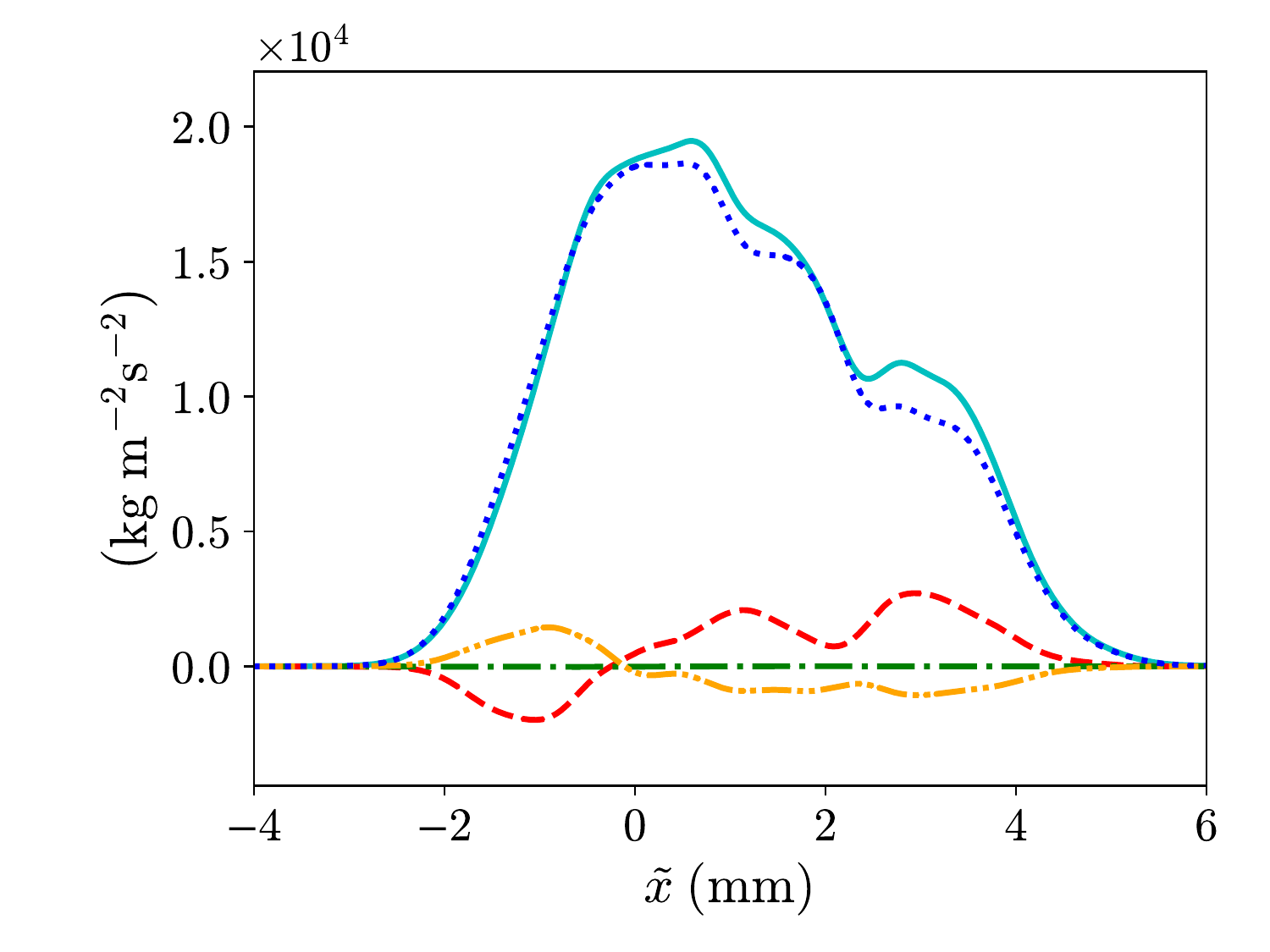}}
\caption{Effect of filtering on the compositions of the production term [term (III)] in the transport equation for the large-scale turbulent mass flux component in the streamwise direction, $\overline{\left< \rho \right>}_{\ell} a_{L,1}$, at $t = 1.40\ \mathrm{ms}$. Cyan solid line: overall production; red dashed line: $b_L \overline{\left< p \right>}_{\ell,1}$; green dash-dotted line: $-b_L \overline{\left< \tau_{11} \right>}_{\ell,1}$; orange dash-dot-dotted line: $b_L \overline{\tau_{11}^{SFS}}_{,1}$; blue dotted line: $-\widetilde{R}_{L,11} \overline{\left< \rho \right>}_{\ell,1}$.}
\label{fig:rho_a1_budget_filtered_production_terms_effect_filtering}
\end{figure*}

\begin{figure*}[!ht]
\centering
\subfigure[$\ \ell \approx 16 \Delta$]{%
\includegraphics[width=0.4\textwidth]{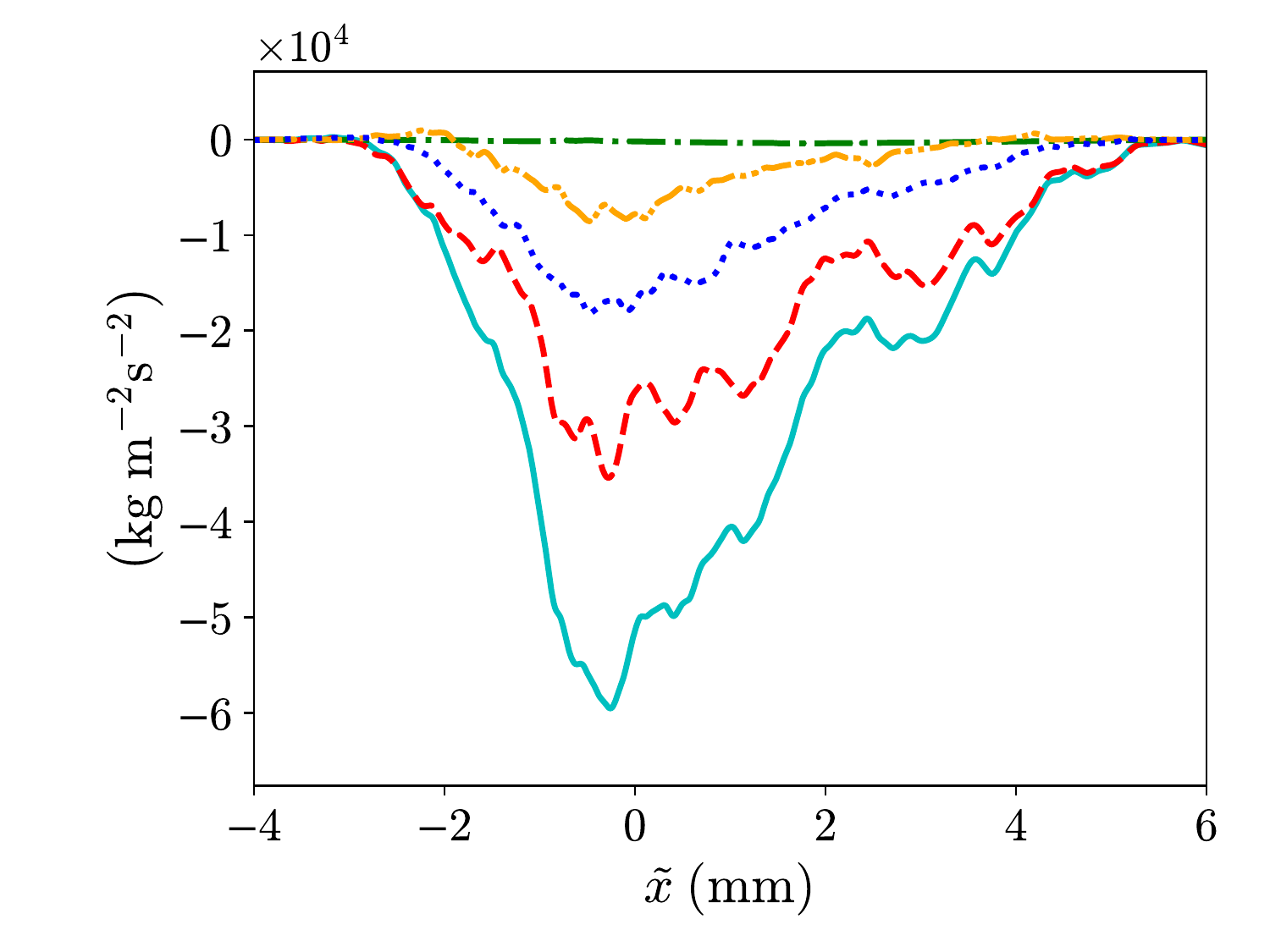}}
\subfigure[$\ \ell \approx 64 \Delta$]{%
\includegraphics[width=0.4\textwidth]{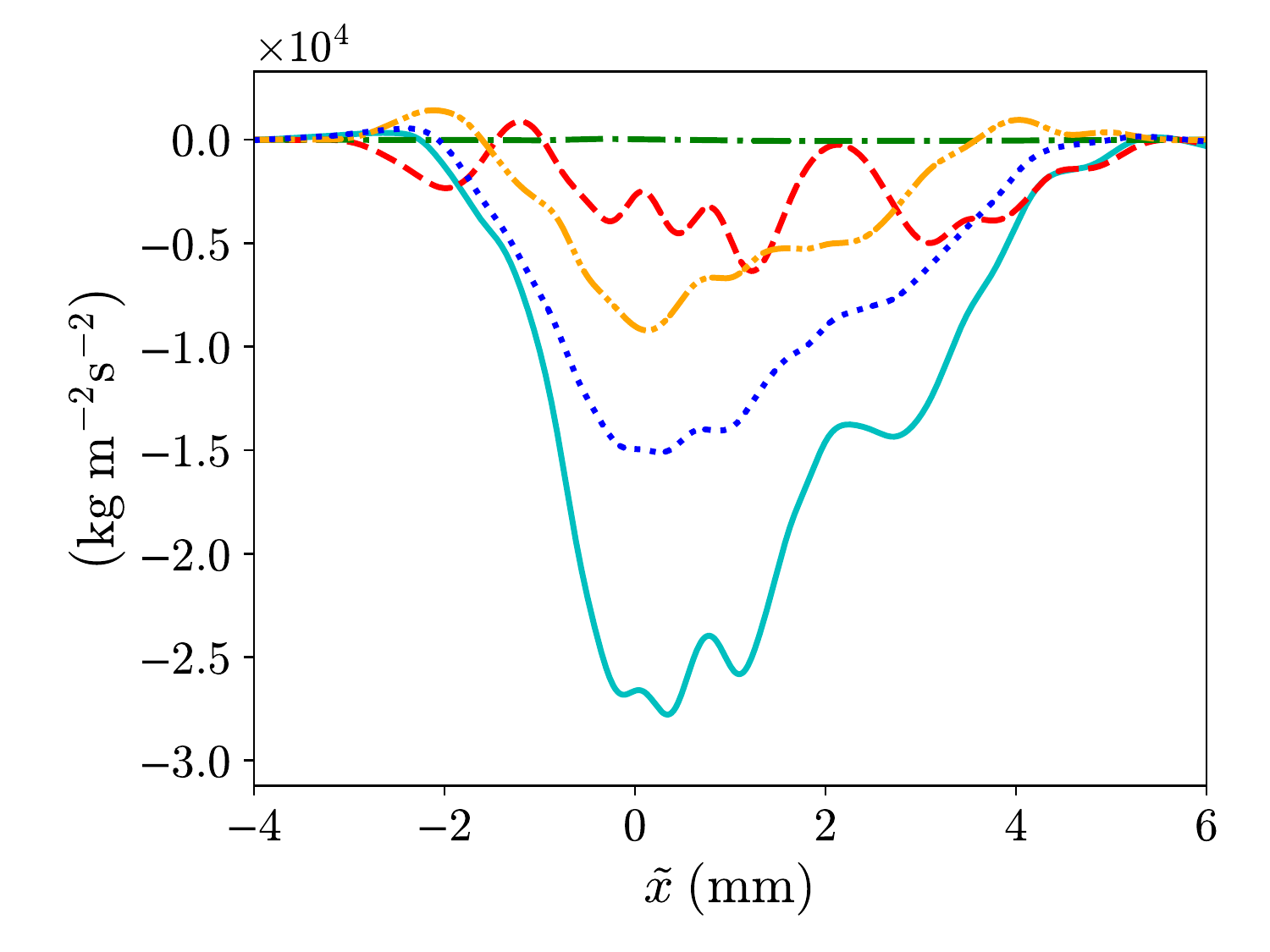}}
\caption{Effect of filtering on the compositions of the destruction term [term (VI)] in the transport equation for the large-scale turbulent mass flux component in the streamwise direction, $\overline{\left< \rho \right>}_{\ell} a_{L,1}$, at $t = 1.40\ \mathrm{ms}$. Cyan solid line: overall destruction; red dashed line: $\overline{\left< \rho \right>}_{\ell} \overline{ \left( 1/\left< \rho \right>_{\ell} \right)^{\prime} \left< p \right>^{\prime}_{\ell,1} }$; green dash-dotted line: $-\overline{\left< \rho \right>}_{\ell} \overline{ \left( 1/\left< \rho \right>_{\ell} \right)^{\prime} \left( \partial \left< \tau_{1i} \right>_{\ell}^{\prime} / \partial x_i \right) }$; orange dash-dot-dotted line: $\overline{\left< \rho \right>}_{\ell} \overline{ \left( 1/\left< \rho \right>_{\ell} \right)^{\prime} \left( \partial {\tau_{1i}^{SFS}}^{\prime} / \partial x_i \right) }$; blue dotted line: $\overline{\left< \rho \right>}_{\ell} \varepsilon_{a_{L,1}}$.}
\label{fig:rho_a1_budget_filtered_destruction_terms_effect_filtering}
\end{figure*}

Figures~\ref{fig:rho_a1_budget_filtered_production_terms_effect_filtering} and \ref{fig:rho_a1_budget_filtered_destruction_terms_effect_filtering} respectively show the effects of filtering on the compositions of the production [term (III)] and destruction [term (VI)] terms in the budgets of the large-scale turbulent mass flux component in the streamwise direction at $t = 1.40\ \mathrm{ms}$. It can be seen from both figures that the magnitudes of the components with the SFS stress in the production and destruction compositions, $b_L \overline{\tau_{11}^{SFS}}_{,1}$ and $\overline{\left< \rho \right>}_{\ell} \overline{ ( 1/\left< \rho \right>_{\ell} )^{\prime} ( \partial {\tau_{1i}^{SFS}}^{\prime} / \partial x_i ) }$, increase when a larger filter width is applied. Examining the production term, while there is a larger effect from the constituent with the SFS stress, $b_L \overline{\tau_{11}^{SFS}}_{,1}$, with a wider filter width,
the magnitude of term $b_L \overline{\left< p \right>}_{\ell,1}$ also becomes larger to offset the increased effect from $b_L \overline{\tau_{11}^{SFS}}_{,1}$.
Thus, the shape of the overall production term remains self-similar with filtering. As for the destruction term, the corresponding component with SFS stress also increases in magnitude to provide more of a destruction effect when the filter width is larger, but another two constituents, $\overline{\left< \rho \right>}_{\ell} \overline{ ( 1/\left< \rho \right>_{\ell} )^{\prime} \left< p \right>^{\prime}_{\ell,1} }$ and $\overline{\left< \rho \right>}_{\ell} \varepsilon_{a_{L,1}}$, adjust (the magnitude of the former decreases and that of the latter increases) and hence the overall destruction term is also self-similar with filtering.

\begin{figure*}[!ht]
\centering
\subfigure[$\ \ell \approx 16 \Delta$]{%
\includegraphics[width=0.4\textwidth]{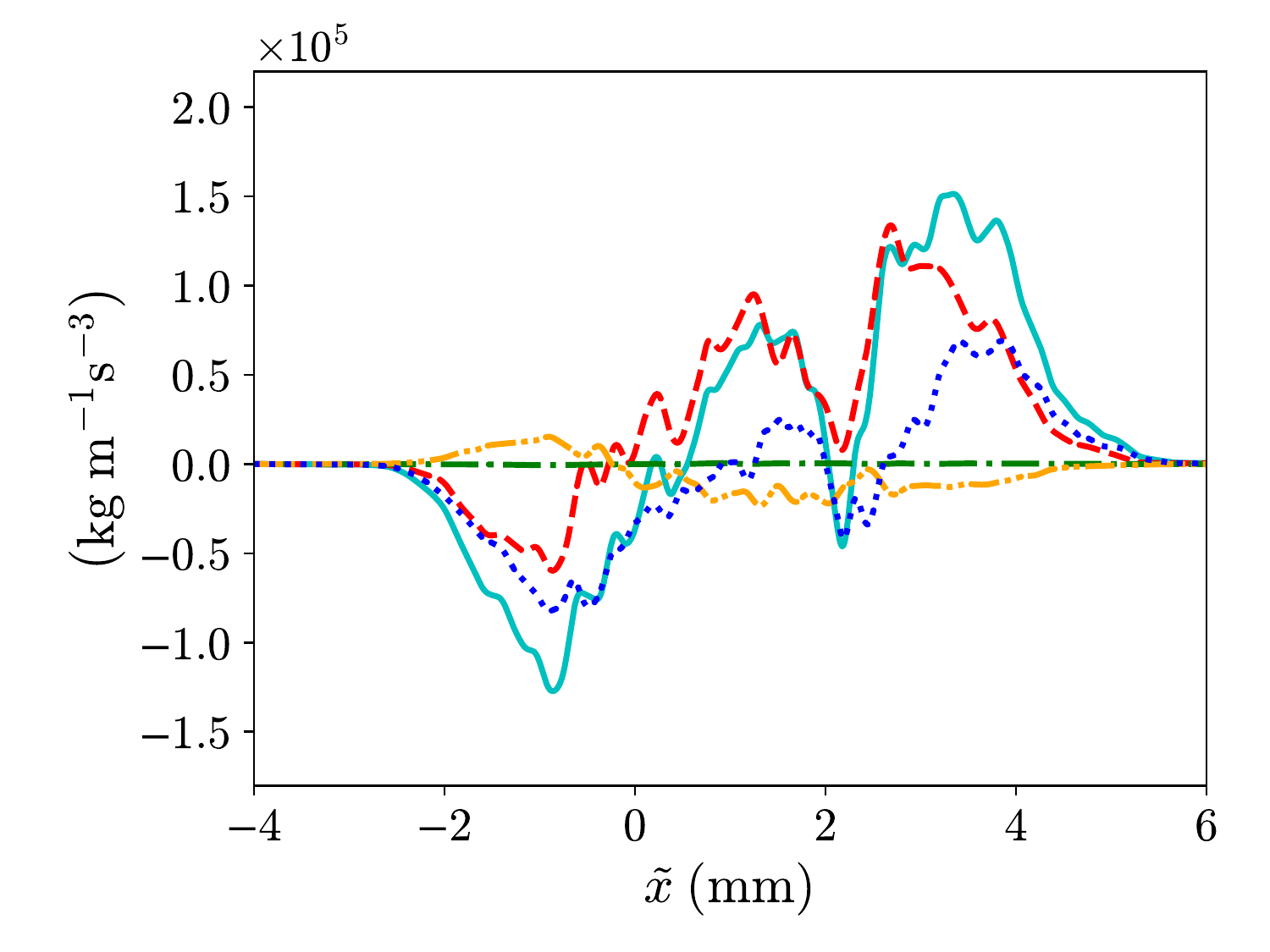}}
\subfigure[$\ \ell \approx 64 \Delta$]{%
\includegraphics[width=0.4\textwidth]{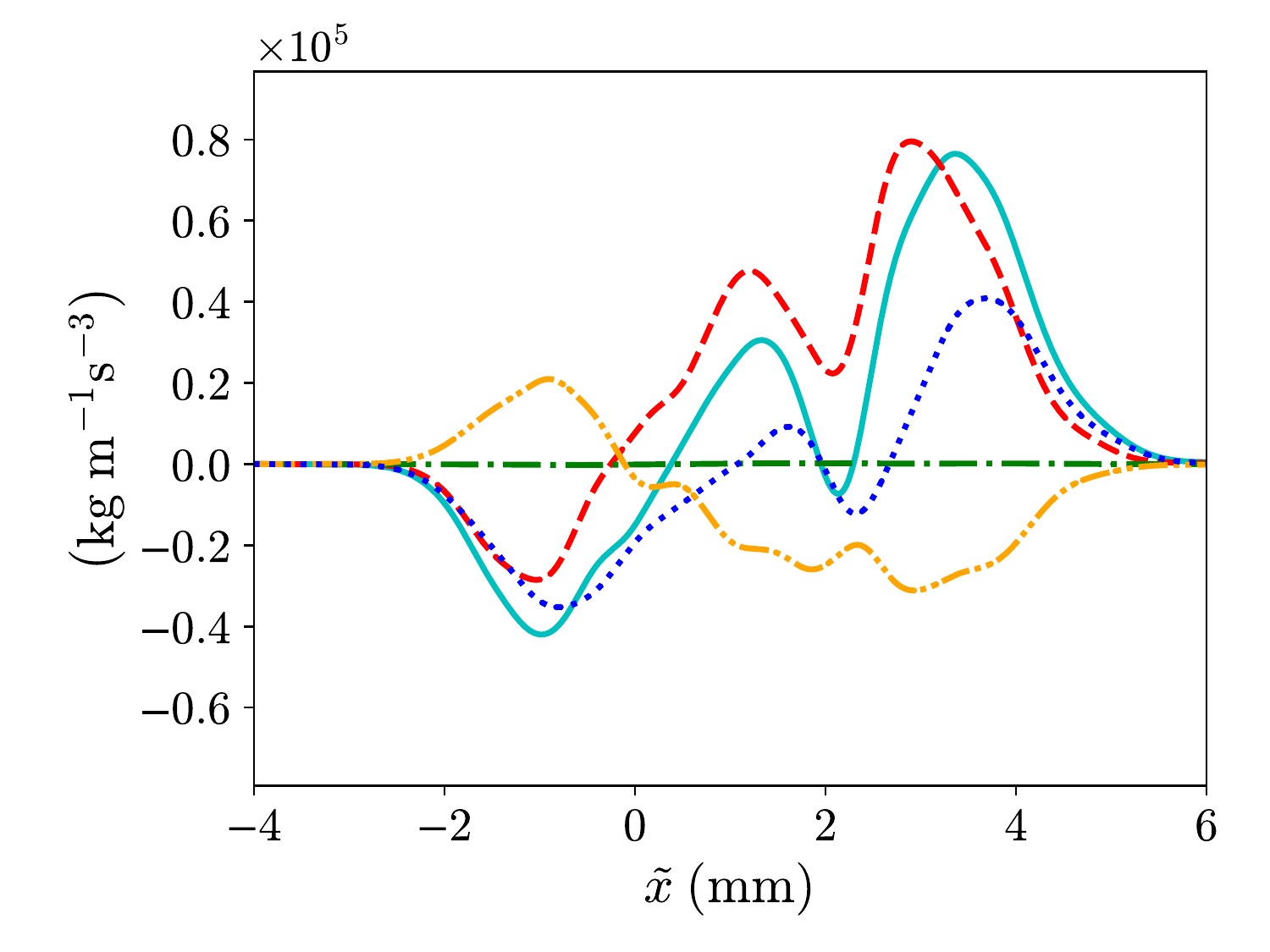}}
\caption{Effect of filtering on the compositions of the production term [term (III)] in the transport equation for the large-scale Favre-averaged Reynolds normal stress component in the streamwise direction multiplied by the mean filtered density, $\overline{\left< \rho \right>}_{\ell} \widetilde{R}_{L,11}$, at $t = 1.40\ \mathrm{ms}$. Cyan solid line: overall production; red dashed line: $2a_{L,1} \overline{\left< p \right>}_{\ell,1}$; green dash-dotted line: $-2a_{L,1} \overline{\left< \tau_{11} \right>}_{\ell,1}$; orange dash-dot-dotted line: $2a_{L,1} {\overline{ \tau_{11}^{SFS} }}_{,1}$; blue dotted line: $-2\overline{\left< \rho \right>}_{\ell} \widetilde{R}_{L,11} \widetilde{\left< u \right>}_{L,1}$.}
\label{fig:rho_R11_budget_filtered_production_terms_effect_filtering}
\end{figure*}

\begin{figure*}[!ht]
\centering
\subfigure[$\ \ell \approx 16 \Delta$]{%
\includegraphics[width=0.4\textwidth]{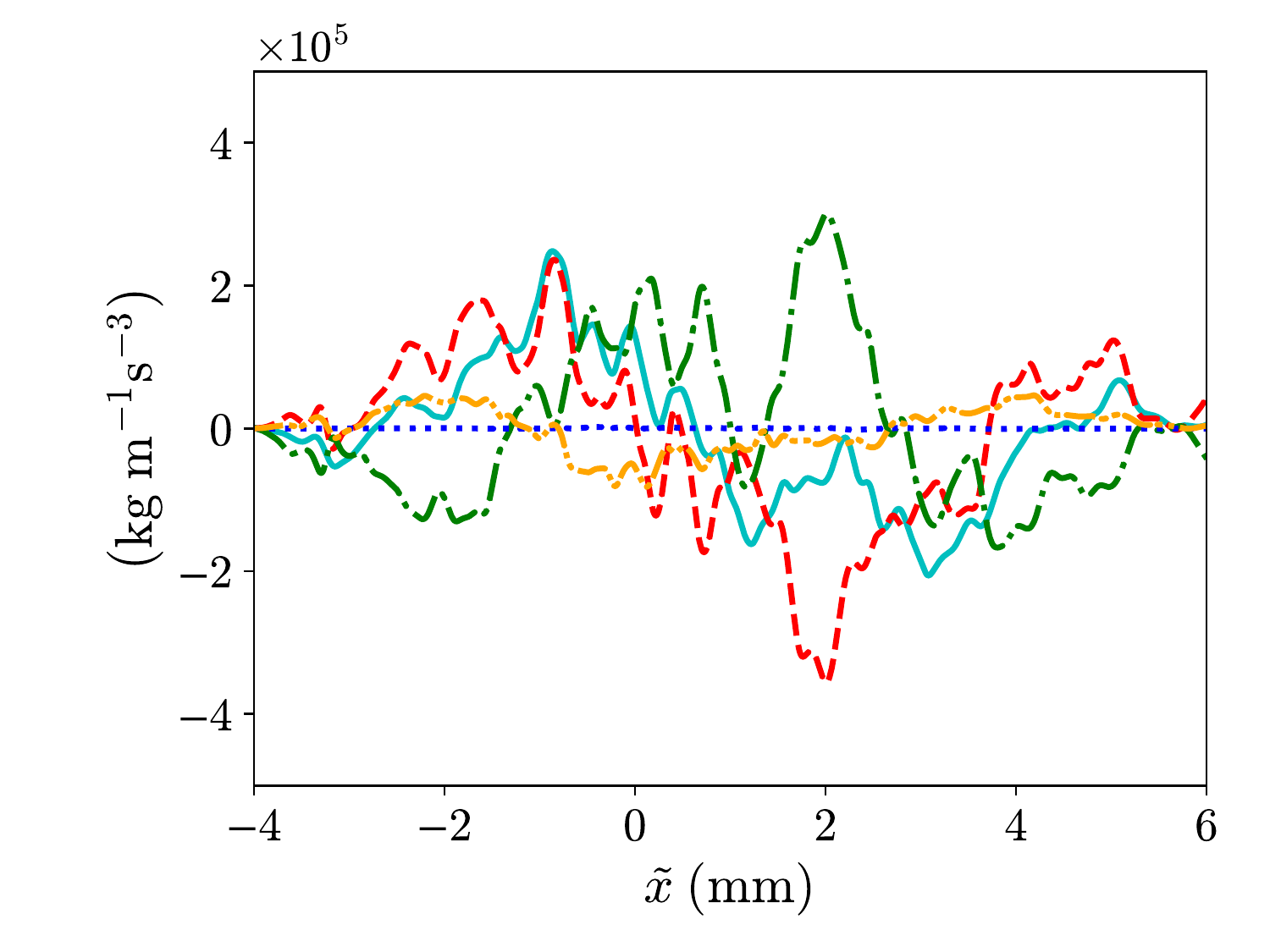}}
\subfigure[$\ \ell \approx 64 \Delta$]{%
\includegraphics[width=0.4\textwidth]{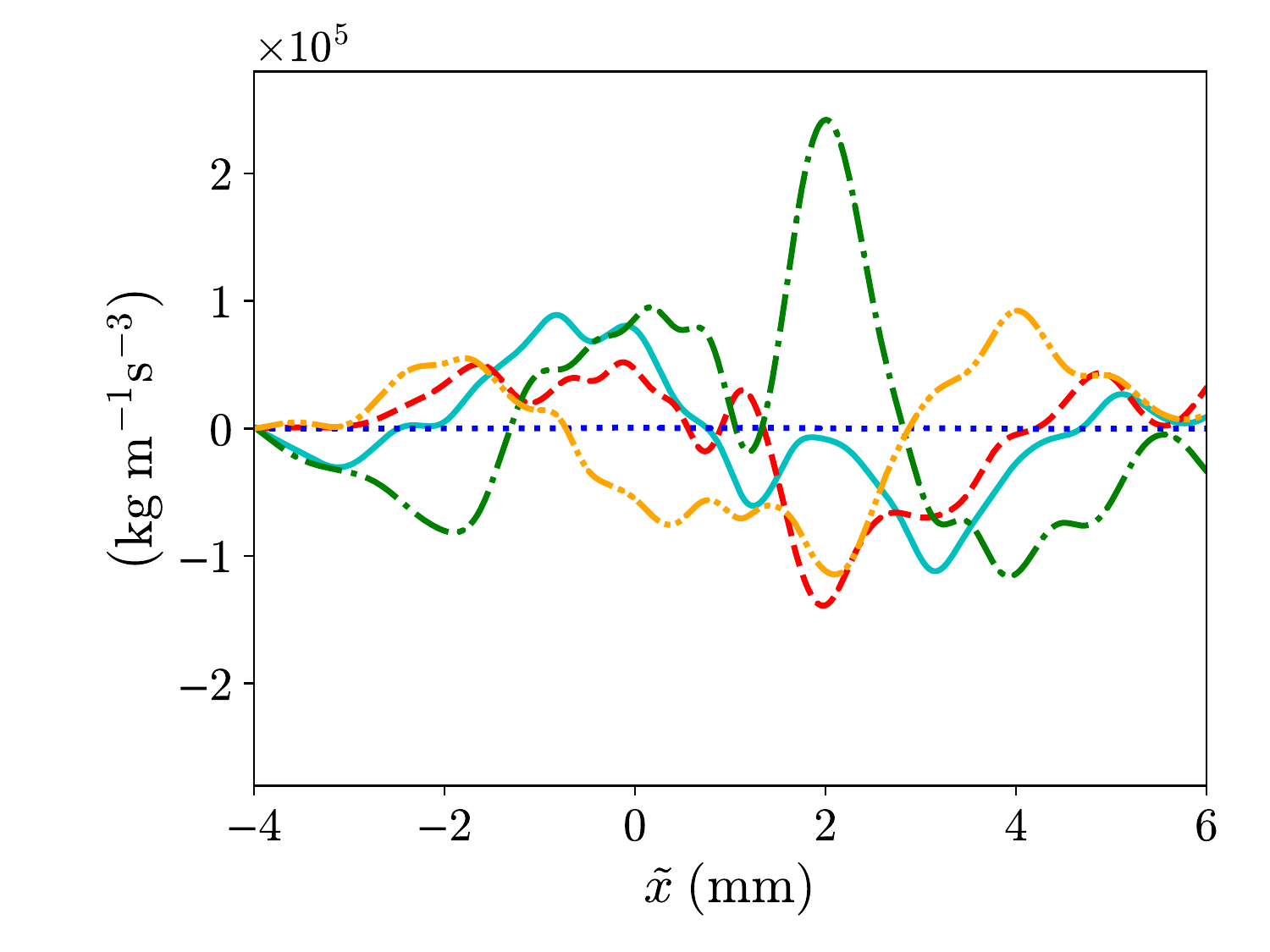}}
\caption{Effect of filtering on the compositions of the turbulent transport term [term (IV)] in the transport equation for the large-scale Favre-averaged Reynolds normal stress component in the streamwise direction multiplied by the mean filtered density, $\overline{\left< \rho \right>}_{\ell} \widetilde{R}_{L,11}$, at $t = 1.40\ \mathrm{ms}$. Cyan solid line: overall turbulent transport; red dashed line: $- ( \overline{ \left< \rho \right>_{\ell} \left< u \right>_{L}^{\prime\prime} \left< u \right>_{L}^{\prime\prime} \left< u \right>_{L}^{\prime\prime} } )_{,1}$; green dash-dotted line: $-2 ( \overline{\left< u \right>_{L}^{\prime} \left< p \right>_{\ell}^{\prime}} )_{,1}$; blue dotted line: $2 ( \overline{ \left< u \right>_{L}^{\prime} \left< \tau_{11} \right>_{\ell}^{\prime} } )_{,1}$; orange dash-dot-dotted line: $-2 ( \overline{ \left< u \right>_{L}^{\prime} {\tau_{11}^{SFS}}^{\prime} } )_{,1}$.}
\label{fig:rho_R11_budget_filtered_turb_transport_terms_effect_filtering}
\end{figure*}

The effects of filtering on the composition of the production [term (III)] and turbulent transport [term (IV)] terms in the budget of the large-scale Reynolds normal stress component in the streamwise direction multiplied by the mean filtered density are studied in figures~\ref{fig:rho_R11_budget_filtered_production_terms_effect_filtering} and \ref{fig:rho_R11_budget_filtered_turb_transport_terms_effect_filtering} respectively at $t = 1.40\ \mathrm{ms}$. The component with the SFS stress in the production term, $2a_{L,1} {\overline{ \tau_{11}^{SFS} }}_{,1}$, has a larger magnitude with more filtering operations but this change is offset by the adjustment of $2a_{L,1} \overline{\left< p \right>}_{\ell,1}$, which also has a larger magnitude but the opposite effect compared with the former. Similarly, the constituent with SFS stress in the turbulent transport term, $-2 ( \overline{ \left< u \right>_{L}^{\prime} {\tau_{11}^{SFS}}^{\prime} } )_{,1}$, has a greater magnitude as the filter width increases, but its larger influence is offset by the change of $-2 ( \overline{\left< u \right>_{L}^{\prime} \left< p \right>_{\ell}^{\prime}} )_{,1}$. In general, the overall shapes of the production and turbulent transport terms are quite similar for different filter widths, but their compositions change as the SFS stress plays a more important role in each of the two terms.

\section{Conclusions}

A second-moment analysis of high Atwood number variable-density mixing induced by RMI was conducted with high-resolution 3D AMR simulation data. In the numerical experiment, the material interface separating $\mathrm{SF_6}$ and air is impulsively accelerated twice and the mixing layer becomes turbulent after re-shock. The roles that the two second-moments, turbulent mass flux and density-specific-volume covariance, play in the development of Favre-averaged Reynolds stress were discussed through the examination the transport equations for the second-moments, including the Favre-averaged Reynolds stress and turbulent kinetic energy. The study of the transport mechanisms of the second-moments can foster the improvement of existing reduced-order models for closing the Favre-averaged Navier--Stokes equations in RANS-based simulations. The quantities of interest, including the second-moments computed with the simulation data, were found to be well grid-converged at the finest grid setting and the study of their time evolution revealed the non-Boussinesq and anisotropic nature of the variable-density flow induced by RMI. The transport equations of the Reynolds stress and second-moments were studied before re-shock when mixing occurs due to the instability. The relative importance of different terms in the budgets of the quantities across the mixing layer was found to vary a lot and the origins of the generation, destruction, and spreading of the quantities of interest over time were traced back to the corresponding budget terms. Unlike the situation where all scales in the flow are well-resolved in the highest resolution simulation before re-shock, the wide span of scales generated due to mixing transition after re-shock leads to under-resolved simulation results. While the budgets of some second-moments, including the Reynolds stress, are unclosed during this time period, the budgets of large-scale Reynolds stress and second-moments at sufficiently large scale were found to be unaffected by the numerical regularization, when the influence of the SFS stress is taken into account. The effects of the SFS stress on the development of large-scale quantities at different filtered scales were studied. Although the SFS stress can significantly contribute to the composition of different budget terms when a large filter width is used, the overall budgets of large-scale Reynolds stress and second-moments remain quite self-similar with filtering as the shapes of different budget terms and their relative magnitudes are similar with different filter widths. This suggests that the budget analysis of large-scale quantities in LESs can be relevant for the development and validation of RANS-based closures that model each budget term as a whole, even when the Reynolds stress and turbulent kinetic energy are not well-resolved, provided that the effects of an accurate representation of the SFS stress are included in the budget terms.
This also addresses the importance of reconstructing the SFS stress in LESs of this type of variable-density flows in order to model the development of the turbulence accurately. Nevertheless, the study of the evolution mechanism of the SFS stress requires the analysis of its transport equation with fully resolved turbulence data. As a result, future research of RMI-induced variable-density turbulence with higher resolution simulations, such as DNSs that resolve all spatio-temporal scales, can largely advance turbulence modeling in LES, RANS, and hybrid RANS-LES approaches.


\section{Acknowledgments}
This work was performed under the auspices of U.S. Department of Energy. M. L. Wong and S. K. Lele were supported by Los Alamos National Laboratory, under Grant No. 431679. Los Alamos National Laboratory is operated by Triad National Security, LLC, for the National Nuclear Security Administration of U.S. Department of Energy (Contract No. 89233218CNA000001). Computational resources were provided by the Los Alamos National Laboratory Institutional Computing Program and the Advanced Simulation and Computation (ASC) Program.

\appendix


\section{\label{sec:spat_conv_second_moments} Grid sensitivity analysis of the spatial profiles of second-moments}

\begin{figure*}[!ht]
\centering
\subfigure[$\ $Before re-shock]{%
\includegraphics[width = 0.4\textwidth]{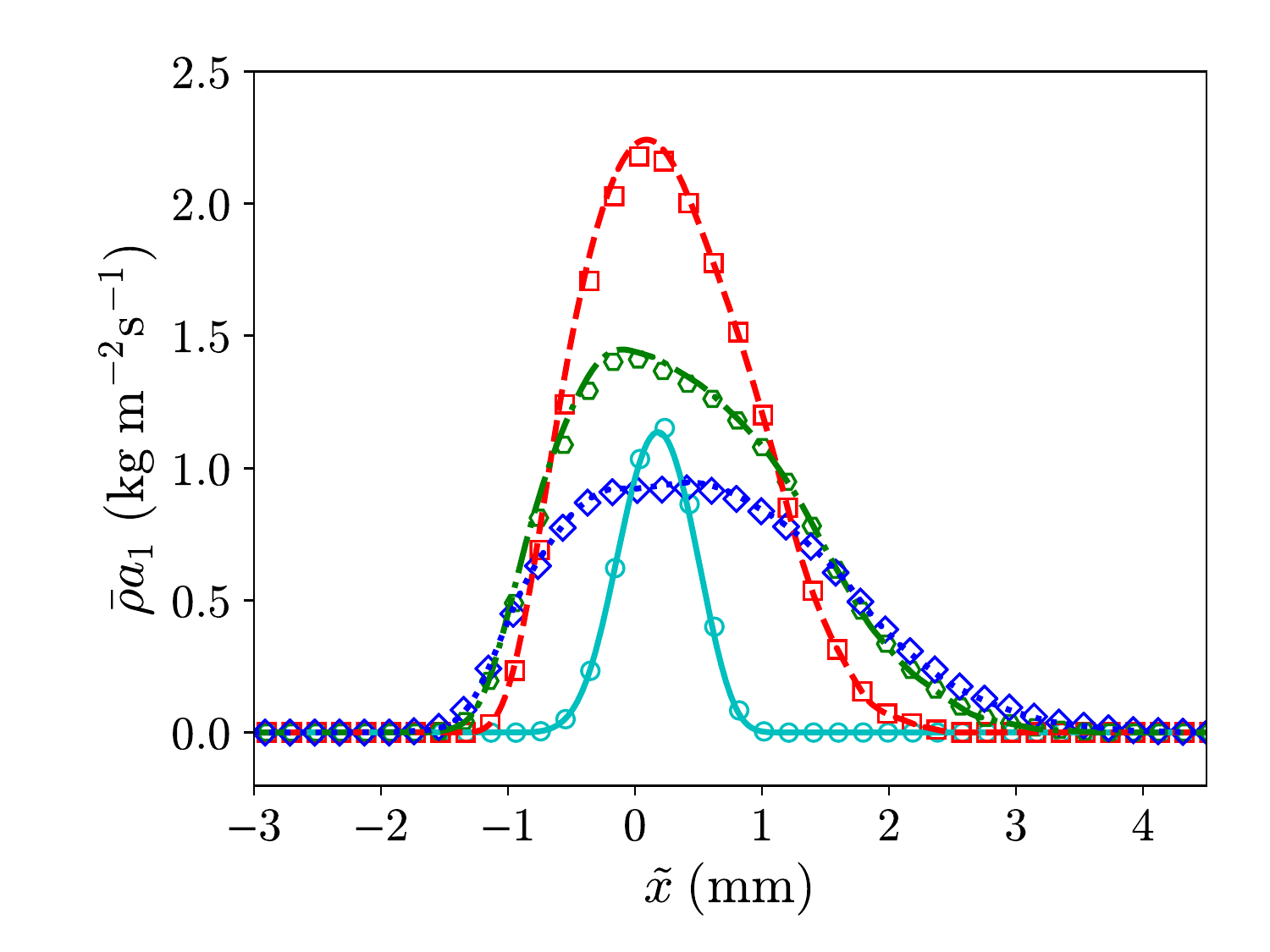}}
\subfigure[$\ $After re-shock]{%
\includegraphics[width = 0.4\textwidth]{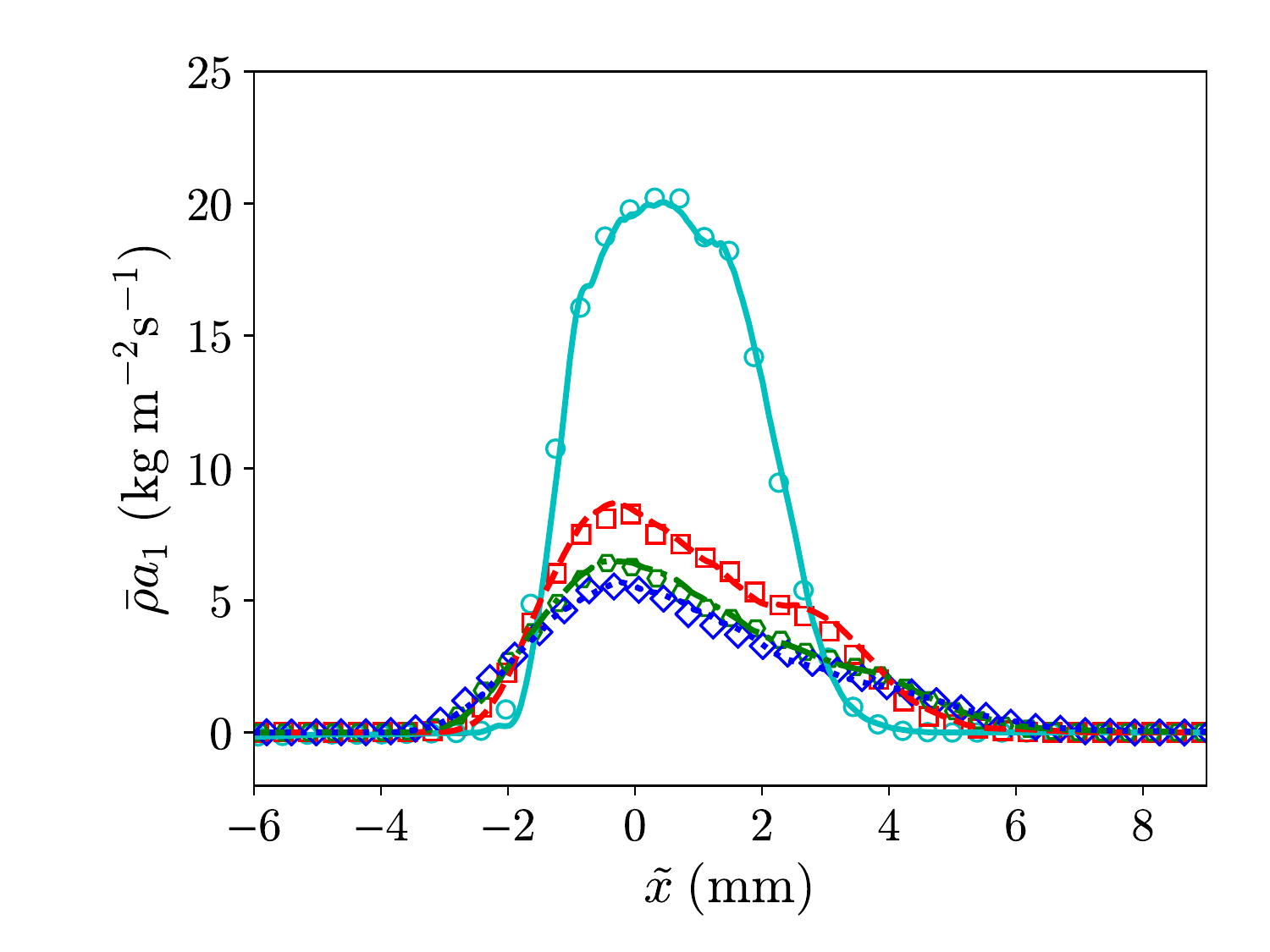}}
\caption{
Grid sensitivities of the profiles of the turbulent mass flux component in the streamwise direction, $\bar{\rho}a_1$, at different times between the grid D and the grid E.
The profiles with the grid D and the grid E are shown with symbols and lines respectively.
Cyan circles or solid line in (a): $t=0.05\ \mathrm{ms}$; red squares or dashed line in (a): $t=0.40\ \mathrm{ms}$; green hexagons or dash-dotted line in (a): $t=0.75\ \mathrm{ms}$; blue diamonds or dotted line in (a): $t=1.10\ \mathrm{ms}$. Cyan circles or solid line in (b): $t=1.20\ \mathrm{ms}$; red squares or dashed line in (b): $t=1.40\ \mathrm{ms}$; green hexagons or dash-dotted line in (b): $t=1.60\ \mathrm{ms}$; blue diamonds or dotted line in (b): $t=1.75\ \mathrm{ms}$.
}
\label{fig:rho_a1_profiles_spat_conv}
\end{figure*}

\begin{figure*}[!ht]
\centering
\subfigure[$\ $Before re-shock]{%
\includegraphics[width=0.4\textwidth]{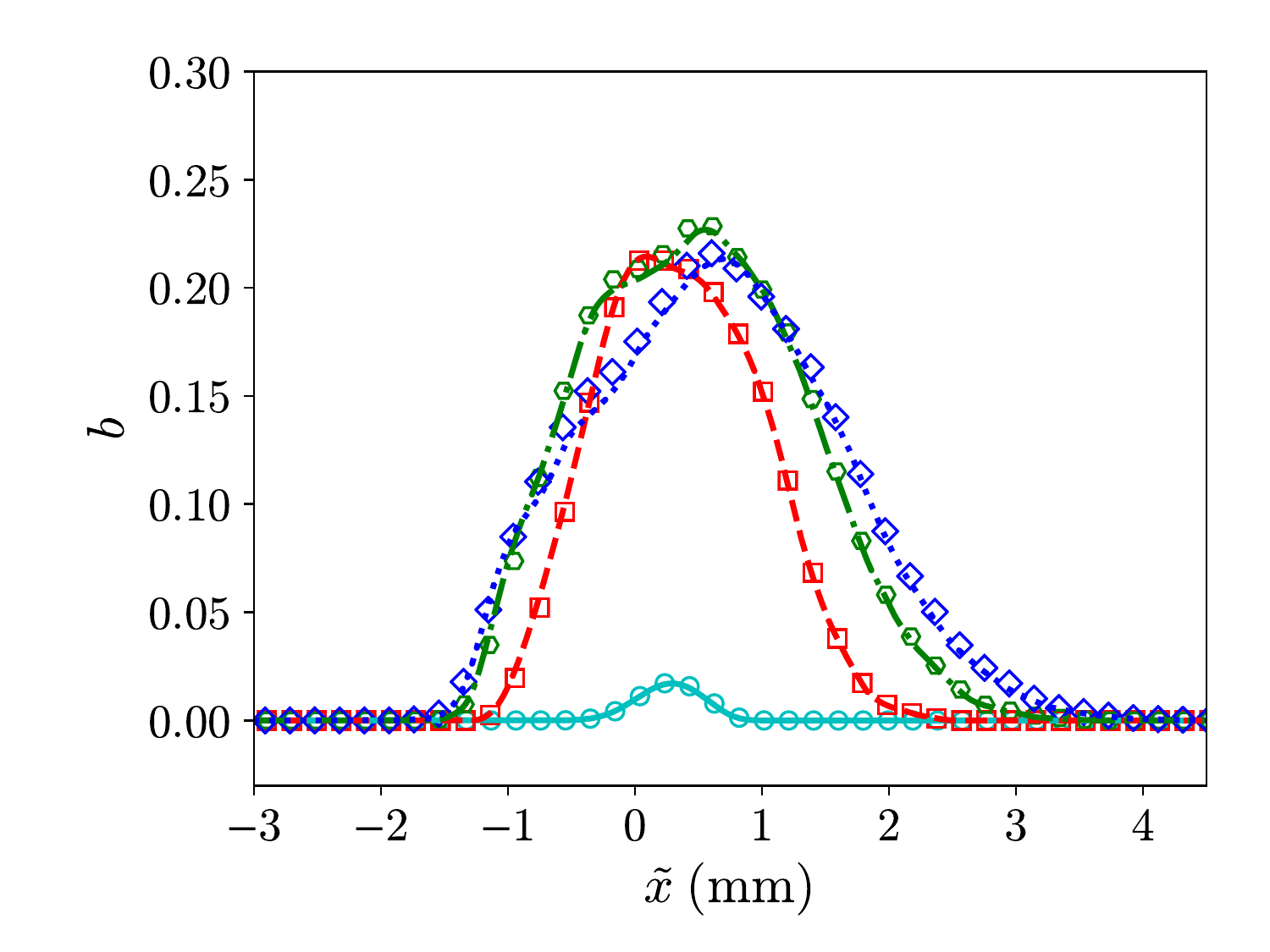}}
\subfigure[$\ $After re-shock]{%
\includegraphics[width=0.4\textwidth]{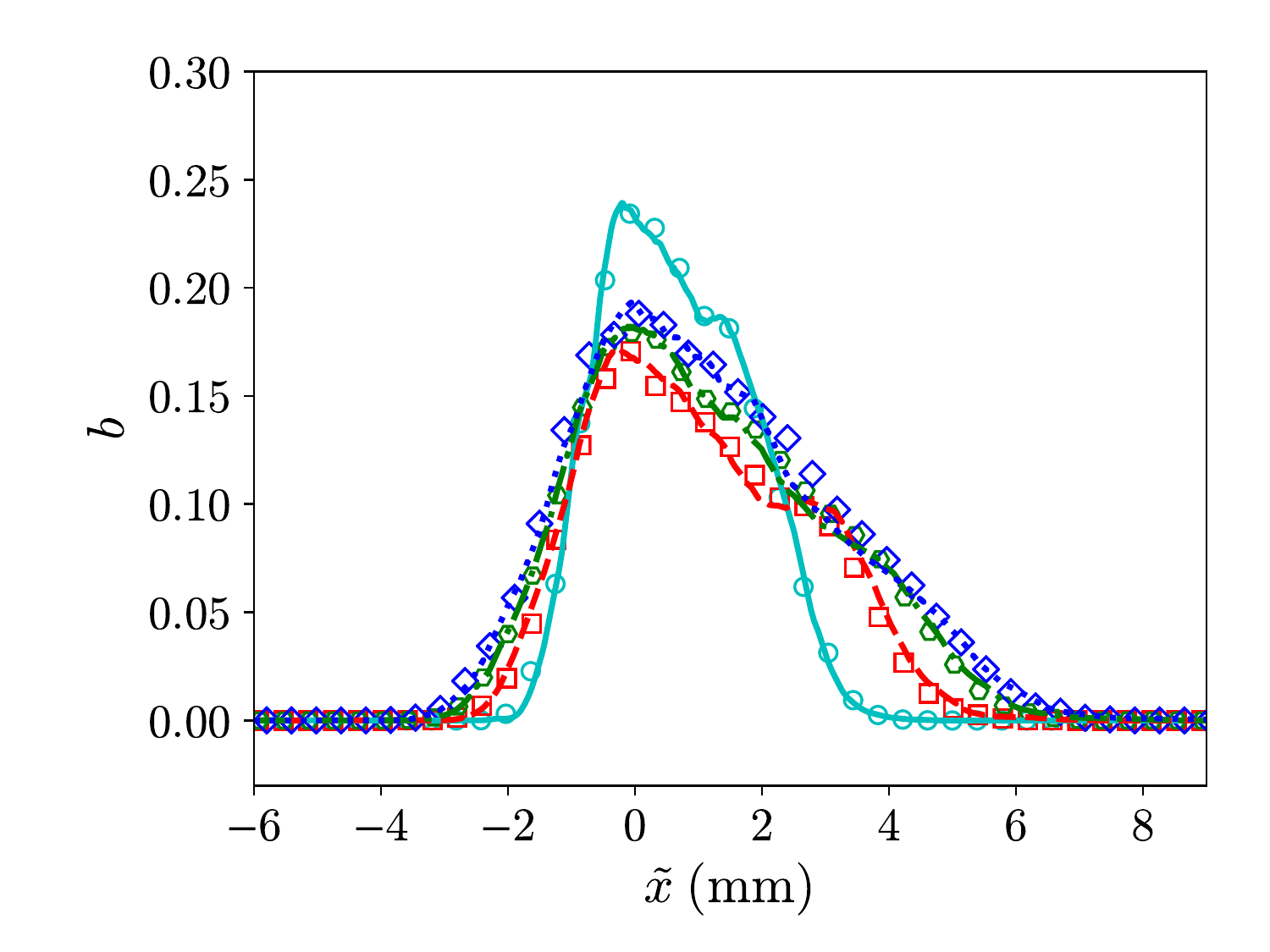}}
\caption{
Grid sensitivities of the profiles of the density-specific-volume covariance, $b$, at different times between the grid D and the grid E.
The profiles with the grid D and the grid E are shown with symbols and lines respectively.
Cyan circles or solid line in (a): $t=0.05\ \mathrm{ms}$; red squares or dashed line in (a): $t=0.40\ \mathrm{ms}$; green hexagons or dash-dotted line in (a): $t=0.75\ \mathrm{ms}$; blue diamonds or dotted line in (a): $t=1.10\ \mathrm{ms}$. Cyan circles or solid line in (b): $t=1.20\ \mathrm{ms}$; red squares or dashed line in (b): $t=1.40\ \mathrm{ms}$; green hexagons or dash-dotted line in (b): $t=1.60\ \mathrm{ms}$; blue diamonds or dotted line in (b): $t=1.75\ \mathrm{ms}$.
}
\label{fig:b_profiles_spat_conv}
\end{figure*}

\begin{figure*}[!ht]
\centering
\subfigure[$\ $Before re-shock]{%
\includegraphics[width=0.4\textwidth]{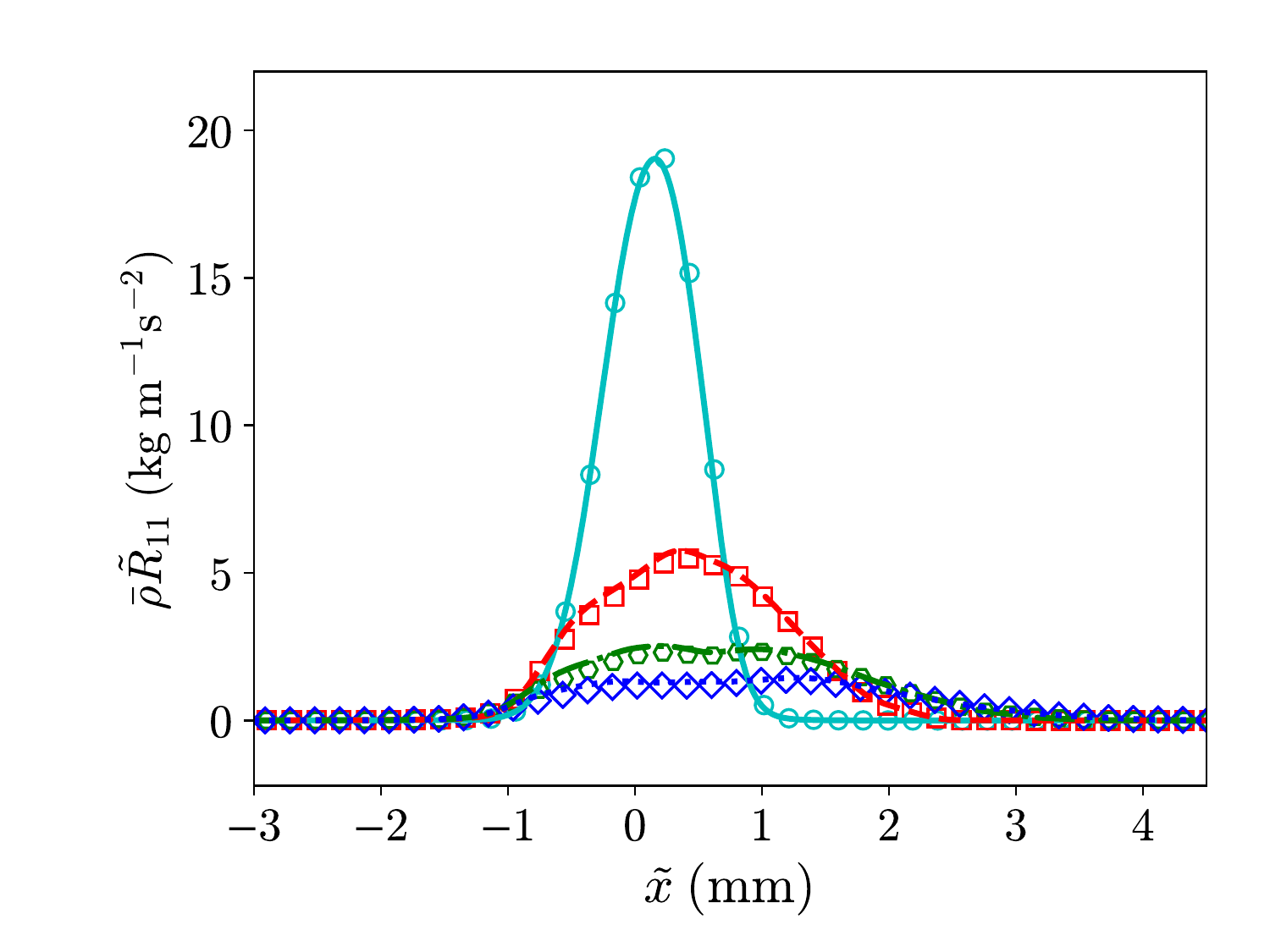}}
\subfigure[$\ $After re-shock]{%
\includegraphics[width=0.4\textwidth]{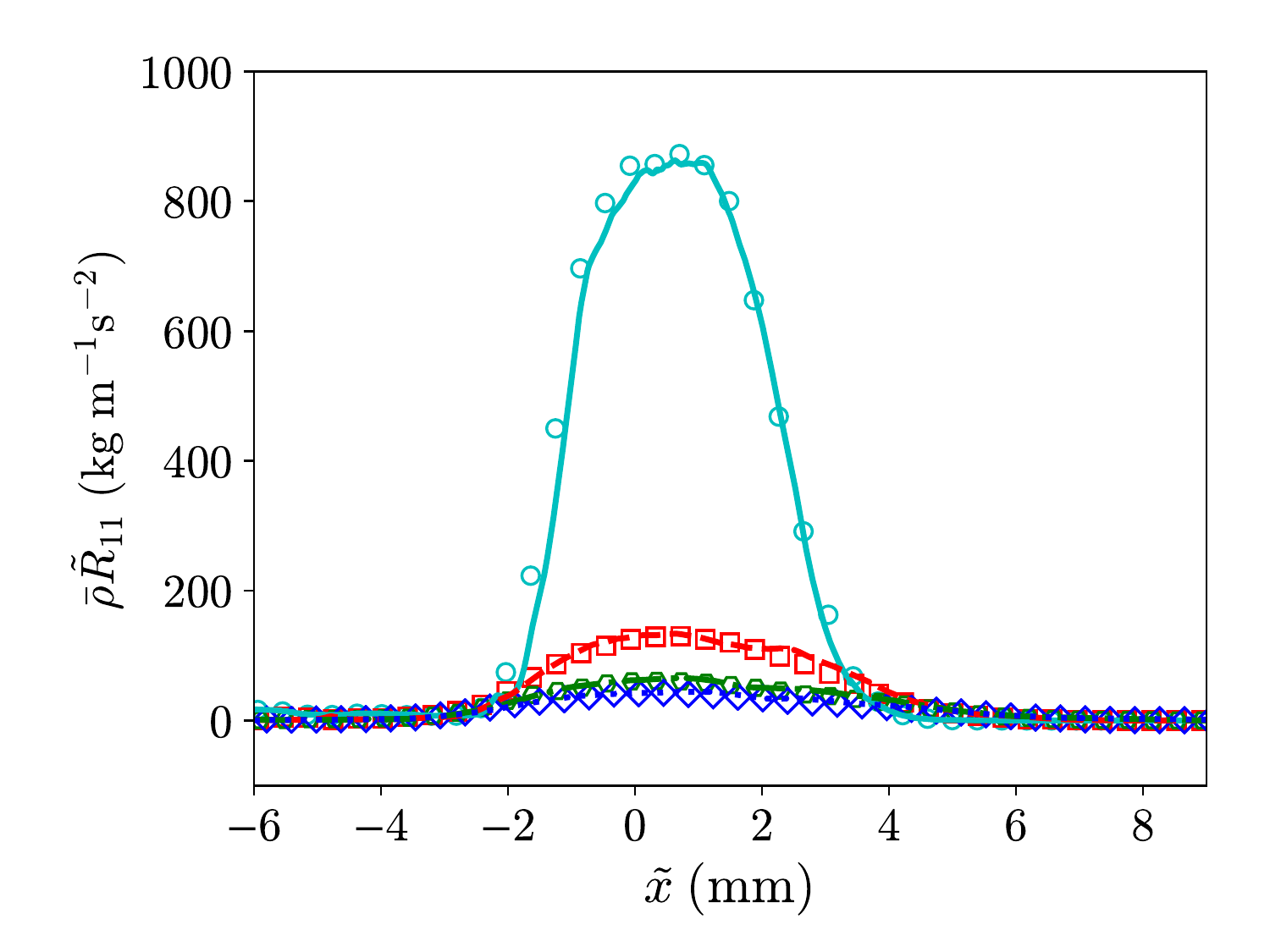}}
\caption{
Grid sensitivities of the profiles of the Reynolds normal stress component in the streamwise direction multiplied by the mean density, $\bar{\rho} \tilde{R}_{11}$, at different times between the grid D and the grid E.
The profiles with the grid D and the grid E are shown with symbols and lines respectively.
Cyan circles or solid line in (a): $t=0.05\ \mathrm{ms}$; red squares or dashed line in (a): $t=0.40\ \mathrm{ms}$; green hexagons or dash-dotted line in (a): $t=0.75\ \mathrm{ms}$; blue diamonds or dotted line in (a): $t=1.10\ \mathrm{ms}$. Cyan circles or solid line in (b): $t=1.20\ \mathrm{ms}$; red squares or dashed line in (b): $t=1.40\ \mathrm{ms}$; green hexagons or dash-dotted line in (b): $t=1.60\ \mathrm{ms}$; blue diamonds or dotted line in (b): $t=1.75\ \mathrm{ms}$.
}
\label{fig:rho_R11_profiles_spat_conv}
\end{figure*}

\begin{figure*}[!ht]
\centering
\subfigure[$\ $Before re-shock]{%
\includegraphics[width=0.4\textwidth]{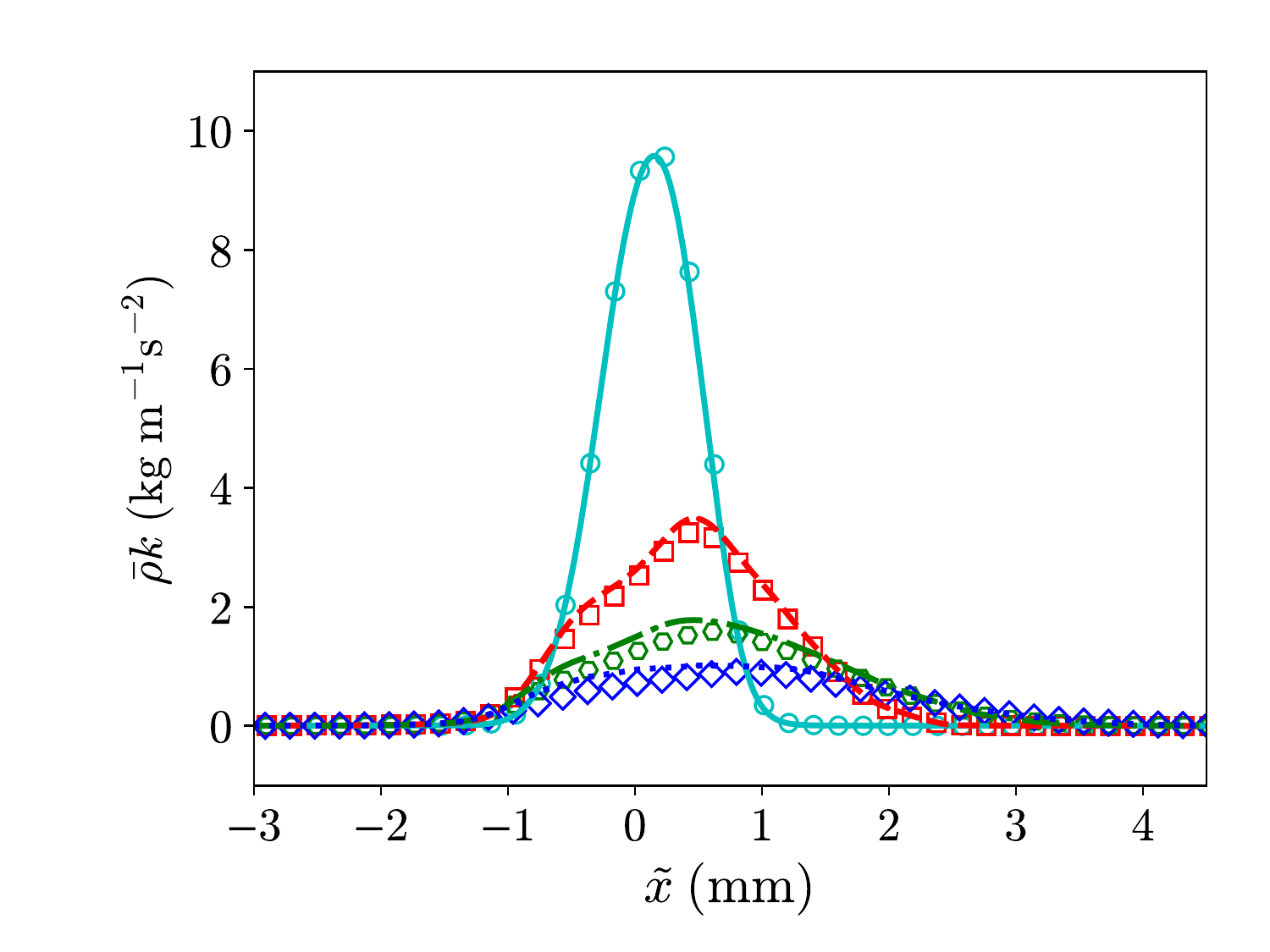}}
\subfigure[$\ $After re-shock]{%
\includegraphics[width=0.4\textwidth]{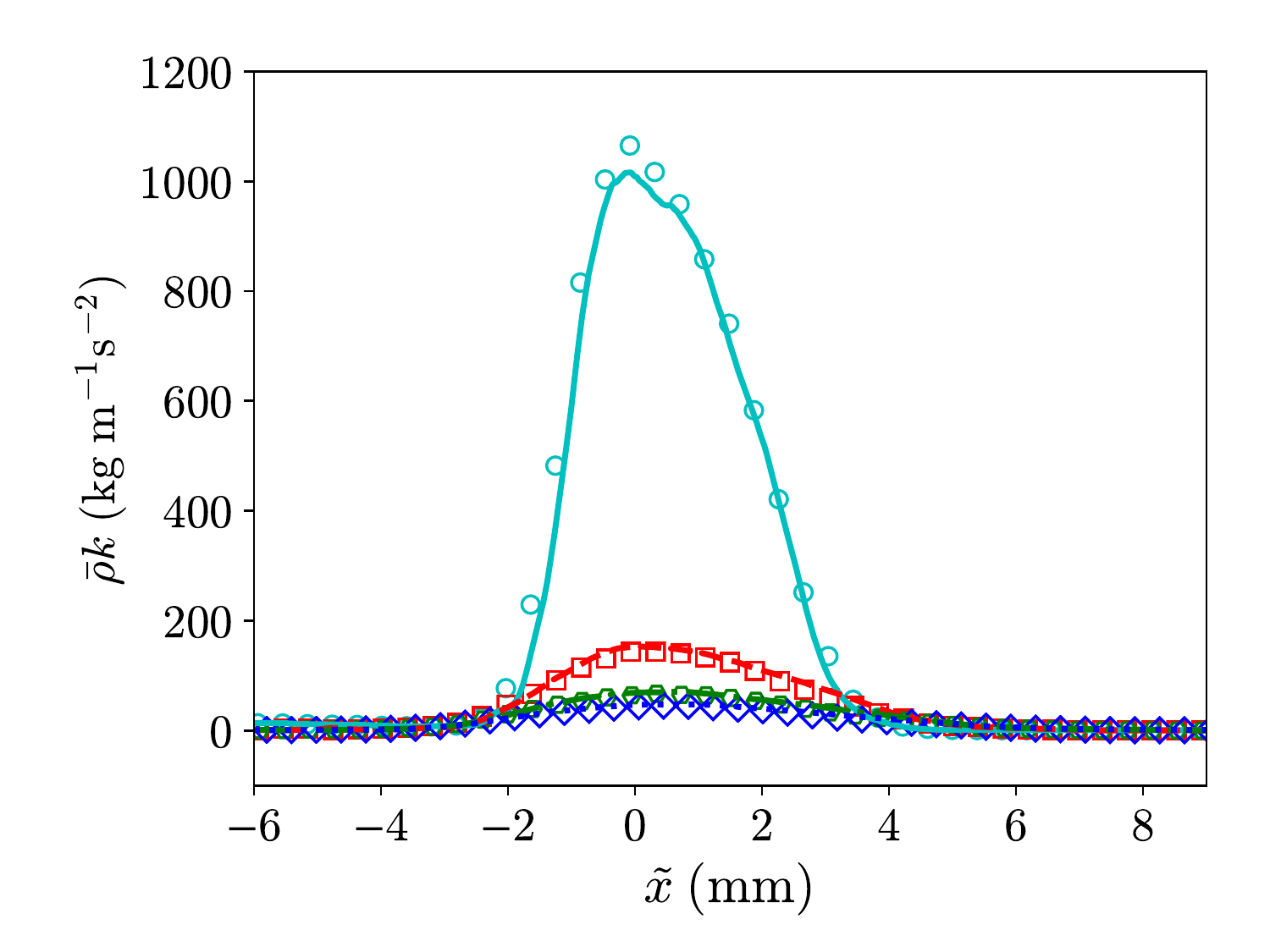}}
\caption{
Grid sensitivities of the profiles of the turbulent kinetic energy, $\bar{\rho} k$, at different times between the grid D and the grid E.
The profiles with the grid D and the grid E are shown with symbols and lines respectively.
Cyan circles or solid line in (a): $t=0.05\ \mathrm{ms}$; red squares or dashed line in (a): $t=0.40\ \mathrm{ms}$; green hexagons or dash-dotted line in (a): $t=0.75\ \mathrm{ms}$; blue diamonds or dotted line in (a): $t=1.10\ \mathrm{ms}$. Cyan circles or solid line in (b): $t=1.20\ \mathrm{ms}$; red squares or dashed line in (b): $t=1.40\ \mathrm{ms}$; green hexagons or dash-dotted line in (b): $t=1.60\ \mathrm{ms}$; blue diamonds or dotted line in (b): $t=1.75\ \mathrm{ms}$.
}
\label{fig:rho_k_profiles_spat_conv}
\end{figure*}

The grid sensitivities of the spatial profiles of $\bar{\rho}a_1$, $b$, $\bar{\rho} \tilde{R}_{11}$, and $\bar{\rho} k$ at different times between the grid D and the grid E are shown respectively in figures~\ref{fig:rho_a1_profiles_spat_conv}, \ref{fig:b_profiles_spat_conv}, \ref{fig:rho_R11_profiles_spat_conv}, and \ref{fig:rho_k_profiles_spat_conv}. Overall, these spatial profiles have small grid sensitivities between the two grid resolutions at different times which are consistent with the grid sensitivities of the time evolution of the domain-integrated values.

\section{\label{sec:cells_count} Time evolution of the numbers of grid cells in the simulations}

\begin{figure*}[!ht]
\centering
\subfigure[$\ $Number of cells]{%
\includegraphics[width=0.4\textwidth]{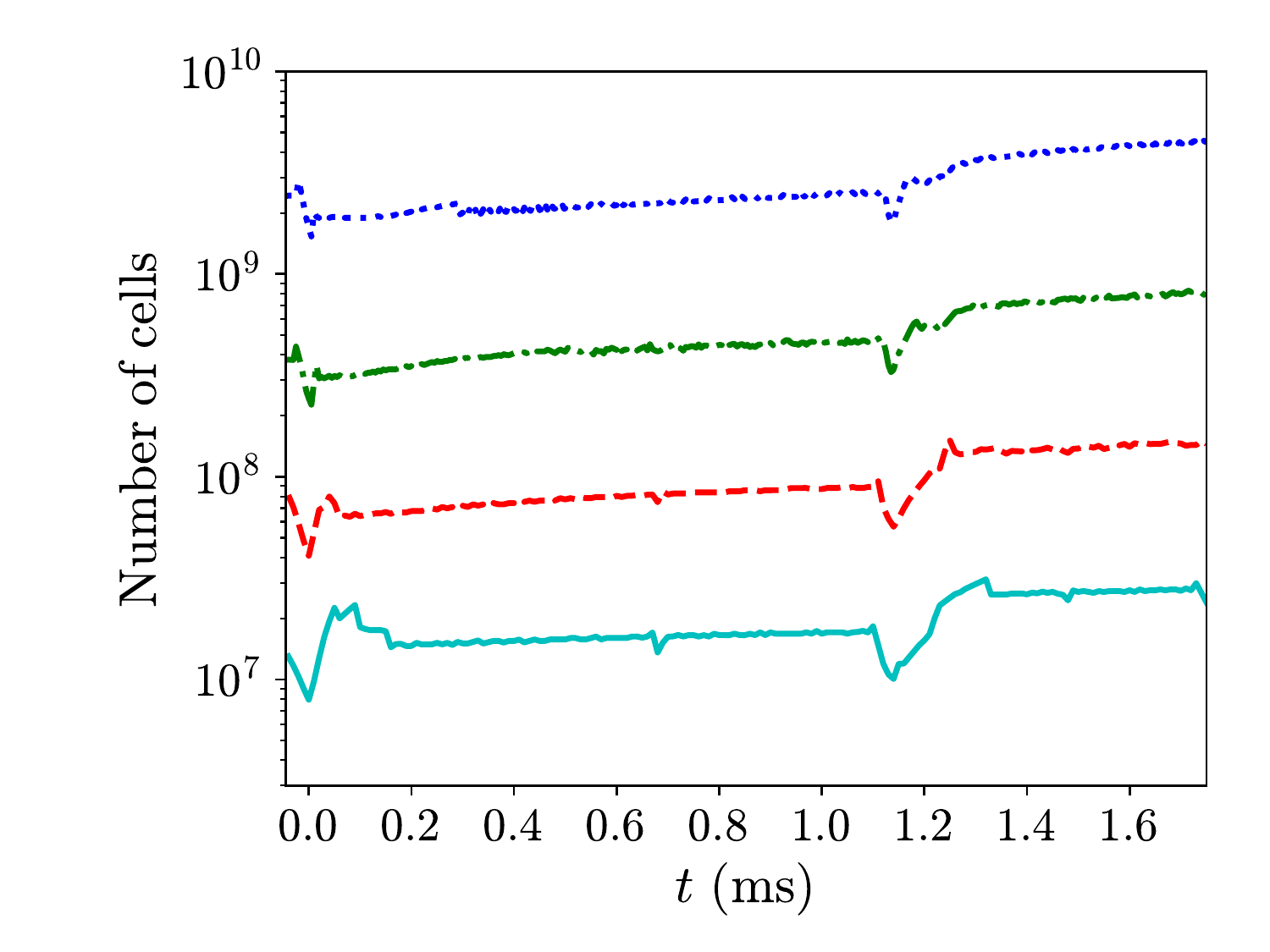}\label{fig:num_cells}}
\subfigure[$\ $Weighted number of cells]{%
\includegraphics[width=0.4\textwidth]{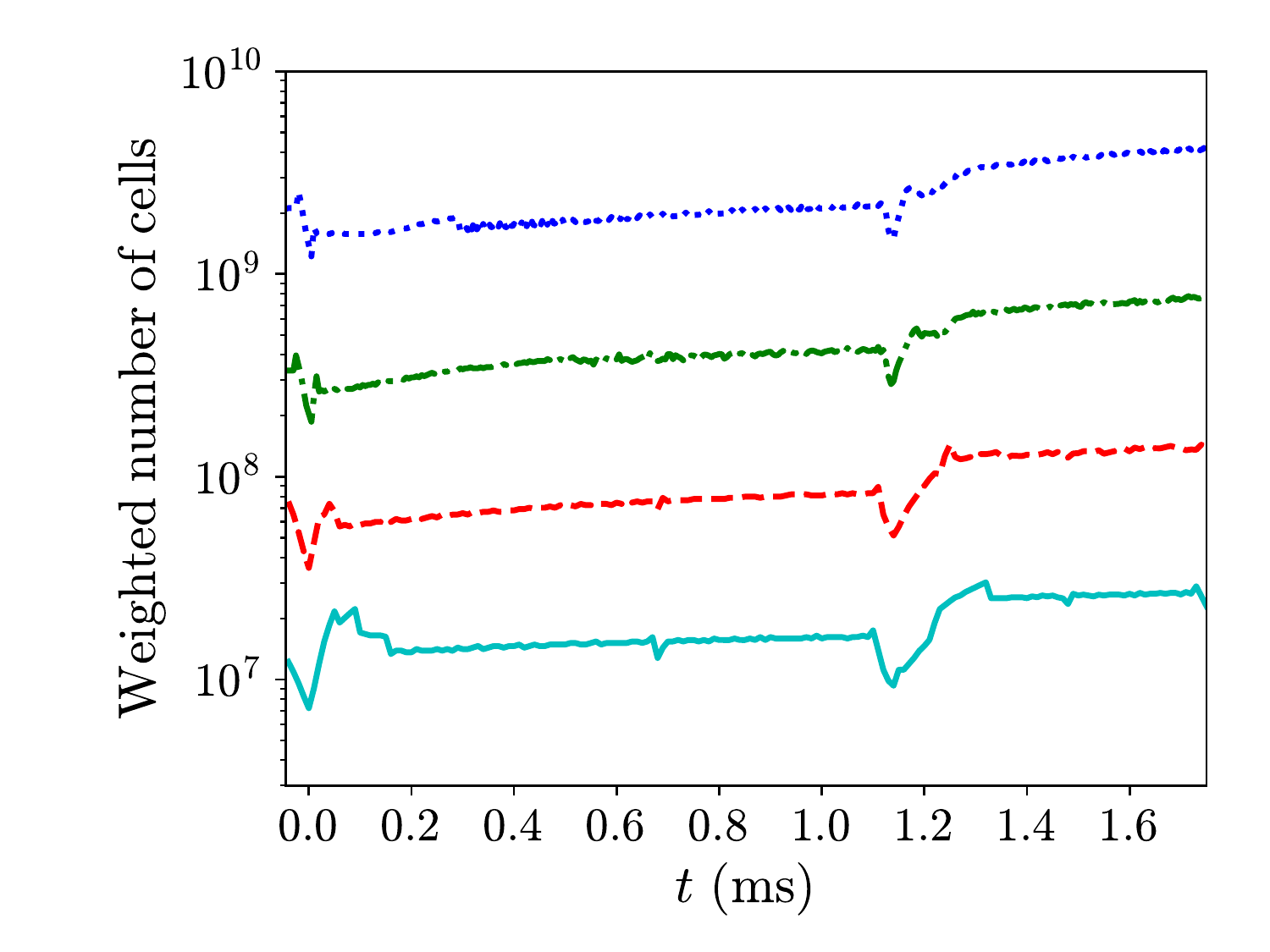}\label{fig:weighted_num_cells}}
\caption{Number of grid cells during the simulations. Cyan solid line: grid B; red dashed line: grid C; green dash-dotted line: grid D; blue dotted line: grid E.}
\label{fig:num_cells_and_weighted_num_cells}
\end{figure*}

Figure~\ref{fig:num_cells_and_weighted_num_cells} shows the number of grid cells and weighted number of grid cells summed over all grid levels for different AMR grid resolutions over time. The weighted number of grid cells is defined as:
\begin{equation}
    \sum^{l_{\mathrm{max}}}_{l=0} \frac{\Delta x_{l_{\mathrm{max}}}}{\Delta x_l} N_l,
\end{equation}
where $N_l$ and $\Delta x_l$ are the number of grid cells and grid spacing respectively at level $l$. The maximum level number $l_{\mathrm{max}}=2$ is used in this work. The weighted number of grid cells accounts for the fact that the time step size is larger for grid cells at the lower grid level from the CFL condition and has less computational cost compared to grid cells at higher grid levels. Since larger time step sizes are used for coarser grid levels in the multi-time stepping (sub-cycling) algorithm of the AMR code, the weighted number of grid cells is a better metric for comparing the computational cost of different AMR simulations. From figure~\ref{fig:num_cells_and_weighted_num_cells}, it can be seen that both the number of cells and weighted number of cells are the largest near the end of simulation for each grid resolution. The maximum number of cells and weighted number of cells for the grid E setting are around 4.55 and 4.19 billions respectively. Both number of cells and weighted number of cells are close to each other over time since most of the grid cells are on the finest level.

\section{\label{sec:appendix_TC} Transport coefficients}

The shear viscosity, $\mu_i$, of species $i$ is given by the Chapman-Enskog's model~\cite{chapman1991mathematical}:
\begin{equation}
	\mu_i = 2.6693 \times 10^{-6} \frac{\sqrt{M_i T}}{\Omega_{\mu,i} \sigma^2_i},
\end{equation}
\noindent where $\sigma_i$ is the collision diameter and $\Omega_{\mu, i}$ is the collision integral of the species given by
\begin{equation}
	\Omega_{\mu, i} = A \left( T^*_{i} \right)^B + C \textnormal{exp} \left( D T^*_{i} \right) + E \textnormal{exp} \left( F T^*_{i} \right),
\end{equation}
\noindent where $T^*_{i} = T/( \epsilon/k )_i$, $A = 1.16145$, $B = -0.14874$, $C = 0.52487$, $D = -0.7732$, $E = 2.16178$, and $F = -2.43787$. $T$ is the temperature of the species. $( \epsilon/k )_i$ is the Lennard-Jones energy parameter and $M_i$ is the molecular mass of the species. The values of $M_i$, $( \epsilon/k )_i$, and $\sigma_i$ are given in Table \ref{tab:fluid_properties}.

The bulk viscosity, $\mu_v$, of air is given by the linear model by \citet{gu2014systematic}:
\begin{equation}
	\mu_{v} = A + B T,
\end{equation}
\noindent where $A=-3.15\times10^{-5}\ {\mathrm{kg \ m^{-1} s^{-1}}}$ and $B = 1.58\times10^{-7}\ {\mathrm{kg \ m^{-1} s^{-1} K^{-1}}}$.

The bulk viscosity, $\mu_v$, of $\mathrm{SF_6}$ is given by Cramer's model~\cite{cramer2012numerical}:
\begin{align}
	\mu_{v} &= \left( \gamma - 1 \right)^2 \left. c_v \right|_{v}  (p \tau_v), \\
	\left. c_v \right|_{v} &= \left( \frac{c_{v}}{R} - \frac{f_{r} + 3}{2} \right), \\
    (p \tau_v) &= A \textnormal{exp} \left( \frac{B}{T^{\frac{1}{3}}} + \frac{C}{T^{\frac{2}{3}}} \right),
\end{align}
\noindent where $f_r = 3$, $A=0.2064\times10^{-5}\ {\mathrm{kg\ m^{-1}s^{-1}}}$, $B=121\ {\mathrm{K^{1/3}}}$, and $C=-339\ {\mathrm{K^{2/3}}}$ for $\mathrm{SF_6}$.

The thermal conductivity of species $i$, $\kappa_i$, is defined by:
\begin{equation}
	\kappa_i = c_{p,i} \frac{\mu_i}{Pr_i},
\end{equation}
\noindent where $Pr_i$ and $c_{p,i}$ are the species-specific Prandtl number and specific heat at constant pressure respectively.

Mass diffusion coefficient of a binary mixture, $D_{ij}$, is given by~\cite{poling2001properties}:
\begin{equation}
	D_{ij} = D_{i} = D_{j} = \frac{0.0266}{\Omega_{D,ij}} \frac{T^{3/2}}{p \sqrt{M_{ij}} \sigma_{ij}^2},
\end{equation}
\noindent where $p$ and $T$ are the pressure and temperature of the mixture. $\Omega_{D,ij}$ is the collision integral for diffusion given by:
\begin{equation}
	\Omega_{D,ij} = A \left( T^*_{ij} \right)^B + C \textnormal{exp} \left( D T^*_{ij} \right) + E \textnormal{exp} \left( F T^*_{ij} \right) + G \textnormal{exp} \left( H T^*_{ij} \right),
\end{equation}
\noindent where $T^*_{ij} = T/T_{\epsilon_{ij}}$, $A = 1.06036$, $B = -0.1561$, $C = 0.19300$, $D = -0.47635$, $E = 1.03587$, $F = -1.52996$, $G = 1.76474$, and $H = -3.89411$. $M_{ij}$, $\sigma_{ij}$, and $T_{\epsilon_{ij}}$ are the effective molecular mass, collision diameter, and Lennard--Jones energy parameter respectively for the mixture:
\begin{align}
	M_{ij} &= \frac{2}{\frac{1}{M_i} + \frac{1}{M_j}}, \\
    \sigma_{ij} &= \frac{\sigma_i + \sigma_j}{2}, \\
    T_{\epsilon_{ij}} &= \sqrt{\left( \frac{\epsilon}{k} \right)_i \left( \frac{\epsilon}{k} \right)_j }.
\end{align}
\noindent The values of $M_i$, $( \epsilon/k )_i$, and $\sigma_i$ of different species are given in table \ref{tab:fluid_properties}.

\begin{table}[!ht]
\caption{\label{tab:fluid_properties}%
Fluid properties.}
\begin{ruledtabular}
\begin{tabular}{ c c c c c c c c c }
Gas & $\gamma_i$ & $c_{p,i}\ (\mathrm{J\ kg^{-1} K^{-1}})$ &
$c_{v,i}\ (\mathrm{J\ kg^{-1} K^{-1}})$ & $M_i\ (\mathrm{g\ mol^{-1}})$ &
$R_i\ (\mathrm{J\ kg^{-1} K^{-1}})$ & $\left( \epsilon/k \right)_i\ (\mathrm{K})$ &
$\sigma_i\ (\mbox{\normalfont\AA})$ & $Pr_i$ \\
\hline
$\mathrm{SF_6}$ & 1.09312 & 668.286 & 611.359 & 146.055 & 56.9269 & 222.1 & 5.128 & 0.79 \\
Air & 1.39909 & 1040.50 & 743.697 & 28.0135 & 296.802 & 78.6 & 3.711 & 0.71
\end{tabular}
\end{ruledtabular}
\end{table}



\section{\label{sec:appendix_mixing_rules} Mixing rules}
With the assumption that all species are at pressure and temperature equilibria, the ratio of specific heats of the mixture follows as
\begin{equation}
    \gamma = \frac{c_p}{c_v}, \quad c_{p} = \sum_{i=1}^{N} Y_i c_{p,i}, \quad c_{v} = \sum_{i=1}^{N} Y_i c_{v,i}. \label{eq:specific_heats}
\end{equation}
The molecular mass of the mixture is given by
\begin{equation}
    M = \left( \sum_{i=1}^{N} \frac{Y_i}{M_i} \right) ^{-1}.
\end{equation}
The mixture shear viscosity, bulk viscosity, and thermal conductivity are given by
\begin{align}
	\mu    &= \frac{\sum^{N}_{i=1} \mu_i Y_i/\sqrt{M_i}}{\sum^{N}_{i=1} Y_i/\sqrt{M_i}}, \\
	\mu_v  &= \frac{\sum^{N}_{i=1} \mu_{v, i} Y_i/\sqrt{M_i}}{\sum^{N}_{i=1} Y_i/\sqrt{M_i}}, \\
	\kappa &= \frac{\sum^{N}_{i=1} \kappa_i Y_i/\sqrt{M_i}}{\sum^{N}_{i=1} Y_i/\sqrt{M_i}}.
\end{align}

\bibliography{references}
  
\end{document}



\title{Supplemental Material: Analysis of second-moments and their budgets for \\
Richtmyer--Meshkov instability with \\
variable-density turbulence induced by re-shock}

\author{Man Long Wong}
 \email{mlwong@alumni.stanford.edu }
 \affiliation{%
  Department of Aeronautics and Astronautics, \& \\
  Center for Turbulence Research, Stanford University \\
  Stanford, CA 94305, USA
 }%

\author{Jon R. Baltzer}
 \affiliation{%
  XTD-IDA, Los Alamos National Laboratory \\
  Los Alamos, NM 87545, USA
 }%

\author{Daniel Livescu}
 \affiliation{%
  CCS-2, Los Alamos National Laboratory \\
  Los Alamos, NM 87545, USA
 }%

\author{Sanjiva K. Lele}
 \affiliation{%
  Department of Aeronautics and Astronautics, \\
  Department of Mechanical Engineering, \& \\
  Center for Turbulence Research, Stanford University \\
  Stanford, CA 94305, USA
 }%

\date{\today}

\begin{abstract}
\end{abstract}

\pacs{Valid PACS appear here}
\maketitle


\section{Phase shifts of the initial perturbation}

The phase shifts of each mode, $\phi_m$ and $\psi_m$, with wavenumber $m$ are given in table~\ref{tab:phase_shifts}.

\begin{table}[!ht]
\caption{\label{tab:phase_shifts}%
Phase shifts $\phi_m$ and $\psi_m$ of each mode with wavenumber $m$.}
\begin{ruledtabular}
\begin{tabular}{ c c c }
 $m$ & $\phi_m$ & $\psi_m$ \\ 
 \hline
  $20$ & $0.99031219901567$ & $2.63387681374906$ \\
  $21$ & $6.06257346743012$ & $3.38548538635238$ \\
  $22$ & $6.01619287992480$ & $2.49296356445290$ \\
  $23$ & $3.43615792623680$ & $2.17122208289517$ \\
  $24$ & $1.74985591686112$ & $1.17030742344034$ \\
  $25$ & $0.61286443954864$ & $0.58018050193692$ \\
  $26$ & $3.97323032474265$ & $0.92209445692414$ \\
  $27$ & $5.73890975922526$ & $1.89961157824218$ \\
  $28$ & $0.79788169834087$ & $0.00071863817185$ \\
  $29$ & $5.69125859039526$ & $4.52593227359734$ \\
  $30$ & $5.11905989575681$ & $2.62022653271778$
\end{tabular}
\end{ruledtabular}
\end{table}


\section{Grid sensitivity analysis of the ratio of the density-specific-volume covariance to the square of density intensity}

\begin{figure*}[!ht]
\centering
\subfigure[$\ $Before re-shock]{%
\includegraphics[width=0.4\textwidth]{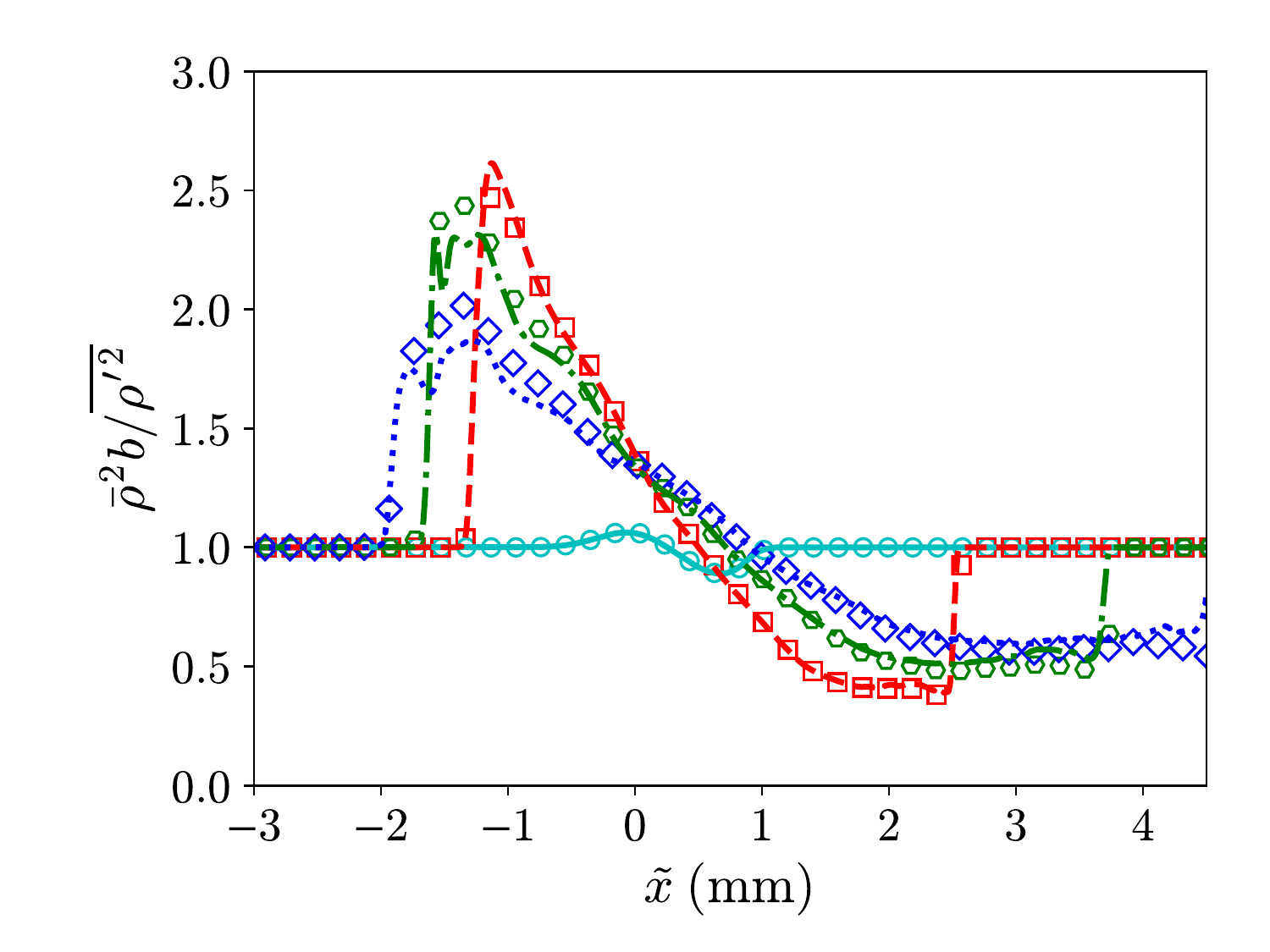}}
\subfigure[$\ $After re-shock]{%
\includegraphics[width=0.4\textwidth]{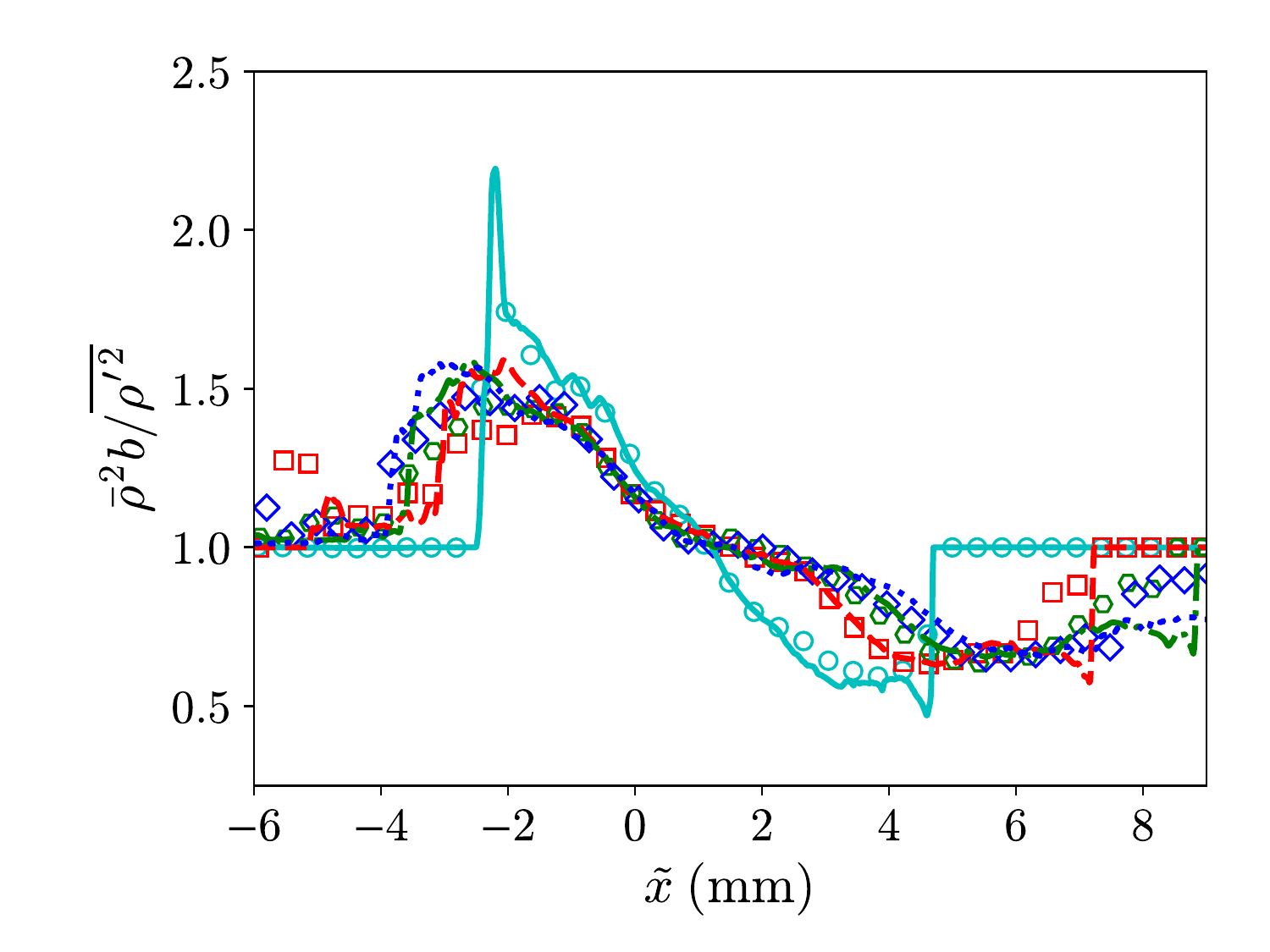}}
\caption{Grid sensitivities of the profiles of the ratio of the density-specific-volume covariance to the square of density intensity at different times between the grid D and the grid E.
The profiles with the grid D and the grid E are shown with symbols and lines respectively.
Cyan circles or solid line in (a): $t=0.05\ \mathrm{ms}$; red squares or dashed line in (a): $t=0.40\ \mathrm{ms}$; green hexagons or dash-dotted line in (a): $t=0.75\ \mathrm{ms}$; blue diamonds or dotted line in (a): $t=1.10\ \mathrm{ms}$. Cyan circles or solid line in (b): $t=1.20\ \mathrm{ms}$; red squares or dashed line in (b): $t=1.40\ \mathrm{ms}$; green hexagons or dash-dotted line in (b): $t=1.60\ \mathrm{ms}$; blue diamonds or dotted line in (b): $t=1.75\ \mathrm{ms}$.}
\label{fig:Bouss_profiles_spat_conv}
\end{figure*}

The grid sensitivities of the ratio of the density-specific-volume covariance to the square of density intensity at different times between the grid D and the grid E are shown in figure~\ref{fig:Bouss_profiles_spat_conv}. The profiles at different times show minor grid sensitivities between the two grid resolutions, which have small effects on the analysis of this quantity in the main article.

\section{Grid sensitivity analysis of the decomposition of the Reynolds normal stress component in the streamwise direction}

\begin{figure*}[!ht]
\centering
\subfigure[$\ t=0.40\ \mathrm{ms}$]{%
\includegraphics[height = 0.33\textwidth]{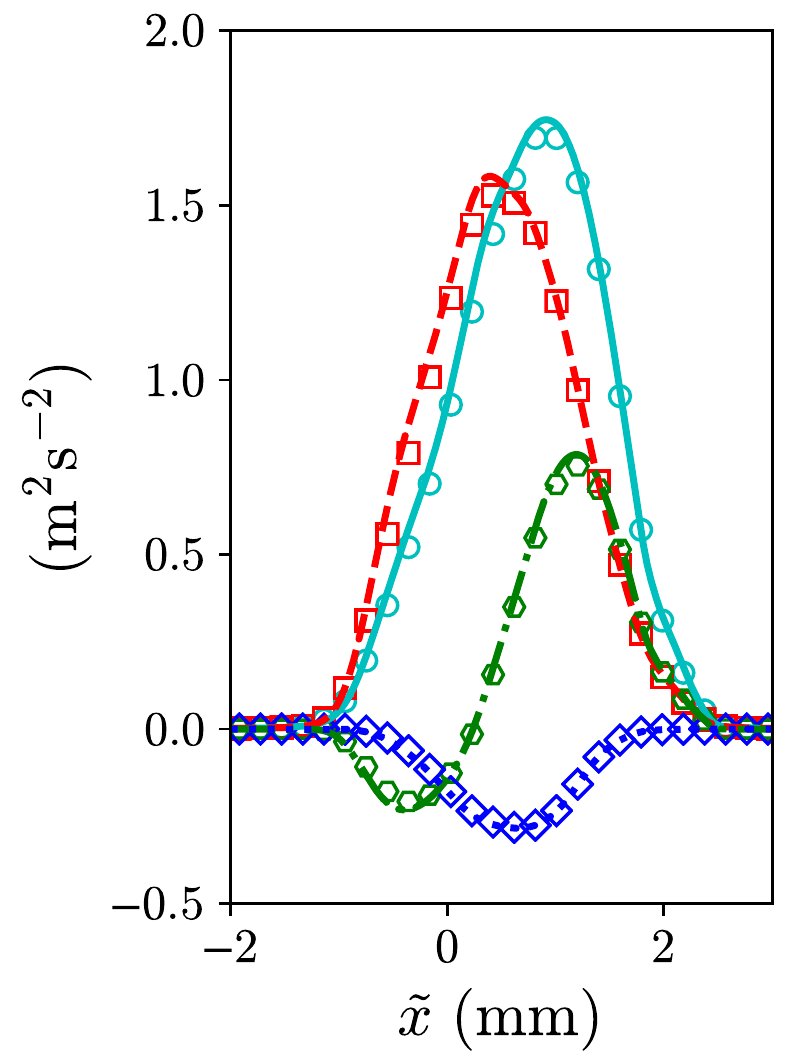}}
\subfigure[$\ t=1.10\ \mathrm{ms}$]{%
\includegraphics[height = 0.33\textwidth]{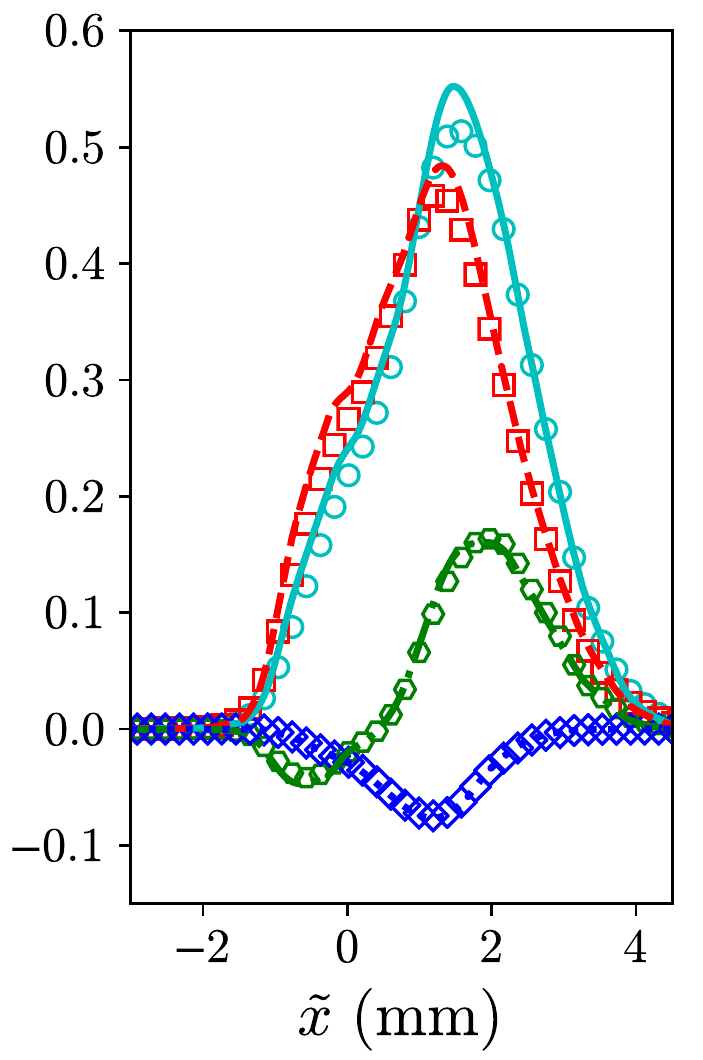}}
\subfigure[$\ t=1.20\ \mathrm{ms}$]{%
\includegraphics[height = 0.33\textwidth]{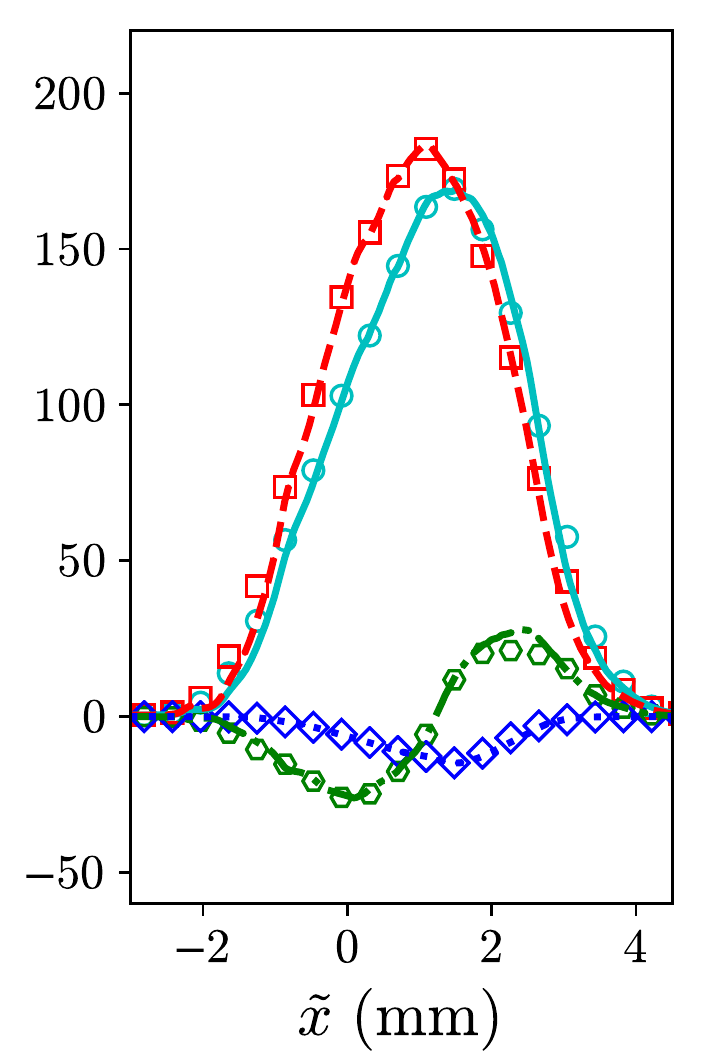}}
\subfigure[$\ t=1.75\ \mathrm{ms}$]{%
\includegraphics[height = 0.33\textwidth]{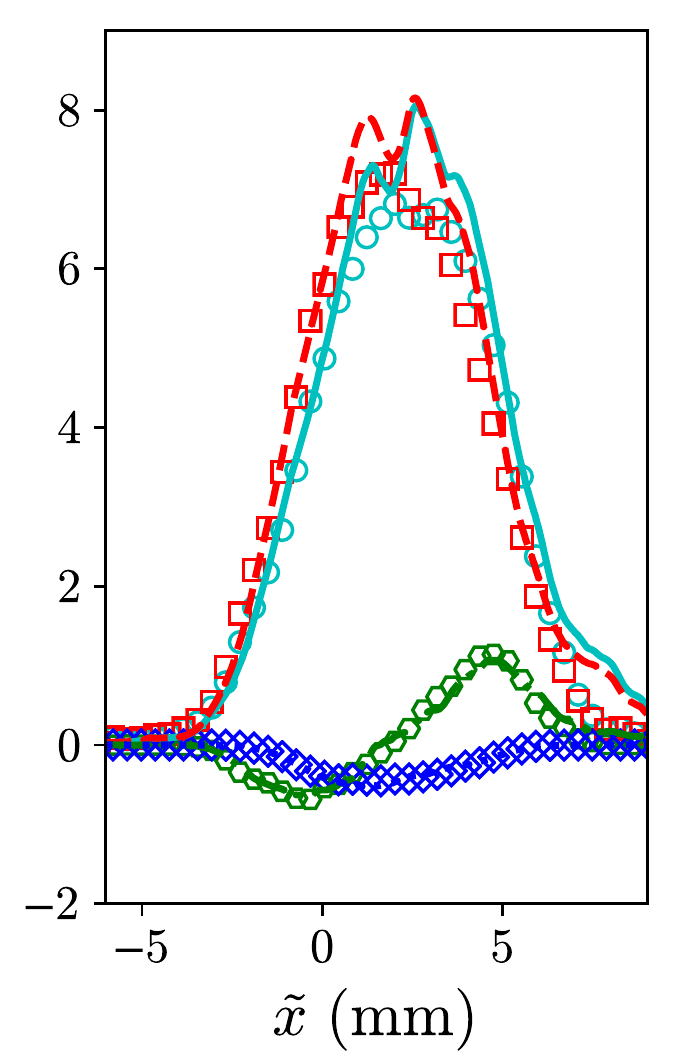}}
\caption{Grid sensitivities of the decomposition of the Reynolds normal stress component in the streamwise direction multiplied by the mean density, $\bar{\rho} \tilde{R}_{11}$, at different times between grid D and grid E.
The profiles with the grid D and the grid E are shown with symbols and lines respectively.
Cyan circles or solid line: $\tilde{R}_{11}$; red squares or dashed line: $\overline{u^{\prime}u^{\prime}}$ [term (I)]; green hexagons or dash-dotted line: $\overline{\rho^{\prime}u^{\prime}u^{\prime}} / \bar{\rho}$ [term (II)]; blue diamonds or dotted line: $-a_1^2$ [term (III)].}
\label{fig:R11_decomposition_spat_conv}
\end{figure*}

The grid sensitivities of the decomposition of the Reynolds normal stress component in the streamwise direction multiplied by the mean density, $\bar{\rho} \tilde{R}_{11}$, at different times between the grid D and the grid E are shown in figure~\ref{fig:R11_decomposition_spat_conv}. The profiles at different times show minor grid sensitivities between the two grid resolutions and have insignificant effects on the analysis in the main article.

\section{Grid sensitivity analysis of the budgets of the second-moments before re-shock}

\begin{figure*}[!ht]
\centering
\subfigure[$\ t=0.40\ \mathrm{ms}$]{%
\includegraphics[width=0.4\textwidth]{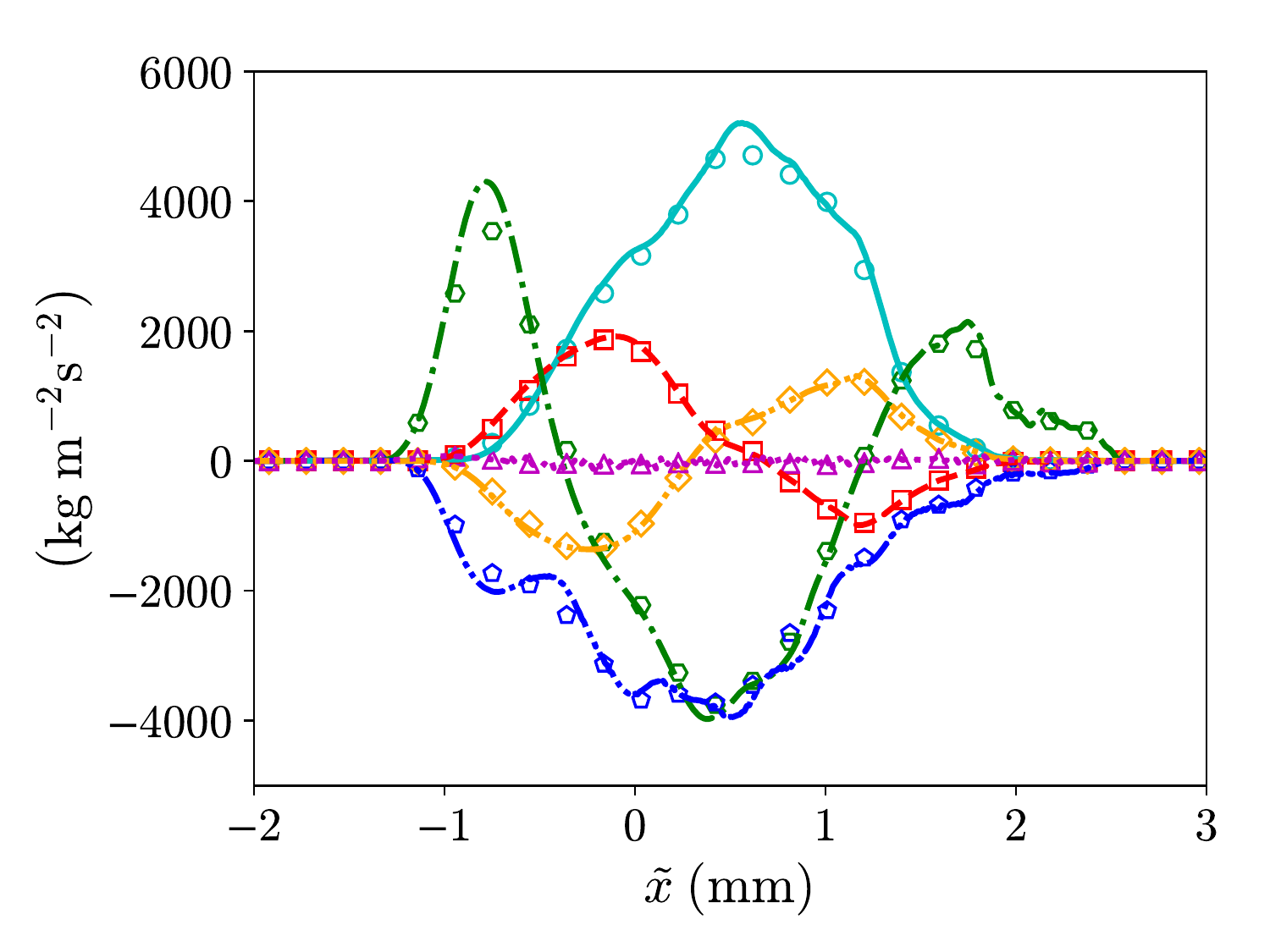}}
\subfigure[$\ t=1.10\ \mathrm{ms}$]{%
\includegraphics[width=0.4\textwidth]{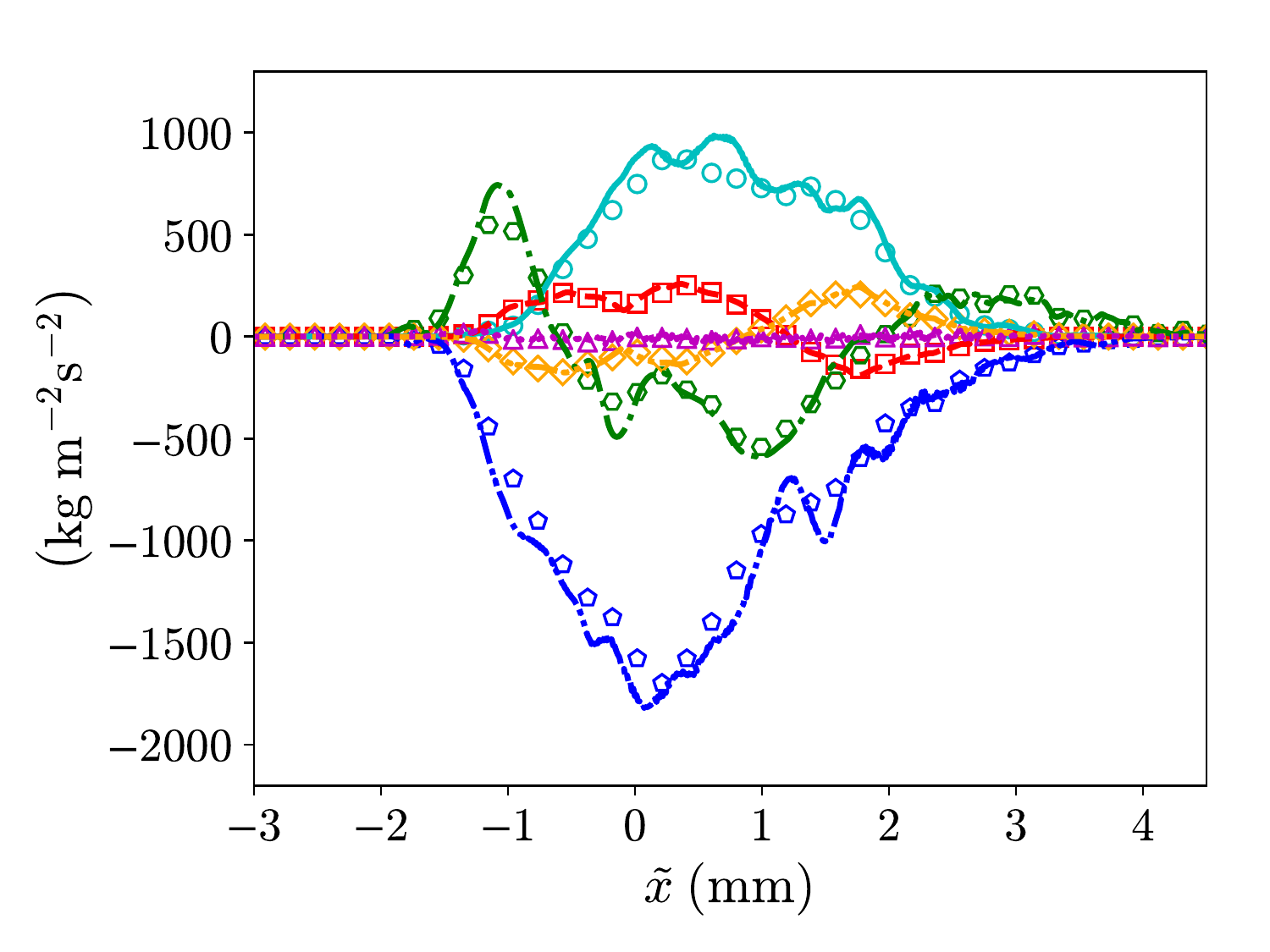}}
\caption{Grid sensitivities of the budgets of the turbulent mass flux component in the streamwise direction, $\bar{\rho} a_1$, at different times before re-shock between the grid D and the grid E.
The budgets with the grid D and the grid E are shown with symbols and lines respectively.
Cyan circles or solid line: production [term (III)]; red squares or dashed line: redistribution [term (IV)]; green hexagons or dash-dotted line: turbulent transport [term (V)]; blue pentagons or dash-dot-dotted line: destruction [term (VI)]; orange diamonds or dash-triple-dotted line: negative of convection due to streamwise velocity associated with turbulent mass flux; magenta triangles or dotted line: residue.}
\label{fig:rho_a1_budget_spat_conv}
\end{figure*}

\begin{figure*}[!ht]
\centering
\subfigure[$\ t=0.40\ \mathrm{ms}$]{%
\includegraphics[width=0.4\textwidth]{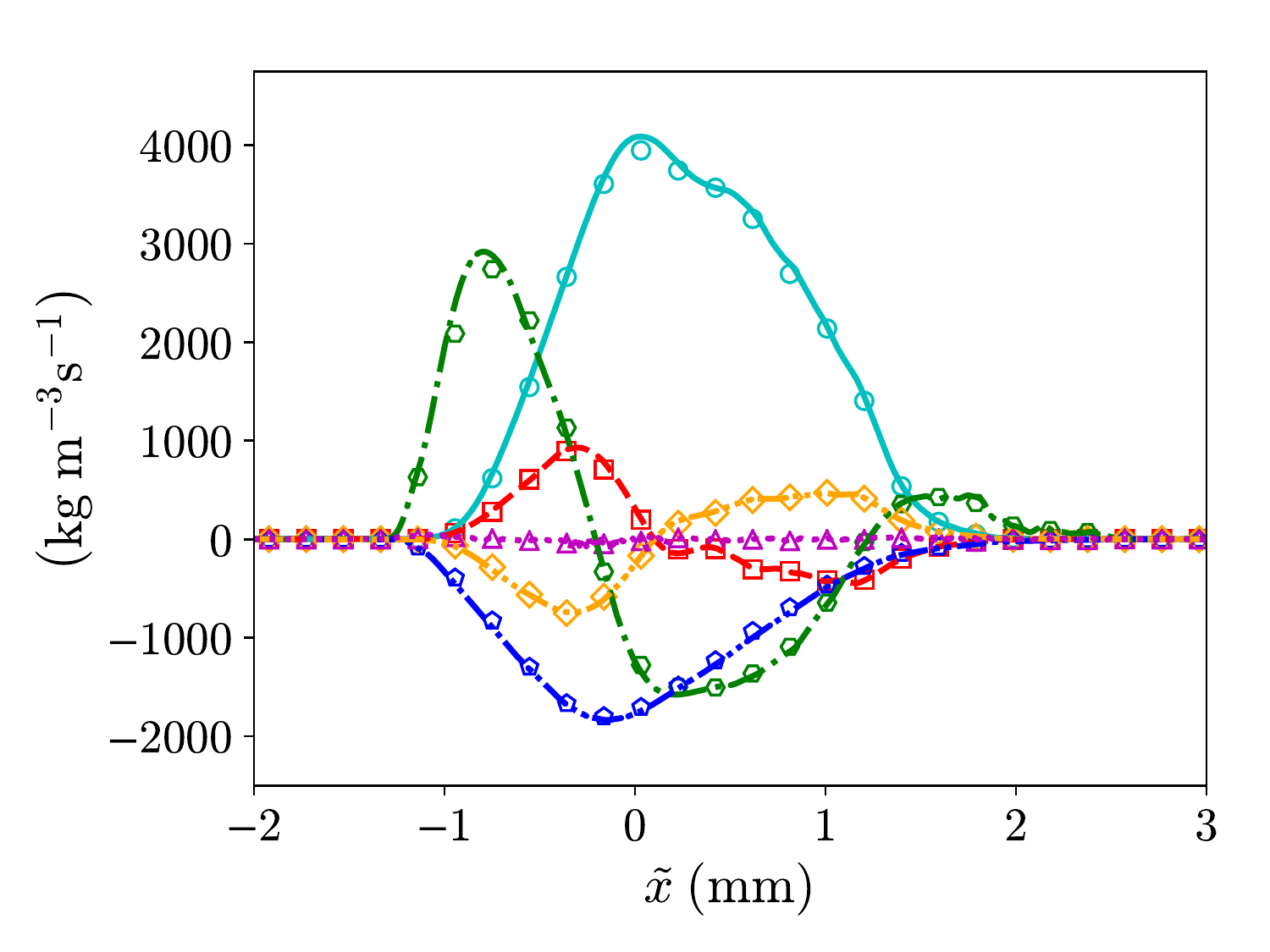}}
\subfigure[$\ t=1.10\ \mathrm{ms}$]{%
\includegraphics[width=0.4\textwidth]{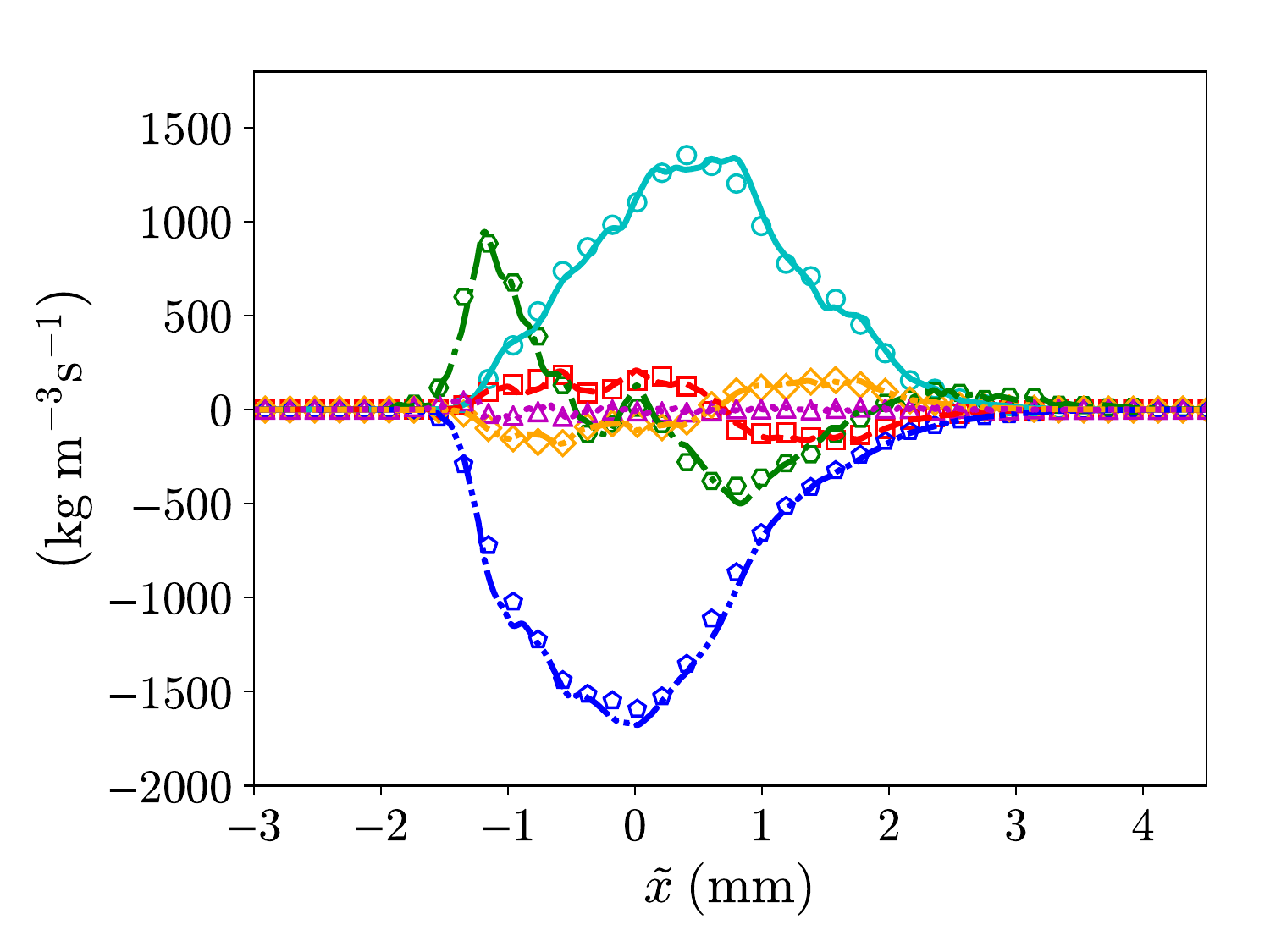}}
\caption{Grid sensitivities of the budgets of the density-specific-volume covariance multiplied by the mean density, $\bar{\rho} b$, at different times before re-shock between the grid D and the grid E.
The budgets with the grid D and the grid E are shown with symbols and lines respectively.
Cyan circles or solid line: production [term (III)]; red squares or dashed line: redistribution [term (IV)]; green hexagons or dash-dotted line: turbulent transport [term (V)]; blue pentagons or dash-dot-dotted line: destruction [term (VI)]; orange diamonds or dash-triple-dotted line: negative of convection due to streamwise velocity associated with turbulent mass flux; magenta triangles or dotted line: residue.}
\label{fig:rho_b_budget_spat_conv}
\end{figure*}

\begin{figure*}[!ht]
\centering
\subfigure[$\ t=0.40\ \mathrm{ms}$]{%
\includegraphics[width=0.4\textwidth]{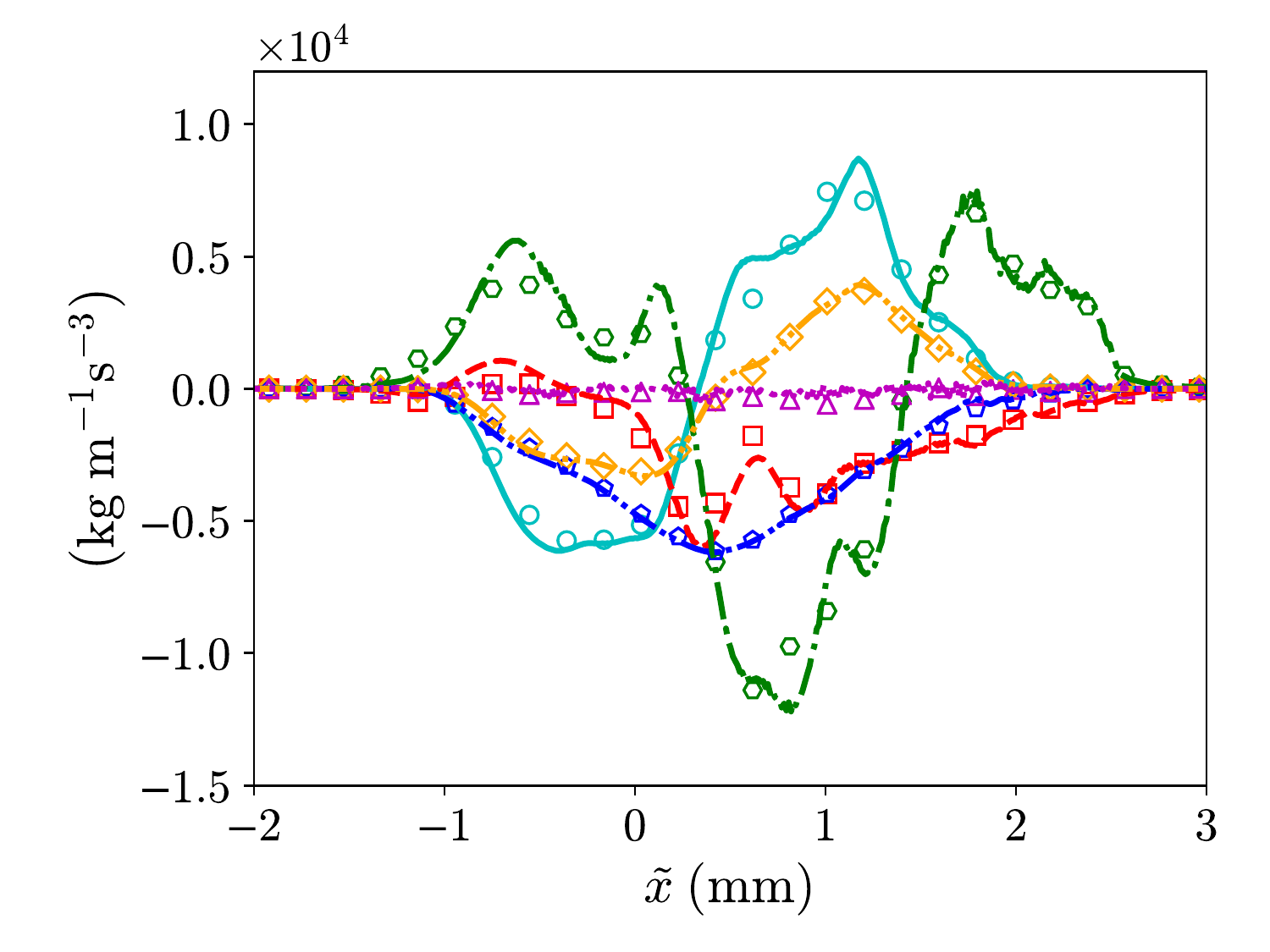}}
\subfigure[$\ t=1.10\ \mathrm{ms}$]{%
\includegraphics[width=0.4\textwidth]{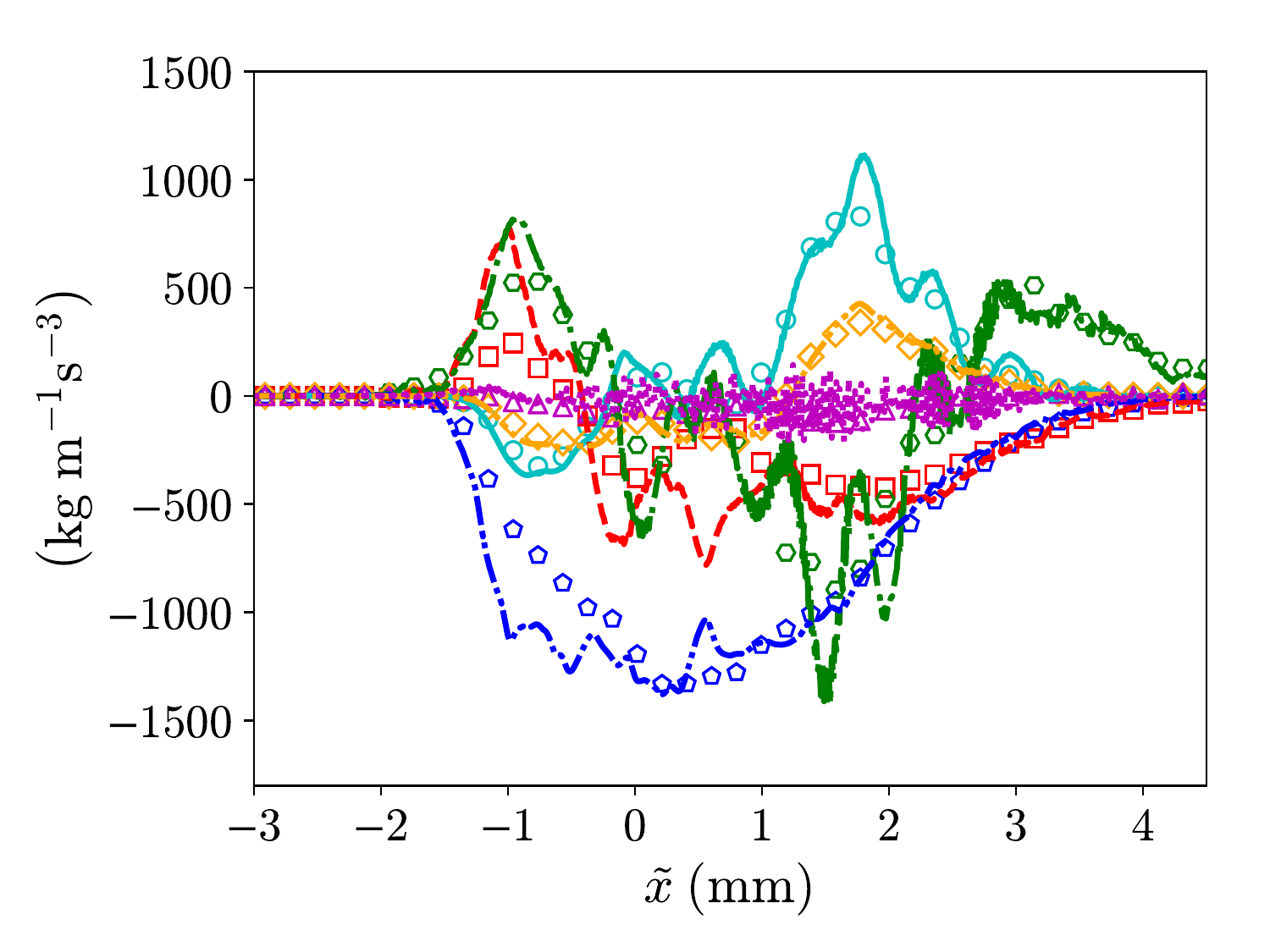}}
\caption{Grid sensitivities of the budgets of the Reynolds normal stress component in the streamwise direction multiplied by the mean density, $\bar{\rho} \tilde{R}_{11}$, at different times before re-shock between the grid D and the grid E. 
The budgets with the grid D and the grid E are shown with symbols and lines respectively.
Cyan circles or solid line: production [term (III)]; red squares or dashed line: press-strain redistribution [term (V)]; green hexagons or dash-dotted line: turbulent transport [term (IV)]; blue pentagons or dash-dot-dotted line: dissipation [term (VI)]; orange diamonds or dash-triple-dotted line: negative of convection due to streamwise velocity associated with turbulent mass flux; magenta triangles or dotted line: residue.}
\label{fig:rho_R11_budget_spat_conv}
\end{figure*}

\begin{figure*}[!ht]
\centering
\subfigure[$\ t=0.40\ \mathrm{ms}$]{%
\includegraphics[width=0.4\textwidth]{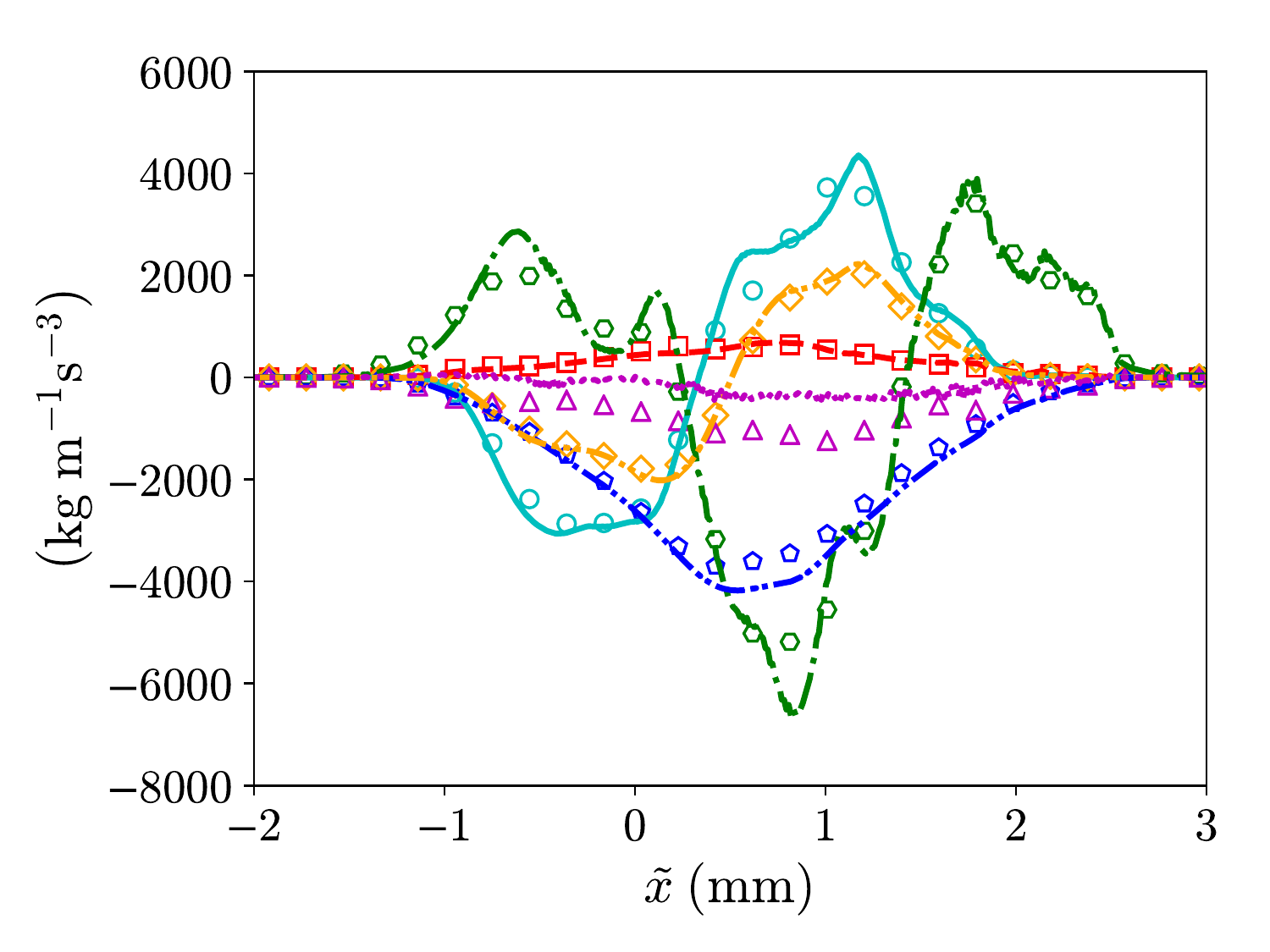}}
\subfigure[$\ t=1.10\ \mathrm{ms}$]{%
\includegraphics[width=0.4\textwidth]{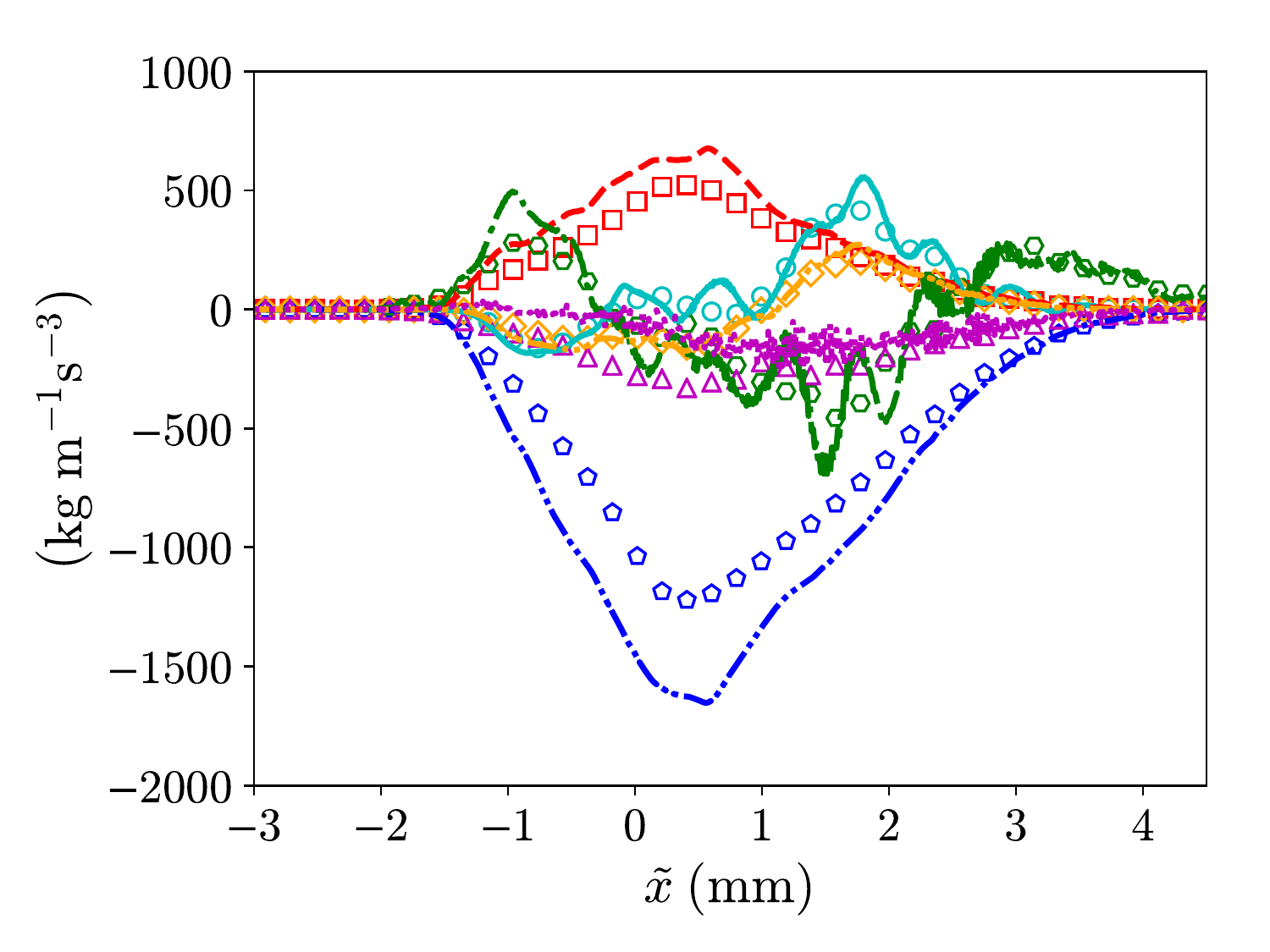}}
\caption{Grid sensitivities of the budgets of the turbulent kinetic energy, $\bar{\rho} k$, at different times before re-shock between the grid D and the grid E.
The budgets with the grid D and the grid E are shown with symbols and lines respectively.
Cyan circles or solid line: production [term (III)]; red squares or dashed line: pressure-dilatation [term (V)]; green hexagons or dash-dotted line: turbulent transport [term (IV)]; blue pentagons or dash-dot-dotted line: dissipation [term (VI)]; orange diamonds or dash-triple-dotted line: negative of convection due to streamwise velocity associated with turbulent mass flux; magenta triangles or dotted line: residue.}
\label{fig:rho_k_budget_spat_conv}
\end{figure*}

The grid sensitivities of $\bar{\rho} a_1$, $\bar{\rho} b$, $\bar{\rho} \tilde{R}_{11}$, and $\bar{\rho} k$ at different times before re-shock between the grid D and the grid E are shown respectively in figures~\ref{fig:rho_a1_budget_spat_conv}, \ref{fig:rho_b_budget_spat_conv}, \ref{fig:rho_R11_budget_spat_conv}, and \ref{fig:rho_k_budget_spat_conv}. Small grid sensitivities of the budgets of $\bar{\rho} a_1$ and $\bar{\rho} b$ can be seen between the two grid resolutions. Reasonably good grid convergence can also be seen for the budgets of $\bar{\rho} \tilde{R}_{11}$. The residues of the budgets of $\bar{\rho} k$ with grid D are not negligible but become smaller when the grid is refined to grid E as a larger amount of dissipation is captured due to finer grid resolution.

\section{Grid sensitivity analysis of the budgets of the large-scale second-moments after re-shock}

\begin{figure*}[!ht]
\centering
\subfigure[$\ t=1.20\ \mathrm{ms}$]{%
\includegraphics[width=0.4\textwidth]{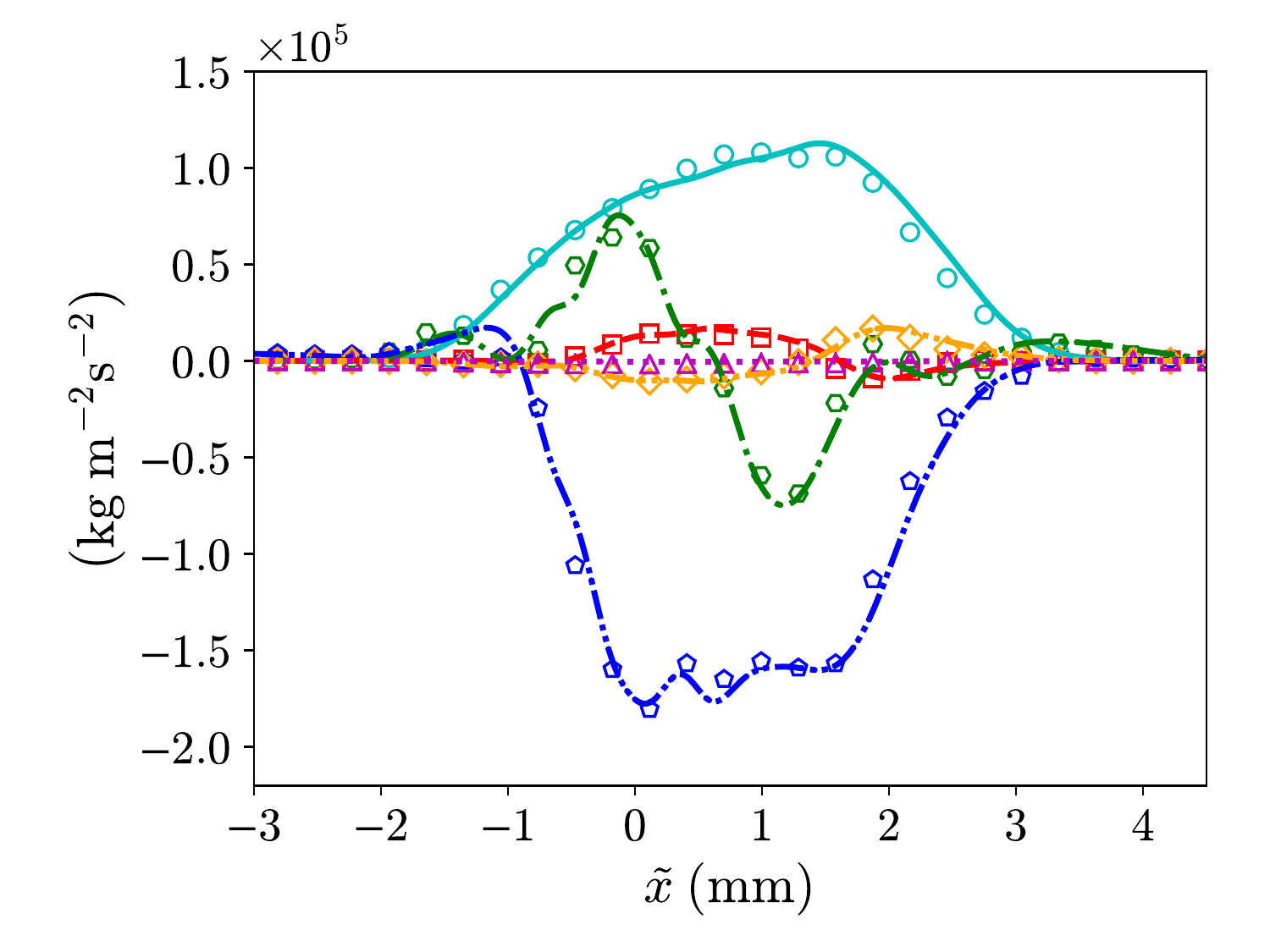}}
\subfigure[$\ t=1.60\ \mathrm{ms}$]{%
\includegraphics[width=0.4\textwidth]{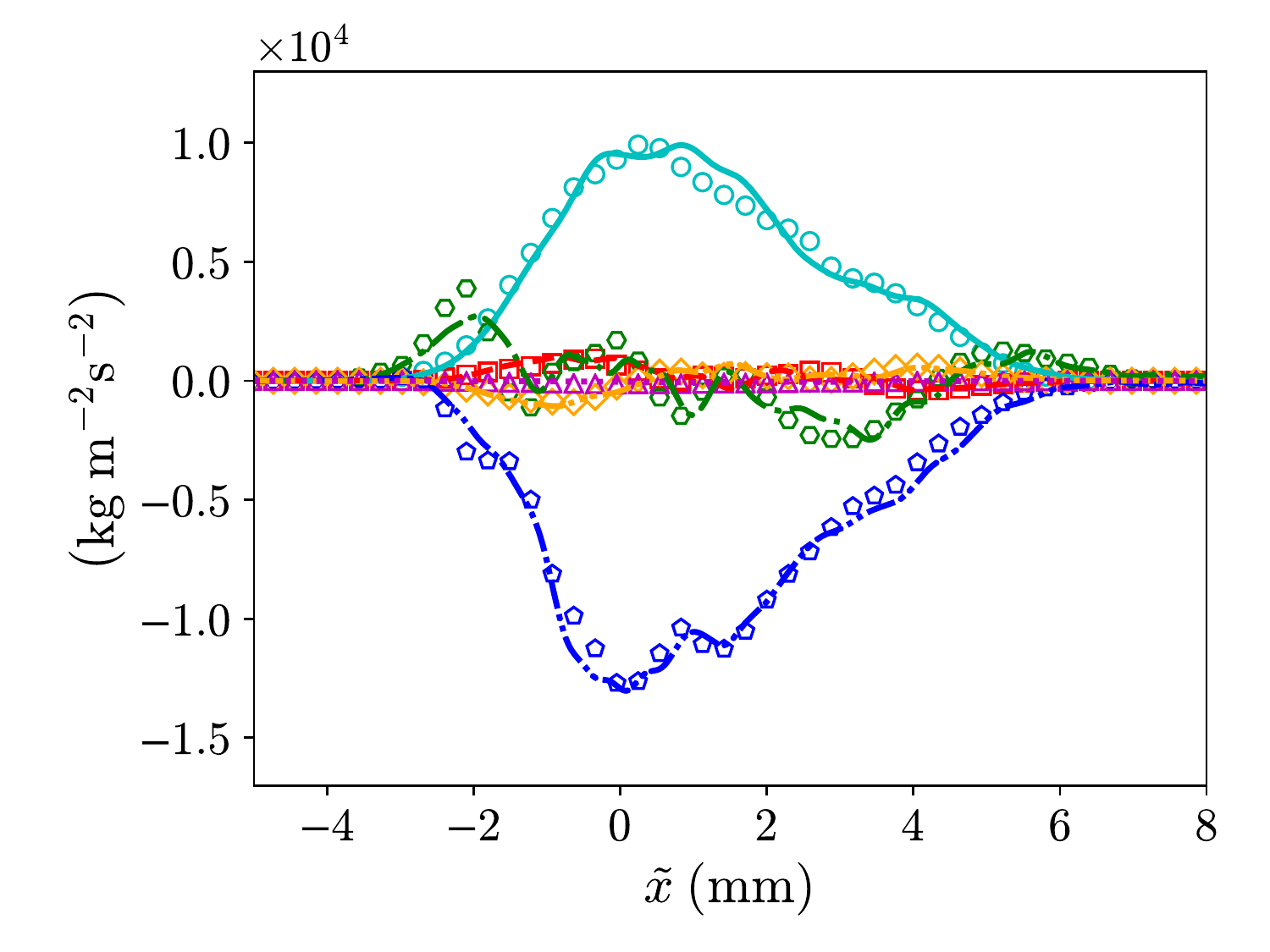}}
\caption{Grid sensitivities of the budgets of the large-scale turbulent mass flux component in the streamwise direction, $\overline{\left< \rho \right>}_{\ell} a_{L,1}$, at different times after re-shock between the grid D and the grid E.
The budgets with the grid D and the grid E are shown with symbols and lines respectively.
Cyan circles or solid line: production [term (III)]; red squares or dashed line: redistribution [term (IV)]; green hexagons or dash-dotted line: turbulent transport [term (V)]; blue pentagons or dash-dot-dotted line: destruction [term (VI)]; orange diamonds or dash-triple-dotted line: negative of convection due to streamwise velocity associated with turbulent mass flux; magenta triangles or dotted line: residue.}
\label{fig:rho_a1_budget_filtered_spat_conv}
\end{figure*}

\begin{figure*}[!ht]
\centering
\subfigure[$\ t=1.20\ \mathrm{ms}$]{%
\includegraphics[width=0.4\textwidth]{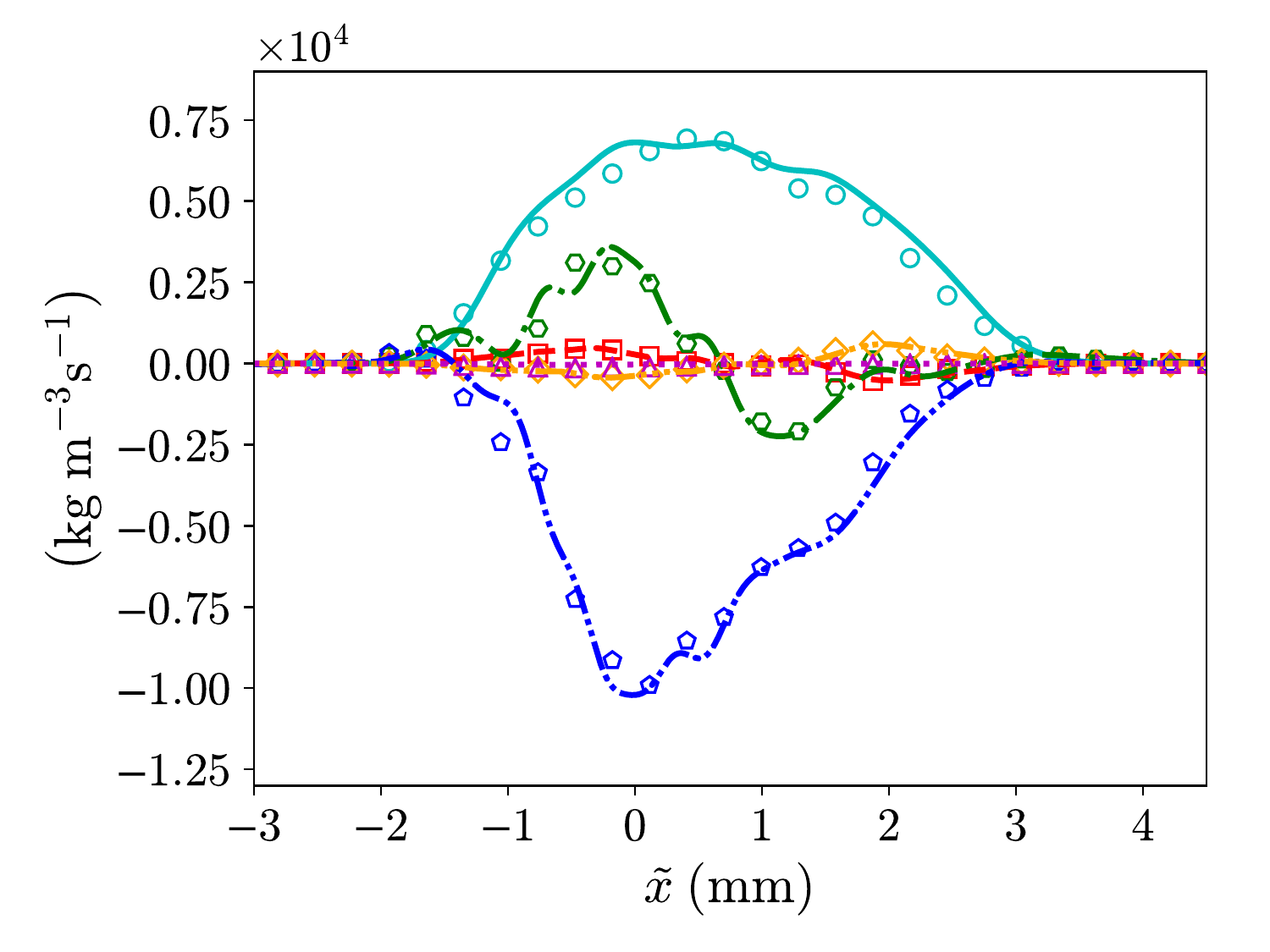}}
\subfigure[$\ t=1.60\ \mathrm{ms}$]{%
\includegraphics[width=0.4\textwidth]{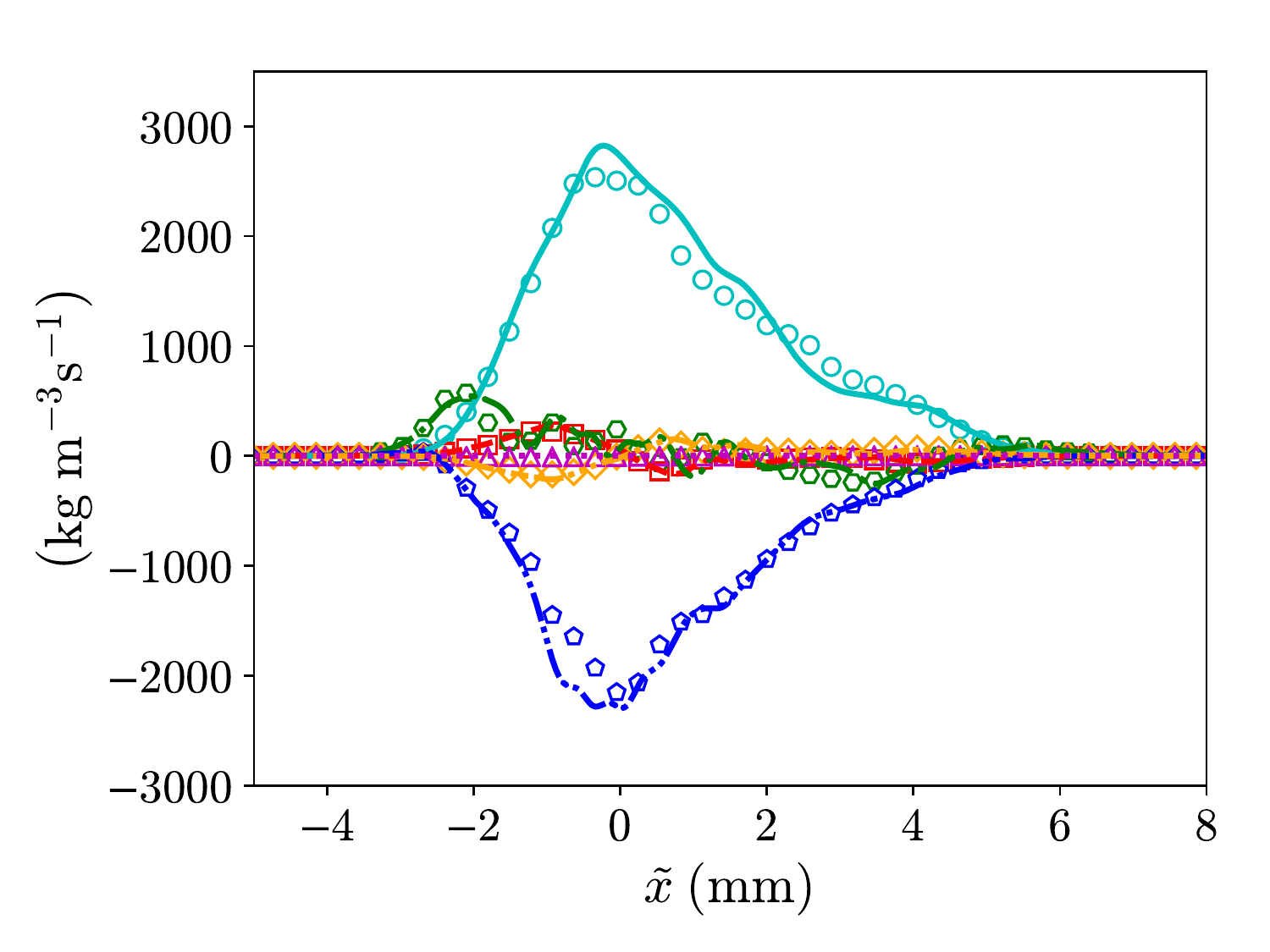}}
\caption{Grid sensitivities of the budgets of the large-scale density-specific-volume covariance multiplied by the mean filtered density, $\overline{\left< \rho \right>}_{\ell} b_L$, at different times after re-shock between the grid D and the grid E.
The budgets with the grid D and the grid E are shown with symbols and lines respectively.
Cyan circles or solid line: production [term (III)]; red squares or dashed line: redistribution [term (IV)]; green hexagons or dash-dotted line: turbulent transport [term (V)]; blue pentagons or dash-dot-dotted line: destruction [term (VI)]; orange diamonds or dash-triple-dotted line: negative of convection due to streamwise velocity associated with turbulent mass flux; magenta triangles or dotted line: residue.}
\label{fig:rho_b_budget_filtered_spat_conv}
\end{figure*}

\begin{figure*}[!ht]
\centering
\subfigure[$\ t=1.20\ \mathrm{ms}$]{%
\includegraphics[width=0.4\textwidth]{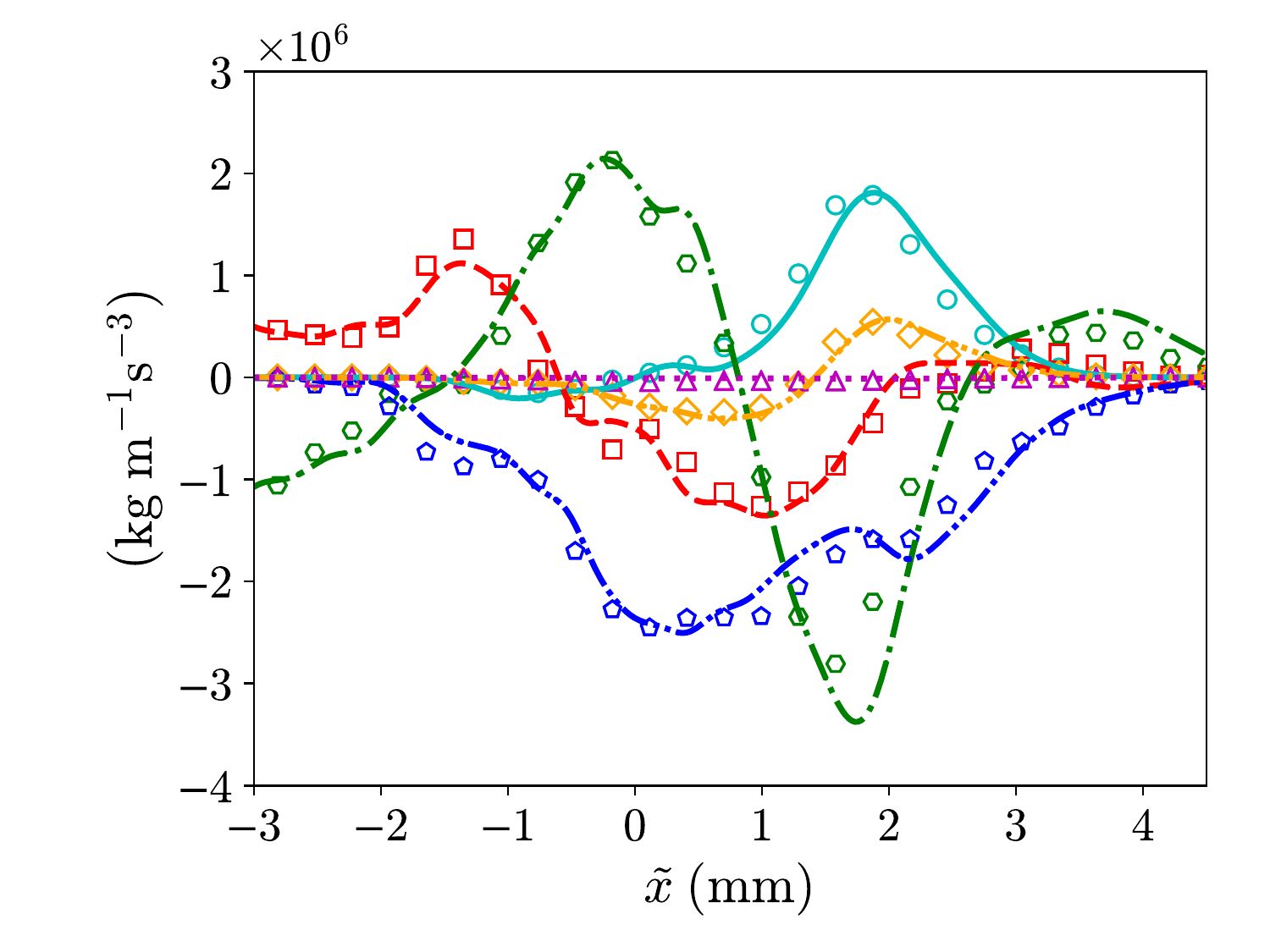}}
\subfigure[$\ t=1.60\ \mathrm{ms}$]{%
\includegraphics[width=0.4\textwidth]{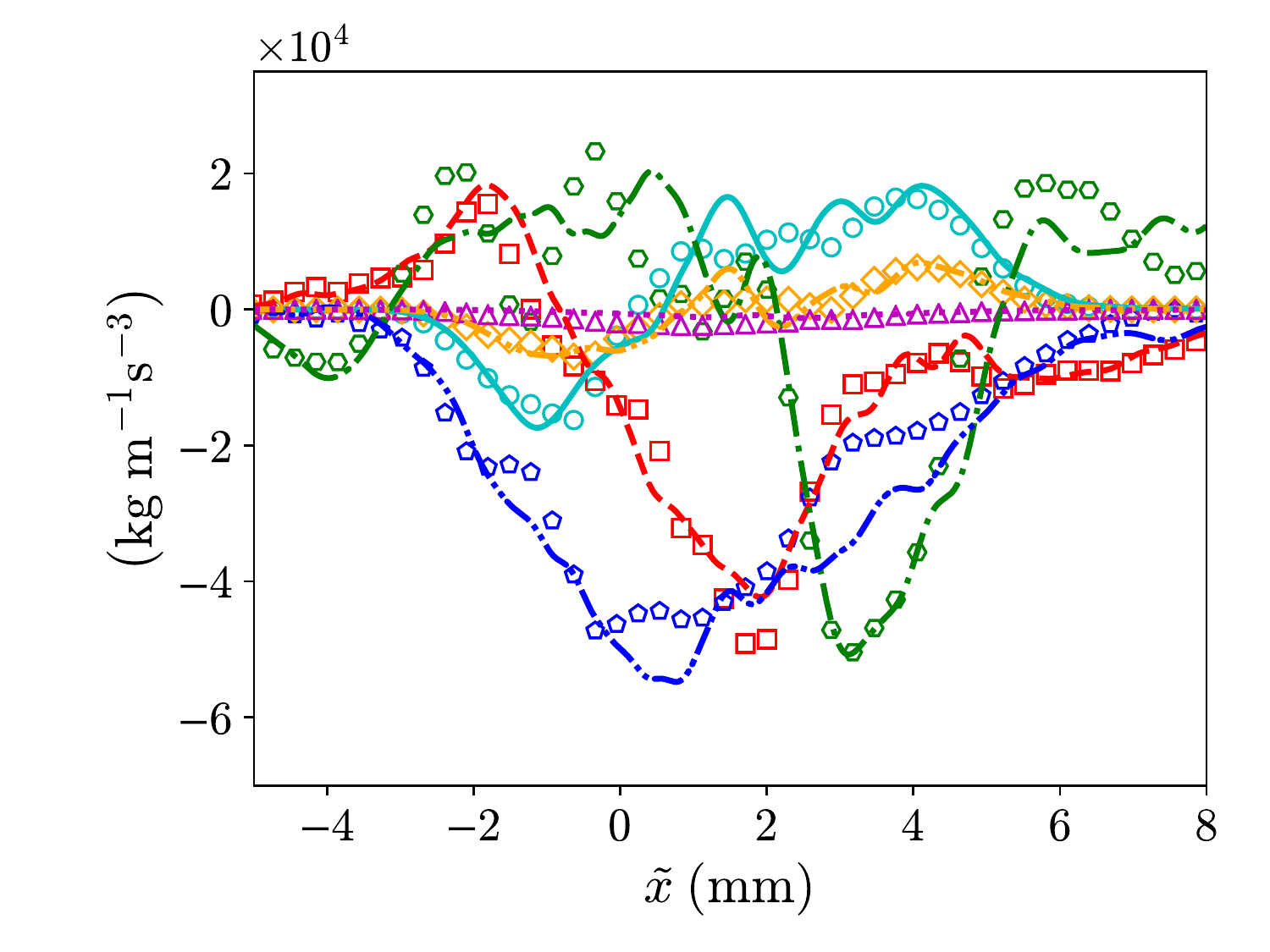}}
\caption{Grid sensitivities of the budgets of the large-scale Favre-averaged Reynolds normal stress component in the streamwise direction multiplied by the mean filtered density, $\overline{\left< \rho \right>}_{\ell} \widetilde{R}_{L,11}$, at different times after re-shock between the grid D and the grid E.
The budgets with the grid D and the grid E are shown with symbols and lines respectively.
Cyan circles or solid line: production [term (III)]; red squares or dashed line: press-strain redistribution [term (V)]; green hexagons or dash-dotted line: turbulent transport [term (IV)]; blue pentagons or dash-dot-dotted line: dissipation [term (VI)]; orange diamonds or dash-triple-dotted line: negative of convection due to streamwise velocity associated with turbulent mass flux; magenta triangles or dotted line: residue.}
\label{fig:rho_R11_budget_filtered_spat_conv}
\end{figure*}

\begin{figure*}[!ht]
\centering
\subfigure[$\ t=1.20\ \mathrm{ms}$]{%
\includegraphics[width=0.4\textwidth]{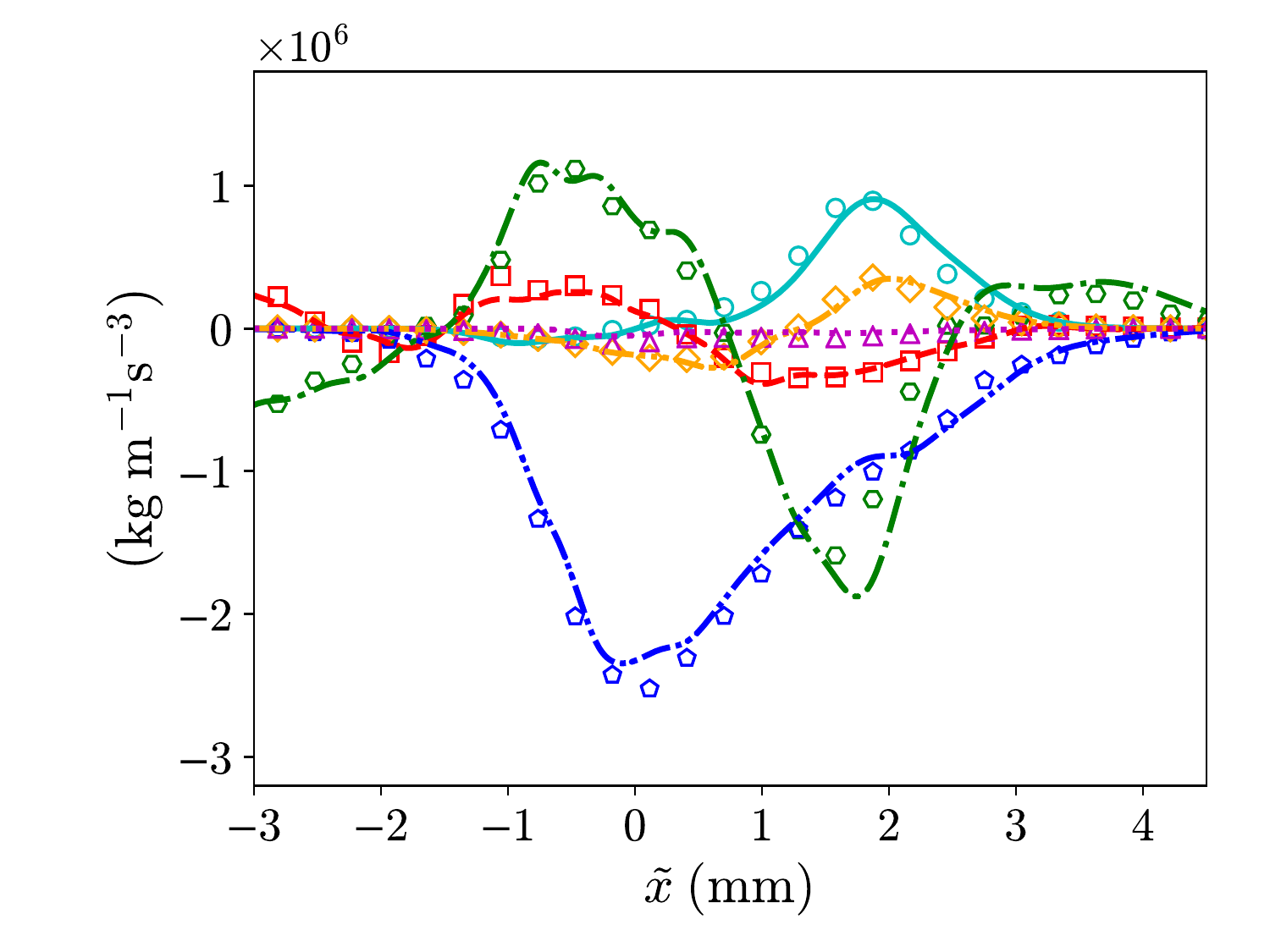}}
\subfigure[$\ t=1.60\ \mathrm{ms}$]{%
\includegraphics[width=0.4\textwidth]{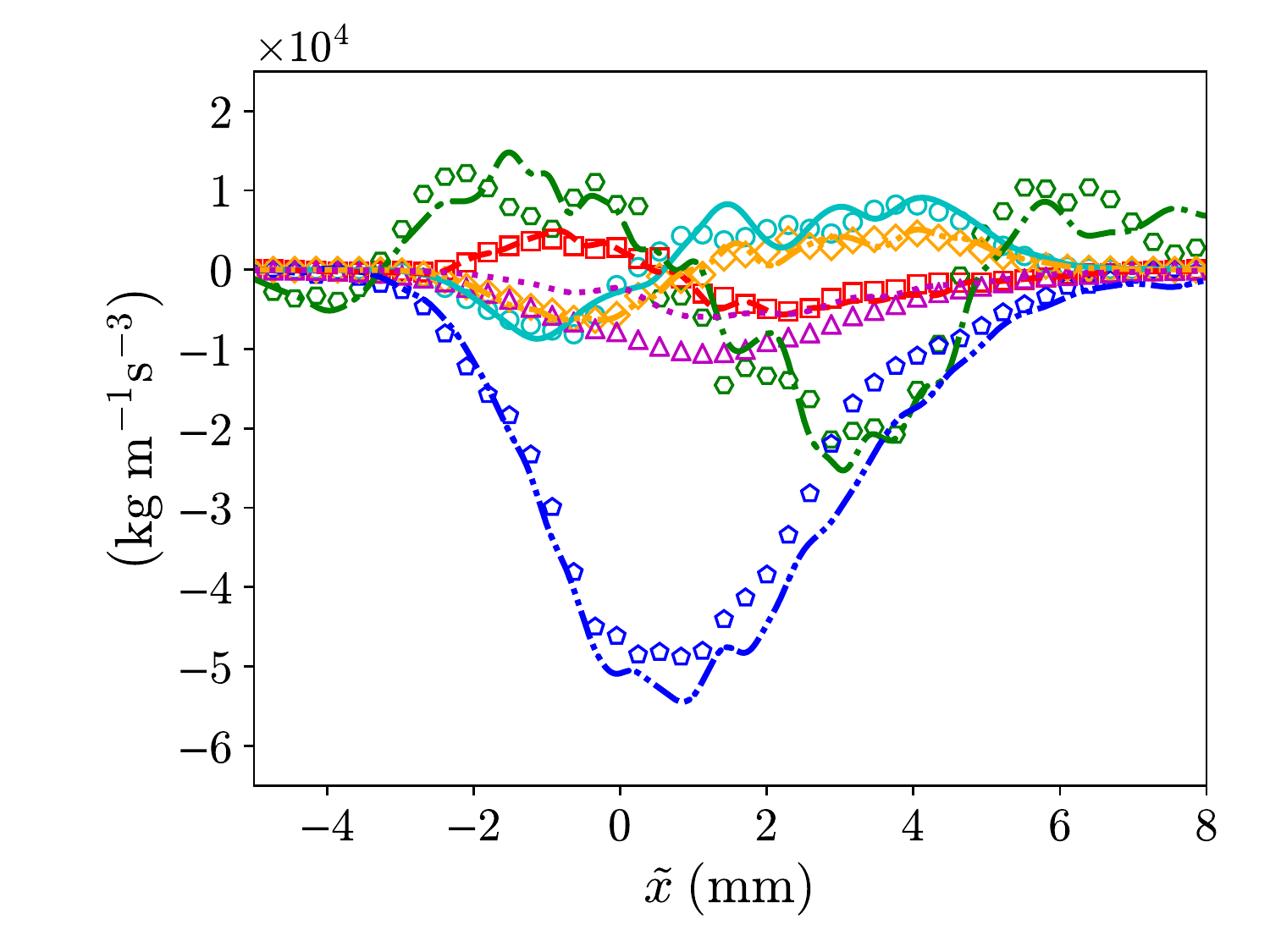}}
\caption{Grid sensitivities of the budgets of the large-scale turbulent kinetic energy, $\overline{\left< \rho \right>}_{\ell} k_L$, at different times after re-shock between the grid D and the grid E.
The budgets with the grid D and the grid E are shown with symbols and lines respectively.
Cyan circles or solid line: production [term (III)]; red squares or dashed line: pressure-dilatation [term (V)]; green hexagons or dash-dotted line: turbulent transport [term (IV)]; blue pentagons or dash-dot-dotted line: dissipation [term (VI)]; orange diamonds or dash-triple-dotted line: negative of convection due to streamwise velocity associated with turbulent mass flux; magenta triangles or dotted line: residue.}
\label{fig:rho_k_budget_filtered_spat_conv}
\end{figure*}

The budgets of $\overline{\left< \rho \right>}_{\ell} a_{L,1}$, $\overline{\left< \rho \right>}_{\ell} b_L$, $\overline{\left< \rho \right>}_{\ell} \widetilde{R}_{L,11}$, and $\overline{\left< \rho \right>}_{\ell} k_L$  with $\ell \approx 0.781\ \mathrm{mm}$ at different times after re-shock between the grid D and the grid E are shown respectively in figures~\ref{fig:rho_a1_budget_filtered_spat_conv}, \ref{fig:rho_b_budget_filtered_spat_conv}, \ref{fig:rho_R11_budget_filtered_spat_conv}, and \ref{fig:rho_k_budget_filtered_spat_conv}. Note that the number of filtering operations are 64 and 256 respectively for grid D and grid E. It can be seen that the budgets of $\overline{\left< \rho \right>}_{\ell} a_{L,1}$ and $\overline{\left< \rho \right>}_{\ell} b_L$ are grid converged reasonably well. The grid sensitivities are small for the budgets of $\overline{\left< \rho \right>}_{\ell} \widetilde{R}_{L,11}$ and $\overline{\left< \rho \right>}_{\ell} k_L$ at $t=1.20\ \mathrm{ms}$. While larger grid sensitivities can be seen at late times after re-shock for these two quantities such as at $t=1.60\ \mathrm{ms}$, the profiles with the grid D and the grid E are in reasonably good agreement. The residue of $\overline{\left< \rho \right>}_{\ell} k_L$ with grid D at $t=1.60\ \mathrm{ms}$ is not negligible but becomes smaller when the grid is refined to grid E.

\section{\label{sec:second_moments_effect_filtering} Effects of the filter width on the large-scale second-moments and the SFS stress at $t=1.20\ \mathrm{ms}$ and $t=1.60\ \mathrm{ms}$ after re-shock}

\begin{figure*}[!ht]
\centering
\subfigure[$\ t=1.20\ \mathrm{ms}$]{%
\includegraphics[width=0.4\textwidth]{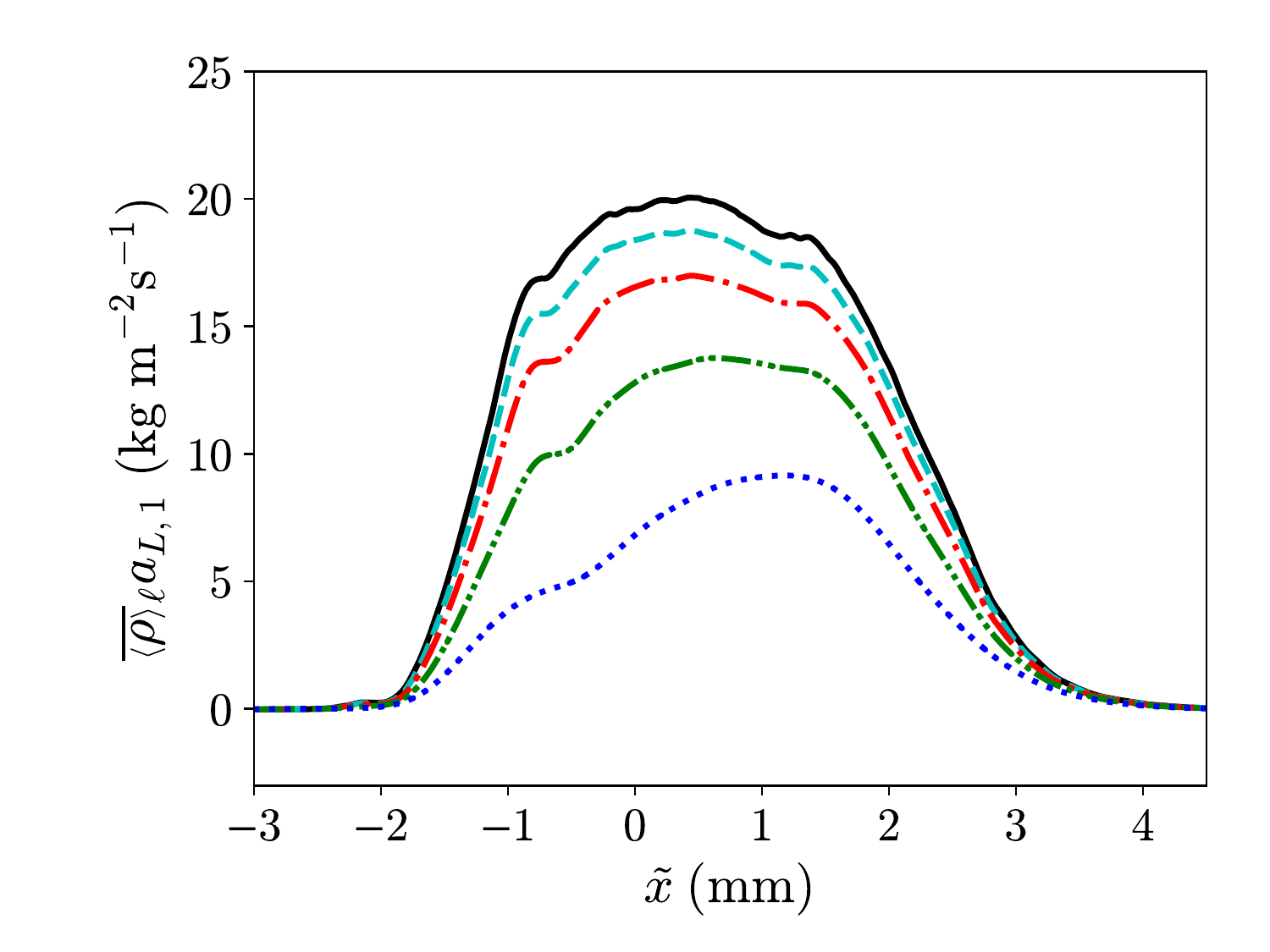}\label{fig:rho_a1_filtered_t_after_reshock}}
\subfigure[$\ t=1.60\ \mathrm{ms}$]{%
\includegraphics[width=0.4\textwidth]{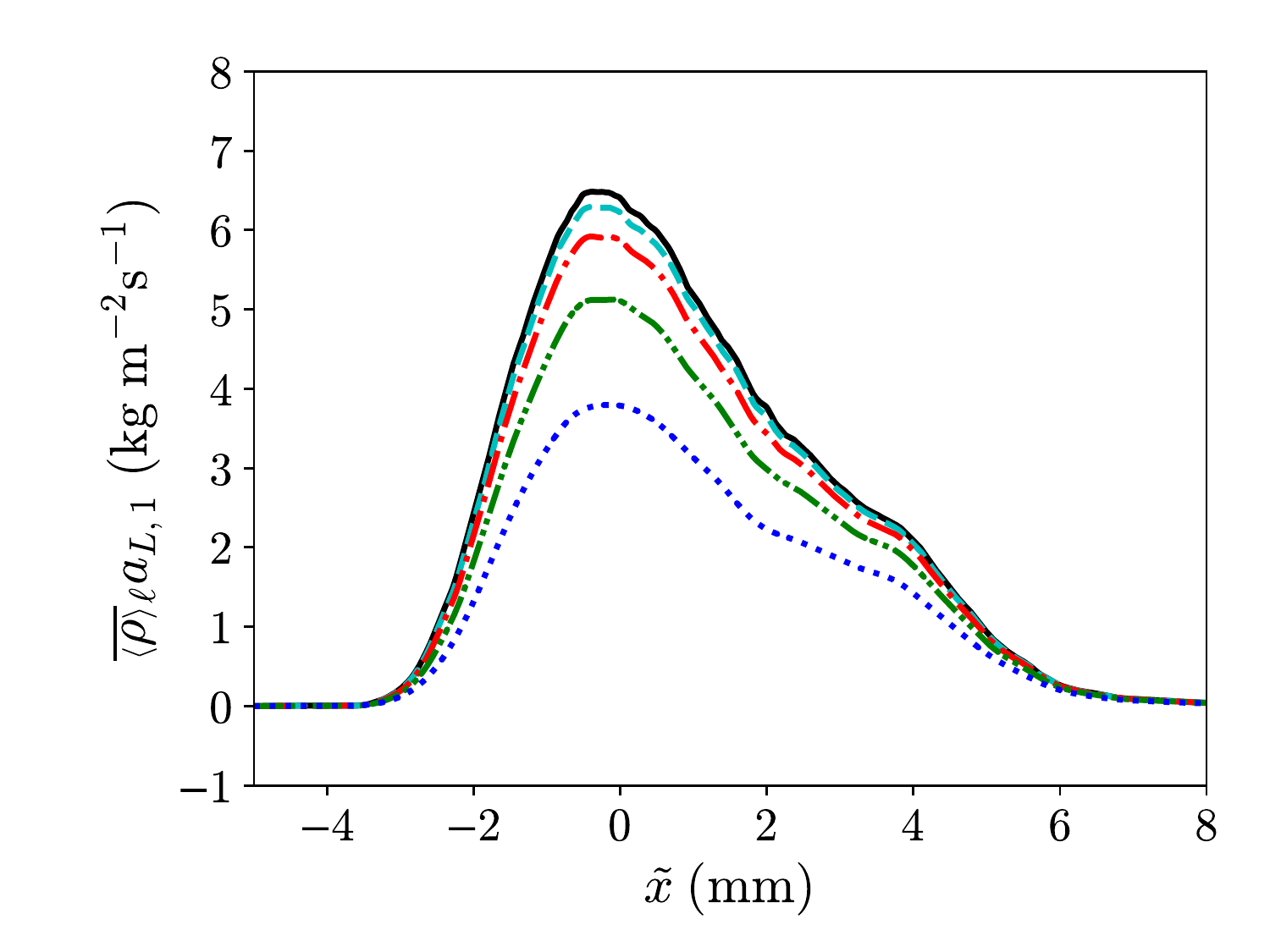}\label{fig:rho_a1_filtered_t_1_60}}
\caption{Effect of filtering on the large-scale turbulent mass flux component in the streamwise direction, $\overline{\left< \rho \right>}_{\ell} a_{L,1}$, at $t=1.20\ \mathrm{ms}$ and $t=1.60\ \mathrm{ms}$ after re-shock. Black solid line: no filtering; cyan dashed line: $\ell \approx 8 \Delta$; red dash-dotted line: $\ell \approx 16 \Delta$; green dash-dot-dotted line: $\ell \approx 32 \Delta$; blue dotted line: $\ell \approx 64 \Delta$.
}
\label{fig:rho_a1_filtered}
\end{figure*}

\begin{figure*}[!ht]
\centering
\subfigure[$\ t=1.20\ \mathrm{ms}$]{%
\includegraphics[width=0.4\textwidth]{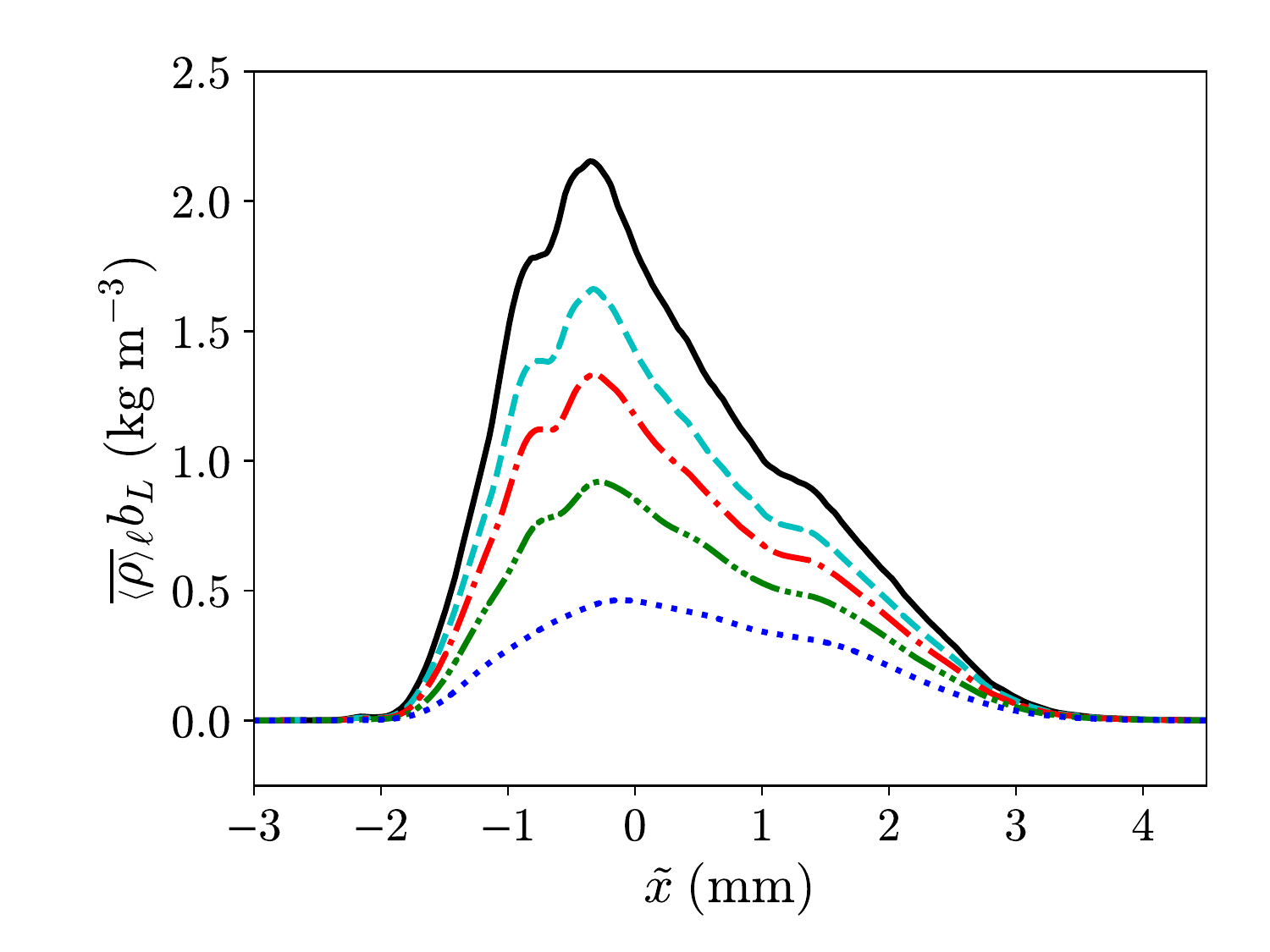}\label{fig:rho_b_filtered_after_reshock}}
\subfigure[$\ t=1.60\ \mathrm{ms}$]{%
\includegraphics[width=0.4\textwidth]{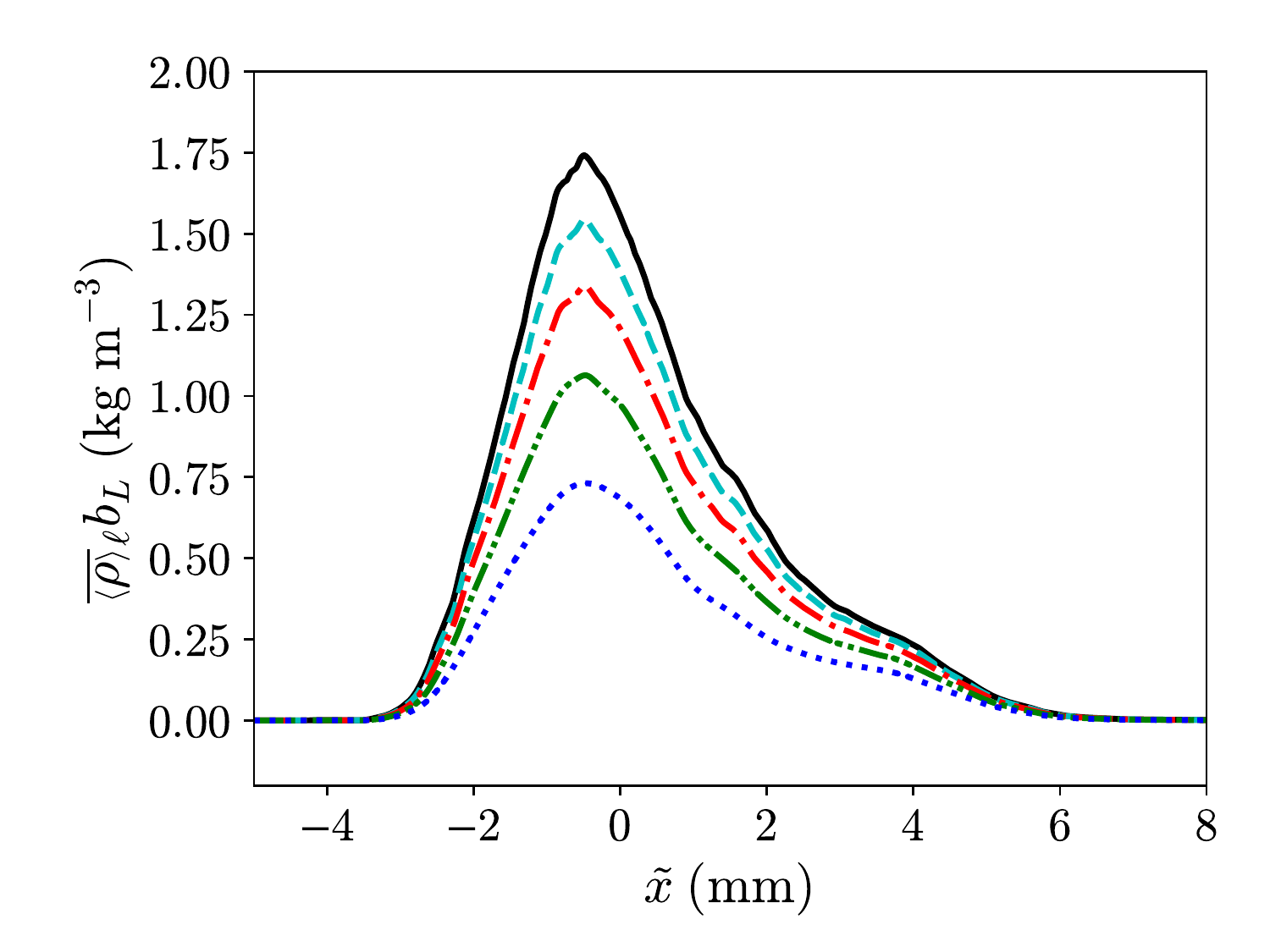}\label{fig:rho_b_filtered_t_1_60}}
\caption{Effect of filtering on the large-scale density-specific-volume covariance multiplied by the mean filtered density, $\overline{\left< \rho \right>}_{\ell} b_L$, at $t=1.20\ \mathrm{ms}$ and $t=1.60\ \mathrm{ms}$ after re-shock. Black solid line: no filtering; cyan dashed line: $\ell \approx 8 \Delta$; red dash-dotted line: $\ell \approx 16 \Delta$; green dash-dot-dotted line: $\ell \approx 32 \Delta$; blue dotted line: $\ell \approx 64 \Delta$.
}
\label{fig:rho_b_filtered}
\end{figure*}

\begin{figure*}[!ht]
\centering
\subfigure[$\ t=1.20\ \mathrm{ms}$]{%
\includegraphics[width=0.4\textwidth]{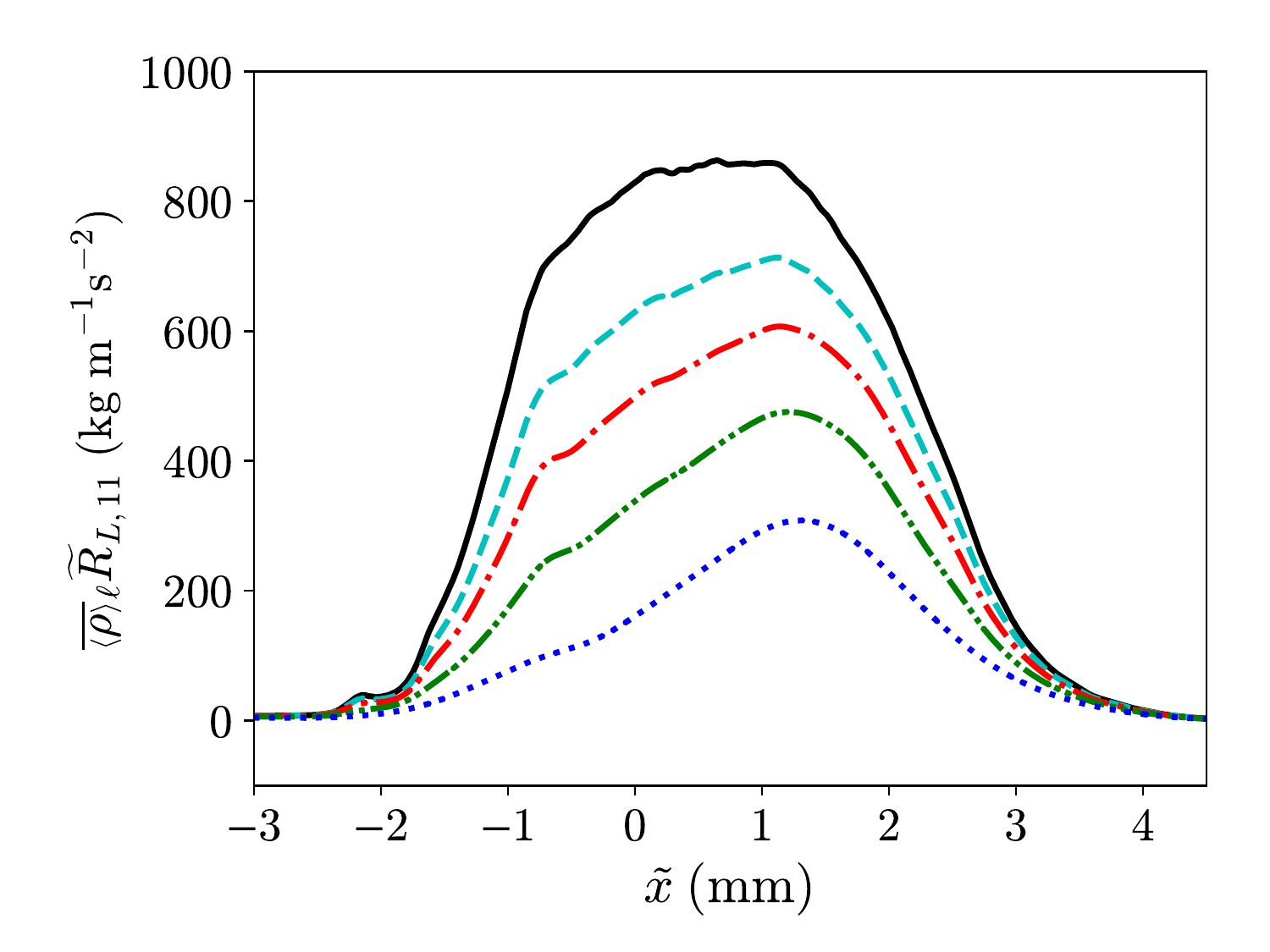}\label{fig:rho_R11_filtered_after_reshock}}
\subfigure[$\ t=1.60\ \mathrm{ms}$]{%
\includegraphics[width=0.4\textwidth]{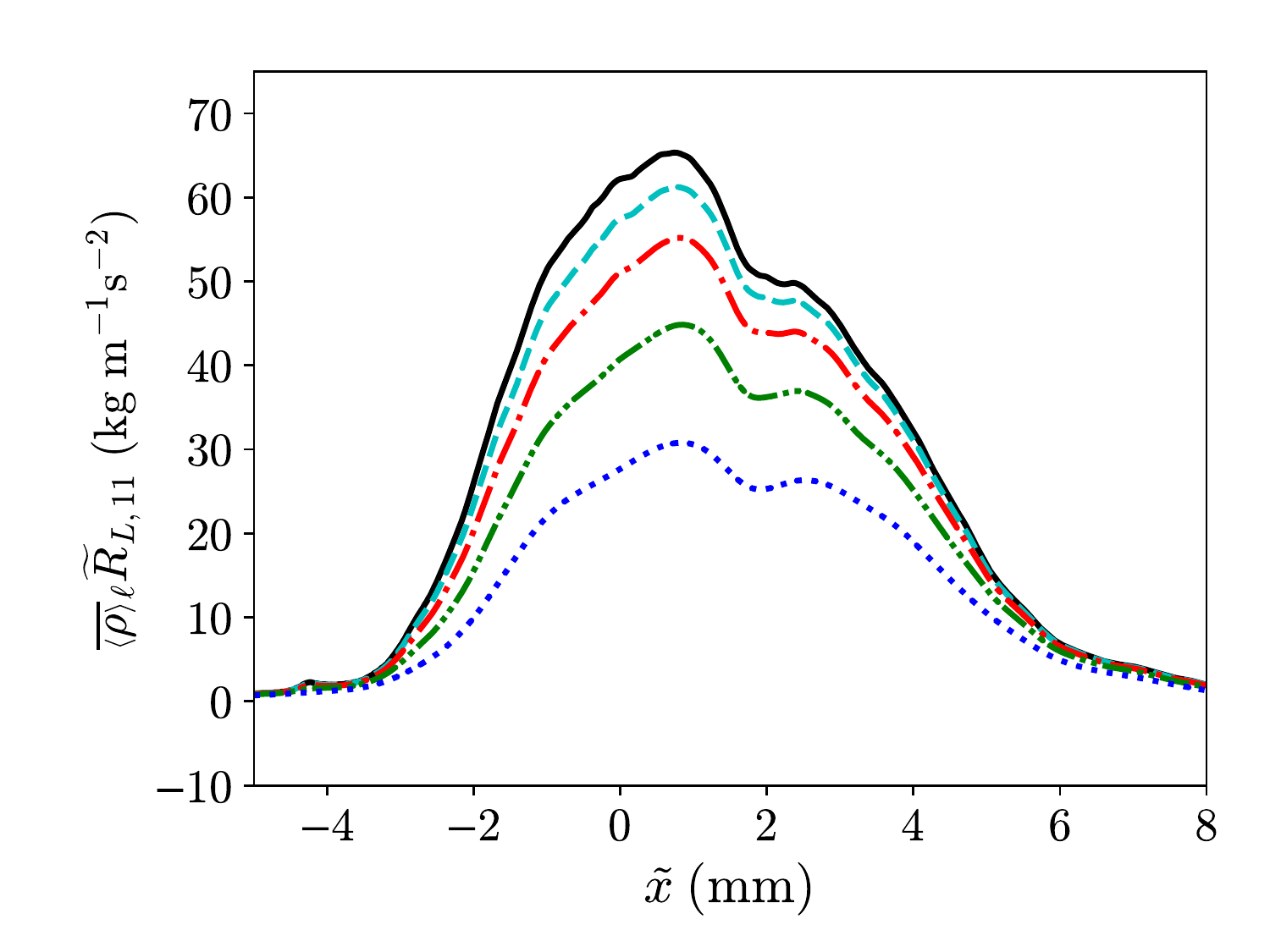}\label{fig:rho_R11_filtered_t_1_60}}
\caption{Effect of filtering on the large-scale Reynolds normal stress component in the streamwise direction multiplied by the mean filtered density, $\overline{\left< \rho \right>}_{\ell} \widetilde{R}_{L,11}$, at $t=1.20\ \mathrm{ms}$ and $t=1.60\ \mathrm{ms}$ after re-shock. Black solid line: no filtering; cyan dashed line: $\ell \approx 8 \Delta$; red dash-dotted line: $\ell \approx 16 \Delta$; green dash-dot-dotted line: $\ell \approx 32 \Delta$; blue dotted line: $\ell \approx 64 \Delta$.
}
\label{fig:rho_R11_filtered}
\end{figure*}

\begin{figure*}[!ht]
\centering
\subfigure[$\ t=1.20\ \mathrm{ms}$]{%
\includegraphics[width=0.4\textwidth]{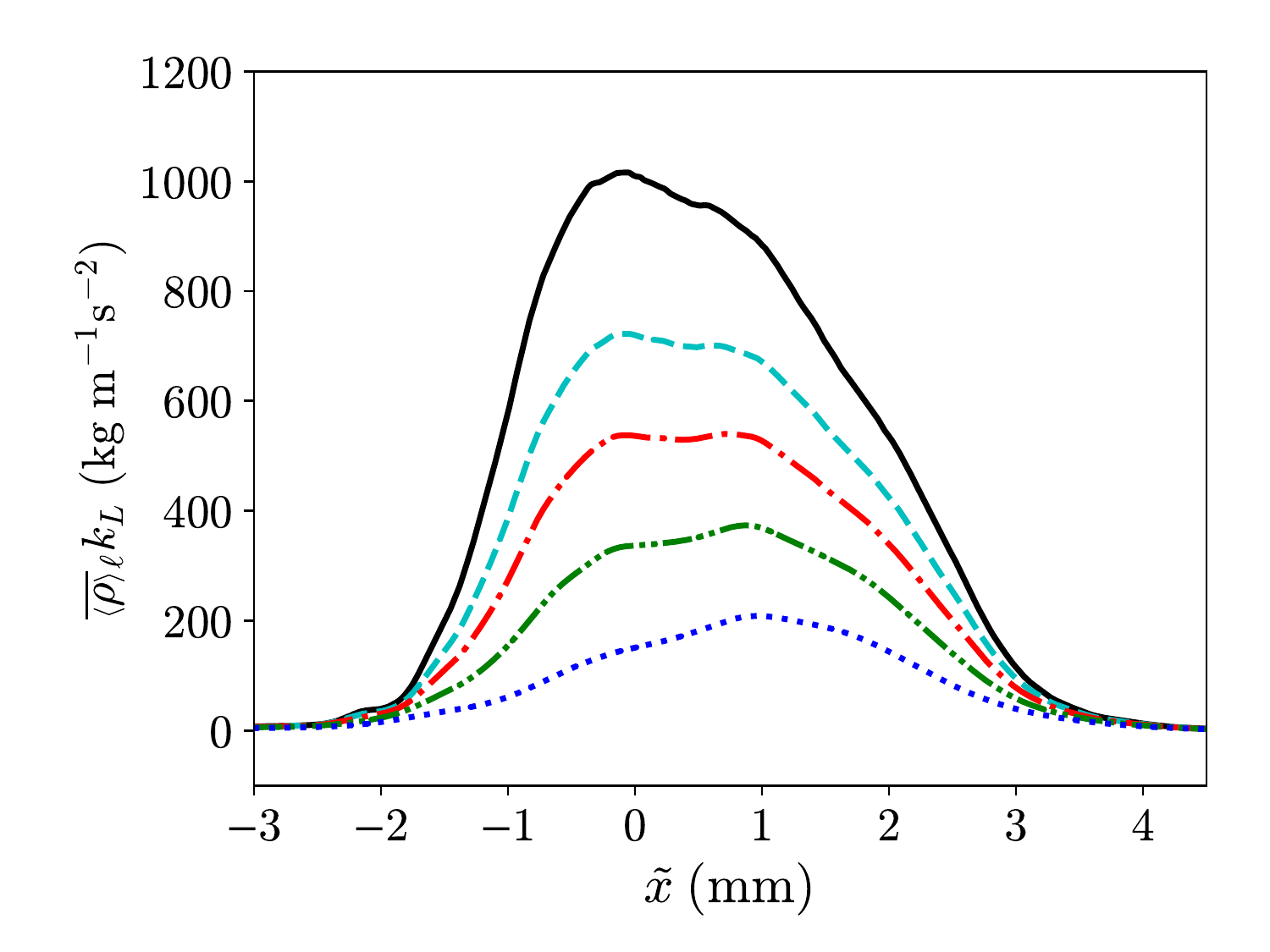}\label{fig:rho_k_filtered_after_reshock}}
\subfigure[$\ t=1.60\ \mathrm{ms}$]{%
\includegraphics[width=0.4\textwidth]{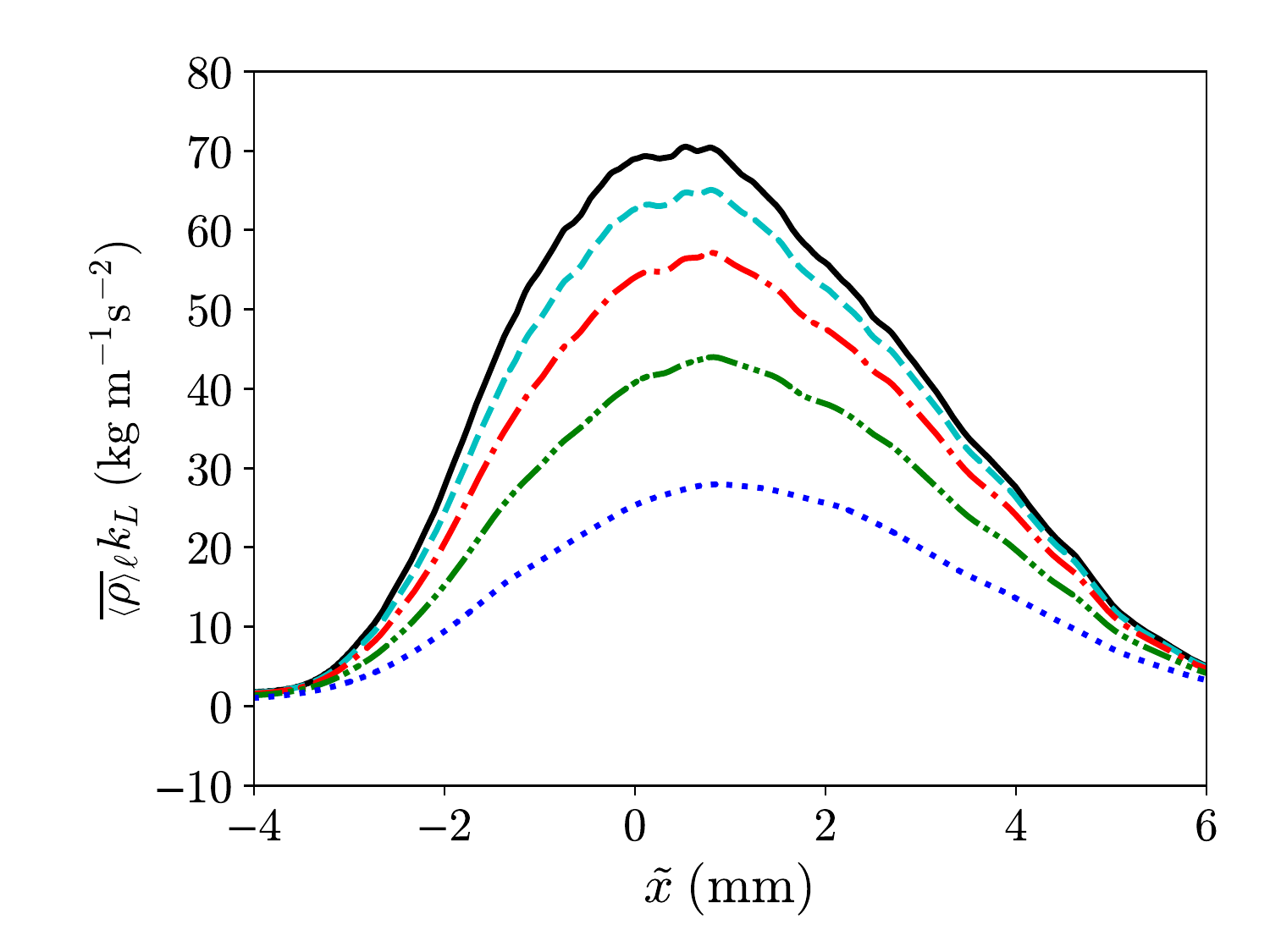}\label{fig:rho_k_filtered_t_1_60}}
\caption{Effect of filtering on the large-scale turbulent kinetic energy, $\overline{\left< \rho \right>}_{\ell} k_L$, at $t=1.20\ \mathrm{ms}$ and $t=1.60\ \mathrm{ms}$ after re-shock. Black solid line: no filtering; cyan dashed line: $\ell \approx 8 \Delta$; red dash-dotted line: $\ell \approx 16 \Delta$; green dash-dot-dotted line: $\ell \approx 32 \Delta$; blue dotted line: $\ell \approx 64 \Delta$.
}
\label{fig:rho_k_filtered}
\end{figure*}

\begin{figure*}[!ht]
\centering
\subfigure[$\ t=1.20\ \mathrm{ms}$]{%
\includegraphics[width=0.4\textwidth]{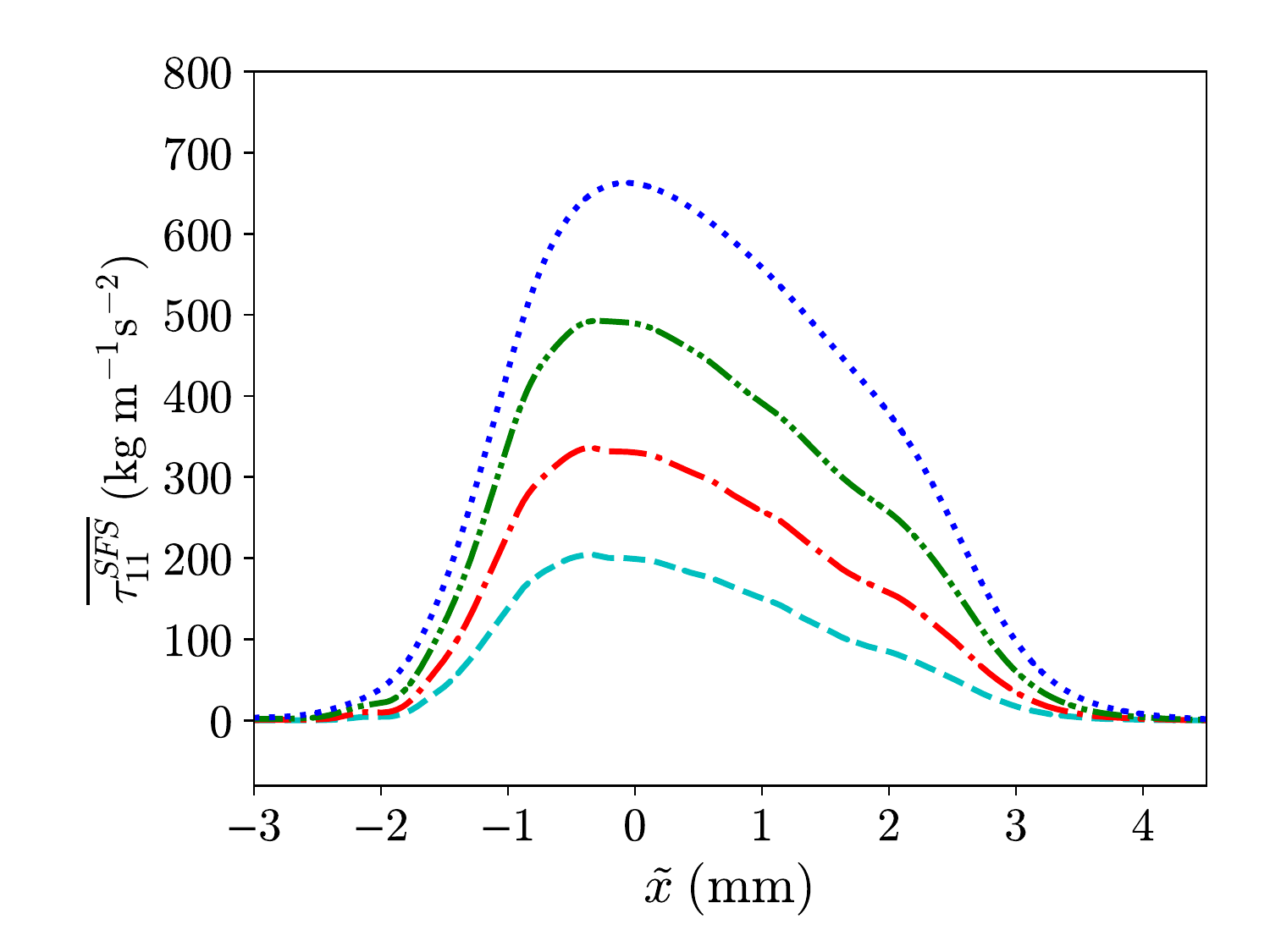}\label{fig:SFS_stress_after_reshock}}
\subfigure[$\ t=1.60\ \mathrm{ms}$]{%
\includegraphics[width=0.4\textwidth]{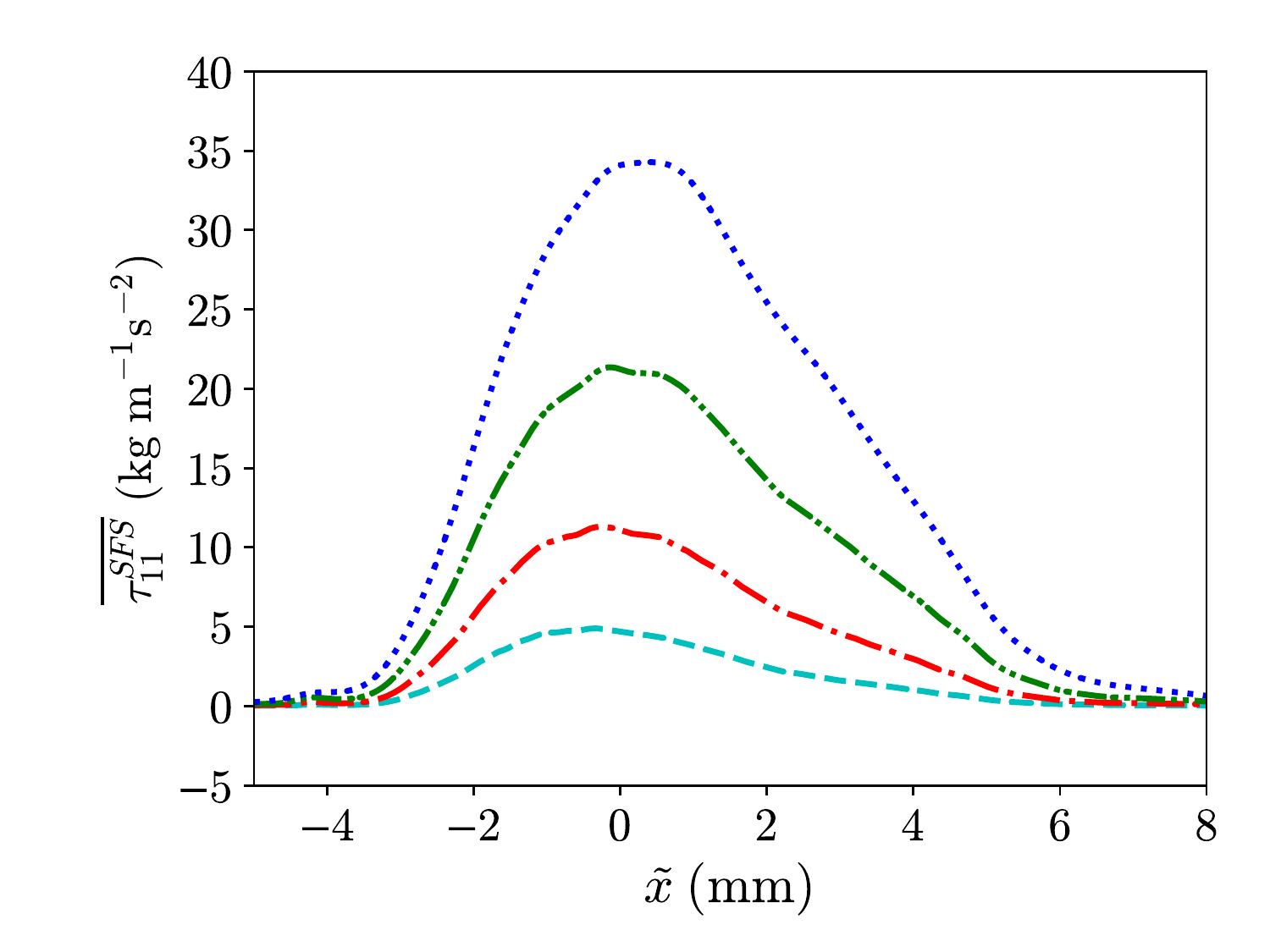}\label{fig:SFS_stress_t_1_60}}
\caption{Effect of filtering on the mean SFS stress component in the streamwise direction, $\overline{\tau_{11}^{SFS}}$, at $t=1.20\ \mathrm{ms}$ and $t=1.60\ \mathrm{ms}$ after re-shock. Cyan dashed line: $\ell \approx 8 \Delta$; red dash-dotted line: $\ell \approx 16 \Delta$; green dash-dot-dotted line: $\ell \approx 32 \Delta$; blue dotted line: $\ell \approx 64 \Delta$.
}
\label{fig:SFS_stress}
\end{figure*}

\begin{figure*}[!ht]
\centering
\subfigure[$\ t=1.20\ \mathrm{ms}$]{%
\includegraphics[width=0.4\textwidth]{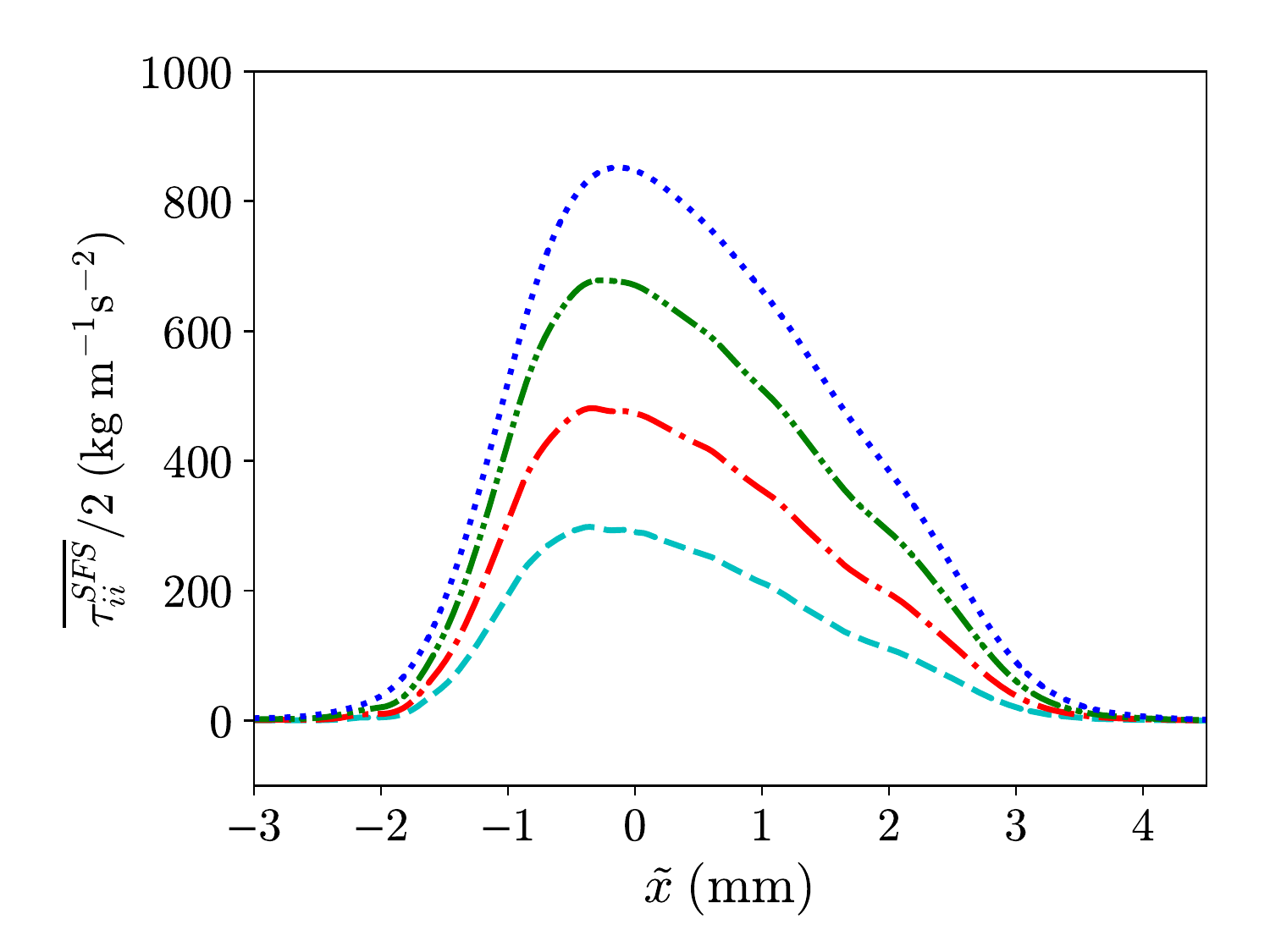}\label{fig:SFS_TKE_after_reshock}}
\subfigure[$\ t=1.60\ \mathrm{ms}$]{%
\includegraphics[width=0.4\textwidth]{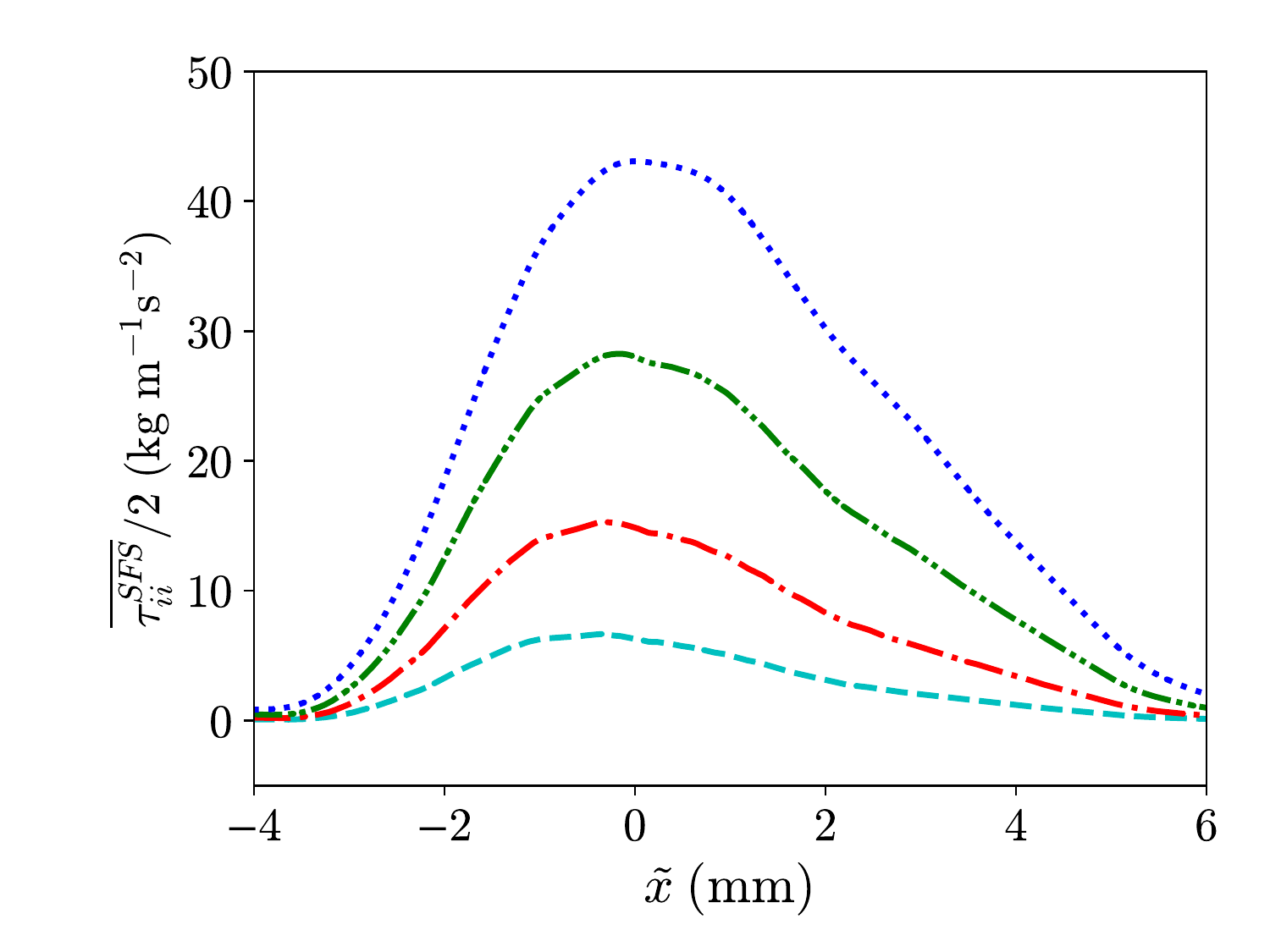}\label{fig:SFS_TKE_t_1_60}}
\caption{Effect of filtering on the mean SFS turbulent kinetic energy, $\overline{\tau_{ii}^{SFS}}/2$, at $t=1.20\ \mathrm{ms}$ and $t=1.60\ \mathrm{ms}$ after re-shock. Cyan dashed line: $\ell \approx 8 \Delta$; red dash-dotted line: $\ell \approx 16 \Delta$; green dash-dot-dotted line: $\ell \approx 32 \Delta$; blue dotted line: $\ell \approx 64 \Delta$.
}
\label{fig:SFS_TKE}
\end{figure*}

The effects of the filter width on the large-scale second-moments, $\overline{\left< \rho \right>}_{\ell} a_{L,1}$, $\overline{\left< \rho \right>}_{\ell} b_L$, and $\overline{\left< \rho \right>}_{\ell} \widetilde{R}_{L,11}$, at $t=1.20\ \mathrm{ms}$ and $t=1.60\ \mathrm{ms}$ obtained with grid E after re-shock
are shown in figures~\ref{fig:rho_a1_filtered}, \ref{fig:rho_b_filtered}, and \ref{fig:rho_R11_filtered} respectively. Figure~\ref{fig:rho_k_filtered} shows the effect of the filter width on the large-scale turbulent kinetic energy, $\overline{\left< \rho \right>}_{\ell} k_L$. The effects of filtering on the mean SFS stress component in streamwise direction, $\overline{\tau_{11}^{SFS}}$, and the mean SFS turbulent kinetic energy, $\overline{\tau_{ii}^{SFS}}/2$, are shown in figures~\ref{fig:SFS_stress} and \ref{fig:SFS_TKE} respectively.

\section{Budgets of the large-scale second-moments at $t=1.40\ \mathrm{ms}$ and $t=1.75\ \mathrm{ms}$ after re-shock}

\begin{figure*}[!ht]
\centering
\subfigure[$\ t=1.40\ \mathrm{ms}$]{%
\includegraphics[width=0.4\textwidth]{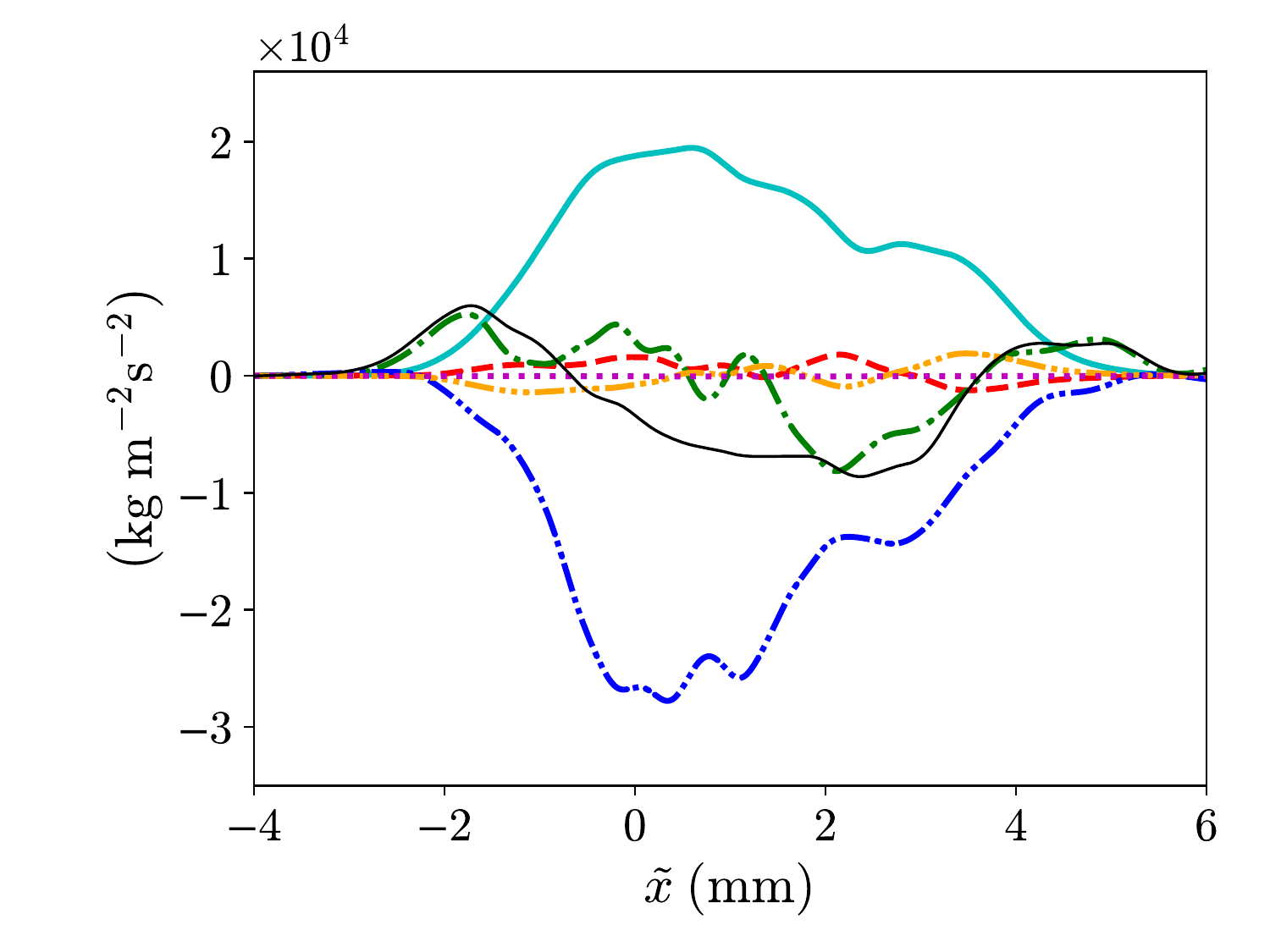}\label{fig:rho_a1_budget_filtered_t_1_40}}
\subfigure[$\ t=1.75\ \mathrm{ms}$]{%
\includegraphics[width=0.4\textwidth]{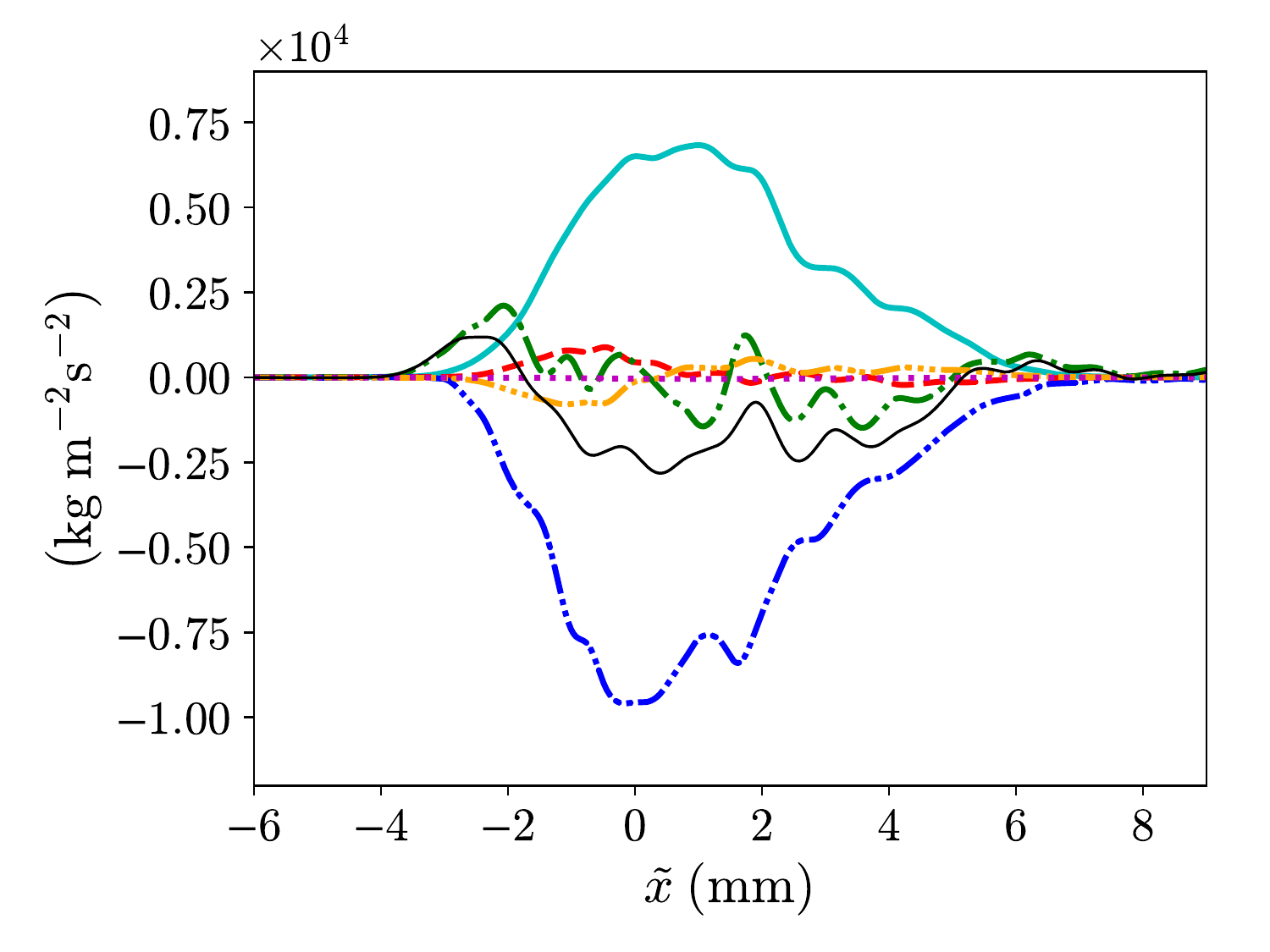}\label{fig:rho_a1_budget_filtered_end}}
\caption{Budgets of the large-scale turbulent mass flux component in the streamwise direction, $\overline{\left< \rho \right>}_{\ell} a_{L,1}$, at $t=1.40\ \mathrm{ms}$ and $t=1.75\ \mathrm{ms}$ after re-shock. Cyan solid line: production [term (III)]; red dashed line: redistribution [term (IV)]; green dash-dotted line: turbulent transport [term (V)]; blue dash-dot-dotted line: destruction [term (VI)]; orange dash-triple-dotted line: negative of convection due to streamwise velocity associated with turbulent mass flux; magenta dotted line: residue; thin black solid line: summation of all terms (rate of change in the moving frame).}
\label{fig:rho_a1_budget_filtered}
\end{figure*}

\begin{figure*}[!ht]
\centering
\subfigure[$\ t=1.40\ \mathrm{ms}$]{%
\includegraphics[width=0.4\textwidth]{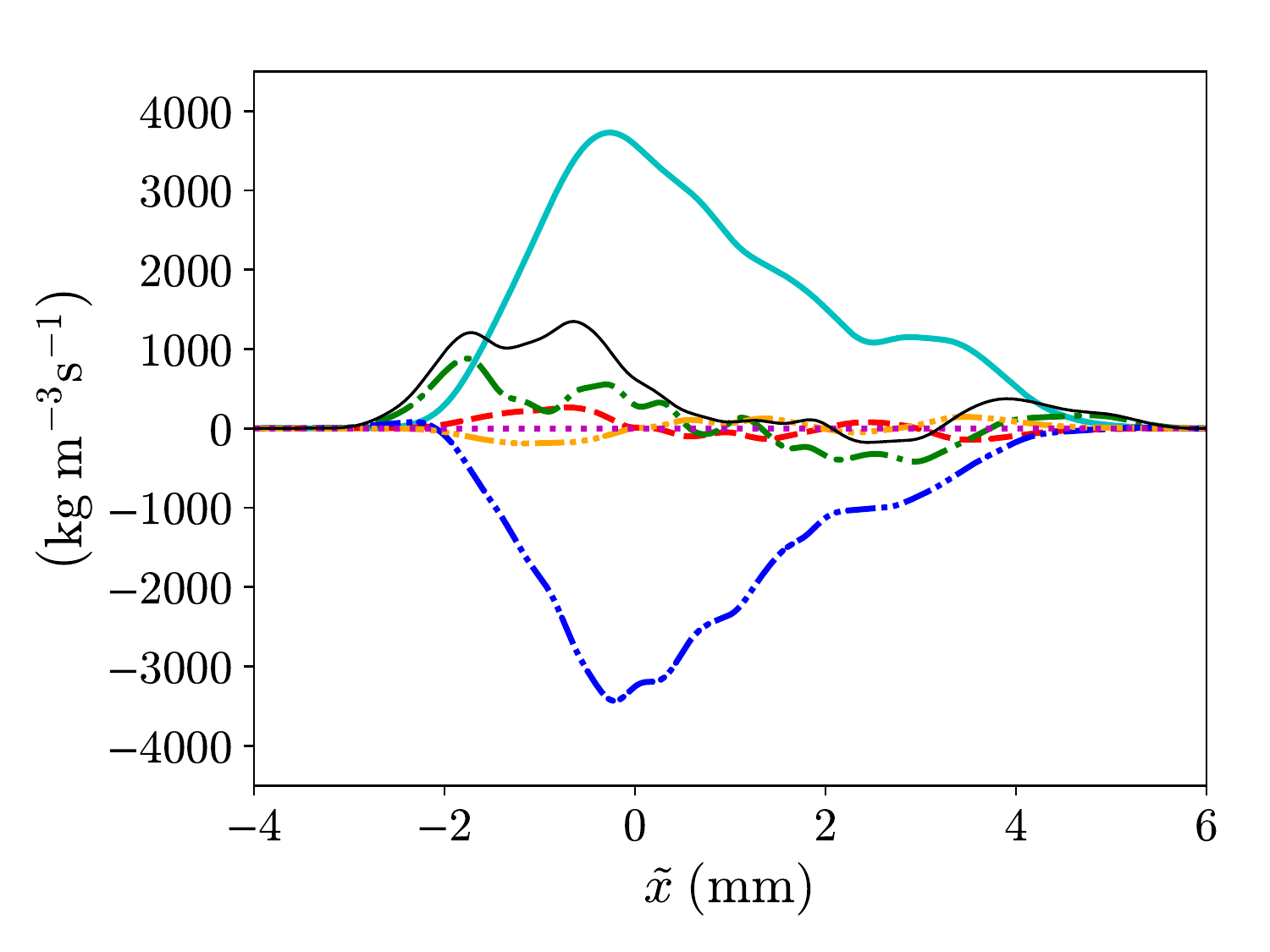}\label{fig:rho_b_budget_filtered_t_1_40}}
\subfigure[$\ t=1.75\ \mathrm{ms}$]{%
\includegraphics[width=0.4\textwidth]{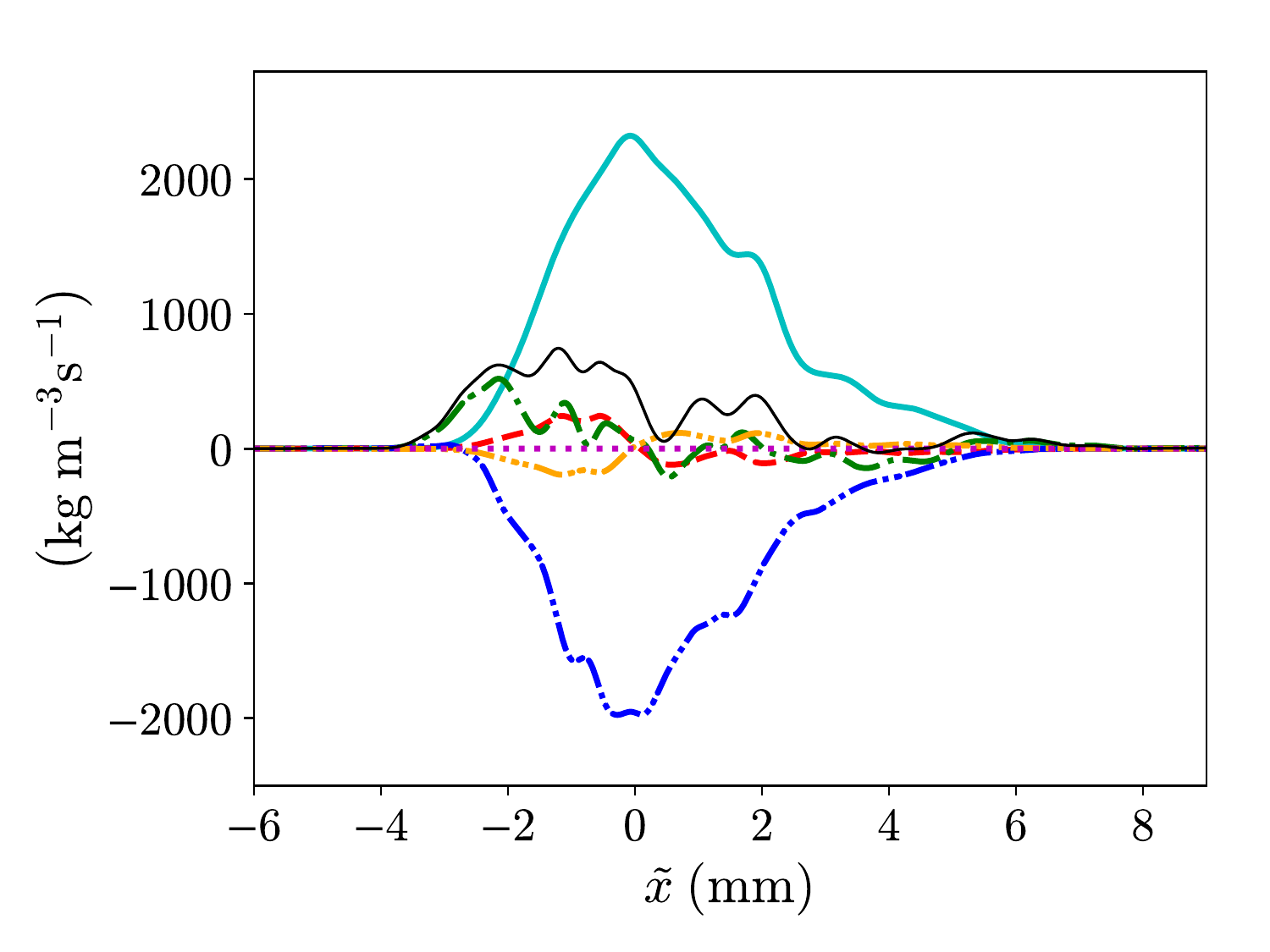}\label{fig:rho_b_budget_filtered_end}}
\caption{Budgets of the large-scale density-specific-volume covariance multiplied by the mean filtered density, $\overline{\left< \rho \right>}_{\ell} b_L$, at $t=1.40\ \mathrm{ms}$ and $t=1.75\ \mathrm{ms}$ after re-shock. Cyan solid line: production [term (III)]; red dashed line: redistribution [term (IV)]; green dash-dotted line: turbulent transport [term (V)]; blue dash-dot-dotted line: destruction [term (VI)]; orange dash-triple-dotted line: negative of convection due to streamwise velocity associated with turbulent mass flux; magenta dotted line: residue; thin black solid line: summation of all terms (rate of change in the moving frame).}
\label{fig:rho_b_budget_filtered}
\end{figure*}

\begin{figure*}[!ht]
\centering
\subfigure[$\ t=1.40\ \mathrm{ms}$]{%
\includegraphics[width=0.4\textwidth]{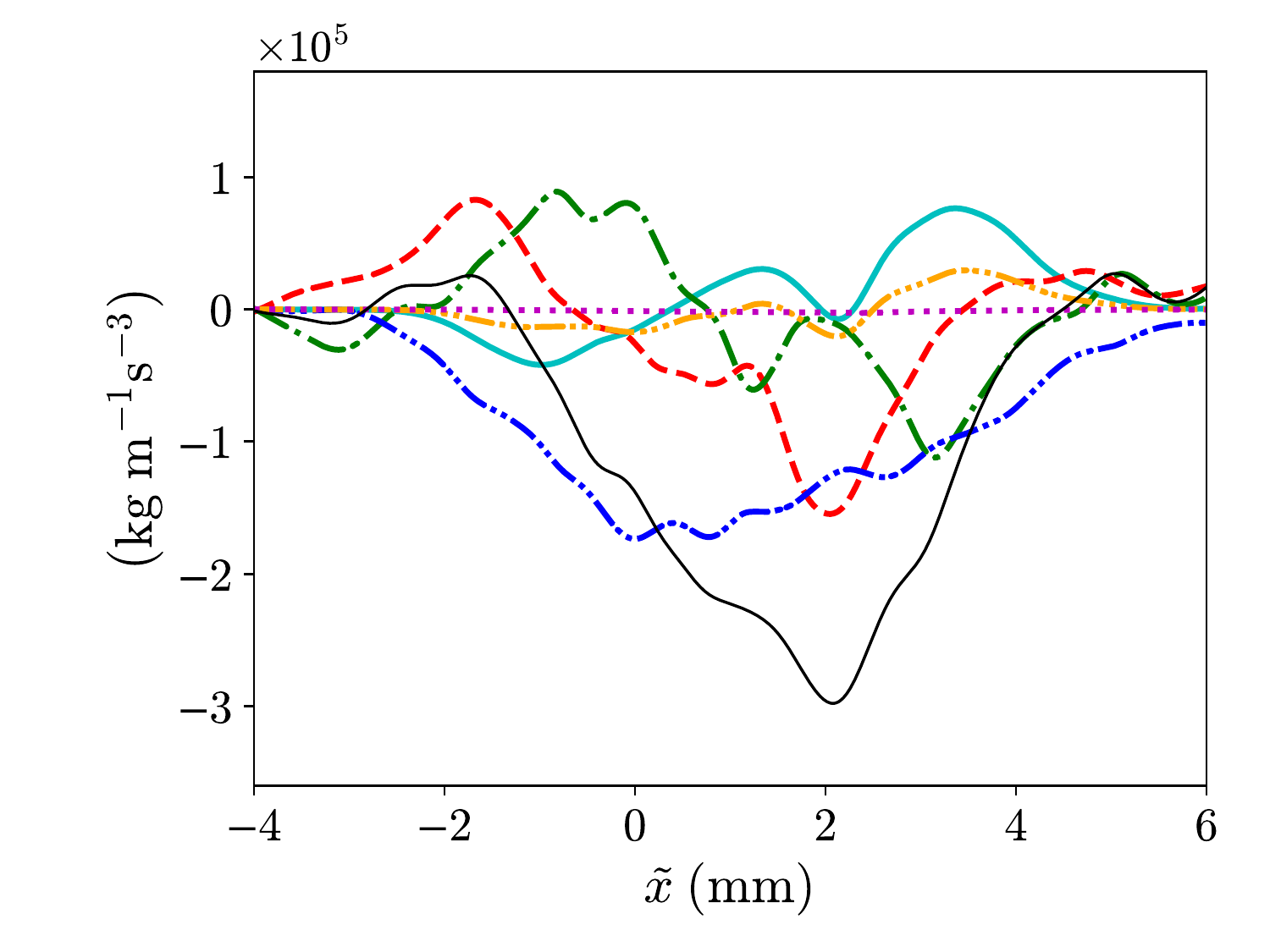}}
\subfigure[$\ t=1.75\ \mathrm{ms}$]{%
\includegraphics[width=0.4\textwidth]{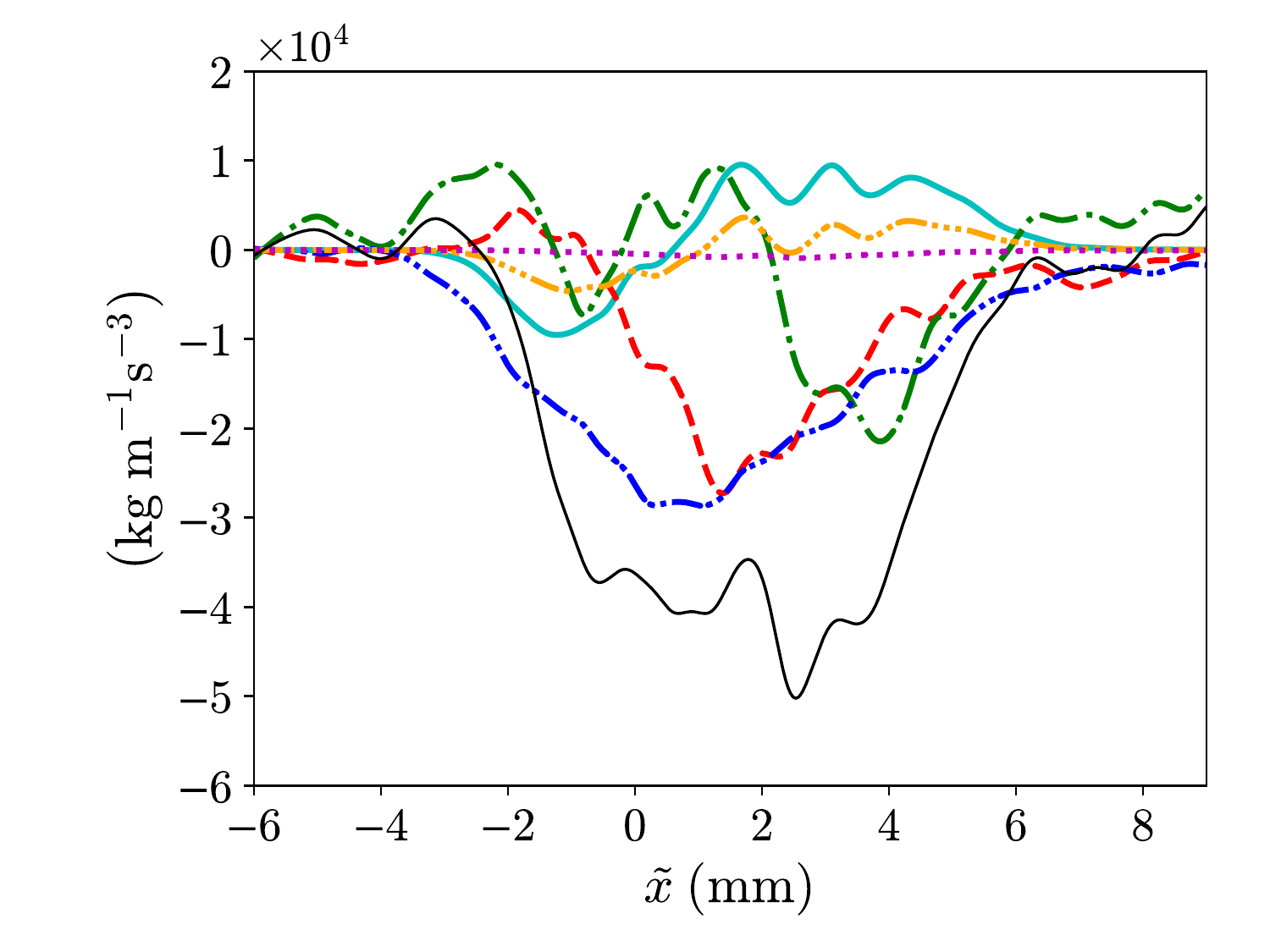}}
\caption{Budgets of the large-scale Favre-averaged Reynolds normal stress component in the streamwise direction multiplied by the mean filtered density, $\overline{\left< \rho \right>}_{\ell} \widetilde{R}_{L,11}$, at $t=1.40\ \mathrm{ms}$ and $t=1.75\ \mathrm{ms}$ after re-shock. Cyan solid line: production [term (III)]; red dashed line: press-strain redistribution [term (V)]; green dash-dotted line: turbulent transport [term (IV)]; blue dash-dot-dotted line: dissipation [term (VI)]; orange dash-triple-dotted line: negative of convection due to streamwise velocity associated with turbulent mass flux; magenta dotted line: residue; thin black solid line: summation of all terms (rate of change in the moving frame).}
\label{fig:rho_R11_budget_filtered}
\end{figure*}

\begin{figure*}[!ht]
\centering
\subfigure[$\ t=1.40\ \mathrm{ms}$]{%
\includegraphics[width=0.4\textwidth]{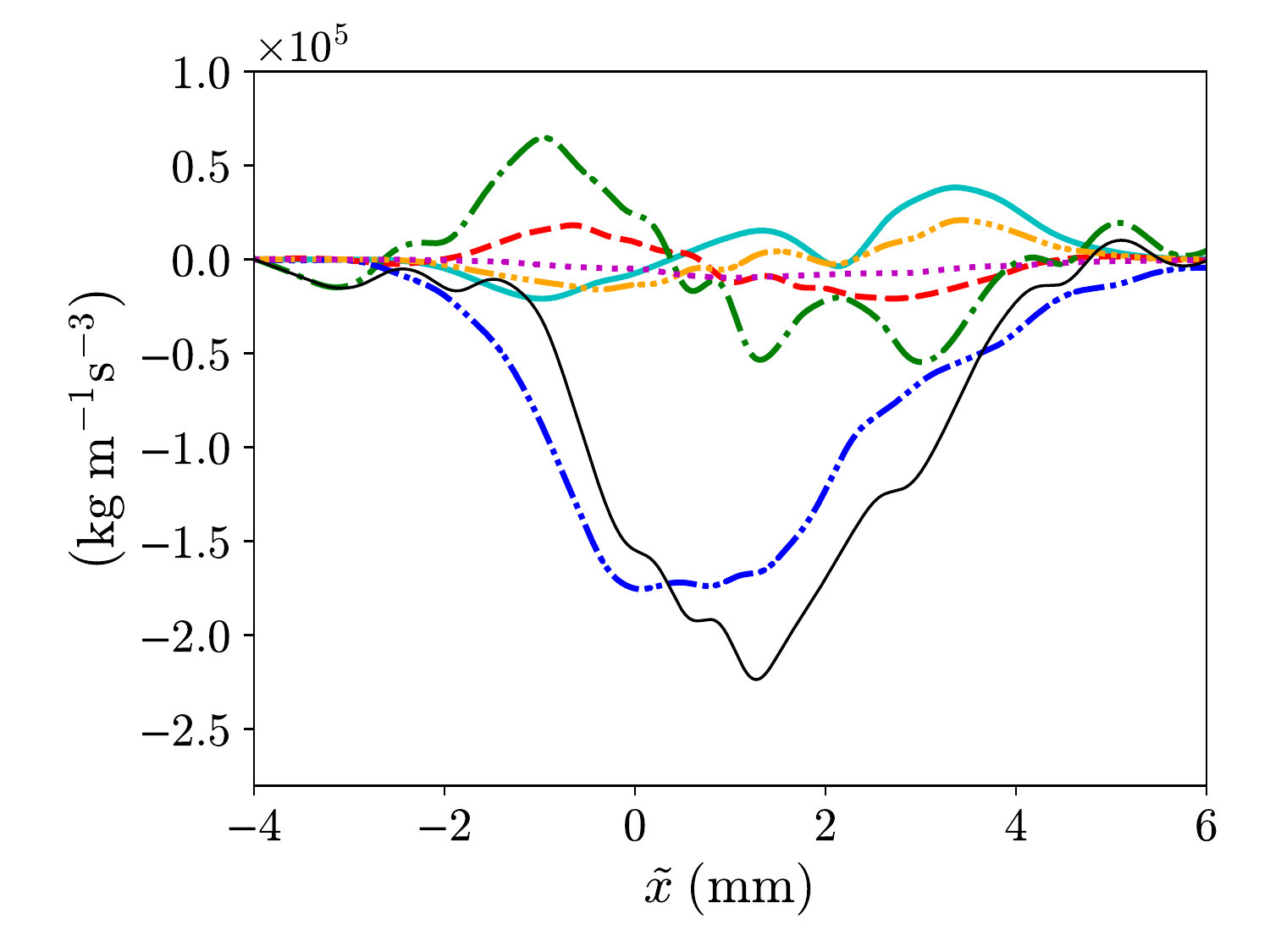}}
\subfigure[$\ t=1.75\ \mathrm{ms}$]{%
\includegraphics[width=0.4\textwidth]{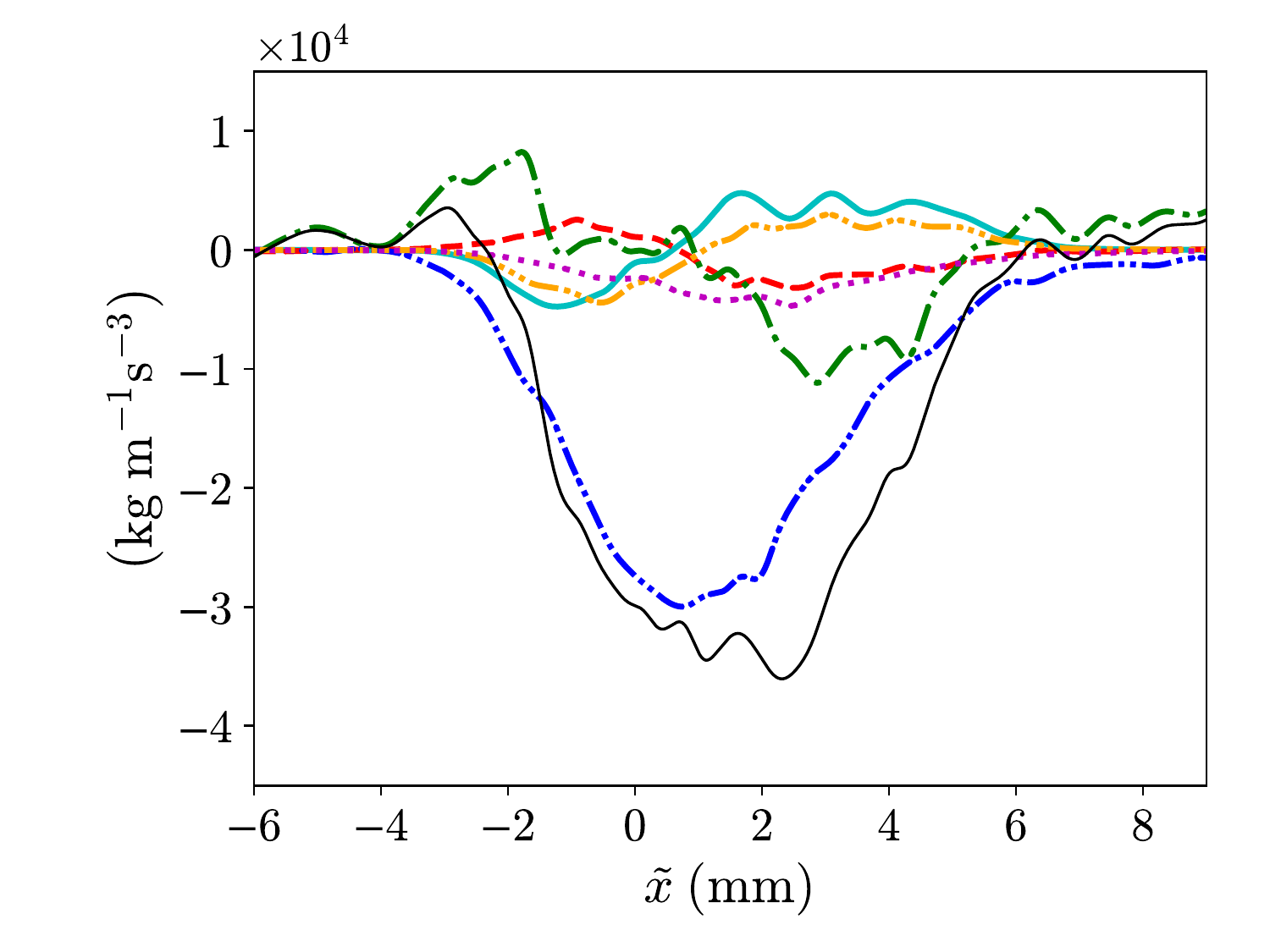}}
\caption{Budgets of the large-scale turbulent kinetic energy, $\overline{\left< \rho \right>}_{\ell} k_L$, at $t=1.40\ \mathrm{ms}$ and $t=1.75\ \mathrm{ms}$ after re-shock. Cyan solid line: production [term (III)]; red dashed line: pressure-dilatation [term (V)]; green dash-dotted line: turbulent transport [term (IV)]; blue dash-dot-dotted line: dissipation [term (VI)]; orange dash-triple-dotted line: negative of convection due to streamwise velocity associated with turbulent mass flux; magenta dotted line: residue; thin black solid line: summation of all terms (rate of change in the moving frame).}
\label{fig:rho_k_budget_filtered}
\end{figure*}

\begin{figure*}[!ht]
\centering
\subfigure[$\ t=1.40\ \mathrm{ms}$]{%
\includegraphics[width=0.4\textwidth]{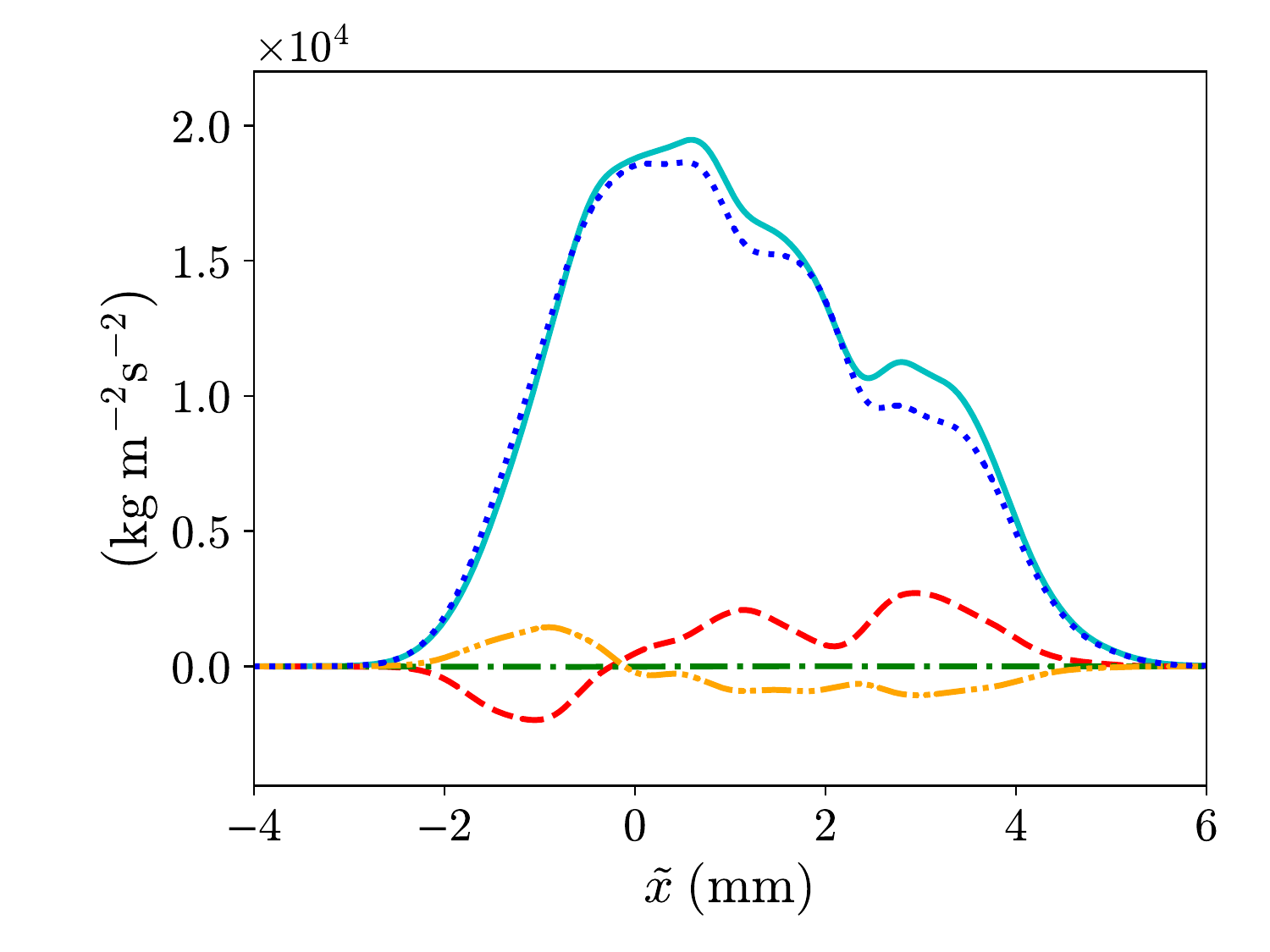}\label{fig:rho_a1_budget_filtered_production_terms_t_1_40}}
\subfigure[$\ t=1.75\ \mathrm{ms}$]{%
\includegraphics[width=0.4\textwidth]{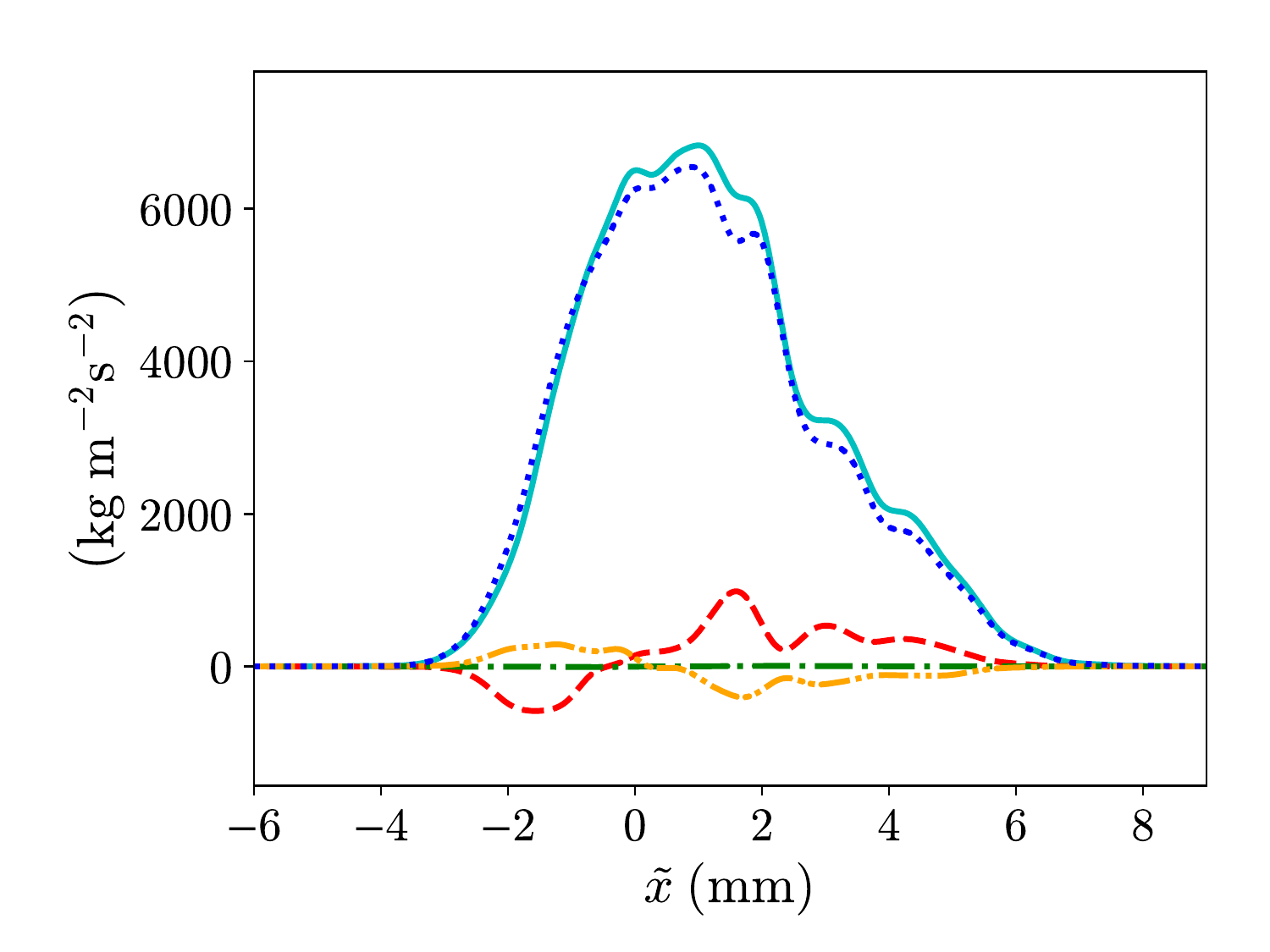}\label{fig:rho_a1_budget_filtered_production_terms_end}}
\caption{Compositions of the production term [term (III)] in the transport equation for the large-scale turbulent mass flux component in the streamwise direction, $\overline{\left< \rho \right>}_{\ell} a_{L,1}$, at $t=1.40\ \mathrm{ms}$ and $t=1.75\ \mathrm{ms}$ after re-shock. Cyan solid line: overall production; red dashed line: $b_L \overline{\left< p \right>}_{\ell,1}$; green dash-dotted line: $-b_L \overline{\left< \tau_{11} \right>}_{\ell,1}$; orange dash-dot-dotted line: $b_L \overline{\tau_{11}^{SFS}}_{,1}$; blue dotted line: $-\widetilde{R}_{L,11} \overline{\left< \rho \right>}_{\ell,1}$.}
\label{fig:rho_a1_budget_filtered_production_terms}
\end{figure*}

\begin{figure*}[!ht]
\centering
\subfigure[$\ t=1.40\ \mathrm{ms}$]{%
\includegraphics[width=0.4\textwidth]{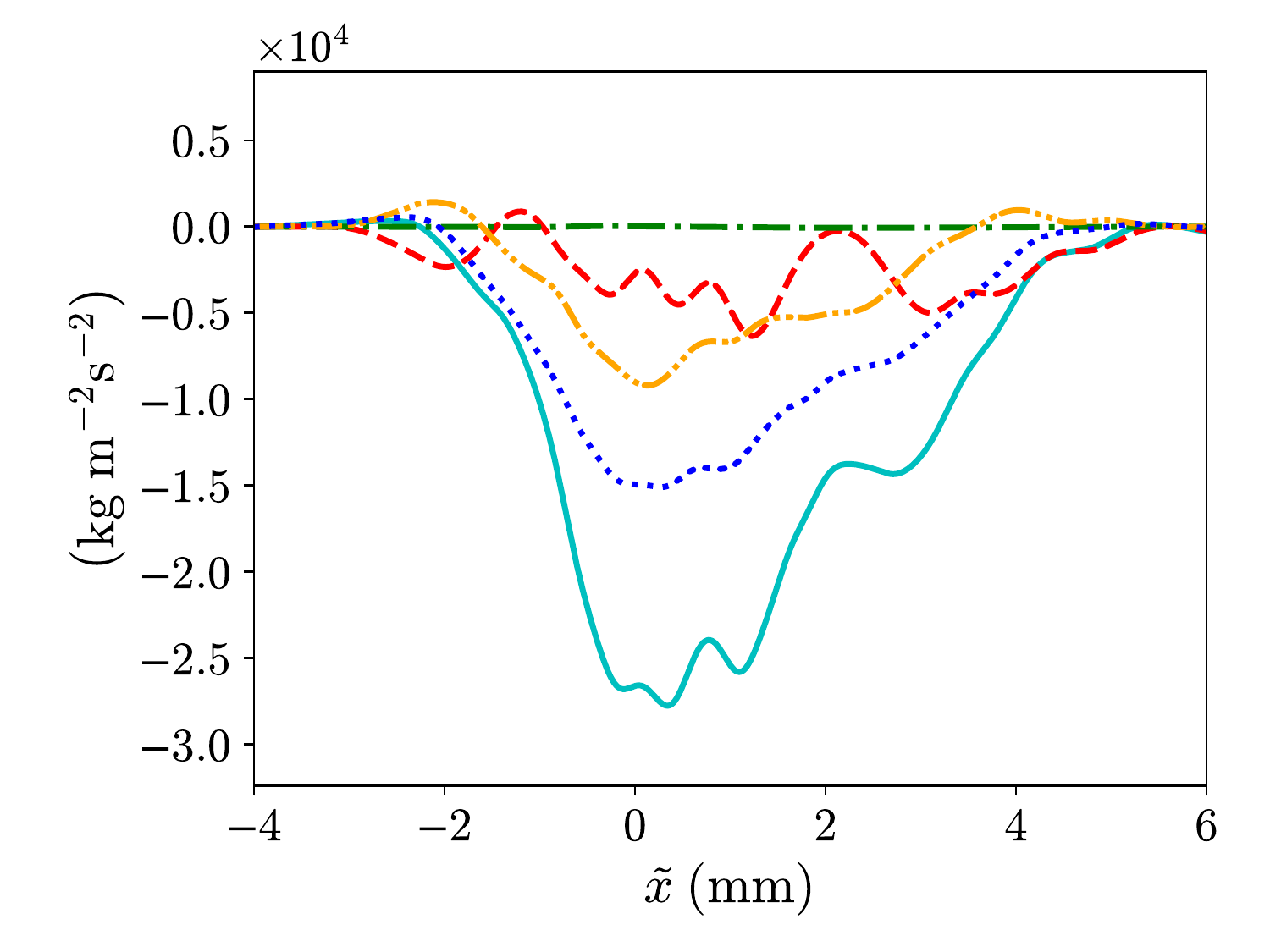}\label{fig:rho_a1_budget_filtered_destruction_terms_t_1_40}}
\subfigure[$\ t=1.75\ \mathrm{ms}$]{%
\includegraphics[width=0.4\textwidth]{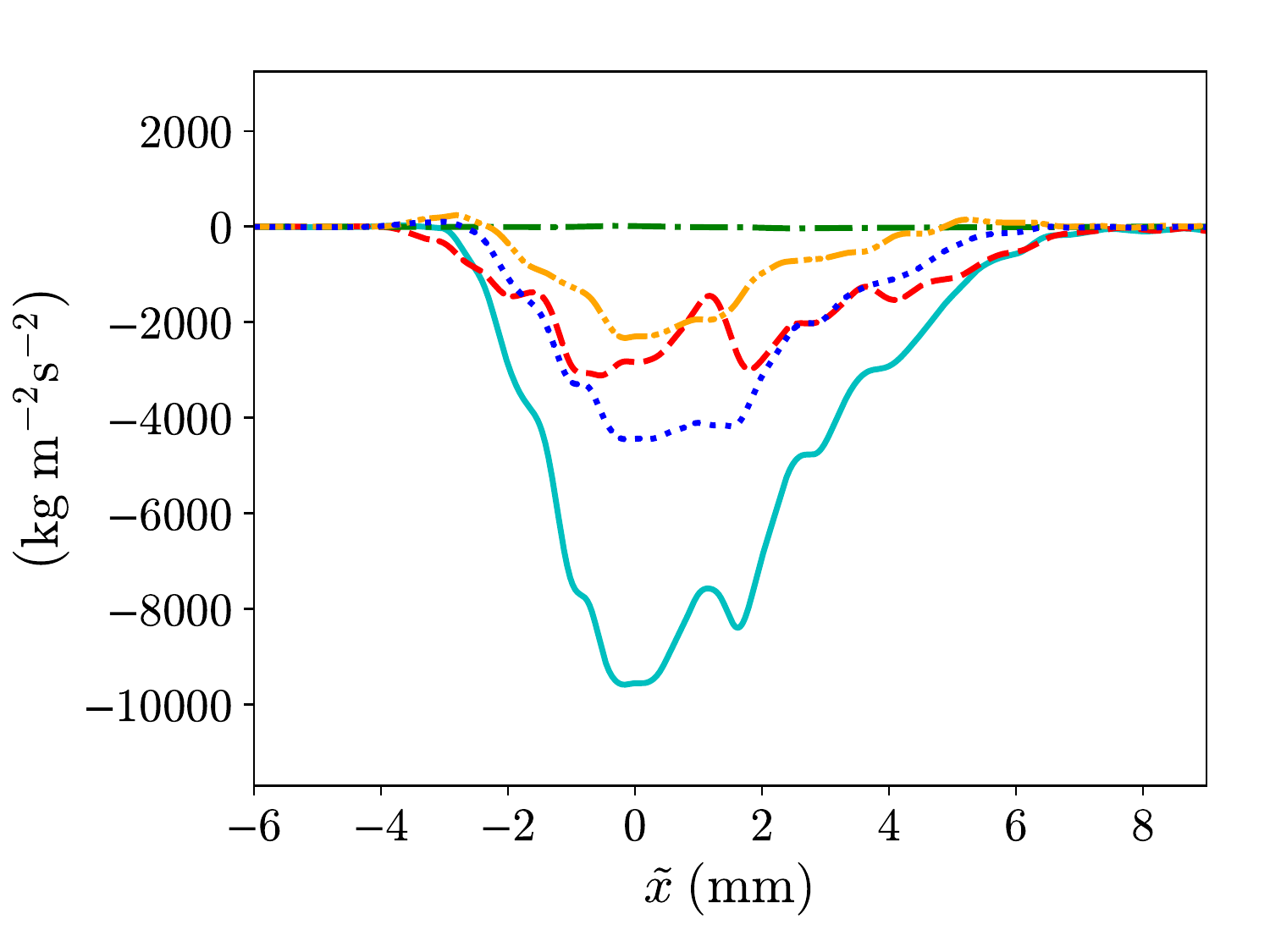}\label{fig:rho_a1_budget_filtered_destruction_terms_end}}
\caption{Compositions of the destruction term [term (VI)] in the transport equation for the large-scale turbulent mass flux component in the streamwise direction, $\overline{\left< \rho \right>}_{\ell} a_{L,1}$, at $t=1.40\ \mathrm{ms}$ and $t=1.75\ \mathrm{ms}$ after re-shock. Cyan solid line: overall destruction; red dashed line: $\overline{\left< \rho \right>}_{\ell} \overline{ \left( 1/\left< \rho \right>_{\ell} \right)^{\prime} \left< p \right>^{\prime}_{\ell,1} }$; green dash-dotted line: $-\overline{\left< \rho \right>}_{\ell} \overline{ \left( 1/\left< \rho \right>_{\ell} \right)^{\prime} \left( \partial \left< \tau_{1i} \right>_{\ell}^{\prime} / \partial x_i \right) }$; orange dash-dot-dotted line: $\overline{\left< \rho \right>}_{\ell} \overline{ \left( 1/\left< \rho \right>_{\ell} \right)^{\prime} \left( \partial {\tau_{1i}^{SFS}}^{\prime} / \partial x_i \right) }$; blue dotted line: $\overline{\left< \rho \right>}_{\ell} \varepsilon_{a_{L,1}}$.}
\label{fig:rho_a1_budget_filtered_destruction_terms}
\end{figure*}

\begin{figure*}[!ht]
\centering
\subfigure[$\ t=1.40\ \mathrm{ms}$]{%
\includegraphics[width=0.4\textwidth]{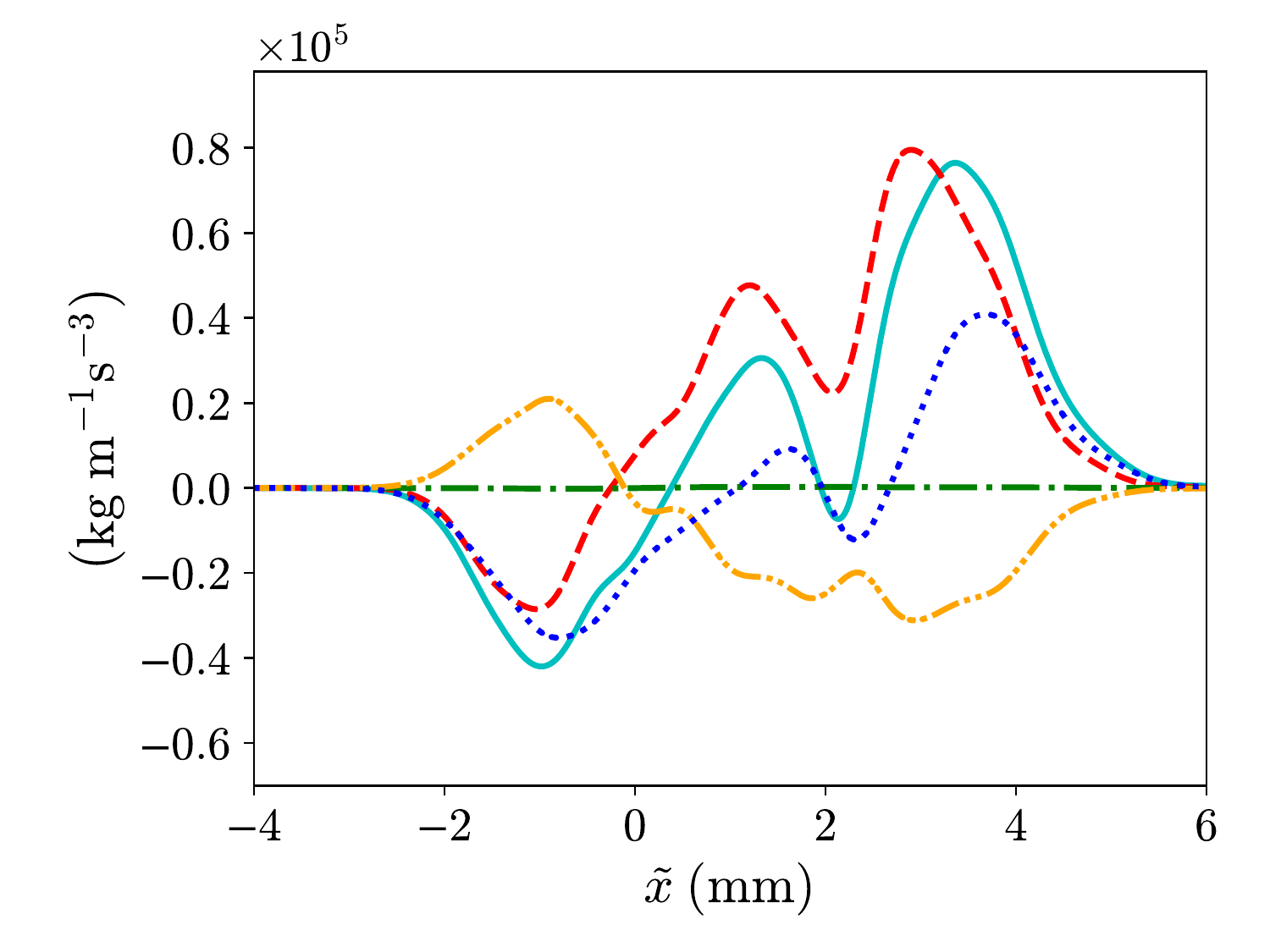}\label{fig:rho_R11_budget_filtered_production_terms_t_1_40}}
\subfigure[$\ t=1.75\ \mathrm{ms}$]{%
\includegraphics[width=0.4\textwidth]{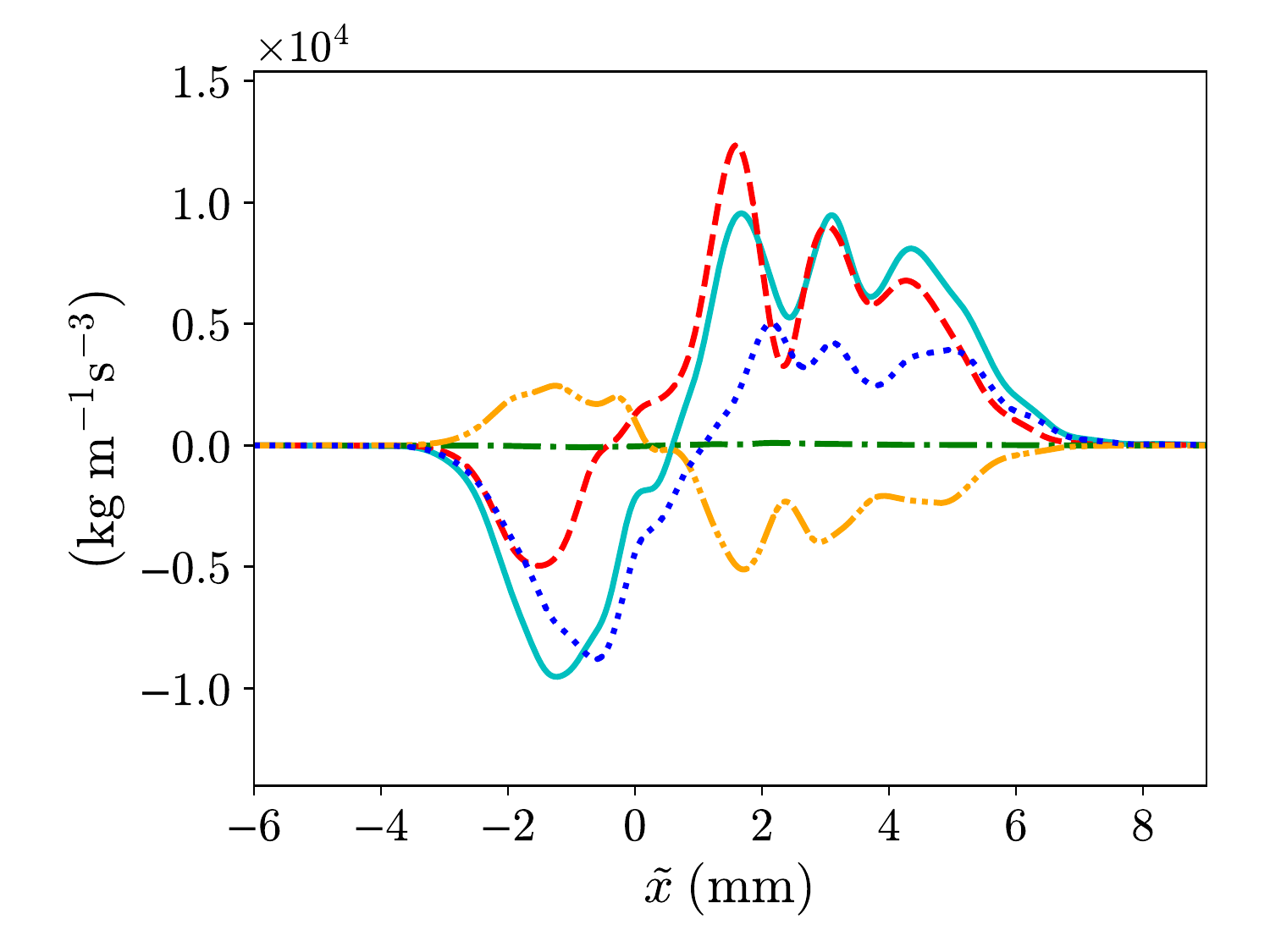}\label{fig:rho_R11_budget_filtered_filtered_production_terms_end}}
\caption{Compositions of the production term [term (III)] in the transport equation for the large-scale Favre-averaged Reynolds normal stress component in the streamwise direction multiplied by the mean filtered density, $\overline{\left< \rho \right>}_{\ell} \widetilde{R}_{L,11}$, at $t=1.40\ \mathrm{ms}$ and $t=1.75\ \mathrm{ms}$ after re-shock. Cyan solid line: overall production; red dashed line: $2a_{L,1} \overline{\left< p \right>}_{\ell,1}$; green dash-dotted line: $-2a_{L,1} \overline{\left< \tau_{11} \right>}_{\ell,1}$; orange dash-dot-dotted line: $2a_{L,1} {\overline{ \tau_{11}^{SFS} }}_{,1}$; blue dotted line: $-2\overline{\left< \rho \right>}_{\ell} \widetilde{R}_{L,11} \widetilde{\left< u \right>}_{L,1}$.}
\label{fig:rho_R11_budget_filtered_production_terms}
\end{figure*}

\begin{figure*}[!ht]
\centering
\subfigure[$\ t=1.40\ \mathrm{ms}$]{%
\includegraphics[width=0.4\textwidth]{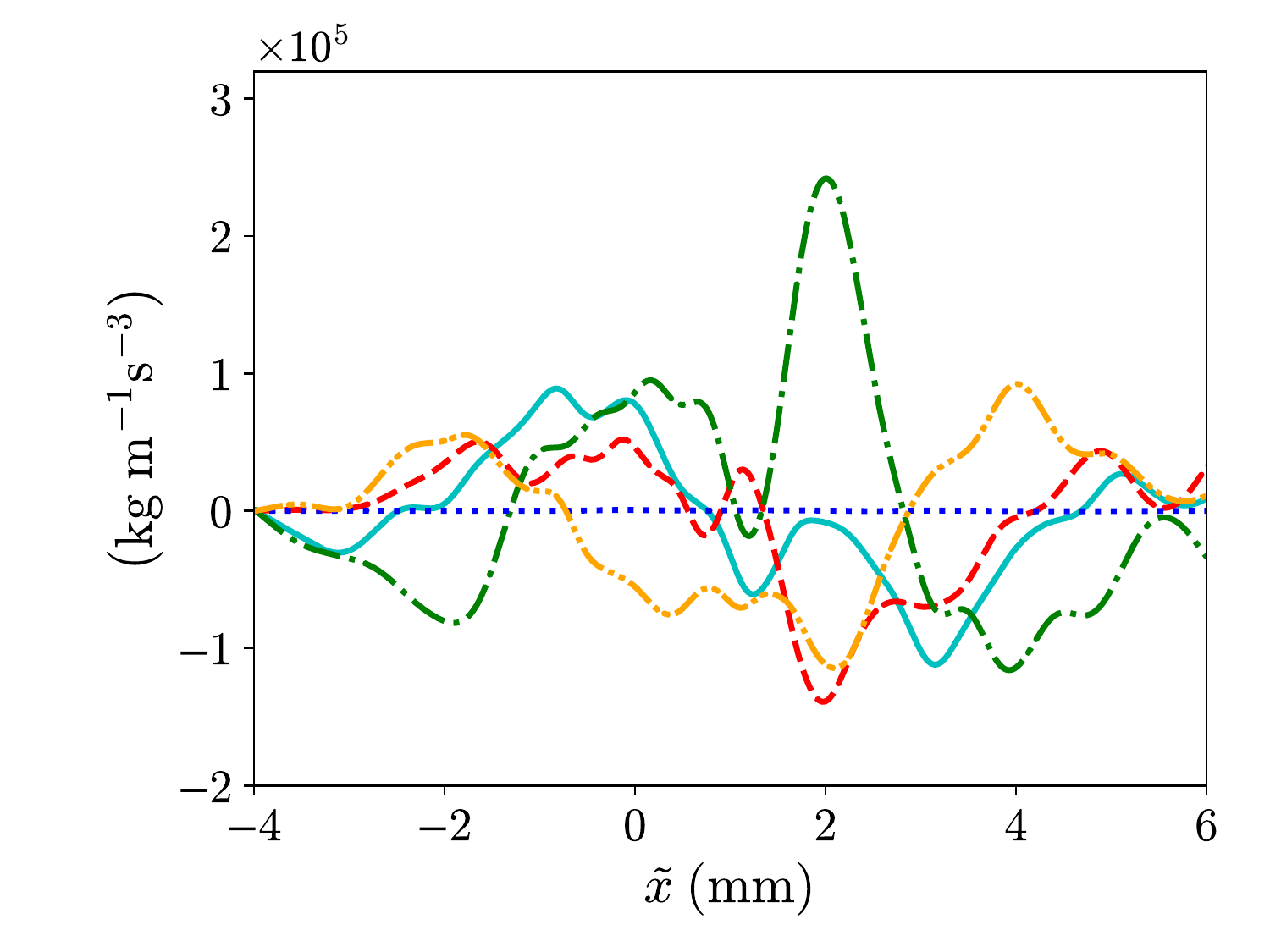}\label{fig:rho_R11_budget_filtered_turb_transport_terms_t_1_40}}
\subfigure[$\ t=1.75\ \mathrm{ms}$]{%
\includegraphics[width=0.4\textwidth]{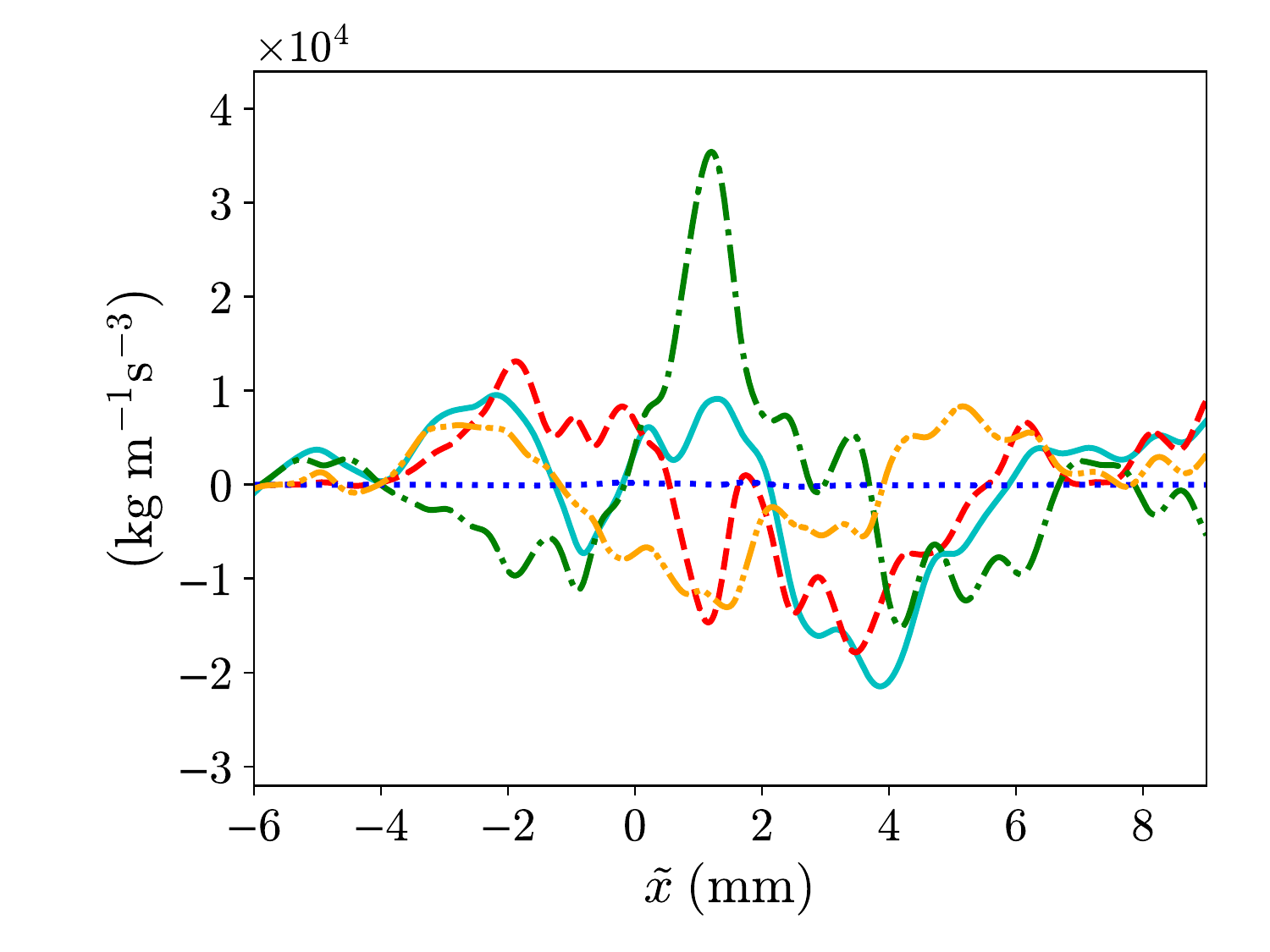}\label{fig:rho_R11_budget_filtered_turb_transport_terms_end}}
\caption{Compositions of the turbulent transport term [term (IV)] in the transport equation for the large-scale Favre-averaged Reynolds normal stress component in the streamwise direction multiplied by the mean filtered density, $\overline{\left< \rho \right>}_{\ell} \widetilde{R}_{L,11}$, at $t=1.40\ \mathrm{ms}$ and $t=1.75\ \mathrm{ms}$ after re-shock. Cyan solid line: overall turbulent transport; red dashed line: $- ( \overline{ \left< \rho \right>_{\ell} \left< u \right>_{L}^{\prime\prime} \left< u \right>_{L}^{\prime\prime} \left< u \right>_{L}^{\prime\prime} } )_{,1}$; green dash-dotted line: $-2 ( \overline{\left< u \right>_{L}^{\prime} \left< p \right>_{\ell}^{\prime}} )_{,1}$; blue dotted line: $2 ( \overline{ \left< u \right>_{L}^{\prime} \left< \tau_{11} \right>_{\ell}^{\prime} } )_{,1}$; orange dash-dot-dotted line: $-2 ( \overline{ \left< u \right>_{L}^{\prime} {\tau_{11}^{SFS}}^{\prime} } )_{,1}$.}
\label{fig:rho_R11_budget_filtered_turb_transport_terms}
\end{figure*}

The budgets of $\overline{\left< \rho \right>}_{\ell} a_{L,1}$, $\overline{\left< \rho \right>}_{\ell} b_L$, $\overline{\left< \rho \right>}_{\ell} \widetilde{R}_{L,11}$, and $\overline{\left< \rho \right>}_{\ell} k_L$  with $\ell \approx 64 \Delta = 0.781\ \mathrm{mm}$ at $t=1.40\ \mathrm{ms}$ and $t=1.75\ \mathrm{ms}$ obtained with grid E after re-shock are shown respectively in figures~\ref{fig:rho_a1_budget_filtered}, \ref{fig:rho_b_budget_filtered}, \ref{fig:rho_R11_budget_filtered}, and \ref{fig:rho_k_budget_filtered}. The compositions of the production and destruction terms in the transport equation for $\overline{\left< \rho \right>}_{\ell} a_{L,1}$ at the two times are displayed in figures~\ref{fig:rho_a1_budget_filtered_production_terms} and \ref{fig:rho_a1_budget_filtered_destruction_terms} respectively. The compositions of the production and turbulent transport terms in the transport equation for $\overline{\left< \rho \right>}_{\ell} \widetilde{R}_{L,11}$ at the two times are shown in figures~\ref{fig:rho_R11_budget_filtered_production_terms} and \ref{fig:rho_R11_budget_filtered_turb_transport_terms} respectively.

\section{Unfiltered budgets and filtered budgets of the second-moments after re-shock with $\ell \approx 32 \Delta$}

\begin{figure*}[!ht]
\centering
\subfigure[$\ $No filtering]{%
\includegraphics[width=0.4\textwidth]{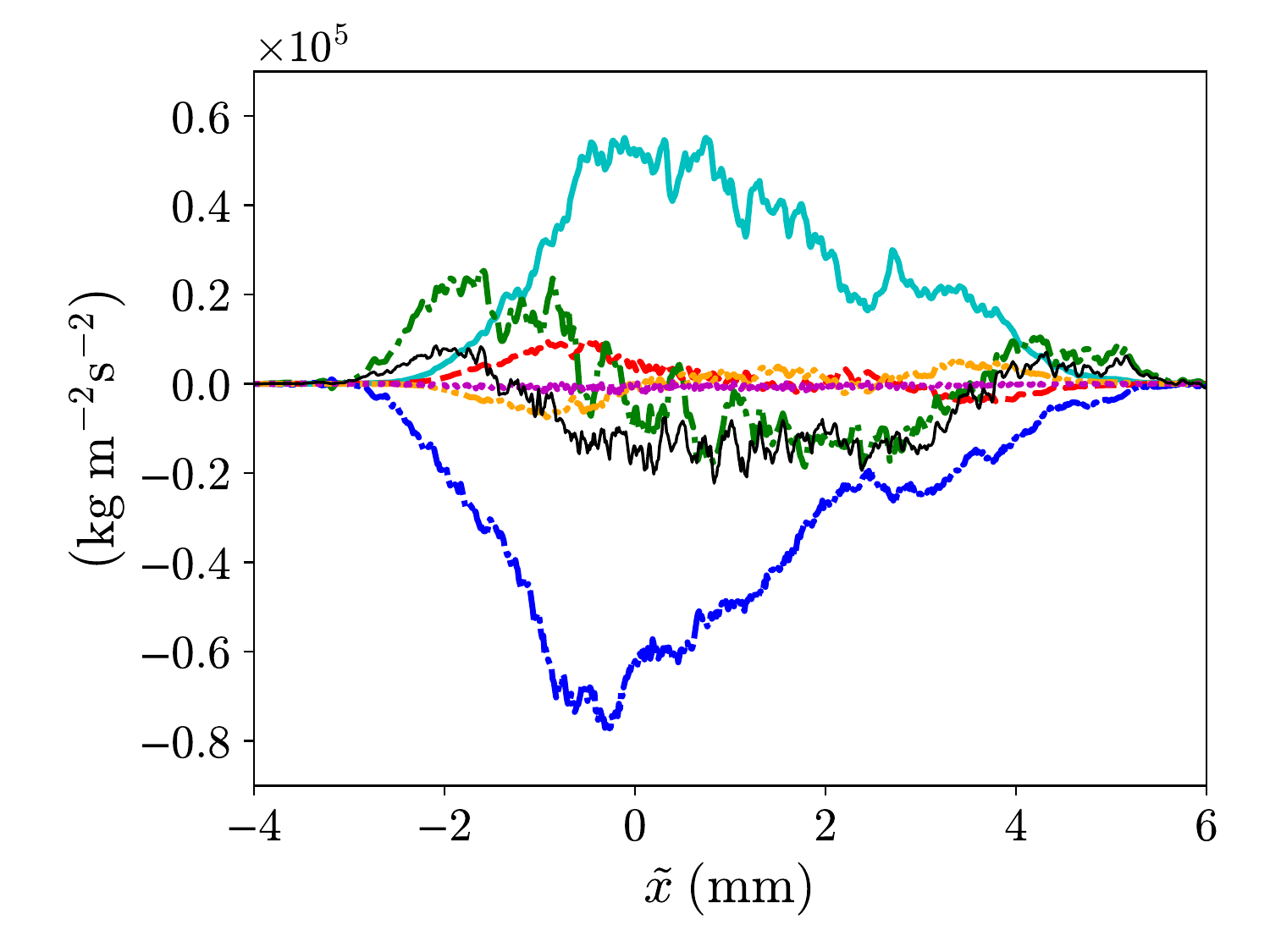}\label{fig:rho_a1_budget_filtered_t_1_40_no_filter}}
\subfigure[$\ \ell \approx 32 \Delta$]{%
\includegraphics[width=0.4\textwidth]{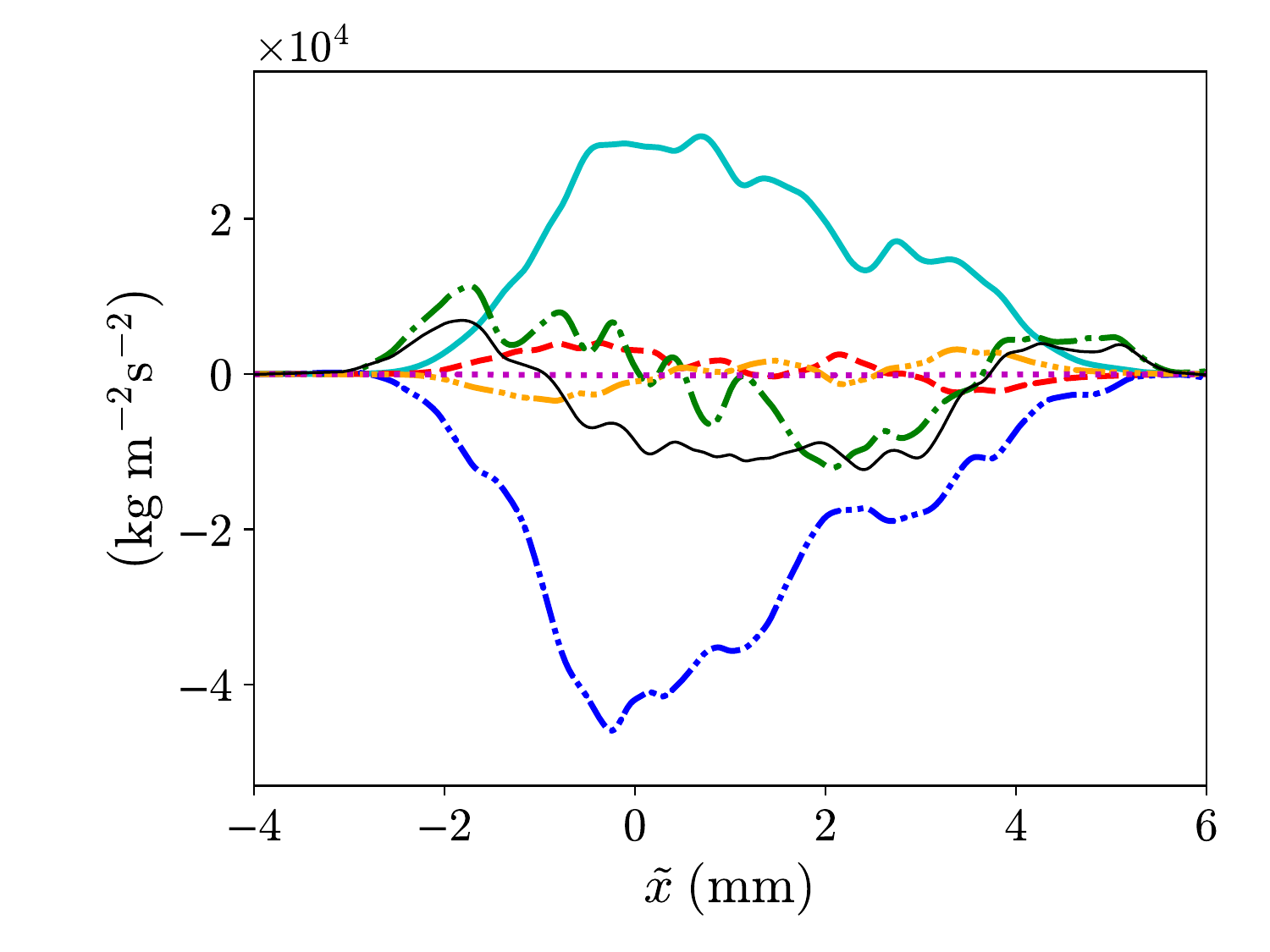}\label{fig:rho_a1_budget_filtered_t_1_40_064x}}
\caption{Effect of filtering on the budgets of the large-scale turbulent mass flux component in the streamwise direction, $\overline{\left< \rho \right>}_{\ell} a_{L,1}$, at $t =1.40\ \mathrm{ms}$. Cyan solid line: production [term (III)]; red dashed line: redistribution [term (IV)]; green dash-dotted line: turbulent transport [term (V)]; blue dash-dot-dotted line: destruction [term (VI)]; orange dash-triple-dotted line: negative of convection due to streamwise velocity associated with turbulent mass flux; magenta dotted line: residue; thin black solid line: summation of all terms (rate of change in the moving frame).}
\label{fig:rho_a1_budget_filtered_effect_filtering}
\end{figure*}

\begin{figure*}[!ht]
\centering
\subfigure[$\ $No filtering]{%
\includegraphics[width=0.4\textwidth]{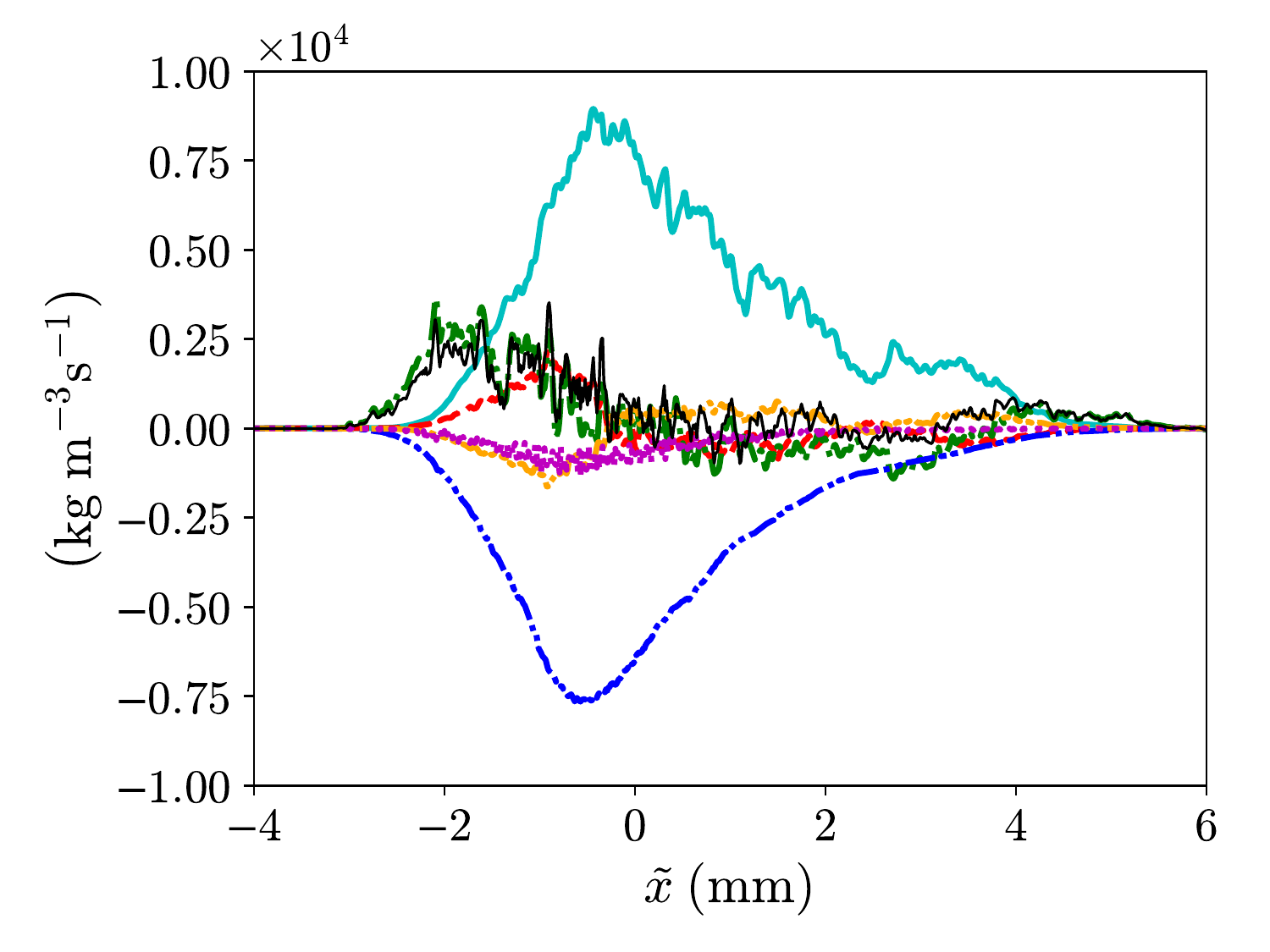}\label{fig:rho_b_budget_filtered_t_1_40_no_filter}}
\subfigure[$\ \ell \approx 32 \Delta$]{%
\includegraphics[width=0.4\textwidth]{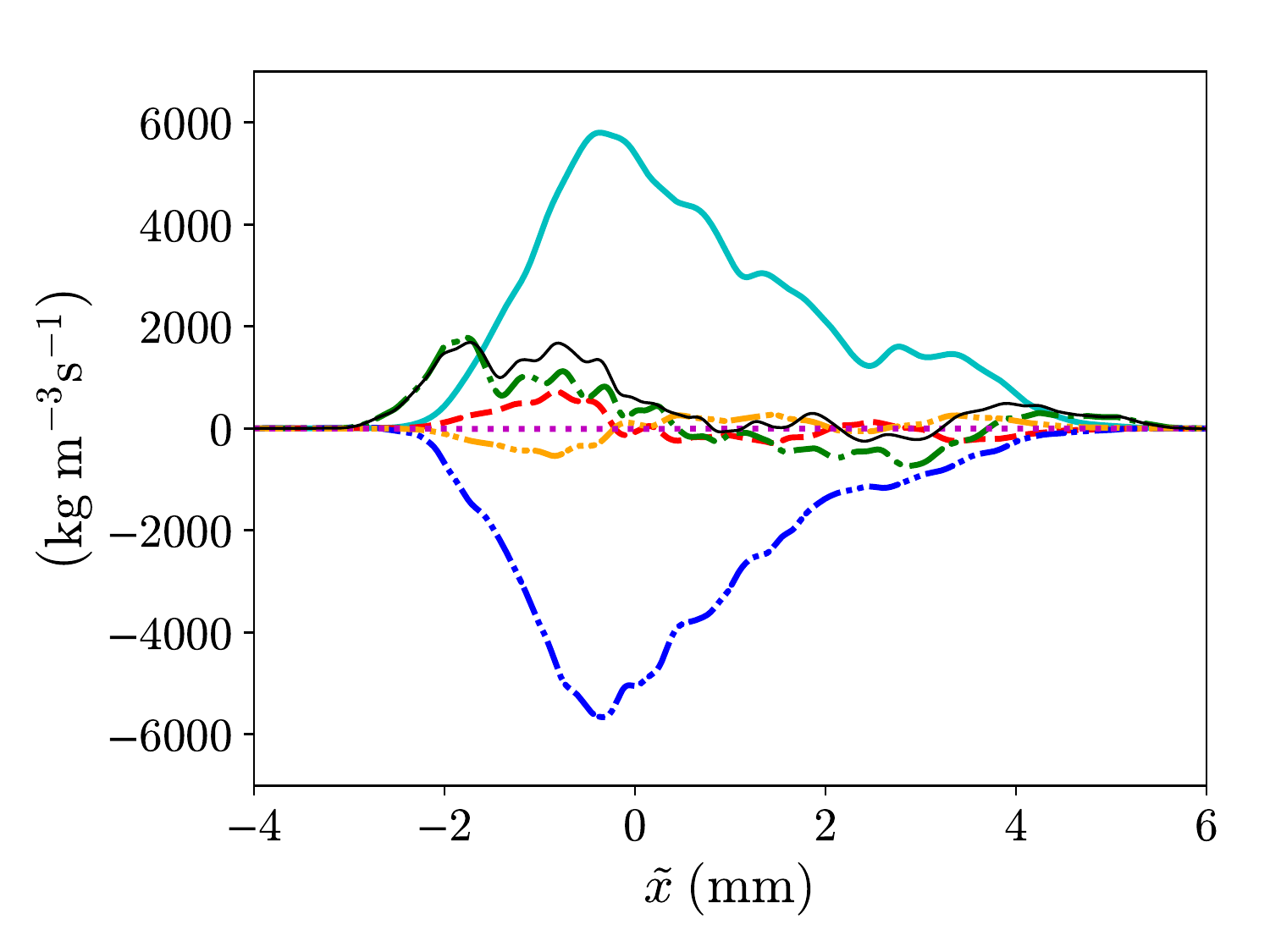}\label{fig:rho_b_budget_filtered_t_1_40_064x}}
\caption{Effect of filtering on the budgets of the large-scale density-specific-volume covariance multiplied by the mean filtered density, $\overline{\left< \rho \right>}_{\ell} b_L$, at $t =1.40\ \mathrm{ms}$. Cyan solid line: production [term (III)]; red dashed line: redistribution [term (IV)]; green dash-dotted line: turbulent transport [term (V)]; blue dash-dot-dotted line: destruction [term (VI)]; orange dash-triple-dotted line: negative of convection due to streamwise velocity associated with turbulent mass flux; magenta dotted line: residue; thin black solid line: summation of all terms (rate of change in the moving frame).}
\label{fig:rho_b_budget_filtered_effect_filtering}
\end{figure*}

\begin{figure*}[!ht]
\centering
\subfigure[$\ $No filtering]{%
\includegraphics[width=0.4\textwidth]{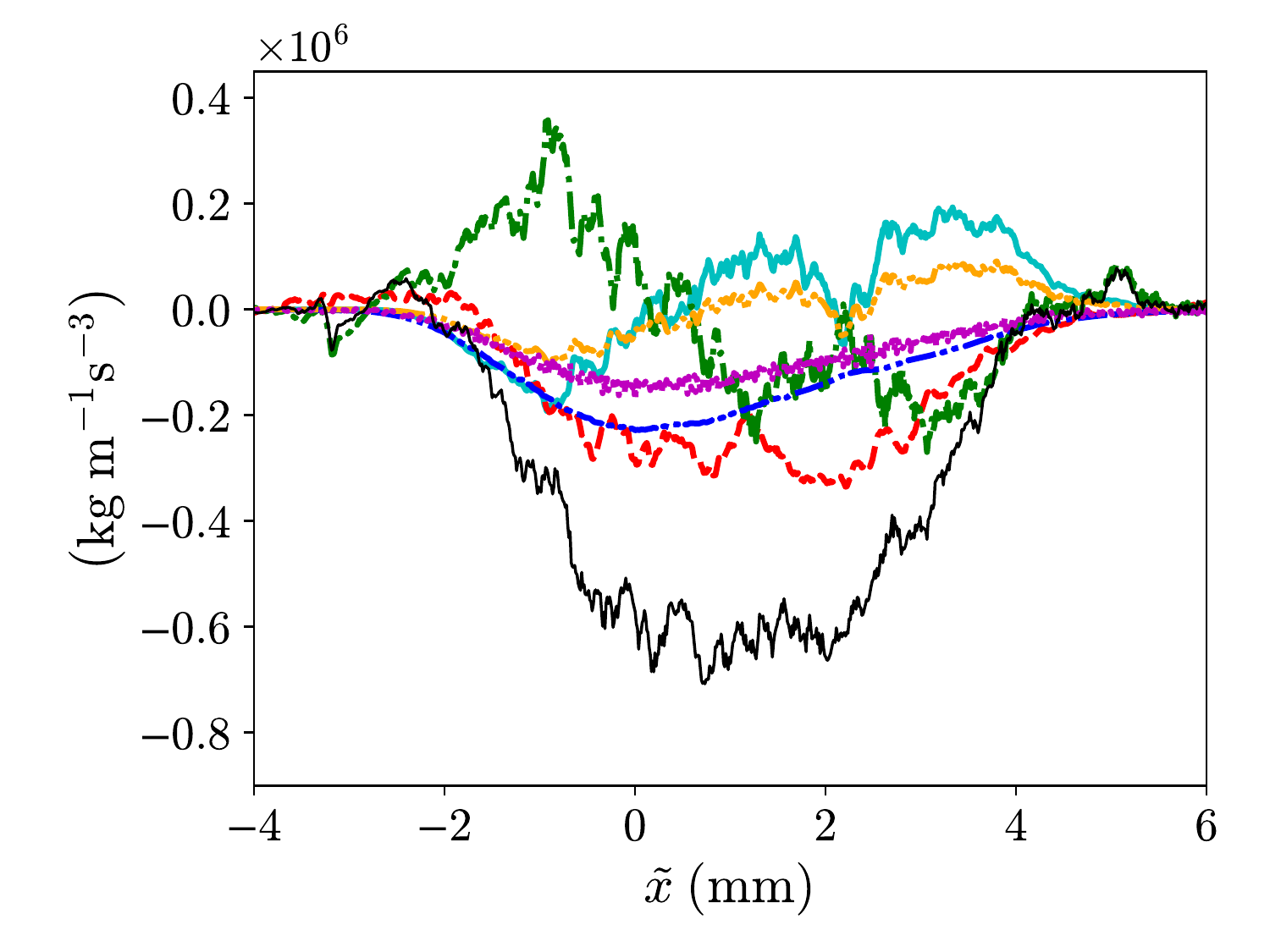}}
\subfigure[$\ \ell \approx 32 \Delta$]{%
\includegraphics[width=0.4\textwidth]{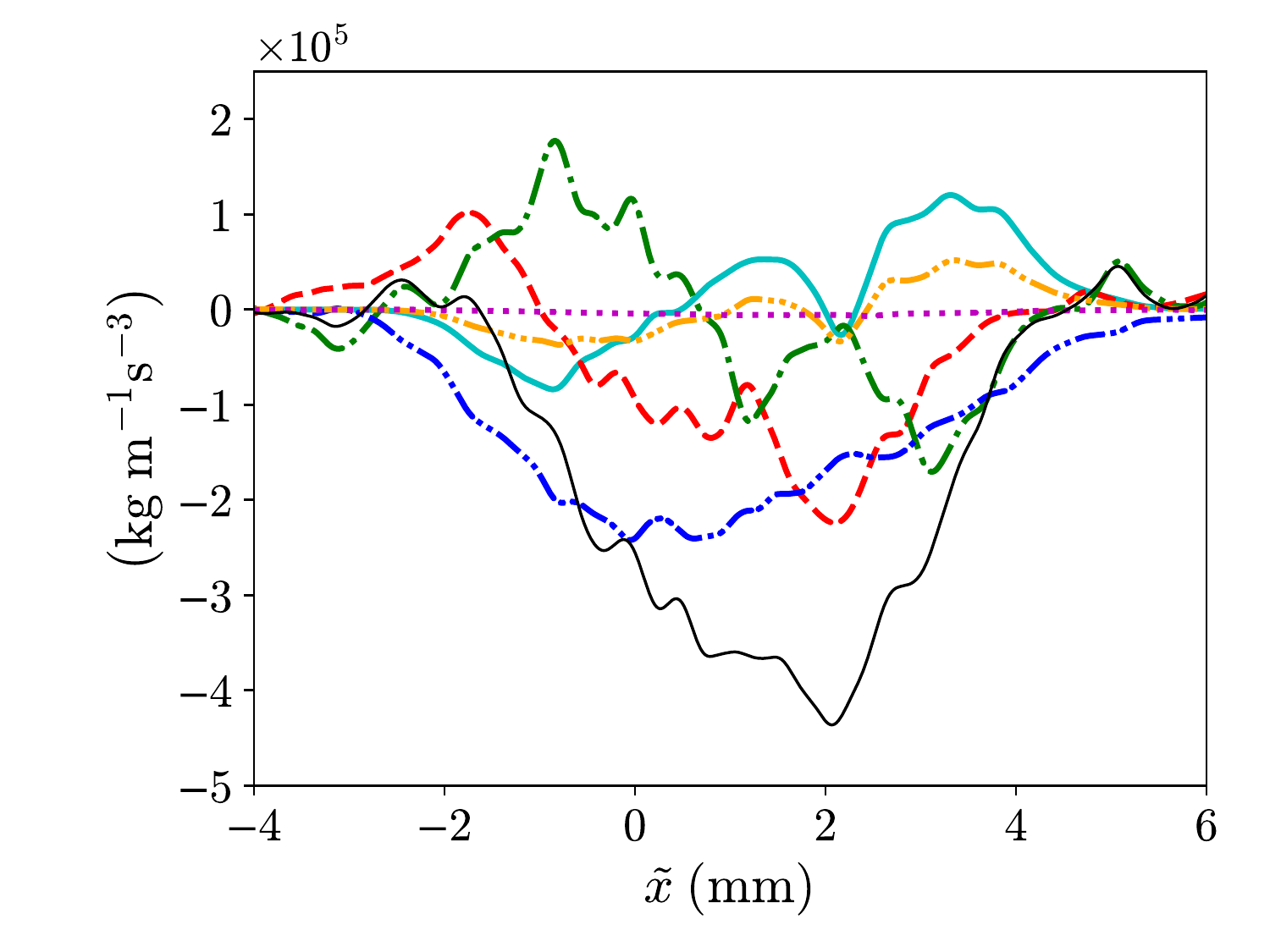}}
\caption{Effect of filtering on the budgets of the large-scale Reynolds normal stress component in the streamwise direction multiplied by the mean filtered density, $\overline{\left< \rho \right>}_{\ell} \widetilde{R}_{L,11}$, at $t =1.40\ \mathrm{ms}$. Cyan solid line: production [term (III)]; red dashed line: press-strain redistribution [term (V)]; green dash-dotted line: turbulent transport [term (IV)]; blue dash-dot-dotted line: dissipation [term (VI)]; orange dash-triple-dotted line: negative of convection due to streamwise velocity associated with turbulent mass flux; magenta dotted line: residue; thin black solid line: summation of all terms (rate of change in the moving frame).}
\label{fig:rho_R11_budget_filtered_effect_filtering}
\end{figure*}

\begin{figure*}[!ht]
\centering
\subfigure[$\ $No filtering]{%
\includegraphics[width=0.4\textwidth]{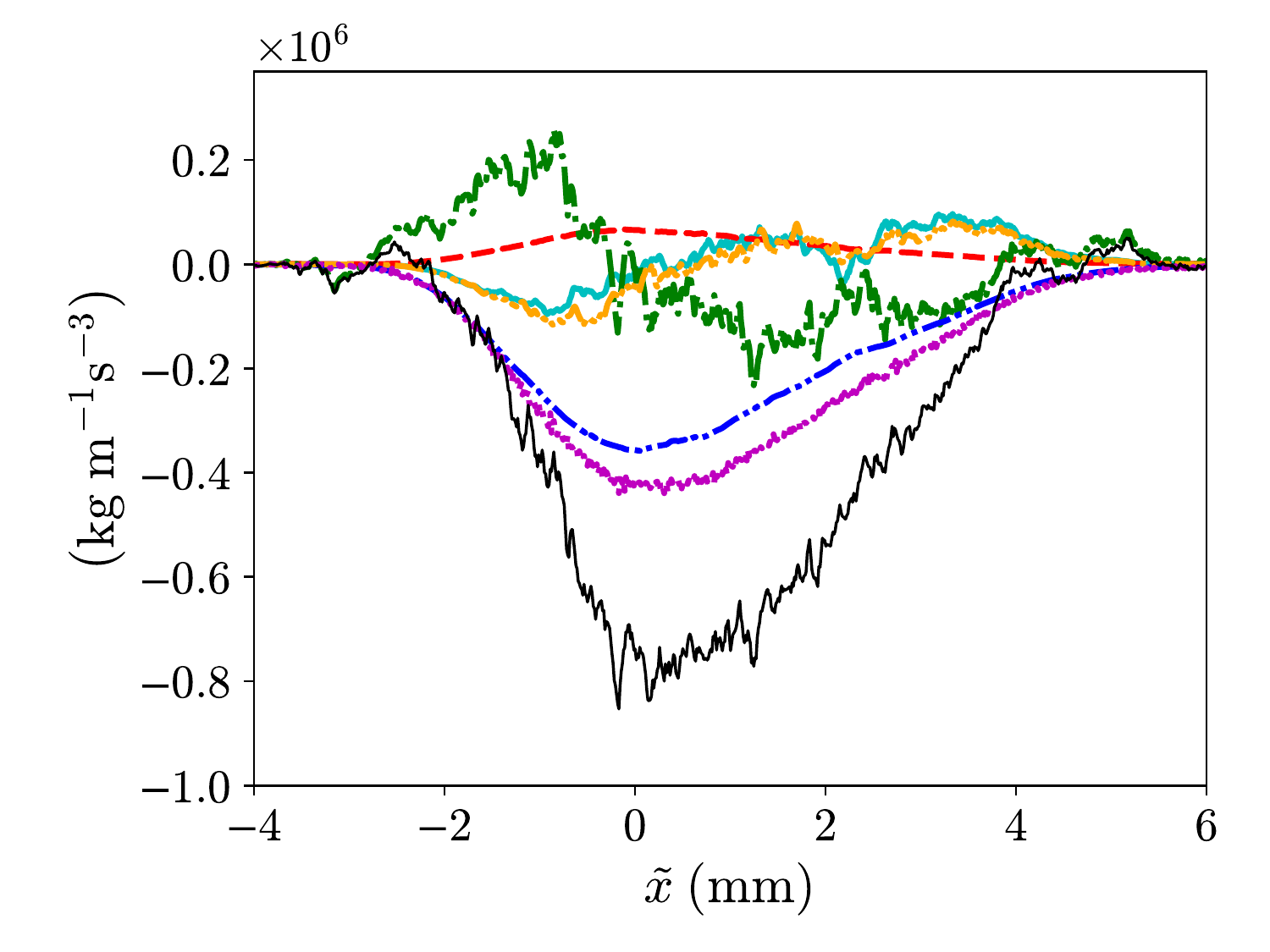}}
\subfigure[$\ \ell \approx 32 \Delta$]{%
\includegraphics[width=0.4\textwidth]{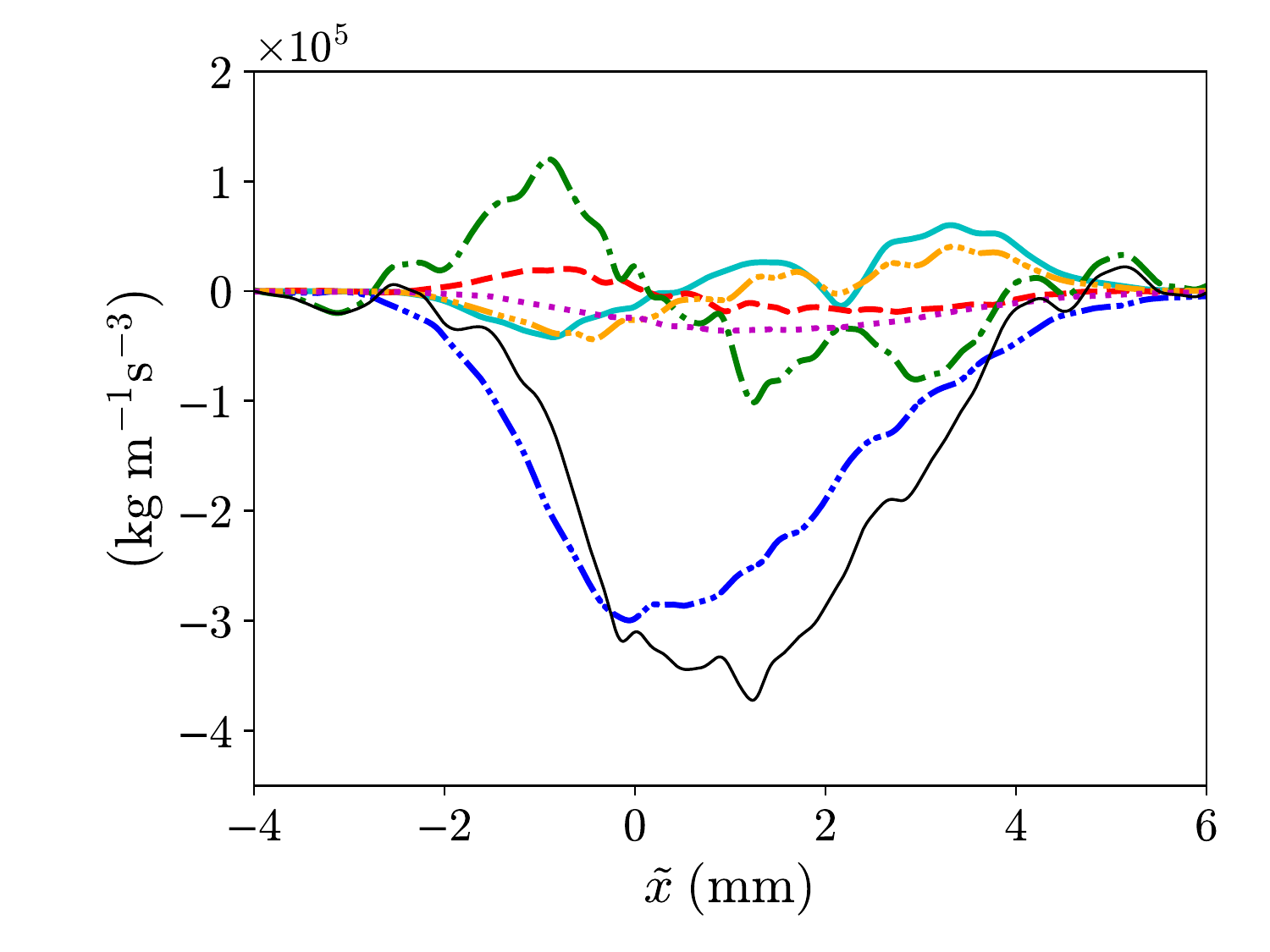}}
\caption{Effect of filtering on the budgets of the large-scale turbulent kinetic energy, $\overline{\left< \rho \right>}_{\ell} k_L$, at $t =1.40\ \mathrm{ms}$. Cyan solid line: production [term (III)]; red dashed line: pressure-dilatation [term (V)]; green dash-dotted line: turbulent transport [term (IV)]; blue dash-dot-dotted line: dissipation [term (VI)]; orange dash-triple-dotted line: negative of convection due to streamwise velocity associated with turbulent mass flux; magenta dotted line: residue; thin black solid line: summation of all terms (rate of change in the moving frame).}
\label{fig:rho_k_budget_filtered_effect_filtering}
\end{figure*}

\begin{figure*}[!ht]
\centering
\subfigure[$\ $No filtering]{%
\includegraphics[width=0.4\textwidth]{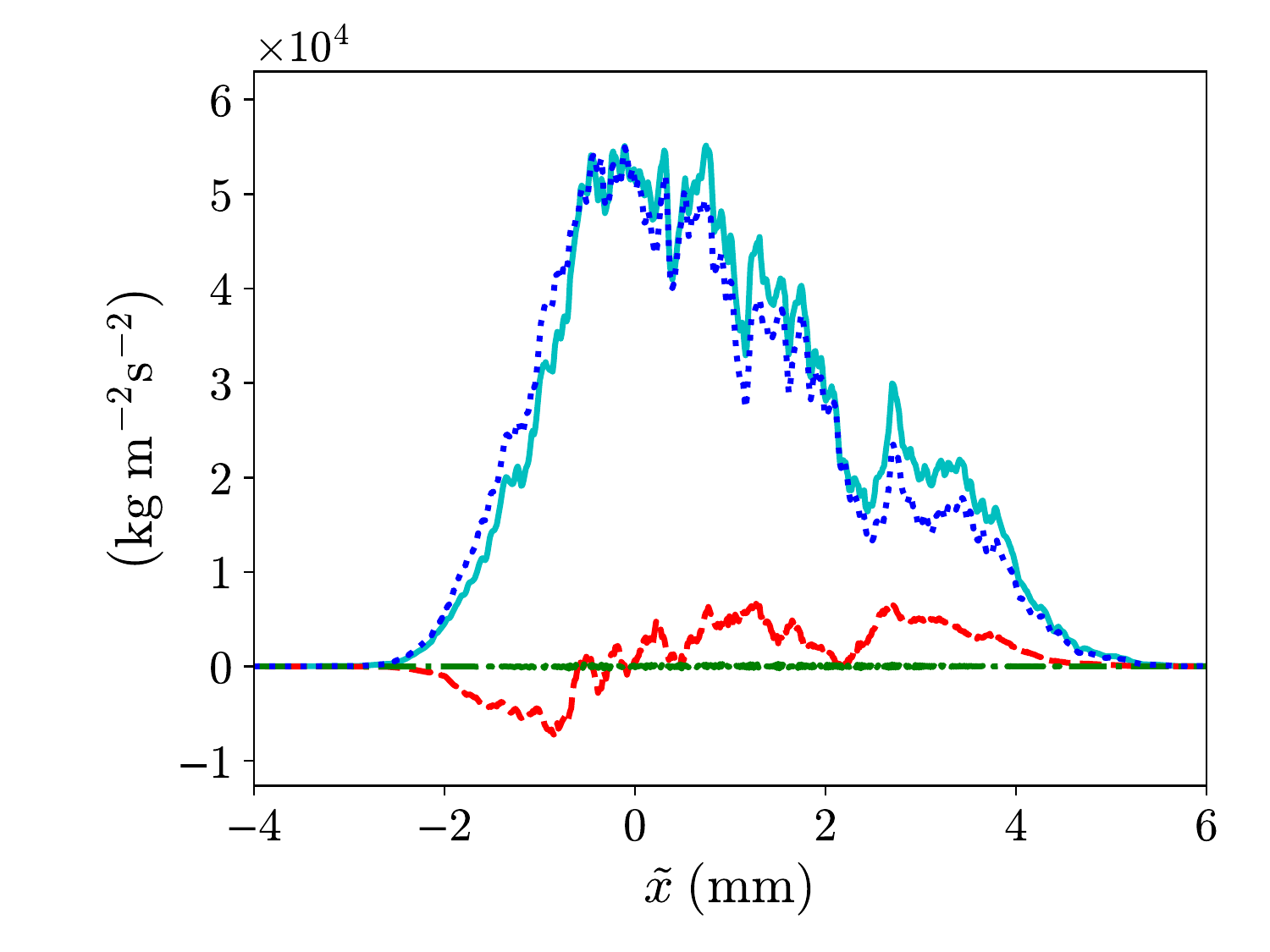}}
\subfigure[$\ \ell \approx 32 \Delta$]{%
\includegraphics[width=0.4\textwidth]{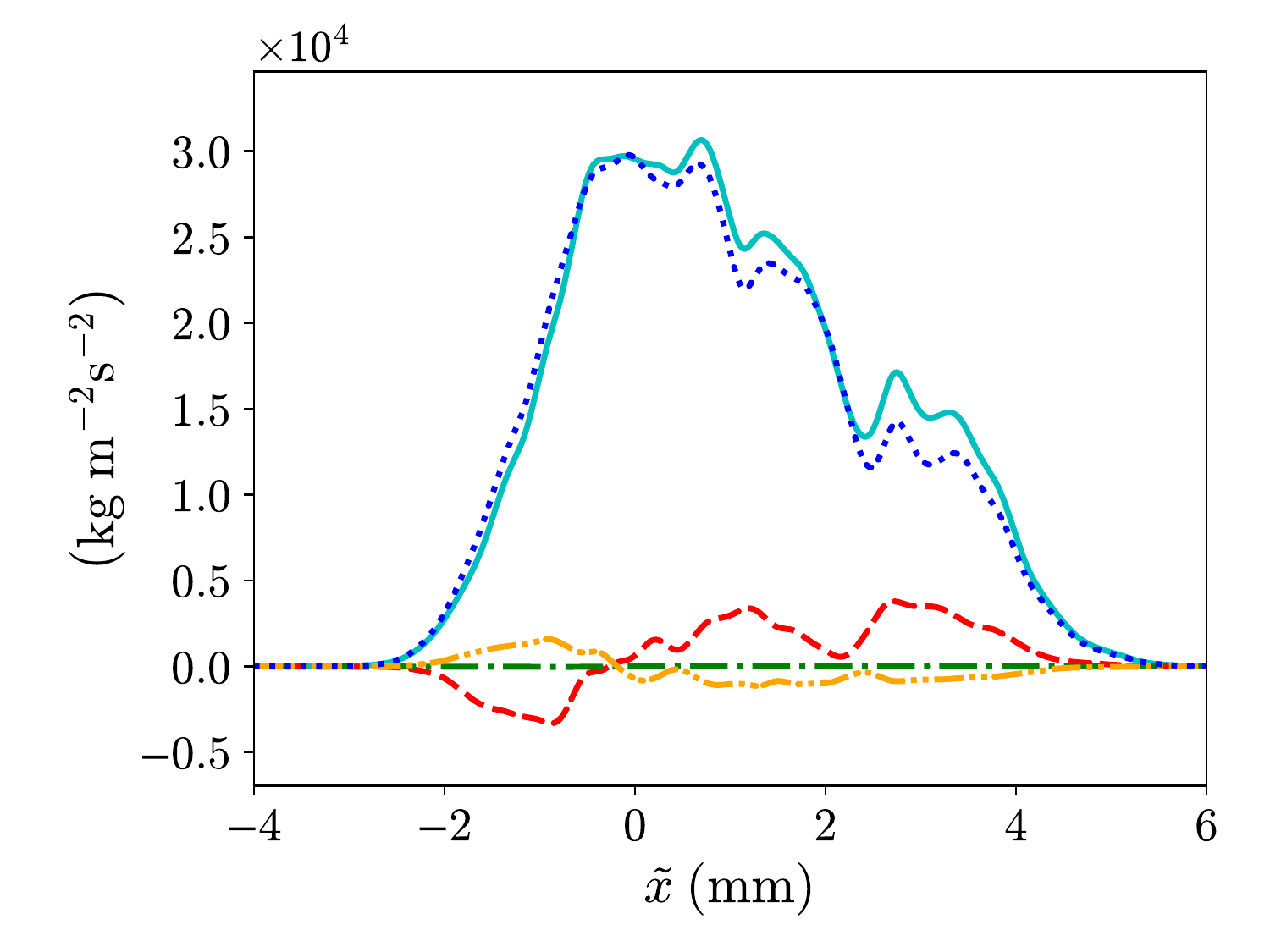}}
\caption{Effect of filtering on the compositions of the production term [term (III)] in the transport equation for the large-scale turbulent mass flux component in the streamwise direction, $\overline{\left< \rho \right>}_{\ell} a_{L,1}$, at $t = 1.40\ \mathrm{ms}$. Cyan solid line: overall production; red dashed line: $b_L \overline{\left< p \right>}_{\ell,1}$; green dash-dotted line: $-b_L \overline{\left< \tau_{11} \right>}_{\ell,1}$; orange dash-dot-dotted line: $b_L \overline{\tau_{11}^{SFS}}_{,1}$; blue dotted line: $-\widetilde{R}_{L,11} \overline{\left< \rho \right>}_{\ell,1}$.}
\label{fig:rho_a1_budget_filtered_production_terms_effect_filtering}
\end{figure*}

\begin{figure*}[!ht]
\centering
\subfigure[$\ $No filtering]{%
\includegraphics[width=0.4\textwidth]{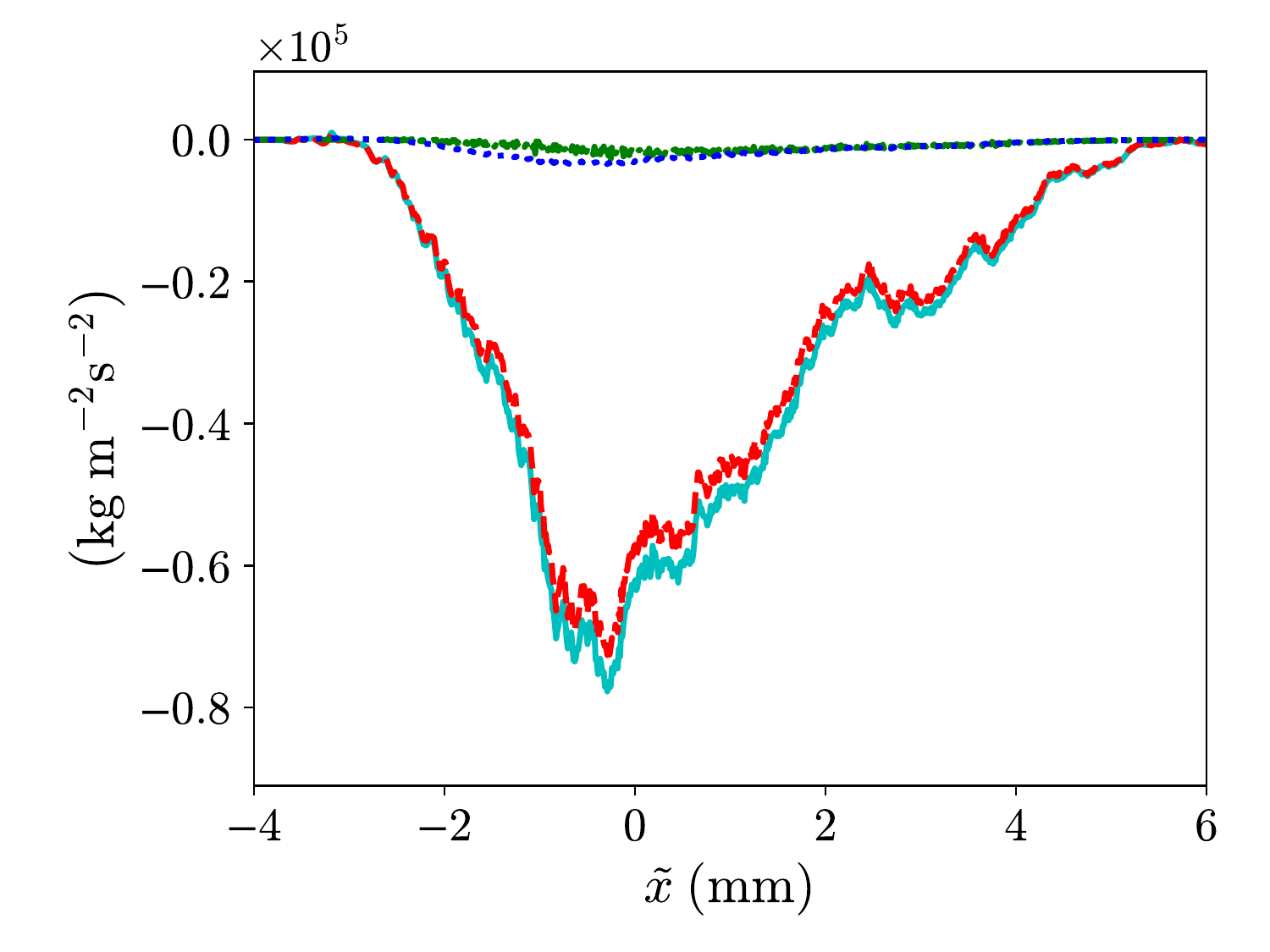}}
\subfigure[$\ \ell \approx 32 \Delta$]{%
\includegraphics[width=0.4\textwidth]{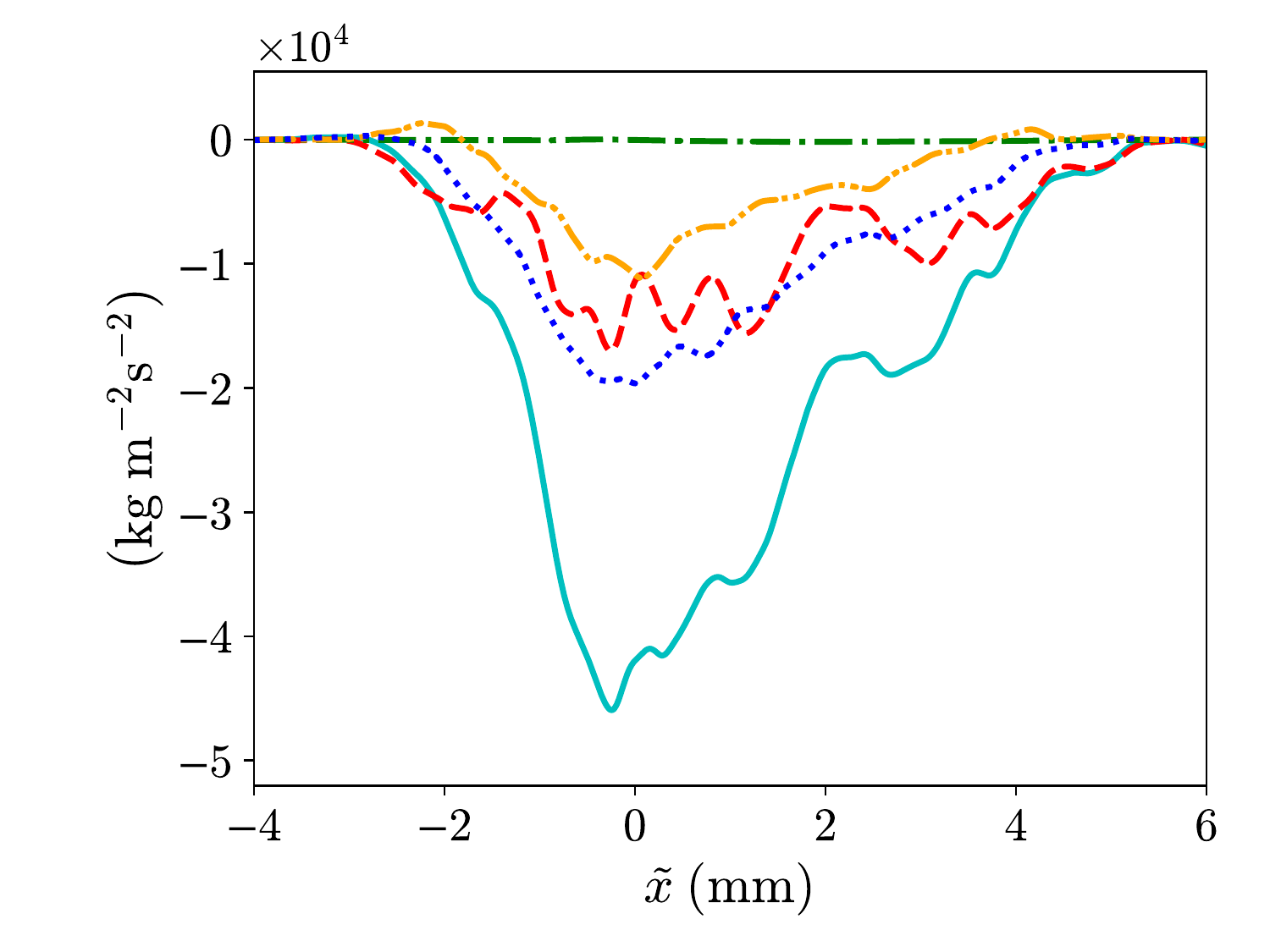}}
\caption{Effect of filtering on the compositions of the destruction term [term (VI)] in the transport equation for the large-scale turbulent mass flux component in the streamwise direction, $\overline{\left< \rho \right>}_{\ell} a_{L,1}$, at $t = 1.40\ \mathrm{ms}$. Cyan solid line: overall destruction; red dashed line: $\overline{\left< \rho \right>}_{\ell} \overline{ \left( 1/\left< \rho \right>_{\ell} \right)^{\prime} \left< p \right>^{\prime}_{\ell,1} }$; green dash-dotted line: $-\overline{\left< \rho \right>}_{\ell} \overline{ \left( 1/\left< \rho \right>_{\ell} \right)^{\prime} \left( \partial \left< \tau_{1i} \right>_{\ell}^{\prime} / \partial x_i \right) }$; orange dash-dot-dotted line: $\overline{\left< \rho \right>}_{\ell} \overline{ \left( 1/\left< \rho \right>_{\ell} \right)^{\prime} \left( \partial {\tau_{1i}^{SFS}}^{\prime} / \partial x_i \right) }$; blue dotted line: $\overline{\left< \rho \right>}_{\ell} \varepsilon_{a_{L,1}}$.}
\label{fig:rho_a1_budget_filtered_destruction_terms_effect_filtering}
\end{figure*}

\begin{figure*}[!ht]
\centering
\subfigure[$\ $No filtering]{%
\includegraphics[width=0.4\textwidth]{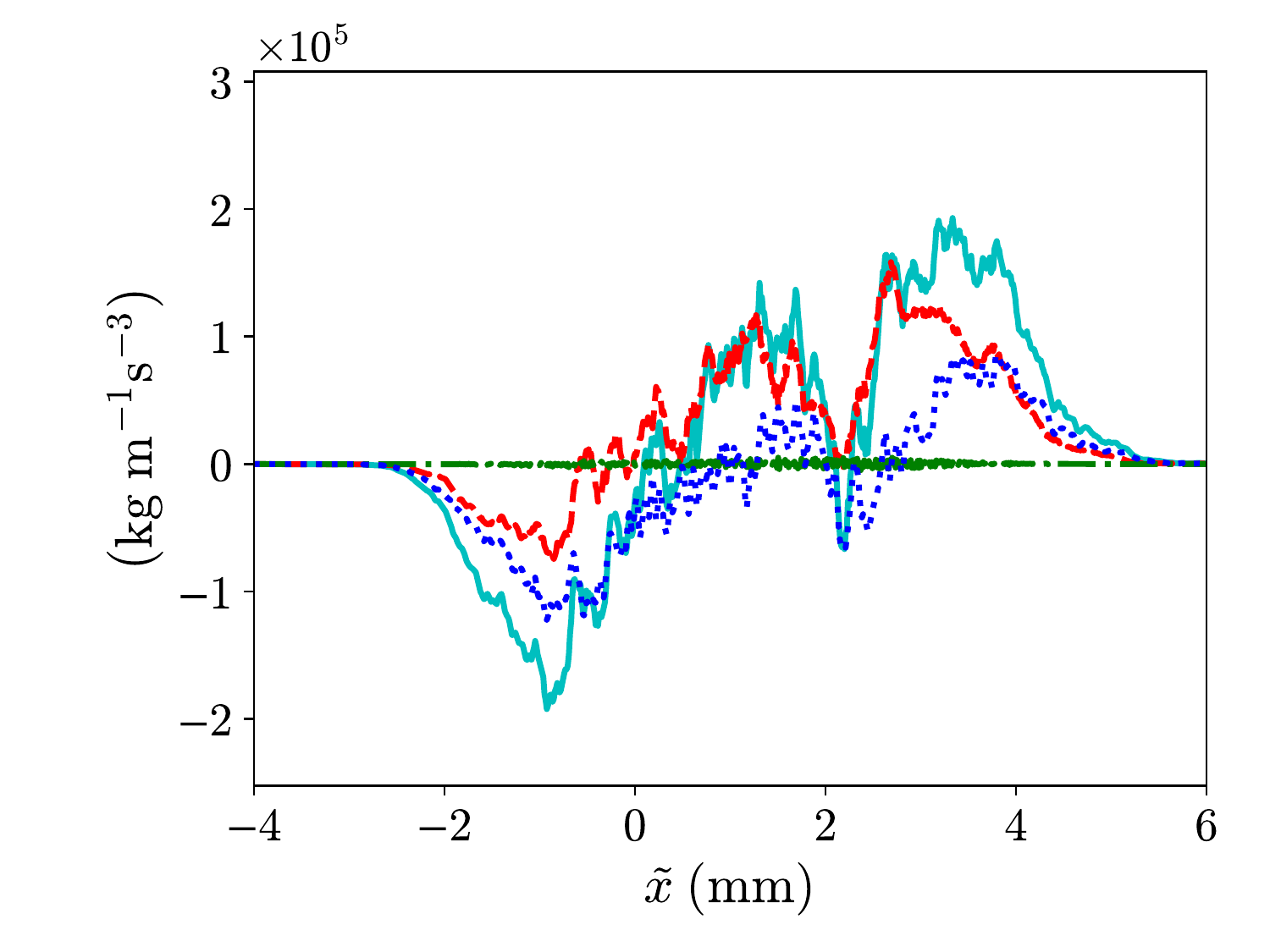}}
\subfigure[$\ \ell \approx 32 \Delta$]{%
\includegraphics[width=0.4\textwidth]{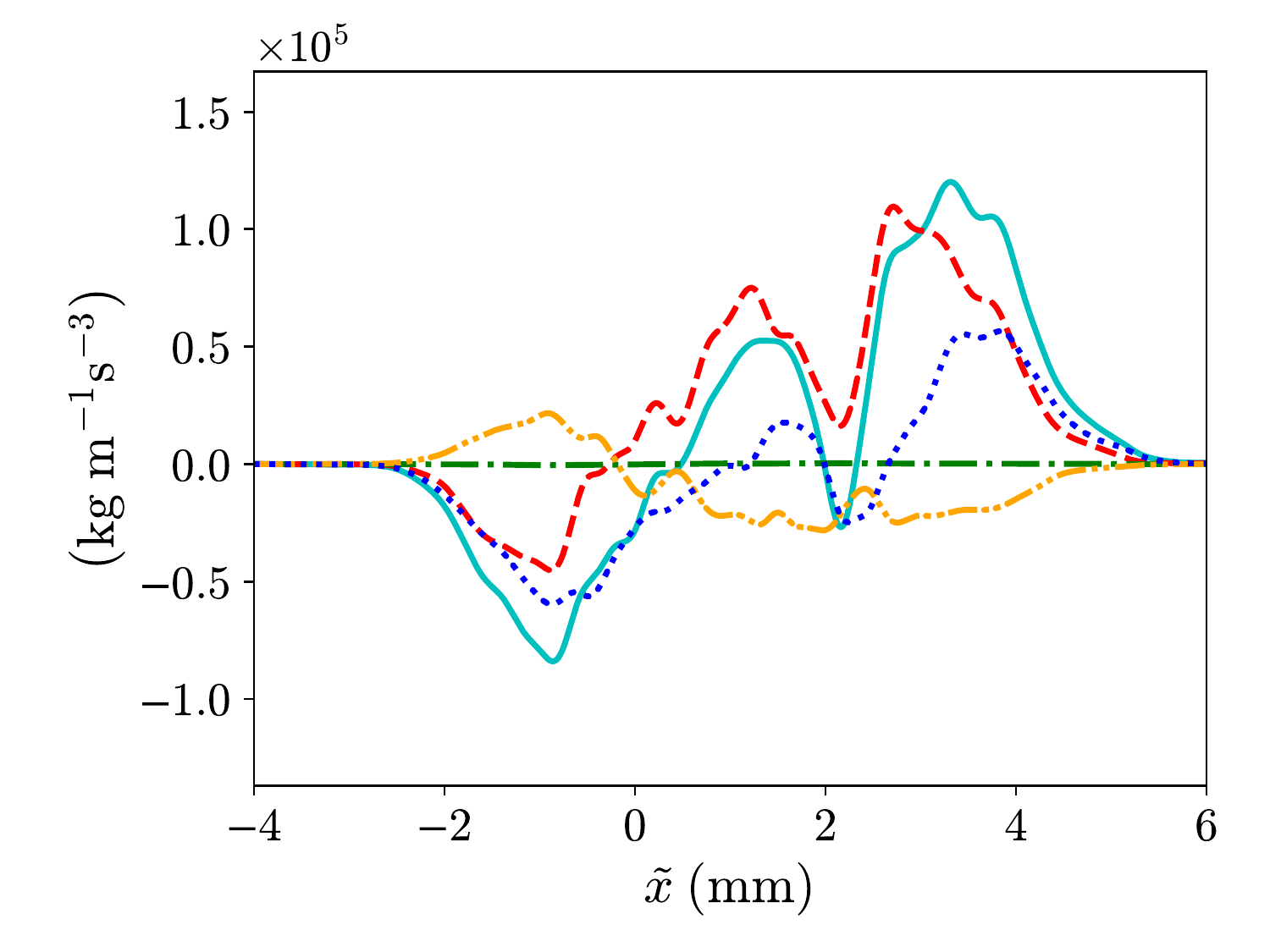}}
\caption{Effect of filtering on the compositions of the production term [term (III)] in the transport equation for the large-scale Favre-averaged Reynolds normal stress component in the streamwise direction multiplied by the mean filtered density, $\overline{\left< \rho \right>}_{\ell} \widetilde{R}_{L,11}$, at $t = 1.40\ \mathrm{ms}$. Cyan solid line: overall production; red dashed line: $2a_{L,1} \overline{\left< p \right>}_{\ell,1}$; green dash-dotted line: $-2a_{L,1} \overline{\left< \tau_{11} \right>}_{\ell,1}$; orange dash-dot-dotted line: $2a_{L,1} {\overline{ \tau_{11}^{SFS} }}_{,1}$; blue dotted line: $-2\overline{\left< \rho \right>}_{\ell} \widetilde{R}_{L,11} \widetilde{\left< u \right>}_{L,1}$.}
\label{fig:rho_R11_budget_filtered_production_terms_effect_filtering}
\end{figure*}

\begin{figure*}[!ht]
\centering
\subfigure[$\ $No filtering]{%
\includegraphics[width=0.4\textwidth]{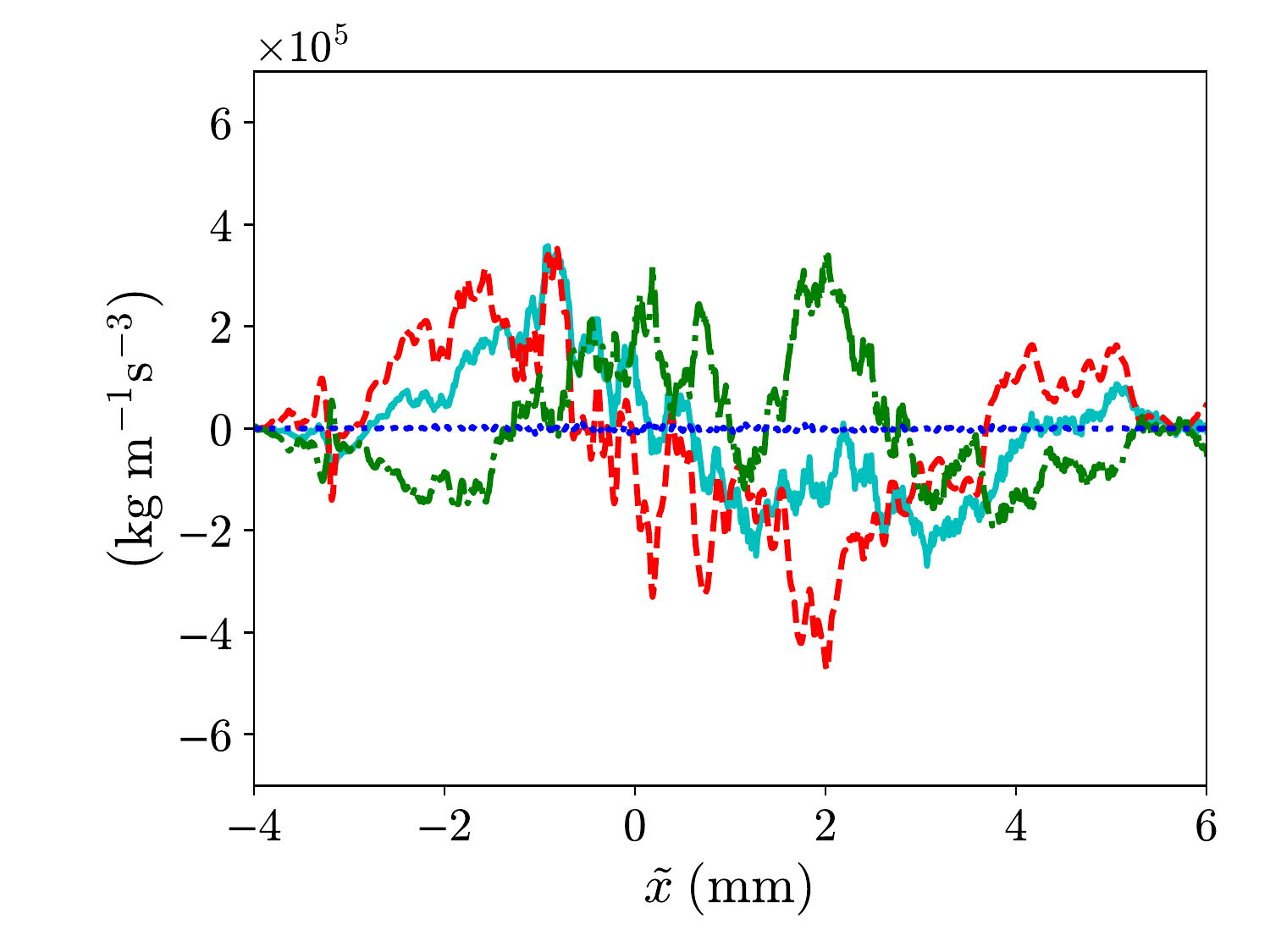}}
\subfigure[$\ \ell \approx 32 \Delta$]{%
\includegraphics[width=0.4\textwidth]{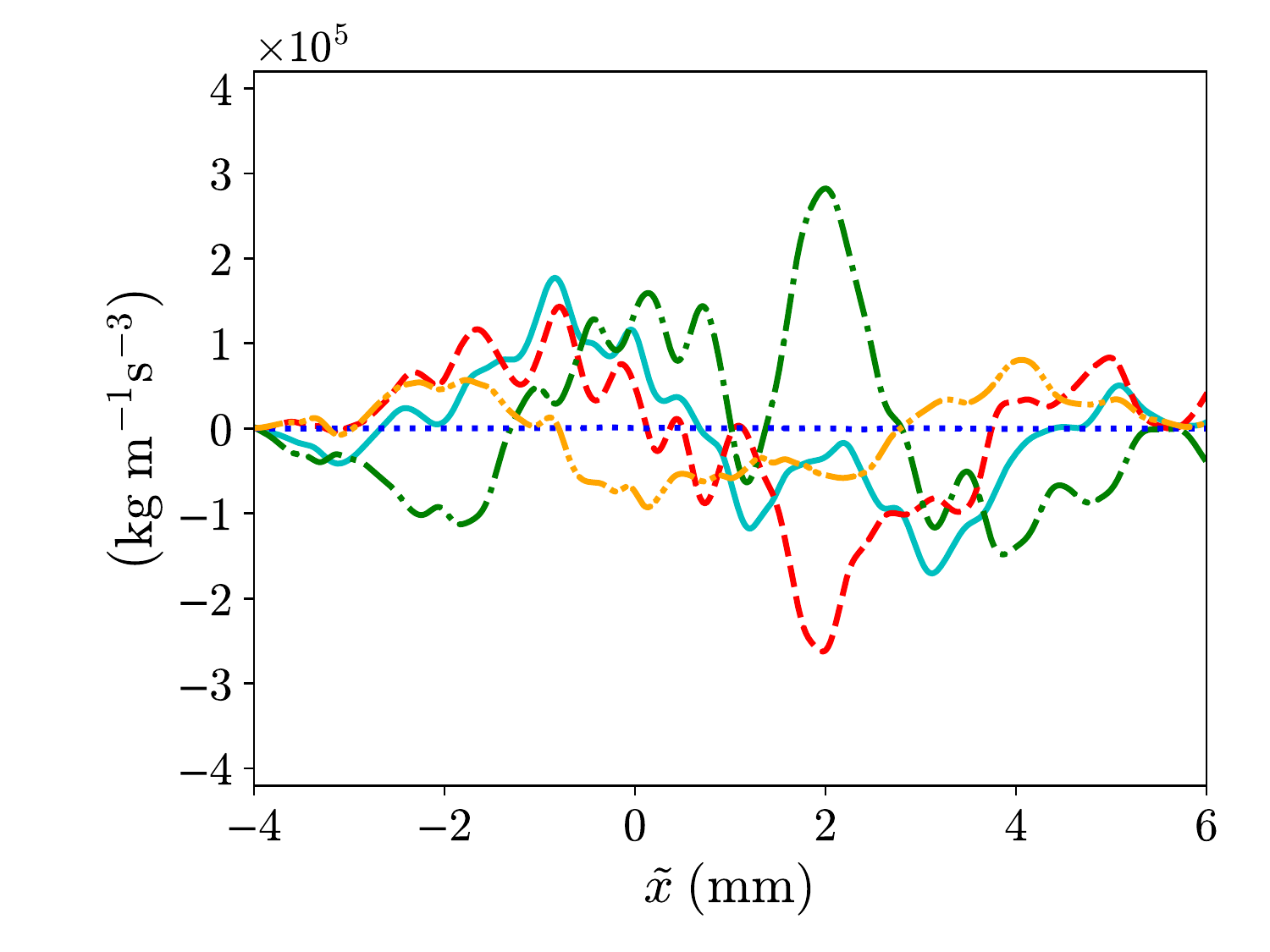}}
\caption{Effect of filtering on the compositions of the turbulent transport term [term (IV)] in the transport equation for the large-scale Favre-averaged Reynolds normal stress component in the streamwise direction multiplied by the mean filtered density, $\overline{\left< \rho \right>}_{\ell} \widetilde{R}_{L,11}$, at $t = 1.40\ \mathrm{ms}$. Cyan solid line: overall turbulent transport; red dashed line: $- ( \overline{ \left< \rho \right>_{\ell} \left< u \right>_{L}^{\prime\prime} \left< u \right>_{L}^{\prime\prime} \left< u \right>_{L}^{\prime\prime} } )_{,1}$; green dash-dotted line: $-2 ( \overline{\left< u \right>_{L}^{\prime} \left< p \right>_{\ell}^{\prime}} )_{,1}$; blue dotted line: $2 ( \overline{ \left< u \right>_{L}^{\prime} \left< \tau_{11} \right>_{\ell}^{\prime} } )_{,1}$; orange dash-dot-dotted line: $-2 ( \overline{ \left< u \right>_{L}^{\prime} {\tau_{11}^{SFS}}^{\prime} } )_{,1}$.}
\label{fig:rho_R11_budget_filtered_turb_transport_terms_effect_filtering}
\end{figure*}

The budgets of the unfiltered second-moments and those of the large-scale second-moments with $\ell \approx 32 \Delta = 0.0.391\ \mathrm{mm}$ obtained with grid E at $t =1.40\ \mathrm{ms}$ are shown in figures~\ref{fig:rho_a1_budget_filtered_effect_filtering}, \ref{fig:rho_b_budget_filtered_effect_filtering}, \ref{fig:rho_R11_budget_filtered_effect_filtering}, and \ref{fig:rho_k_budget_filtered_effect_filtering} respectively. The compositions of the production and destruction terms in the transport equation for $\overline{\left< \rho \right>}_{\ell} a_{L,1}$ at the same time are displayed in figures~\ref{fig:rho_a1_budget_filtered_production_terms_effect_filtering} and \ref{fig:rho_a1_budget_filtered_destruction_terms_effect_filtering} respectively. The compositions of the production and turbulent transport terms in the transport equation for $\overline{\left< \rho \right>}_{\ell} \widetilde{R}_{L,11}$ are shown in figures~\ref{fig:rho_R11_budget_filtered_production_terms_effect_filtering} and \ref{fig:rho_R11_budget_filtered_turb_transport_terms_effect_filtering} respectively.

